\pdfoutput=1

\documentclass[11pt]{article}
\usepackage{jheppub}

\usepackage{amsmath}
\usepackage{amssymb}
\usepackage{graphicx,color,slashed}
\usepackage{ifpdf}
\usepackage[english]{babel}
\usepackage{url}
\usepackage{booktabs}

\usepackage{float}

\hypersetup{bookmarksnumbered}

\allowdisplaybreaks[1]

\DeclareFontFamily{U}{mathx}{\hyphenchar\font45}
\DeclareFontShape{U}{mathx}{m}{n}{
      <5> <6> <7> <8> <9> <10>
      <10.95> <12> <14.4> <17.28> <20.74> <24.88>
      mathx10
      }{}
\DeclareSymbolFont{mathx}{U}{mathx}{m}{n}
\DeclareFontSubstitution{U}{mathx}{m}{n}
\DeclareMathAccent{\widecheck}{0}{mathx}{"71}

\font\tenshuffle=shuffle10 \font\sevenshuffle=shuffle7 \font\fiveshuffle=shuffle7 at 5pt
\def\shuffle{{%
\def\Dshuffle{\mathbin{\hbox{\tenshuffle\char'001}}}%
\def\Sshuffle{\mathbin{\hbox{\sevenshuffle\char'001}}}%
\def\SSshuffle{\mathbin{\hbox{\fiveshuffle\char'001}}}%
\mathchoice{\Dshuffle}{\Dshuffle}{\Sshuffle}{\SSshuffle}}}

\restylefloat{figure}
\definecolor{dgreen}{rgb}{0,0.70,0.30}
\definecolor{gold}{rgb}{0.85,.66,0}
\definecolor{purple}{rgb}{1.0,0.3,0.6}

\usepackage{nicefrac}
\newcommand{\tauh}{\nicefrac{\tau}{2}}

\usepackage{tikz}
\usetikzlibrary{calc} \usetikzlibrary{patterns,snakes} \usetikzlibrary{decorations.pathreplacing} \usetikzlibrary{decorations.markings} \usetikzlibrary{decorations.pathmorphing} \usetikzlibrary{positioning} \usetikzlibrary{arrows.meta}
\def\be{\begin{equation}}
\def\ee{\end{equation}}
\def\ba{\begin{array}}
\def\ea{\end{array}}

\newcommand{\bea}{\begin{eqnarray}}
\newcommand{\eea}{\end{eqnarray}}

\def\beq{\begin{equation}}
\def\eeq{\end{equation}}
\let\Re\relax
\let\Im\relax
\DeclareMathOperator{\Re}{Re}
\DeclareMathOperator{\Im}{Im}

\newcommand{\vecb}{\left(\begin{array}{c}}
\newcommand{\vece}{\end{array}\right)}
\newcommand{\ccb}{\left(\begin{array}{cc}}
\newcommand{\cce}{\end{array}\right)}
\newcommand{\cccb}{\left(\begin{array}{ccc}}
\newcommand{\ccce}{\end{array}\right)}
\newcommand{\ccccb}{\left(\begin{array}{cccc}}
\newcommand{\cccce}{\end{array}\right)}
\newcommand{\cccccb}{\left(\begin{array}{ccccc}}
\newcommand{\ccccce}{\end{array}\right)}

\newcommand{\dd}{\mathrm{d}}
\newcommand{\te}{\textrm}
\newcommand{\ap}{{\alpha'}}

\newcommand{\ep}{\epsilon}

\newcommand{\RR}{\mathbb R}
\newcommand{\CC}{\mathbb C}
\newcommand{\NN}{\mathbb N}
\newcommand{\ZZ}{\mathbb Z}

\newcommand{\QQ}{\mathbb Q}

\usepackage[format=plain,
            labelfont=bf,
            textfont=it]{caption}

\title{One-loop open-string integrals from differential equations: all-order $\ap$-expansions at $n$ points}
\author[a]{Carlos R.\ Mafra}
\author[b]{$\! \!$, Oliver Schlotterer}

\affiliation[a]{STAG Research Centre and Mathematical Sciences, University of Southampton,
Highfield, Southampton SO17 1BJ, UK}
\affiliation[b]{Department of Physics and Astronomy, Uppsala University, 75108 Uppsala, Sweden}

\emailAdd{c.r.mafra@soton.ac.uk}
\emailAdd{oliver.schlotterer@physics.uu.se}

\date{\today}

\abstract{We study generating functions of moduli-space integrals at genus one that are expected to
form a basis for massless $n$-point one-loop amplitudes of open superstrings and open bosonic strings.
These integrals are shown to satisfy the same type of linear and homogeneous first-order differential equation
w.r.t.\ the modular parameter $\tau$ which is known from the A-elliptic Knizhnik--Zamolodchikov--Bernard associator.
The expressions for their $\tau$-derivatives take a universal form for the integration cycles in
planar and non-planar one-loop open-string amplitudes.
These differential equations manifest the uniformly transcendental appearance of iterated
integrals over holomorphic Eisenstein series in the low-energy expansion w.r.t.\ the inverse string tension $\ap$.
In fact, we are led to conjectural matrix representations of certain derivations dual to Eisenstein series. Like this,
also the $\ap$-expansion of non-planar integrals is manifestly expressible in terms of iterated Eisenstein
integrals without referring to twisted elliptic multiple zeta values.
The degeneration of the moduli-space integrals at $\tau \rightarrow i\infty$ is expressed in terms
of their genus-zero analogues -- $(n{+}2)$-point Parke--Taylor integrals over disk boundaries.
Our results yield a compact formula for $\ap$-expansions of $n$-point integrals over boundaries of
cylinder- or M\"obius-strip worldsheets, where any desired order is accessible from elementary operations.}

\preprint{UUITP--36/19}

\begin{document}
\maketitle{}

\setcounter{tocdepth}{2}

\numberwithin{equation}{section}



\section{Introduction}
\label{sec:1}

Recent studies of scattering amplitudes in string theories have extended our computational reach
into several directions and led to a variety of structural insights. Most of these developments were
driven by a separation of string amplitudes into kinematic factors and moduli-space integrals. While
kinematic factors carry the entire dependence on the external polarizations, the accompanying integrals
over moduli spaces of punctured worldsheets depend on dimensionless combinations of the inverse string
tension $\ap$ and external momenta.

In this work, we describe a new method to integrate over open-string punctures in generating functions of
genus-one integrals in one-loop amplitudes of bosonic strings and superstrings. These
integrations will be performed in a Laurent expansion
w.r.t.\ $\ap$ -- for any number of open-string punctures in both
planar and non-planar one-loop amplitudes. A brief introduction of the method
in a letter format has been given in \cite{lettercomp}.

In general, $\ap$-expansions of string amplitudes have been identified as rewarding laboratory to encounter the
periods of the underlying moduli spaces in a simple context, leading to fruitful crosstalk between
string theory, particle phenomenology and number theory. Open- and closed-string tree amplitudes for instance
are governed by disk and sphere worldsheets, and the associated periods are multiple zeta values (MZVs)
\cite{Terasoma, Brown:2009qja, Schlotterer:2012ny, Broedel:2013aza} and their single-valued\footnote{The
notion of single-valued MZVs was introduced in \cite{Schnetz:2013hqa, Brown:2013gia}.} analogues
\cite{Schlotterer:2012ny, Stieberger:2013wea, Stieberger:2014hba, Schlotterer:2018zce, Brown:2018omk,
Vanhove:2018elu, Brown:2019wna}, respectively\footnote{See for instance
\cite{Barreiro:2005hv, Oprisa:2005wu, Stieberger:2006te, Stieberger:2009rr} for earlier work on tree-level
$\ap$-expansions at $n\leq 7$ points, in particular \cite{Oprisa:2005wu, Stieberger:2006te,
Boels:2013jua, Puhlfuerst:2015gta} for synergies with hypergeometric-function representations.}.

One-loop open-string amplitudes in turn were shown \cite{Broedel:2014vla,
Broedel:2017jdo} to yield elliptic multiple zeta values (eMZVs) defined by
Enriquez \cite{Enriquez:Emzv} upon integration over punctures on a cylinder or
M\"obius-strip worldsheet. Both of these worldsheet topologies are captured by
more general integrals over $A$-cycles of a torus by different specializations
of its complex modular parameter $\tau$. The main target of this work is the
$\ap$-expansion of such $A$-cycle integrals, and we will unravel the patterns
of eMZVs therein for any number of punctures.

The techniques in this work are driven by first-order differential equations for generating functions of
$A$-cycle integrals w.r.t.\ the modular parameter $\tau$ of the torus. Moreover,
the degeneration of the $n$-point genus-one integrals at the cusp $\tau \rightarrow i \infty$ is
expressed in terms of $(n{+}2)$-point genus-zero integrals whose $\ap$-expansions can be imported from
\cite{Schlotterer:2012ny, Broedel:2013aza, wwwMZV, Mafra:2016mcc}.
The desired $\ap$-expansions at genus one are generated by solving the differential equations (along with an initial value
at $\tau \rightarrow i \infty$) via standard Picard iteration. Accordingly, the accompanying eMZVs
will appear as iterated integrals over holomorphic Eisenstein series of ${\rm SL}_2(\ZZ)$
\cite{Enriquez:Emzv, Broedel:2015hia, Broedel:2019vjc}, ``iterated Eisenstein integrals'' for short.
In this way, all relations among eMZVs are automatically built in\footnote{This relies
on the linear-independence result of \cite{Nilsnewarticle} on iterated Eisenstein integrals.},
and the $\ap$-expansions of $A$-cycle integrals are available in a maximally simplified form.

One can view this work as a one-loop generalization of the following
structural results on tree-level $\ap$-expansions:
\begin{itemize}
\item First, the coefficients of Riemann zeta values $\zeta_{w}$ in the $\ap$-expansion of disk integrals
are related to those of products or higher-depth MZVs \cite{Schlotterer:2012ny}. These relations
are most succinct in the $f$-alphabet description of (motivic) MZVs \cite{Brown:2011ik, BrownTate}
and imply \cite{Drummond:2013vz} that disk integrals are stable under the motivic coaction
\cite{Goncharov:2005sla, BrownTate, Brown:2011ik, Brown:ICM14}.
Based on the differential equations at genus one, we will identify similar patterns among the coefficients
of different eMZVs or iterated Eisenstein integrals in the $\ap$-expansion. Accordingly, the $A$-cycle integrals
under investigation are preserved by the coaction of iterated Eisenstein integrals
\cite{Francislecture, Broedel:2018iwv} once we take a suitable quotient by their degeneration at $\tau \rightarrow i \infty$.
\item Second, $\ap$-expansions of $n$-point disk integrals can be recursively generated from
matrix representations of the Drinfeld associator \cite{Broedel:2013aza}. The specific representations
follow from Knizhnik--Zamolodchikov (KZ) equations of higher-multiplicity disk integrals with an
additional puncture and line up with the construction of more general braid matrices in
\cite{Mizera:2019gea}. As a genus-one generalization, the $A$-cycle integrals under investigation
are shown to obey the same type of differential equation in $\tau$ as the elliptic
Knizhnik--Zamolodchikov--Bernard (KZB) associator \cite{KZB, EnriquezEllAss, Hain}.
In particular, $n$-point $A$-cycle integrals
induce $(n{-}1)! \times (n{-}1)!$ matrix representations of certain derivations dual to Eisenstein
series \cite{Tsunogai} which accompany the iterated Eisenstein integrals in the $\ap$-expansions.
\item Third, only $(n{-}3)!$ choices for $n$-point disk integrands are inequivalent under integration by parts, i.e.\ the
so-called twisted cohomology at genus zero has dimension $(n{-}3)!$.
This counting follows from the work of Aomoto in the mathematics literature \cite{Aomoto87} and
has been independently conjectured and applied in the context of string tree amplitudes
\cite{Mafra:2011nv, Zfunctions, Huang:2016tag, Azevedo:2018dgo}. As a tentative one-loop analogue,
the $A$-cycle integrands of this work are proposed to furnish representatives of an $(n{-}1)!$-dimensional
twisted cohomology at genus one\footnote{We are grateful to Albert Schwarz for discussions on twisted
cohomologies at genus one. In collaboration with Dmitry Fuchs he calculated  and analyzed the twisted
homology of configuration spaces for surfaces of arbitrary genus.}. Our $(n{-}1)!$-family of $A$-cycle
integrals is stable under $\tau$-derivatives and should capture massless $n$-point one-loop amplitudes
of both superstrings and open bosonic strings.
\end{itemize}
The differential-equation setup outlined above will be used to manifest the following
properties of open-string $\ap$-expansions at genus one:
\begin{itemize}
\item {\bf Uniform transcendentality:} In conventional basis choices for disk integrals
\cite{Mafra:2011nv, Schlotterer:2012ny, Zfunctions}, the order in the $\ap$-expansion is correlated
with the transcendental weight of the accompanying MZVs, as can for instance be seen from \cite{Broedel:2013aza}.
The same kind of ``uniform transcendentality''\footnote{See \cite{Kotikov:1990kg, ArkaniHamed:2010gh, Henn:2013pwa, Adams:2018yfj, Broedel:2018qkq} for a discussion of uniform transcendentality
in the context of Feynman integrals in dimensional regularization, where the regularization parameter
$\varepsilon$ takes the role of $\ap$.} becomes manifest for the $A$-cycle integrals in this work when
generating their $\ap$-expansions from the differential equations: Among other things, we will exploit that
the $(n{-}1)! \times (n{-}1)!$ matrix representations of the derivations are linear in $\ap$,
and that the initial value at $\tau \rightarrow i \infty$ is built from uniformly transcendental genus-zero integrals.
Our open-string results are complemented by recent investigations of transcendentality properties of one-loop
amplitudes of heterotic strings \cite{Gerken:2018jrq} and type-II superstrings \cite{DHoker:2019blr}.
\item {\bf No twisted eMZVs in non-planar amplitudes:} The earlier expansion method for non-planar
open-string amplitudes \cite{Broedel:2017jdo, Broedel:2019vjc} introduces
twisted eMZVs or cyclotomic analogues of eMZVs in intermediate steps. To the $\ap$-orders considered,
however, the end results were found to be expressible in terms of conventional (i.e.\ untwisted) eMZVs.
The dropout of twisted eMZVs was not at all obvious from the techniques of \cite{Broedel:2017jdo, Broedel:2019vjc},
and only some of the cases admitted an explanation by factorization on closed-string poles. The differential equations in
this work will expose the absence of twisted eMZVs for any partition of $n$ punctures on the two cylinder boundaries.
The appearance of holomorphic Eisenstein series of ${\rm SL}_2(\ZZ)$ in $\tau$-derivatives turns out to be universal for
$A$-cycle integrals referring to planar and non-planar cylinders, respectively.
\end{itemize}

\begin{figure}
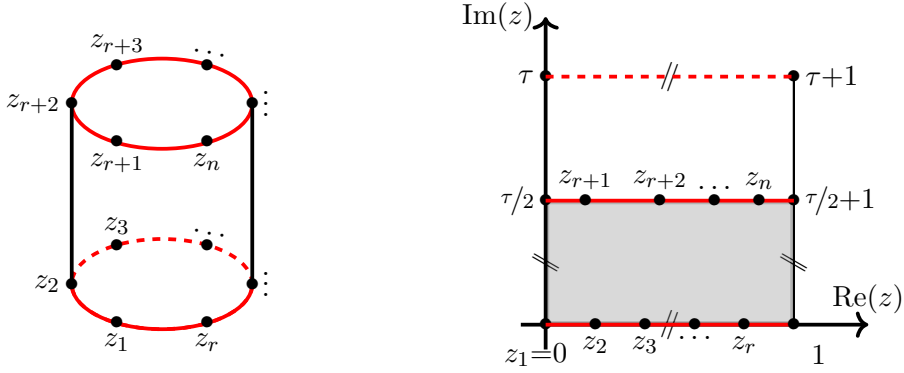

  \begin{center}
\tikzpicture [scale=0.6,line width=0.50mm]
\draw[red] (2,-4) ellipse (2cm and 1cm);
\draw[white,fill=white] (0,-2.8) rectangle (4,-3.9);
\draw[red,dashed] (2,-4) ellipse (2cm and 1cm);
\draw[red] (2,0) ellipse (2cm and 1cm);
\draw (0,0) -- (0,-4);
\draw (4,0) -- (4,-4);
\draw (1,0.85)node{$\bullet$}node[above]{$z_{r{+}3}$};
\draw (3,0.85)node{$\bullet$}node[above,rotate=-10]{$\ldots$};
\draw (4,0) node{$\bullet$};
\draw (4.3,0.18) node{$\vdots$};
\draw (0,0) node{$\bullet$}node[left]{$z_{r{+}2}$};
\draw (1,-0.85) node{$\bullet$}  node[below]{$z_{r{+}1}$};
\draw (3,-0.85) node{$\bullet$}  node[below]{$z_n$};
\scope[yshift=-4cm]
\draw (1,0.85)node{$\bullet$}node[above]{$z_3$};
\draw (3,0.85)node{$\bullet$}node[above,rotate=-10]{$\ldots$};
\draw (4,0) node{$\bullet$} ;
\draw (4.3,0.18) node{$\vdots$};
\draw (0,0) node{$\bullet$}node[left]{$z_2$};
\draw (1,-0.85) node{$\bullet$}  node[below]{$z_1$};
\draw (3,-0.85) node{$\bullet$}  node[below]{$z_r$};
\endscope
\scope[xshift=10.5cm,yshift=-4.9cm,scale=1.1]
\draw[->](-0.5,0) -- (6.5,0) node[above]{${\rm Re}(z)$};
\draw[->](0,-0.5) -- (0,6.2) node[left]{${\rm Im}(z)$};
\draw[line width=0.3mm](5,0) -- (5,5) ;
\draw(0,0)node{$\bullet$};
\draw(-0.2,-0.6)node{$z_1{=}0$};
\draw(0,1.25)node[rotate=60]{$| \! |$};
\draw(5,1.25)node[rotate=60]{$| \! |$};
\draw(2.5,0)node[rotate=-20]{$| \; \! \! |$};
\draw(2.5,5)node[rotate=-20]{$| \; \! \! |$};
\draw (0,5)node{$\bullet$}node[left]{$\tau$};
\draw (0,2.5)node{$\bullet$}node[left]{$\tauh$};
\draw (5,2.5)node{$\bullet$}node[right]{$\tauh{+}1$};
\draw (5,5)node{$\bullet$}node[right]{$\tau{+}1$};
\draw[fill=gray, opacity=0.3,line width=0.3mm] (0.05,0.05) rectangle (4.95,2.45);
\draw[red](0,0) -- (5,0);
\draw[red](0,2.5) -- (5,2.5);
\draw[red,dashed](0,5) -- (5,5);
\draw (5,0)node{$\bullet$};
\draw (5.5,-0.6)node{$1$};
\draw(1,0)node{$\bullet$}node[below]{$z_2$};
\draw(2,0)node{$\bullet$}node[below]{$z_3$};
\draw(3,0)node{$\bullet$}node[below]{$\ldots$};
\draw(4,0)node{$\bullet$}node[below]{$z_r$};
\draw(0.8,2.5)node{$\bullet$}node[above]{$z_{r+1}$};
\draw(2.3,2.5)node{$\bullet$}node[above]{$z_{r+2}$};
\draw(3.4,2.5)node{$\bullet$}node[above]{$\ldots$};
\draw(4.3,2.5)node{$\bullet$}node[above]{$z_n$};
\endscope
\endtikzpicture
    \caption{The cylinder worldsheet for one-loop open-string amplitudes seen in the left panel is
    parameterized by the gray rectangle in the right panel, where the periodic direction is reflected by
    the identification of horizontal lines. The cylinder boundaries drawn in red are the integration domain ${\cal C}(\ast)$ of
    the integrals in (\ref{intro2}) where a cyclic ordering of the punctures is prescribed within both boundaries.
    The asterisk is a placeholder for various distributions of the punctures,
    say $z_1,z_2,\ldots,z_r$ on the lower boundary and $z_{r+1},\ldots,z_n$ on the upper one for some $0\leq r \leq n$.}
    \label{basiccyl}
  \end{center}
\end{figure}


\subsection{Summary of the main results}
\label{sec:1schem}

We will now give a more detailed summary of the main results and display
the key equations. First of all, the $(n{-}1)!$-family of $A$-cycle integrals to be investigated in
this work is given by the generating functions
\begin{align}
&Z^\tau_{\vec{\eta}}( \ast | 1,2,\ldots,n) = \int_{{\cal C}(\ast)} \dd z_2 \, \dd z_3 \, \ldots \, \dd z_n  \,
\exp \Big( \sum_{i<j}^n s_{ij} {\cal G}(z_{ij},\tau) \Big) \label{intro2}
\\
&\ \ \times
 \Omega( z_{12},\eta_{2}{+}\eta_3{+}\ldots {+}\eta_{n} ,\tau)
  \Omega( z_{23},\eta_3{+}\ldots {+}\eta_{n} ,\tau)
  \ldots
   \Omega( z_{n-1,n},\eta_{ n}  ,\tau) \notag
\end{align}
and their permutations in the external-state labels $2,3,\dots,n$.
The integration domain ${\cal C}(\ast)$ may refer to any planar or non-planar arrangement of
open-string punctures $z_j$ on the two boundaries of the cylinder depicted in figure \ref{basiccyl}.
The two cylinder boundaries are parameterized by the unit interval $z_j \in (0,1)$ and its translate by $\tauh$,
i.e.\ parallel to the $A$-cycle of an auxiliary torus. While the cylinder in figure \ref{basiccyl} is obtained from a
rectangular torus with a purely imaginary modular parameter $\tau \in i\RR_+$, our results for $A$-cycle
integrals (\ref{intro2}) will hold for general tori with $\Re \tau \neq 0$. In particular, integrals over the boundary of a
M\"obius strip can be obtained by specializing the $A$-cycle integrals in a planar ordering to
$\tau \in \frac{1}{2}+i\RR_+$ \cite{Green:1984ed}.

The one-loop Green function
${\cal G}(z,\tau)$ and the doubly-periodic Kronecker--Eisenstein series $\Omega(z,\eta,\tau)$ in (\ref{intro2})
with first arguments $z_{ij}=z_i{-}z_j$ and formal expansion variables $\eta_j$ will be introduced below.
The Mandelstam variables are taken to be dimensionless,
\beq
s_{ij}=-2\ap k_i\cdot k_j  \, , \ \ \ \ \ \ 1 \leq i < j \leq n \, ,
 \label{intro0}
\eeq
i.e.\ the $\ap$-expansions refer to simultaneous Laurent expansion w.r.t.\ all of (\ref{intro0}).

Their dependence on $n{-}1$ bookkeeping variables $\eta_2,\eta_3,\ldots,\eta_n$ promotes the $Z^\tau_{\vec{\eta}}$
in (\ref{intro2}) to generating series of the $A$-cycle integrals that enter specific one-loop string amplitudes.
Open-superstring amplitudes with maximal supersymmetry give rise to component integrals at homogeneity
degree $-3$ in the $\eta_j$, and also at degrees $-7,-9,\ldots$ in case of $n\geq 8$ external legs
\cite{Tsuchiya:1988va, Broedel:2014vla}. Similarly, one-loop superstring amplitudes with reduced supersymmetry
additionally involve degree-$\eta^{-1}$ and $\eta^{-5}$ parts of (\ref{intro2})
\cite{Bianchi:2015vsa, Berg:2016wux}, while degree-$\eta^{+1}$ parts only enter bosonic-string amplitudes
or chiral halves of the heterotic string \cite{Dolan:2007eh}. Higher orders in
the $\eta_j$ are likely to be relevant for one-loop amplitudes involving massive states.

The $A$-cycle integrals in (\ref{intro2}) can be viewed as the one-loop generalization of the
Parke--Taylor integrals over disk boundaries at genus zero \cite{Zfunctions}
\begin{align}
Z^{\rm tree}(a_1,a_2,\ldots,a_n|1,2,\ldots ,n) = \! \! \! \!  \! \! \! \! \int \limits_{-\infty <z_{a_1} < z_{a_2} < \ldots < z_{a_n}< \infty} \! \! \! \!   \! \! \! \! \frac{ \dd z_1 \, \dd z_2\, \ldots \, \dd z_n }{\te{vol} \ \te{SL}_2(\RR)} \, \frac{ \prod^n_{i<j} |z_{ij}|^{-s_{ij}} }{ z_{12} z_{23} \ldots z_{n-1,n} z_{n,1}} \, .
\label{cocyc1}
\end{align}
The labelling of both $Z^{\rm tree}(\cdot|\cdot)$ and $Z^\tau_{\vec{\eta}}(\cdot | \cdot)$ by two slots ``$\cdot$''
emphasizes that moduli-space integrals in open-string amplitudes are pairings of integration cycles and
differential forms. In the tree-level case (\ref{cocyc1}), the cycle is labelled by a permutation
$a_1,a_2,\ldots,a_n \in S_n$ of the external legs $1,2,\ldots,n$ in the first slot which does not need to correlate with
the arrangements of the factors $z_{ij}$ in the Parke--Taylor denominator
$z_{12} z_{23}\ldots z_{n,1}$. The product of Kronecker--Eisenstein series $\Omega(\ldots)$ in (\ref{intro2})
takes the role of the Parke--Taylor forms, though the second slot of $Z^\tau_{\vec{\eta}}(\cdot | 1,2,\ldots,n)$
is not cyclically identified with $2,\ldots,n,1$ by the absence of $\Omega(z_{n,1},\eta,\tau)$ in the integrand.

The permutation symmetric products of $|z_{ij}|^{-s_{ij}}$ and $\exp ( s_{ij} {\cal G}(z_{ij},\tau) )$
in (\ref{cocyc1}) and (\ref{intro2}) are referred to as the Koba--Nielsen factors at genus zero and one,
respectively. They introduce the $\ap$-dependence into the moduli-space integrals and suppress
boundary terms when integrating total derivatives w.r.t.\ the punctures. The resulting integration-by-parts
relations define twisted cohomologies at the respective genus \cite{Mizera:2017cqs}. At genus zero, the twisted cohomology
entering the second slot of $Z^{\rm tree}(\cdot|\cdot)$ in (\ref{cocyc1}) is known to be spanned by $(n{-}3)!$ Parke--Taylor
factors \cite{Aomoto87}.

At genus one, we propose that $(n{-}1)!$ permutations of $2,3,\ldots,n$
in the second slot of (\ref{intro2}) yield valid representatives of the twisted cohomology.
A piece of evidence stems from the fact that these $(n{-}1)!$
permutations are closed under $\tau$-derivatives,
\beq
2\pi i \partial_\tau Z^\tau_{\vec{\eta}}( \ast | 1,a_2,a_3,\ldots,a_n)
= \sum_{B \in S_{n-1}} D^\tau_{\vec{\eta}}(A|B) Z^\tau_{\vec{\eta}}( \ast | 1,b_2,b_3,\ldots,b_n) \, .
 \label{intro3}
\eeq
The matrix $D^\tau_{\vec{\eta}}(A|B)$ is indexed by permutations $A=a_2,a_3,\ldots,a_n$
and $B=b_2,b_3,\ldots,b_n$ in $S_{n-1}$ that act on the labels $2,3,\ldots,n$ of both the $\eta_j$ and
$s_{ij}$ in (\ref{intro2}). As we will see, each entry of $D^\tau_{\vec{\eta}}(A|B)$ is linear in $s_{ij}$
which will be used to demonstrate uniform transcendentality of the $\ap$-expansion of $Z^\tau_{\vec{\eta}}$.
Moreover, $D^\tau_{\vec{\eta}}(A|B)$ is composed of second derivatives $\partial_{\eta_i}\partial_{\eta_j}$
and Weierstrass functions evaluated at sums of $\eta_{j}$.
Given that the Weierstrass function is the generating series of
holomorphic Eisenstein series ${\rm G}_{2m}$ including ${\rm G}_0=-1$,
\beq
\wp(\eta,\tau) = - \frac{ {\rm G}_0 }{\eta^2} + \sum_{m=2}^{\infty} (2m{-}1) \eta^{2m-2} {\rm G}_{2m}(\tau) \, ,
 \label{intro3a2}
 \eeq
the $(n{-}1)!\times (n{-}1)!$ matrix $D^\tau_{\vec{\eta}} $ in (\ref{intro3}) can be
expanded as follows
\beq
D^\tau_{\vec{\eta}} = \sum_{m=0}^{\infty} (1{-}2m) {\rm G}_{2m}(\tau) r_{\vec{\eta}}(\ep_{2m}) \, .
 \label{intro3b}
\eeq
The notation and normalization for the coefficients of the Eisenstein series
is motivated by the expectation that $r_{\vec{\eta}}(\ep_{2m})$ are $(n{-}1)!\times (n{-}1)!$ matrix representations of
the derivations $\ep_0,\ep_2,\ep_4,\ldots$ firstly studied by Tsunogai \cite{Tsunogai}. Each entry of
$r_{\vec{\eta}}(\ep_{2m>0})$ is of homogeneity degree $\eta^{2m-2}$, and $r_{\vec{\eta}}(\ep_{0})$
contains derivatives $\partial_{\eta_i}\partial_{\eta_j}$ on top of degree-$\eta^{-2}$ and
degree-$\eta^0$ terms. The dependence
of $r_{\vec{\eta}}(\ep_{2m})$ on $\eta_j$ and $\partial_{\eta_j}$ should preserve the commutation relations of the
derivations \cite{LNT, Pollack, Broedel:2015hia}, that is why they are referred to as a conjectural representation.

Note that differential equations of the form in (\ref{intro3}) have also been found for Feynman integrals
that evaluate to elliptic polylogarithms, see e.g.\ \cite{Bloch:2013tra, Bloch:2014qca, Adams:2017ejb, Ablinger:2017bjx, Remiddi:2017har, Bourjaily:2017bsb, Broedel:2017kkb, Broedel:2017siw, Adams:2018yfj, Broedel:2018iwv, Adams:2018bsn, Adams:2018kez, Broedel:2018rwm, Blumlein:2018aeq, Broedel:2019hyg, Bogner:2019lfa, Broedel:2019kmn}
and references therein. In particular, since the differential operator $D^\tau_{\vec{\eta}}(A|B)$ is linear in
$\ap$, (\ref{intro3}) can be viewed as the string-theory analogue of the $\varepsilon$-form of the differential equations
for Feynman integrals in the elliptic case \cite{Adams:2018yfj}. At genus zero, the KZ equations of higher-multiplicity
disk integrals studied in \cite{Broedel:2013aza} exhibit the same linearity of the right-hand side in $\ap$
and thereby furnish the string-theory analogue of the $\varepsilon$-form for Feynman integrals that evaluate
to multiple polylogarithms~\cite{Henn:2013pwa}.

Once we solve the differential equation (\ref{intro3}) via Picard iteration, the expansion (\ref{intro3b})
of the differential operator $D^\tau_{\vec{\eta}}$ naturally leads to iterated Eisenstein integrals
 $\gamma(k_1,\ldots,k_r|\tau)$ to be reviewed below. Since the initial values $ Z^{i\infty}_{\vec{\eta}}( \ast | 1,B)
 = \lim_{\tau \rightarrow i\infty} Z^{\tau}_{\vec{\eta}}( \ast | 1,B) $
 will be assembled from known functions of $\eta_j$ and genus-zero integrals (\ref{cocyc1}),
 we can extract the $\ap$-expansion of the $A$-cycle integrals from
\begin{align}
Z^\tau_{\vec{\eta}}( \ast | 1,A) &= \sum_{r=0}^{\infty}  \sum_{k_1,k_2,\ldots,k_r \atop{=0,4,6,8,\ldots} }  \prod_{j=1}^r (k_j{-}1)\, \gamma(k_1,k_2,\ldots,k_r|\tau)   \label{intro1}  \\
&\ \ \ \ \times \sum_{B \in S_{n-1}} r_{\vec{\eta}}(\ep_{k_r} \ep_{k_{r-1}} \ldots \ep_{k_2} \ep_{k_1} )_A{}^B
Z^{i\infty}_{\vec{\eta}}( \ast | 1,B) \, .
\notag
\end{align}
In an expansion w.r.t.\ $\eta_j$ and $\ap$, each order can be assembled from a finite number of terms
in the sum over $\{k_1,k_2,\ldots,k_r\}$ on the right-hand side.
Hence, the $\ap$-expansion of $A$-cycle integrals in open-string
amplitudes that occur at specific orders of $Z^\tau_{\vec{\eta}}( \ast | 1,A)$ in $\eta_j$
follows from (\ref{intro1}) via elementary operations -- matrix multiplication and differentiation w.r.t.\ $\eta_j$.
Since the eMZVs are represented via iterated
Eisenstein integrals $\gamma(\ldots)$, the $\ap$-expansions in (\ref{intro1}) are available in their minimal
form, i.e.\ all relations among eMZVs are already incorporated. This is analogous to expressing (motivic) MZVs in the
$f$-alphabet \cite{Broedel:2015hia}, so one can think of (\ref{intro1}) as generalizing the structure of
the tree-level $\ap$-expansion unraveled in \cite{Schlotterer:2012ny} to genus one.

On top of the structural insights provided by (\ref{intro1}), it has practical advantages in the explicit evaluation
of $\ap$-expansions. In contrast to earlier expansion methods for $A$-cycle
integrals \cite{Broedel:2014vla, Broedel:2017jdo},
there is no need to rearrange elliptic iterated integrals via so-called ``$z$-removal'' techniques\footnote{The
``$z$-removal'' techniques of \cite{Broedel:2014vla} have been recently described from the perspective of
elliptic symbol calculus \cite{Broedel:2018iwv}.}
when integrating over one puncture after the other. Moreover, all the kinematic poles of the $A$-cycle integrals
such as $s_{12}^{-1}$ are determined by the initial value $Z^{i\infty}_{\vec{\eta}}( \ast | 1,B) $
and do not require any subtraction scheme \cite{Broedel:2019vjc} when integrating over the punctures.

It will turn out to be convenient in this work to consider all the $\frac{n}{2}(n{-}1)$ Mandelstam invariants
in an $n$-point integral (\ref{intro2}) as independent. This may appear surprising since the
primary application of the $\ap$-expansion of this work are massless string amplitudes
with phase-space constraints $\sum_{i=1 \atop{i\neq j}}^n s_{ij} \ \forall \ j=1,2,\ldots,n$
and $\sum_{1\leq i<j}^{n-1} s_{ij}=0$. As a practical benefit of an enlarged momentum phase
space\footnote{The idea of relaxing momentum conservation goes back to Minahan
\cite{Minahan:1987ha} and was instrumental to describe the structure of one-loop string- and field-theory integrands
with half-maximal supersymmetry \cite{Berg:2016wux, Berg:2016fui, He:2017spx}.},
one can already define non-trivial two- and three-point building blocks which can be recycled to
simplify non-planar four and five-point string amplitudes. Moreover, having non-vanishing sums like
$s_{12}+s_{13}+s_{23}$ at four points regularizes $\frac{0}{0}$ indeterminates which would otherwise
arise from the method of this work.


\subsection{Outline}
\label{sec:1out}

The main body of this paper is organized as follows: After a review of basic material on eMZVs and
iterated Eisenstein integrals in section \ref{sec:2}, we illustrate the
main results and techniques of this work by means of two-point examples in section~\ref{sec:3}.
Section \ref{sec:4} is dedicated to the differential equation
(\ref{intro3}), with rigorous derivations up to five points and an $n$-point conjecture for the
explicit form of the differential operator $D^\tau_{\vec{\eta}}$. The associated initial values for planar and non-planar
$A$-cycle integrals at $\tau \rightarrow i \infty$ are expressed via linear combinations of disk
integrals in section \ref{sec:5} and \ref{sec:6},
respectively. In section \ref{sec:others}, the results of this work are shown to imply that $A$-cycle
integrals (\ref{intro2}) exhibit uniformly transcendental $\ap$-expansions and are preserved under the coaction.

The key equations of this work include the $n$-point proposal for $D^\tau_{\vec{\eta}}$ in (\ref{B460a})
as well as the relations (\ref{CC10G}), (\ref{rePT12}) and (\ref{s0jL}) between their initial values and
$(n{+}2)$-point disk integrals.


\section{Basics of eMZVs and iterated Eisenstein integrals}
\label{sec:2}

This section aims to review various properties of elliptic analogues of MZVs and iterated integrals
over holomorphic Eisenstein series. Our conventions for MZVs are fixed by ($n_j \in \NN$)
\beq
\zeta_{n_1,n_2,\ldots,n_r} \equiv \sum_{0<k_1<k_2<\ldots <k_r} k_1^{-n_1} k_2^{-n_2} \ldots k_r^{-n_r} \, , \ \ \ \ \ \ n_r \geq 2\, ,
\label{convMZV}
\eeq
where $r$ and $n_1{+}n_2{+}\ldots{+}n_r$ are referred to as their depth and transcendental weight, respectively.
The customary terminology ``transcendental weight'' used throughout this work is highly abusive since none of the
$\zeta_{n}$ with odd $n$ have been proven to be transcendental so far.


\subsection{Kronecker--Eisenstein series}
\label{sec:2.1}

\begin{figure}
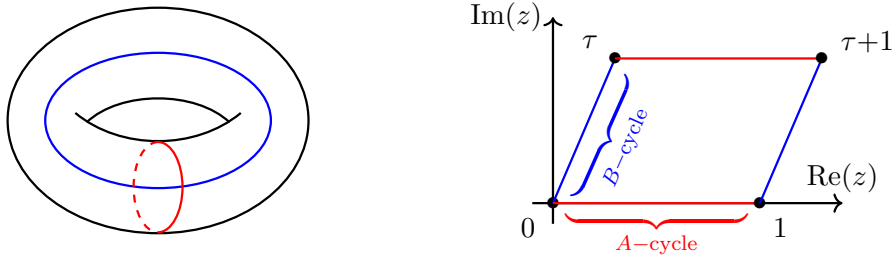

\begin{center}
\tikzpicture[scale=0.5,line width=0.30mm]
\draw(0,0) ellipse  (4cm and 3cm);
\draw(-2.2,0.2) .. controls (-1,-0.8) and (1,-0.8) .. (2.2,0.2);
\draw(-1.9,-0.05) .. controls (-1,0.8) and (1,0.8) .. (1.9,-0.05);
\draw[blue](0,0) ellipse  (3cm and 1.8cm);
\draw[red] (0,-2.975) arc (-90:90:0.65cm and 1.2cm);
\draw[red,dashed] (0,-0.575) arc (90:270:0.65cm and 1.2cm);
\scope[xshift=10.5cm,yshift=-2.2cm,scale=1.1]
\draw[->](-0.5,0) -- (7,0) node[above]{${\rm Re}(z)$};
\draw[->](0,-0.5) -- (0,4.5) node[left]{${\rm Im}(z)$};
\draw(0,0)node{$\bullet$};
\draw(-0.6,-0.6)node{$0$};
\draw[blue](0,0) -- (1.5,3.5);
\draw[blue](1.4,1.5)node[rotate=66]{$\textcolor{blue}{\underbrace{\phantom{xxxxxxxx}}_{B-\te{cycle}}}$};
\draw (1.5,3.5)node{$\bullet$} ;
\draw(0.9,3.9)node{$\tau$};
\draw[red](0,0) -- (5,0);
\draw[red](2.5,-0.6)node{$\textcolor{red}{\underbrace{\phantom{xxxxxxxxxxx}}_{A-\te{cycle}}}$};
\draw (5,0)node{$\bullet$};
\draw (5.5,-0.6)node{$1$};
\draw[red](1.5,3.5) -- (6.5,3.5);
\draw[blue](5,0) -- (6.5,3.5);
\draw(6.5,3.5)node{$\bullet$};
\draw(7.6,3.9)node{$\tau{+}1$};
\endscope
\endtikzpicture
\caption{The torus will be parameterized through a complex coordinate $z$, where the
$A$-cycle and $B$-cycle translate into periodicities $z \cong z{+}1$ and $z \cong z{+}\tau$,
respectively.}
\label{figureone}
\end{center}
\end{figure}

The integrands under investigation in (\ref{intro2}) are built from a non-holomorphic version
of the Kronecker--Eisenstein series \cite{Kronecker, BrownLev},
\begin{align}
\Omega(z,\eta,\tau) \equiv \exp \Big( 2\pi i \eta\frac{ \Im z }{\Im \tau} \Big) \frac{ \theta_1'(0,\tau) \theta_1(z+\eta,\tau) }{\theta_1(z,\tau)  \theta_1(\eta,\tau)} \, ,
\label{1.2}
\end{align}
where $\Im \tau >0$, and the tick of $\theta_1'(0,\tau)$ denotes a derivative w.r.t.\ the first argument.
Our conventions for the modular parameter and the odd Jacobi theta function are $q = e^{2\pi i \tau}$ and
\begin{align}
\theta_1(z,\tau) \equiv 2 q^{1/8} \sin(\pi z) \prod_{n=1}^{\infty} (1-q^n) (1-e^{2\pi i z} q^n) (1-e^{-2\pi i z} q^n) \, .
\label{1.theta}
\end{align}
The double periodicity of the Kronecker--Eisenstein series (\ref{1.2})
\beq
 \Omega(z,\eta,\tau) = \Omega(z{+}1,\eta,\tau) = \Omega(z{+}\tau,\eta,\tau)
\label{1.2doub}
\eeq
qualifies the first argument $z$ to live on a torus $\frac{  \CC }{ \ZZ + \tau \ZZ}$, see figure \ref{figureone}
for the parameterization used in this work.

Upon Laurent expansion in the second argument $\eta \in \CC$,
the expression for $\Omega(z,\eta,\tau)$ in (\ref{1.2}) defines an
infinite family of doubly-periodic functions $f^{(k)}$ with $k \in \NN_0$,
\beq
\Omega(z,\eta,\tau) =  \sum_{k=0}^{\infty} \eta^{k-1} f^{(k)}(z,\tau)\, , \ \ \ \ \ \ \
f^{(k)}(z,\tau)=f^{(k)}(z{+}1,\tau)=f^{(k)}(z{+}\tau,\tau)\, ,
\label{A2}
\eeq
starting with $f^{(0)}(z,\tau)=1$ and $f^{(1)}(z,\tau)=\partial_z \log \theta_1(z,\tau)+2\pi i \frac{ \Im z }{\Im \tau}$.
While $f^{(1)}(z,\tau)$ has simple poles on the lattice $z \in \ZZ + \tau \ZZ$, the remaining $f^{(k\neq 1)}(z,\tau)$
are regular for any $z \in \CC$.

It is worthwhile to highlight two properties of the doubly-periodic Kronecker--Eisenstein
series (\ref{1.2}) that will be used extensively in this work: Its bilinears obey Fay
identities \cite{mumford1984tata, BrownLev}
\beq
\Omega(z_1,\eta_1,\tau)\Omega(z_2,\eta_2,\tau) =
\Omega(z_1,\eta_1{+}\eta_2,\tau) \Omega(z_2{-}z_1,\eta_2,\tau)
+\Omega(z_2,\eta_1{+}\eta_2,\tau) \Omega(z_1{-}z_2,\eta_1,\tau)
\, ,
\label{A21}
\eeq
and its derivatives are related through the mixed heat equation
\beq
2\pi i \partial_{\tau} \Omega(u\tau{+} v,\eta,\tau) = \partial_v \partial_\eta \Omega(u\tau{+} v,\eta,\tau)
\, , \ \ \ \ \ \ u,v \in \RR \, .
\label{Aheat}
\eeq
The $\tau$-derivative in (\ref{Aheat}) is taken at a fixed value of the real coordinates $u,v \in \RR$
of the first argument $z= u\tau + v$ in (\ref{1.2}) which will be shown to be a natural choice
in the context of one-loop open-string amplitudes.


\subsection{eMZVs and twisted eMZVs}
\label{sec:2.2}

Enriquez defined eMZVs as the iterated integrals of the above Kronecker--Eisenstein
coefficients $f^{(k)}(z,\tau)$ over the homology cycles of a torus \cite{Enriquez:Emzv}.
We will only consider $A$-cycle eMZVs due to integration over $z\in (0,1)$ in this work
and refer the reader to \cite{Enriquez:Emzv, Broedel:2018izr, Zerbini:2018sox, Zerbini:2018hgs}
for discussions of $B$-cycle eMZVs (with $\frac{z}{\tau} \in (0,1)$
 as an integration path). Based on the recursively defined elliptic iterated integrals \cite{BrownLev, Broedel:2014vla}
\beq
\Gamma\left( \begin{smallmatrix} k_1 &k_2 &\ldots &k_r \\ a_1 &a_2 &\ldots &a_r  \end{smallmatrix} ;  z |\tau\right)
= \int^z_0 \dd t \, f^{(k_1)}(t{-}a_1,\tau) \, \Gamma\left( \begin{smallmatrix} k_2 &\ldots &k_r \\ a_2 &\ldots &a_r  \end{smallmatrix} ;  t |\tau\right) \, , \ \ \ \ \ \ z \in \RR, \ \ \ a_j \in \CC
\label{elliter}
\eeq
with $\Gamma\left( \begin{smallmatrix} \emptyset \\ \emptyset \end{smallmatrix} ;  z |\tau\right) =1$,
Enriquez' $A$-cycle eMZVs can be obtained by evaluation at $z=1$ and specialization to $a_j=0$,
\beq
\omega(k_1,k_2,\ldots,k_r|\tau) \equiv \Gamma\left( \begin{smallmatrix} k_r &k_{r-1} &\ldots &k_2 &k_1 \\
0 &0 &\ldots &0 &0  \end{smallmatrix} ;  1 |\tau\right)\, .
\label{A2emzv}
\eeq
While the elliptic iterated integrals (\ref{elliter}) are not homotopy invariant by themselves,
it is known how to generate homotopy invariant uplifts by adding similar iterated integrals with
simpler integration kernels \cite{BrownLev}. Following the demands of the integrals over $A$-cycles
in figure \ref{basiccyl}, we will always take the interval $(0,z)$ with $z\in \RR$ as the integration path
for (\ref{elliter}).

The integers $r$ and $k_1{+}k_2{+}\ldots{+}k_r$ characterizing the eMZV (\ref{A2emzv}) are
referred to as its length and (transcendental) weight. By the simple pole of $f^{(1)}(z,\tau)$ at lattice points
$z \in \ZZ {+} \tau \ZZ$, both of $\omega(1,\ldots|\tau)$ and $\omega(\ldots,1|\tau)$ generically diverge.
We will follow the regularization prescription in section 2.2.1 of \cite{Broedel:2014vla} which assigns
$\lim_{\tau \rightarrow i \infty} \omega(0,1|\tau) = \frac{i \pi }{2}$ and preserves the reflection and shuffle identities.

When the shifts $a_j$ of the elliptic iterated integrals (\ref{elliter}) are taken to be rational points
on the torus, evaluation at $z=1$ gives rise to twisted eMZVs \cite{Broedel:2017jdo}
\beq
\omega\Big(  \begin{smallmatrix}k_1, &k_2, &\ldots \ , &k_r\\ b_1, &b_2, &\ldots \ , &b_r\end{smallmatrix}  \big| \,  \tau\Big)
\equiv \Gamma\left( \begin{smallmatrix} k_r &k_{r-1} &\ldots &k_2 &k_1 \\
b_r &b_{r-1} &\ldots &b_2 &b_1  \end{smallmatrix} ;  1 |\tau\right) \, , \ \ \ \ \ \
b_j \in \QQ + \tau \QQ
\label{A2temzv}
\eeq
of length $r$ and weight $k_{1}{+}k_2{+}\ldots{+}k_r$.
Again, integration over $f^{(k_j=1)}$ may cause divergences if some
of the twists $b_j$ are real, see section 2.1 of \cite{Broedel:2017jdo} for a possible regularization.
The only twists we encounter in this work are $b_j=0$ and $b_j = \tauh$,
where the latter only occur in intermediate steps and do not cause any divergence
when integrating $f^{(1)}(z{-}\tauh,\tau)$ over the $A$-cycle $z\in (0,1)$. Note that double periodicity
implies that $f^{(k)}(z{-}\tauh,\tau)= f^{(k)}(z{+}\tauh,\tau)$ $\forall \ k \in \NN_0$, so one cannot distinguish
shifts $a_j= \pm \tauh$ in (\ref{elliter}) or $b_j= \pm \tauh$ in (\ref{A2temzv}).

The functional dependence of twisted eMZVs on $\tau$ can be inferred by solving first-order differential
equations along with a known initial value at $\tau\rightarrow i\infty$ \cite{Broedel:2017jdo}: By the mixed
heat equation (\ref{Aheat}), the $\tau$-derivatives of twisted eMZVs of length $r$ are expressible via those
of length $(r{-}1)$ multiplied by holomorphic Eisenstein series of congruence subgroups of ${\rm SL}_2(\ZZ)$.
At the cusp $\tau\rightarrow i\infty$, twisted eMZVs with $\Re b_j =0$ degenerate to $\QQ[(2\pi i)^{-1}]$-linear
combinations of MZVs. For untwisted eMZVs, the $\tau$-derivative solely introduces Eisenstein series of ${\rm SL}_2(\ZZ)$
\cite{Enriquez:Emzv, Broedel:2015hia}. This leads to an expansion around the cusp of the form
\beq
\omega(k_1,k_2,\ldots,k_r|\tau) = \sum_{n=0}^{\infty} c_n(k_1,k_2,\ldots,k_r) q^n \, ,
\label{A2emzvq}
\eeq
where the Fourier coefficients $c_n(\ldots)$ are $\QQ[(2\pi i)^{-1}]$-linear
combinations of MZVs \cite{Enriquez:Emzv, Broedel:2015hia}. Twisted eMZVs
with $b_j \in \{0,\tauh\}$ allow for a similar expansion,
\beq
\omega\Big(  \begin{smallmatrix}k_1, &k_2, &\ldots \ , &k_r\\ b_1, &b_2, &\ldots \ , &b_r\end{smallmatrix}  \big| \,  \tau\Big) = \sum_{n=0}^{\infty} d_n \Big(  \begin{smallmatrix}k_1, &k_2, &\ldots \ , &k_r\\ b_1, &b_2, &\ldots \ , &b_r\end{smallmatrix}  \Big) q^{n/2} \, ,
\label{A2temzvq}
\eeq
where the powers of $q$ include half-odd integers, and $d_n(\ldots)$ are again $\QQ[(2\pi i)^{-1}]$-linear
combinations of MZVs \cite{Broedel:2017jdo}.

As a main result of this work, we will derive and solve differential equations for eMZVs at the level of the
$A$-cycle integrals (\ref{intro2}). Hence, we provide a unified all-order treatment of the $\ap$-dependence
introduced by the Koba--Nielsen factor $\sim e^{s_{ij} {\cal G}(z_{ij},\tau)}$. Even for the integration cycles of non-planar
open-string amplitudes, the $\tau$-derivatives will be shown to solely introduce Eisenstein
series of ${\rm SL}_2(\ZZ)$, manifesting the dropout of twisted eMZVs at all orders in $\ap$.
The $\tau \rightarrow i\infty$ degenerations of the $A$-cycle integrals yield series in
$s_{ij}$ with $\QQ[(2\pi i)^{-1}]$-linear combinations MZVs as coefficients. These series
in $s_{ij}$ and MZVs will be inferred from disk integrals with known $\ap$-expansion
\cite{Schlotterer:2012ny, Broedel:2013aza, wwwMZV, Mafra:2016mcc} which elegantly
resums the degeneration of all the eMZVs in the $\ap$-expansion of the $A$-cycle integrals
(see \cite{Enriquez:Emzv, Broedel:2015hia} for a method to determine the individual
$\omega(k_1,\ldots,k_r|\tau \rightarrow i \infty)$).


\subsection{Open-string Green functions}
\label{sec:2.3}

We will now review the connection between elliptic iterated integrals and the open-string
Green function ${\cal G}(z,\tau)$ that enters the $A$-cycle integrals (\ref{intro2}) through
the Koba--Nielsen factor
\beq
 \te{KN}^{\tau}_{12\ldots n} \equiv \exp \Big( \sum_{i<j}^n s_{ij} {\cal G}(z_i{-}z_j,\tau) \Big) \, .
 \label{dfkn}
 \eeq
According to the parameterization of the cylinder in figure \ref{basiccyl}, placing the
punctures $z_j{=}u_j\tau{+}v_j$ on the cylinder boundaries amounts to fixing the first
real coordinate to $u_j=0$ or $u_j= \frac{1}{2}$. The integrations in (\ref{intro2}) along the boundaries -- i.e.\
along $A$-cycles of a torus -- only concern the second coordinate $v_j \in (0,1)$ with $\dd z_j = \dd v_j$.

When the Green function connects punctures on the same boundary, the first coordinate of
$z_{ij} = z_i - z_j=u_{ij}\tau +v_{ij}$ is set to $u_{ij}=0$, whereas $z_i$ and $z_j$ on different boundaries
give rise to $u_{ij}= \frac{1}{2}$. In both cases, the dependence on $v_{ij} \in (-1,1)$
can be expressed via elliptic iterated integrals (\ref{elliter}) \cite{Broedel:2014vla, Broedel:2017jdo}
(with additional simplifications due to our choice $v_1=0$):
\begin{itemize}
\item both punctures on the same boundary (``planar Green function'')
\begin{align}
{\cal G}(v_{1j},\tau) &=    \omega(1,0|\tau)
- \Gamma\left( \begin{smallmatrix}1 \\ 0
\end{smallmatrix} ;  v_{j}|\tau\right)  \label{B1} \\
{\cal G}(v_{ij},\tau)  &= \omega(1,0|\tau)
-  \Gamma\left( \begin{smallmatrix}1 \\v_j \end{smallmatrix} ;  v_{i} |\tau\right)
- \Gamma\left( \begin{smallmatrix} 1 \\ 0 \end{smallmatrix} ;  v_{j} |\tau\right)
\notag
\end{align}
\item punctures on different boundaries (``non-planar Green function'')
\begin{align}
{\cal G}(v_{1j}{+}\tauh,\tau) &=  \omega \Big(  \begin{smallmatrix}1, &0\\ \tauh, &0 \end{smallmatrix}  \big| \, \tau \Big)
 - \Gamma\left( \begin{smallmatrix}1 \\ \tauh
\end{smallmatrix} ;  v_{j}  \big| \,\tau\right) + \frac{i \pi \tau }{4} \label{B2} \\
{\cal G}(v_{ij}{+}\tauh,\tau) &=  \omega \Big(  \begin{smallmatrix}1, &0\\ \tauh, &0 \end{smallmatrix}  \big| \, \tau \Big)
-  \Gamma\left( \begin{smallmatrix}1 \\v_j+\tauh
\end{smallmatrix} ;  v_{i} \big| \,\tau\right) - \Gamma\left( \begin{smallmatrix}
1 \\ \tauh
\end{smallmatrix} ;  v_{j} \big| \,\tau\right) + \frac{i \pi \tau }{4}
 \notag
\end{align}
\end{itemize}
When restricted to the on-shell kinematics $\sum_{1\leq i<j}^n s_{ij}=0$ of massless $n$-point string
amplitudes, the Koba--Nielsen factor (\ref{dfkn})
is invariant under shifts ${\cal G}(z_{ij},\tau) \rightarrow {\cal G}(z_{ij},\tau) {+} {\cal F}(\tau)$
of the Green function, where ${\cal F}(\tau)$ does not depend on the punctures. In this work, the $A$-cycle integrals
(\ref{intro2}) will be studied in the enlarged phase space with unconstrained $\sum_{1\leq i<j}^n s_{ij} $,
so we committed to a choice of ${\cal F}(\tau)$ in specifying the Green functions in (\ref{B1}) and (\ref{B2}).

As firstly noticed in \cite{Broedel:2018izr}, the benefit of the additive term
$\omega(1,0|\tau) $ in (\ref{B1}) is that the planar Green function integrates to zero along the $A$-cycle
$\int^1_0 \dd v_j  \, {\cal G}(v_{1j},\tau)=0$. Since the planar and non-planar Green functions have to descend
from the same Koba--Nielsen factor (\ref{dfkn}), this leaves no further freedom in the non-planar Green
function (\ref{B2}) which integrates to $\int^1_0 \dd v_j  \, {\cal G}(v_{1j}{+}\tauh,\tau)=\frac{i\pi \tau}{4}$.
The detailed connection of the above ${\cal G}(\ldots,\tau)$ with suitable restrictions of the closed-string
Green function $- \log \big| \frac{ \theta_1(z,\tau) }{\theta_1'(0,\tau)}\big|^2 + \frac{2\pi }{\Im \tau}(\Im z)^2$
is explained in section 4.2 of \cite{Broedel:2017jdo}. Note that
in section 5 of \cite{Broedel:2018izr}, the term $\frac{i \pi \tau }{4}$ in (\ref{B2}) has been absorbed into
a redefinition of the Koba--Nielsen factor by products of $q^{s_{ij}/8}$, and we will similarly peel off these
products in several examples in sections \ref{sec:3.5} and \ref{sec:6}.

The mixed heat equation (\ref{Aheat}) and the recursive definition (\ref{elliter}) of elliptic iterated integrals
can be used to compute the $v_j$- and $\tau$-derivatives of the Green function.
For both choices of $u_{ij} \in \{0,\tfrac{1}{2}\}$ in $z_{ij}=u_{ij}\tau + v_{ij}$, one can show that\footnote{The following
identities have been used in intermediate steps:
\begin{align*}
2\pi i \partial_\tau \Gamma\left( \begin{smallmatrix}1 \\ 0
\end{smallmatrix} ;  z |\tau \right)  &=   f^{(2)}(z,\tau) -   f^{(2)}(0,\tau)  \, , \ \ \ \ \ \
2\pi i \partial_\tau \omega(1,0|\tau)   = - f^{(2)}(0,\tau) - 2 \zeta_2 \, , \ \ \ \ \ \
2\pi i \partial_\tau \frac{ i \pi \tau}{4} = -3 \zeta_2 \,  \\
2\pi i \partial_\tau \Gamma\left( \begin{smallmatrix}1 \\ \tauh
\end{smallmatrix} ;  z |\tau \right) &=  f^{(2)}(z{-}\tauh,\tau) -   f^{(2)}(\tauh,\tau)
 \, , \ \ \ \ \ \
2\pi i \partial_\tau  \omega \Big(  \begin{smallmatrix}1, &0\\ \tauh, &0 \end{smallmatrix} |\tau  \Big)
 = - f^{(2)}(\tauh,\tau) + \zeta_2
\end{align*}}
\begin{align}
\partial_{v_i} {\cal G}(z_{ij} ,\tau)  = - f^{(1)}(z_{ij},\tau) \, , \ \ \ \ \ \
2\pi i \partial_\tau {\cal G}(z_{ij},\tau)  = - f^{(2)}(z_{ij},\tau) -2 \zeta_2 \, ,
\label{B5}
\end{align}
which implies the following differential equations for the Koba--Nielsen factor (\ref{dfkn}):
\beq
\partial_{v_i} \te{KN}^{\tau}_{12\ldots n}  = -  \sum_{j=1 \atop{j\neq i}}^n s_{ij} f^{(1)}_{ij}  \te{KN}^{\tau}_{12\ldots n}
\, , \ \ \ \ \
2\pi i \partial_{\tau} \te{KN}^{\tau}_{12\ldots n} = - \! \sum_{1\leq i <j}^n \! s_{ij}( f^{(2)}_{ij} + 2 \zeta_2 )
\te{KN}^{\tau}_{12\ldots n}  \, .
\label{dkn}
\eeq
Here and in later sections, we employ the shorthands
\beq
f^{(k)}_{ij} = f^{(k)}(z_{ij},\tau) \, ,
\label{shfij}
\eeq
and we emphasize that (\ref{dkn}) is universally valid for any distribution of the $n$ punctures
over the two cylinder boundaries in figure \ref{basiccyl}, i.e.\ for any integration cycle ${\cal C}(\ast)$ in
the $A$-cycle integral (\ref{intro2}).


\subsection{Iterated Eisenstein integrals}
\label{sec:2.4}

By repeatedly integrating the $\tau$-derivative of $A$-cycle eMZVs,
one is lead to iterated integrals of holomorphic Eisenstein series ${\rm G}_k(\tau)$
of $\te{SL}_2(\ZZ)$ \cite{Enriquez:Emzv, Broedel:2015hia}.
We will use conventions where ${\rm G}_0(\tau) = -1$ and
\beq
{\rm G}_k(\tau) = \sum_{(m,n) \in \ZZ^2 \atop{(m,n) \neq (0,0) }} \frac{1}{(m\tau + n)^k} =  \left\{
\begin{array}{cl}
0 &: \ k>0 \ \te{odd} \\ \displaystyle
 2\zeta_k+\frac{2 (2\pi i)^k}{(k-1)!} \sum_{m,n=1}^{\infty} m^{k-1} q^{mn} &: \ k>0 \ \te{even}
\end{array}  \right.
\, ,
\label{A14}
\eeq
with Eisenstein summation prescription in case of $k=2$. Iterated Eisenstein
integrals in the normalization of \cite{Broedel:2015hia} are defined by
\beq
\gamma(k_1,k_2,\ldots,k_r|\tau) = \frac{(-1)^r}{(2\pi i)^{2r}}  \! \! \! \! \int \limits_{0<q_1<q_2<\ldots <q_r <q}   \! \! \! \! \frac{ \dd q_1 }{q_1} \,  \frac{ \dd q_2 }{q_2} \ldots  \frac{ \dd q_r }{q_r} \, {\rm G}_{k_1}(\tau_1) {\rm G}_{k_2}(\tau_2) \ldots {\rm G}_{k_r}(\tau_r) \, .\label{preB5}
\eeq
Integration over ${\rm G}_0$ or the zero mode $2\zeta_k$ in the
$q$-expansions of (\ref{A14}) may cause endpoint divergences from the integration region where $q_1\rightarrow 0 $
in (\ref{preB5}). These divergences will be regularized through the tangential-base-point
prescription \cite{Brown:mmv} with net effect $\int^q_0 \frac{ \dd q_1}{q_1} = \log q$.
This leads to regularized values such as
\beq
\gamma(0|\tau) = \frac{\tau}{2\pi i} \ , \ \ \ \ \ \
\gamma(4|\tau) = \frac{ i \pi^3 \tau }{90} + \frac{4\pi^2}{3} \sum_{m,n=1}^{\infty} \frac{ m^2 }{n} q^{mn}
\, ,
\label{A14reg}
\eeq
that preserve the shuffle relations of $\gamma(k_1,\ldots,k_r|\tau)$. Alternatively, one can bypass the
endpoint divergences by subtracting the zero modes of the non-constant holomorphic Eisenstein series (\ref{A14}),
\beq
{\rm G}^0_{k}(\tau) =  \left\{
\begin{array}{cl} {\rm G}_{k}(\tau)  - 2 \zeta_k  &: \ k>0 \ \te{even} \\
0 &: \ k>0 \ \te{odd} \\
-1 &: \ k=0 \end{array} \right.  \, .\label{preB5GG}
\eeq
These integration kernels give rise to modified versions of
iterated Eisenstein integrals
\beq
\gamma_0(k_1,k_2,\ldots,k_r|\tau) = \frac{(-1)^r}{(2\pi i)^{2r}}  \! \! \! \! \int \limits_{0<q_1<q_2<\ldots <q_r <q}   \! \! \! \! \frac{ \dd q_1 }{q_1} \,  \frac{ \dd q_2 }{q_2} \ldots  \frac{ \dd q_r }{q_r} \, {\rm G}^0_{k_1}(\tau_1) {\rm G}^0_{k_2}(\tau_2) \ldots {\rm G}^0_{k_r}(\tau_r) \, ,\label{preB5con}
\eeq
that converge if $k_1>0$ \cite{Broedel:2015hia}. The $q$-expansion resulting from (\ref{A14})
takes the following compact form in these cases (with $k_j\neq 0$ and $p_j \geq 0$),
\begin{align}
&\gamma_0(k_1,0^{p_1-1},k_2,0^{p_2-1},\ldots,k_d,0^{p_d-1}|\tau) =(-2)^d
 \Big( \prod_{j=1}^{d} \frac{  (2\pi i)^{k_j - 2 p_j} }{(k_j-1)!} \Big)
\label{qgamma1}\\
& \ \ \ \ \ \ \ \times \sum_{m_1,m_2,\ldots,m_d=1 \atop{ n_1,n_2,\ldots,n_d=1}}^{\infty}
\frac{m_1^{k_1-1} m_2^{k_2-1} \ldots m_d^{k_d-1}  q^{m_1n_1+m_2n_2+\ldots +m_dn_d}}{(m_1 n_1)^{p_1} (m_1n_1+m_2n_2)^{p_2} \ldots (m_1n_1+m_2n_2+\ldots +m_dn_d)^{p_d}}  \ ,
\notag
\end{align}
and the analogues of (\ref{A14reg}) are $\gamma_0(0|\tau)=\frac{\tau}{2\pi i} $ and
$\gamma_0(4|\tau)= \frac{4\pi^2}{3} \sum_{m,n=1}^{\infty} \frac{ m^2 }{n} q^{mn}$. Both
of $\gamma(k_1,k_2,\ldots,k_r|\tau)$ and $\gamma_0(k_1,k_2,\ldots,k_r|\tau)$ are said to have
length $r$, and the number of nonzero entries $k_j\neq 0$ (i.e.\ the integer $d$ in (\ref{qgamma1}))
is referred to as their depth. The transcendental weight of both $\gamma(k_1,k_2,\ldots,k_r|\tau)$
and $\gamma_0(k_1,k_2,\ldots,k_r|\tau)$ is $k_1{+}k_2{+}\ldots{+}k_r{-}r$.
This follows from the differential equation of eMZVs as well as the transcendental
weights $k$ of ${\rm G}_k(\tau),{\rm G}^0_k(\tau)$ and $k_1{+}k_2{+}\ldots{+}k_r$ of
$\omega(k_1,k_2,\ldots,k_r|\tau)$.

The decomposition of $A$-cycle eMZVs into iterated Eisenstein integrals
usually involves fewer terms when employing $\gamma_0(\ldots)$ instead of $\gamma(\ldots)$, e.g.\
\cite{Broedel:2015hia}
\begin{align}
  \omega(0,2k{-}1|\tau)&=\delta_{1,k}\frac{\pi i}{2}+(2k{-}1)\gamma_0(2k|\tau)
\label{eqn:gamma1} \\
 \omega(0,0,2k|\tau)&=-\frac{1}{3}\zeta_{2k} - 2k(2k{+}1) \gamma_0(2k{+}2,0|\tau) \, , \notag
\end{align}
see appendix B of the reference for further examples.
Note that eMZVs of length one are constant ($\omega(2k|\tau)=-2\zeta_{2k}$ and $\omega(2k{+}1|\tau)=0$
for $k\in \NN_0$), and eMZVs of length $r$ yield iterated Eisenstein integrals of length $\leq r{-}1$.


\subsection{Derivations}
\label{sec:2.5}

Not all the iterated Eisenstein integrals $\gamma(k_1,k_2,\ldots,k_r|\tau)$ are realized in the decomposition
of $A$-cycle eMZVs. The selection rules on whether
a given combination of $\gamma(k_1,k_2,\ldots,k_r|\tau)$ descends from eMZVs are encoded in a family
of derivations $\ep_{2m}, \ m\geq 0$ firstly studied by Tsunogai \cite{Tsunogai}. These derivations appear
in the differential equation of the elliptic KZB associator whose $A$-cycle
component $A(x,y,\tau)$ satisfies \cite{KZB, EnriquezEllAss, Hain}
\begin{align}
2\pi i \partial_\tau A(x,y,\tau) &= \sum_{m=0}^{\infty} (1{-}2m){\rm G}_{2m}(\tau)\ep_{2m}  A(x,y,\tau) \, .
\label{DER1}
\end{align}
The derivations $\ep_{2m}$ act on two non-commutative variables $x,y$ via ($m>0$)
\begin{align}
  \ep_{2m}(x)&=(\te{ad}_x)^{2m}(y) \label{howep} \\
  \ep_{2m}(y)&=[y,(\te{ad}_x)^{2m-1}(y)]
  +\sum_{j=1}^{m-1} (-1)^j[(\te{ad}_x)^j(y),(\te{ad}_x)^{2m-1-j}(y)] \, , \notag
\end{align}
where $\te{ad}_x(y)\equiv [x,y]$, and $\ep_0$ acts via
\beq
 \ep_{0}(x)  = y \, , \ \ \ \ \ \   \ep_0(y)=0 \, . \label{howep0}
\eeq
These definitions imply a variety of relations in the derivation algebra \cite{LNT, Pollack, Broedel:2015hia},
starting with the fact that $\ep_2$ is a central element,
\beq
[\ep_{2m},\ep_2]=0  \,, \ \ \ \ \ \  m\geq 0 \ .
\label{poll1}
\eeq
Furthermore, representation theory of ${\rm SL}_2$ implies that $[ \ep_0, [\ep_0, [ \ep_0,\ep_4]]] = 0$
and more generally
\beq
({\rm ad}_{\ep_0})^{2m-1}(\ep_{2m}) = 0\, , \ \ \ \ \ \ m>0 \, .
\label{DER3}
\eeq
Moreover, irreducible relations\footnote{A relation is said to be irreducible if it cannot be written as
${\rm ad}_{\ep_{n_1}}  {\rm ad}_{\ep_{n_2}}  \ldots {\rm ad}_{\ep_{n_r}}  N(\ep) =0$ with $r>0$
and $N(\ep)$ denoting some vanishing expression built from (possibly nested) commutators
of $\ep_{2m}$.} at various depths are related to cusp forms \cite{Pollack}, starting with
\begin{align}
0 &=[\ep_{10},\ep_4]-3[\ep_{8},\ep_6] \notag \\
0&=2 [\ep_{14},\ep_4] - 7[\ep_{12},\ep_6] + 11 [\ep_{10},\ep_8]   \label{DER4} \\
0&=80[\ep_{12},[\ep_4,\ep_{0}]] + 16 [\ep_4,[\ep_{12},\ep_0]] - 250 [\ep_{10},[\ep_6,\ep_0]] \notag \\
& \ \ \ \  - 125 [\ep_6,[\ep_{10},\ep_0]] + 280 [\ep_8,[\ep_8,\ep_0]]- 462 [\ep_4,[\ep_4,\ep_8]] - 1725 [\ep_6,[\ep_6,\ep_4]] \, ,
\notag
\end{align}
and further examples can for instance be found in \cite{Pollack, Broedel:2015hia} or downloaded
from \cite{WWWe}. As detailed in section 4 of \cite{Broedel:2015hia}, the vanishing of
$[\ep_{10},\ep_4]-3[\ep_{8},\ep_6]$ for instance constrains the relative factors of
$\gamma(4,10|\tau),\gamma(10,4|\tau),\gamma(6,8|\tau)$ and $\gamma(8,6|\tau)$
in the decomposition of eMZVs.

At each multiplicity $n\geq2$, the $A$-cycle integrals (\ref{intro2}) will induce conjectural $(n{-}1)!\times (n{-}1)!$
matrix representations $r_{\vec{\eta}}(\ep_{2m})$ of the above derivations that preserve
all their commutator relations including (\ref{poll1}) to (\ref{DER4}), see section \ref{sec:4.5}.
We will later on spell out the $(n{\leq}4)$-examples of $r_{\vec{\eta}}(\ep_{2m})$ from the differential
equation (\ref{intro3}) of the $A$-cycle integrals by isolating the coefficient of ${\rm G}_{2m}(\tau)$
in the differential operator, cf.\ (\ref{intro3b}), and the all-multiplicity generalization is generated
by (\ref{B460a}).


\section{Two-point warm-up}
\label{sec:3}

The purpose of this section is to illustrate the main results in a two-point context,
where the basis of integrands for the $Z^\tau_{\vec{\eta}}(\cdot|\cdot)$ in (\ref{intro2}) is one-dimensional.
Once we fix translation variance by setting $z_1=0$, the planar and non-planar arrangement
of the two cylinder punctures in figure \ref{basiccyl} amount to integration cycles
\beq
{\cal C}(1,2) = \big\{ z_2\in (0,1) \big\} \, , \ \ \  \ \ \
{\cal C}(\begin{smallmatrix} 2 \\ 1 \end{smallmatrix} ) = \big\{ z_2 = \tauh + v_2 , \ v_2 \in (0,1) \big\}\,,
\label{2cycl}
\eeq
cf.\ section \ref{sec:2.3}. The two-point instances of (\ref{intro2}) are then given by
\begin{align}
Z^\tau_{\eta_2}(1,2|1,2) &=
\int_{{\cal C}(1,2)} \dd z_2 \,\Omega(z_{12},\eta_2,\tau) \,e^{s_{12}{\cal G}(z_{12},\tau)}
= \int_0^1 \dd v_2 \,\Omega(v_{12},\eta_2,\tau) \,e^{s_{12}{\cal G}(v_{12},\tau)}  \label{B11} \\
Z^\tau_{\eta_2}\big( \begin{smallmatrix} 2 \\ 1 \end{smallmatrix}|1,2\big) &=
\int_{{\cal C}\big(\begin{smallmatrix} 2 \\ 1 \end{smallmatrix} \big)} \dd z_2 \,\Omega(z_{12},\eta_2,\tau) \,e^{s_{12}{\cal G}(z_{12},\tau)}
= \int_0^1 \dd v_2 \,\Omega(v_{12}{+}\tauh,\eta_2,\tau) \,e^{s_{12}{\cal G}(v_{12}{+}\tauh,\tau)} \, ,
\notag
\end{align}
see (\ref{1.2}) for the Kronecker--Eisenstein series $\Omega(\ldots)$ as well as (\ref{B1}) and (\ref{B2})
for the planar and non-planar Green functions ${\cal G}(\ldots)$, respectively. By the expansion methods
of \cite{Broedel:2014vla} and \cite{Broedel:2017jdo}, the $\tau$-dependence at any $\ap$-order of
$Z^\tau_{\eta_2}(1,2|1,2)$ and $q^{-s_{12}/8}Z^\tau_{\eta_2}\big( \begin{smallmatrix} 2 \\ 1 \end{smallmatrix}|1,2\big) $
is carried by eMZVs and twisted eMZVs with Fourier expansions (\ref{A2emzvq}) and (\ref{A2temzvq}), respectively.
As we will see, the half-odd integer powers $q^{m}, \ m \in \NN{-}\frac{1}{2}$ in (\ref{A2temzvq}) drop out.


\subsection{The $\tau$-derivative}
\label{sec:3.1}

Given that the derivatives (\ref{B5}) of the Green functions w.r.t.\ $v_j$ and $\tau$ take a universal
form for the planar and non-planar case, we will evaluate the $\tau$-derivative of both integrals in (\ref{B11}) for
an unspecified integration cycle ${\cal C}(\ast)$,
\begin{align}
2\pi i \partial_\tau Z^\tau_{\eta_2}(\ast|1,2)&= 2\pi i  \!    \int \limits_{{\cal C}(\ast)} \!  \dd z_2 \,\Big( \partial_\tau \Omega(z_{12},\eta_2,\tau) \,e^{s_{12}{\cal G}(z_{12},\tau)}+ \Omega(z_{12},\eta_2,\tau)  \partial_\tau  \,e^{s_{12}{\cal G}(z_{12},\tau)} \Big) \notag
\\
&=  \!    \int \limits_{{\cal C}(\ast)} \!  \dd z_2 \,\Big( \partial_z \partial_{\eta_2} \Omega(z_{12},\eta_2,\tau)
- s_{12} (f^{(2)}_{12} {+} 2 \zeta_2) \Omega(z_{12},\eta_2,\tau)   \Big)\,e^{s_{12}{\cal G}(z_{12},\tau)}
\label{B12} \\
&= s_{12}  \!    \int \limits_{{\cal C}(\ast)} \!  \dd z_2 \,\Big( f^{(1)}_{12} \partial_{\eta_2} \Omega(z_{12},\eta_2,\tau)
-  (f^{(2)}_{12} + 2 \zeta_2) \Omega(z_{12},\eta_2,\tau)   \Big) \,e^{s_{12}{\cal G}(z_{12},\tau)} \, . \notag
\end{align}
We have used (\ref{Aheat}) and (\ref{B5}) in passing to the second line and integrated
the $z$-derivative of $\partial_{\eta_2} \Omega(z_{12},\eta_2,\tau) $ by parts in the last step,
see (\ref{shfij}) for the shorthand $f_{ij}^{(k)}$. Throughout this work, boundary terms w.r.t.\ the punctures
are discarded since the Koba--Nielsen factor always leads to a suppression of the form\footnote{In order to see
this, one first exploits the local behavior ${\cal G}(z,\tau) \rightarrow - \log|z|$ of the Green function around
the origin to write $\lim_{z_{ij}\rightarrow 0 }e^{s_{ij}{\cal G}(z_{ij},\tau)} =  \lim_{z_{ij}\rightarrow 0 }e^{-s_{ij} \log|z_{ij}|}
=\lim_{z_{ij}\rightarrow 0 } |z_{ij}|^{-s_{ij}}$. Then, the vanishing of $\lim_{z_{ij}\rightarrow 0 } |z_{ij}|^{-s_{ij}}$ is
obvious in case of $\Re(s_{ij})<0$ and otherwise follows from analytic continuation.}
$\lim_{z_{ij}\rightarrow 0 }e^{s_{ij}{\cal G}(z_{ij},\tau)} =0$ or $\lim_{z_{ij}\rightarrow 1 }e^{s_{ij}{\cal G}(z_{ij},\tau)} =0$.

In order to relate the right-hand side of (\ref{B12})
to the original integral $Z^\tau_{\eta_2}(\ast|1,2)$, we
rewrite the combination $ f^{(1)}_{12} \partial_{\eta_2}\Omega(z_{12},\eta_2,\tau)
 - f^{(2)}_{12} \Omega(z_{12},\eta_2,\tau)$ using
\beq
f^{(1)}_{ij} = - \Omega(z_{ji},\xi,\tau) \, \big|_{\xi^0} \, , \ \ \ \ \ \
f^{(2)}_{ij} = \partial_\xi \Omega(z_{ji},\xi,\tau) \, \big|_{\xi^0} \, .
\label{B14}
\eeq
The notation $\big|_{\xi^0}$ instructs to pick up the zero$^{\rm th}$ order in the Laurent expansion w.r.t.\
an auxiliary variable $\xi \in \CC$. Moreover, we will need the key identity
\cite{BrownLev,Enriquez:Emzv}
\beq
(\partial_{\eta_2}+\partial_{\xi})
\Omega(z_{12},\eta_2,\tau)  \Omega(z_{21},\xi,\tau) = \big(\wp(\eta_2,\tau) - \wp(\xi,\tau) \big)
\Omega(z_{12},\eta_2-\xi,\tau)
\label{A25DP}
\eeq
whose derivation is reviewed in appendix \ref{app:0.1}. The Weierstrass functions on the right-hand side
generate holomorphic Eisenstein series (\ref{A14}) upon Laurent expansion in $\eta$,
\beq
\wp(\eta,\tau) =  - \partial^2_\eta \log \theta_1(\eta,\tau) - {\rm G}_2(\tau)
 = - \frac{ {\rm G}_0 }{\eta^2} + \sum_{m=2}^{\infty} (2m{-}1) \eta^{2m-2} {\rm G}_{2m}(\tau) \, .
 \label{intro3a}
 \eeq
Based on (\ref{B14}) and (\ref{A25DP}), the last line of (\ref{B12}) can be rewritten as
\begin{align}
&f^{(1)}_{12} \partial_{\eta_2}\Omega(z_{12},\eta_2,\tau)  {-} f^{(2)}_{12} \Omega(z_{12},\eta_2,\tau)
\notag \\
& \ \  = - \Omega(z_{21},\xi,\tau)  \partial_{\eta_2}\Omega(z_{12},\eta_2,\tau)
 {-}  \Omega(z_{12},\eta_2,\tau) \partial_\xi \Omega(z_{21},\xi,\tau) \, \big|_{\xi^0}  \notag \\
&\ \ = - (\partial_{\eta_2} +\partial_\xi)\Omega(z_{12},\eta_2,\tau)  \Omega(z_{21},\xi,\tau) \, \big|_{\xi^0}  \label{B15} \\
&\ \  =  \big(\wp(\xi,\tau) - \wp(\eta_2,\tau) \big) \Omega(z_{12},\eta_2-\xi,\tau) \, \big|_{\xi^0} \notag \\
&\ \ = \Big( \frac{1}{2} \partial_{\eta_2}^2 - \wp(\eta_2,\tau)  \Big) \Omega(z_{12},\eta_2,\tau) \ .
\notag
 \end{align}
The derivative  $\frac{1}{2} \partial_{\eta_2}^2$ in the
last line stems from the interplay of the double pole $\wp(\xi,\tau)=\frac{1}{\xi^2}+{\cal O}(\xi^2)$
with the Taylor expansion of $\Omega(z_{12},\eta_2-\xi,\tau)$ around $\xi=0$. At fixed order in $\eta_2$, one can
extract relations such as $f^{(2)}_{12} f^{(2)}_{12}  - 2 f^{(3)}_{12} f^{(1)}_{12} = 3 {\rm G}_4 - 2 f^{(4)}_{12} $
from (\ref{B15}), see (\ref{B13}) for further examples and (\ref{B13gen}) for a general formula. More
importantly, (\ref{B15}) allows to rewrite the $\tau$-derivative of the two-point integrals (\ref{B11}) as
\begin{align}
2\pi i \partial_\tau Z^\tau_{\eta_2}(\ast|1,2)&= s_{12} \int_{{\cal C}(\ast)} \dd z_2 \,  \Big( \frac{1}{2} \partial_{\eta_2}^2 - \wp(\eta_2,\tau)  - 2 \zeta_2 \Big) \Omega(z_{12},\eta_2,\tau)  \, e^{s_{12}{\cal G}(z_{12},\tau)}  \, ,  \label{B16}
\end{align}
i.e.\ in terms of the original integral
\begin{align}
2\pi i \partial_\tau Z^\tau_{\eta_2}(\ast|1,2)&= s_{12}  \Big( \frac{1}{2} \partial_{\eta_2}^2 - \wp(\eta_2,\tau)  - 2 \zeta_2 \Big) Z^\tau_{\eta_2}(\ast|1,2)  \, .  \label{B16alt}
\end{align}
As we will see, this linear and homogeneous first-order differential equation in $\tau$ has
a variety of structural implications and can be applied to efficiently perform $\ap$-expansions.


\subsection{Derivations and iterated Eisenstein integrals}
\label{sec:3.2}

The upshot (\ref{B16alt}) of the previous section is the two-point instance of our
central result (\ref{intro3}): The $A$-cycle integrals in (\ref{B11})
close under $\tau$-derivatives, and the differential operator in
\beq
2\pi i \partial_\tau Z^\tau_{\eta_2}(\ast|1,2) = D^\tau_{\eta_2} Z^\tau_{\eta_2}(\ast|1,2)\, , \ \ \ \ \ \
D^\tau_{\eta_2} = s_{12}  \Big( \frac{1}{2} \partial_{\eta_2}^2 - \wp(\eta_2,\tau)  - 2 \zeta_2 \Big)
\label{2ptA}
\eeq
takes a universal form for the planar and non-planar integration cycles ${\cal C}(\ast)$ in (\ref{2cycl}).
Once we expand the Weierstrass function via (\ref{intro3a}), this can be lined up with the differential
equation (\ref{DER1}) of the KZB associator,
\beq
D^\tau_{\eta_2} = \sum_{m=0}^{\infty} (1{-}2m) {\rm G}_{2m}(\tau) r_{\eta_2}(\ep_{2m})\, ,
 \label{2ptB}
\eeq
and one can read off a scalar representation $r_{\eta_2}(\cdot)$ of the derivations reviewed in section \ref{sec:2.5},
\begin{align}
 r_{\eta_2}(\ep_{0})  &= s_{12} \Big( \frac{ 1}{\eta_2^{2}} +  2 \zeta_2 - \frac{1}{2} \partial_{\eta_2}^2\Big) \, , \ \ \ \ \ \
  r_{\eta_2}(\ep_{2})  = 0 \notag \\
  r_{\eta_2}(\ep_{2m})  &= s_{12} \eta_2^{2m-2} \, , \ \ \ \ \ \ m>1\, .
 \label{2ptC}
\end{align}
The differential equation (\ref{2ptA}) can be solved via standard Picard iteration once an initial value
at some reference value $\tau_0$ is available (which is chosen as $\tau_0 \rightarrow i\infty$ for
convenience),
\begin{align}
Z^\tau_{\eta_2}(\ast|1,2) &= \Big(1 + \frac{1}{2\pi i } \int^\tau_{i\infty} \dd \tau_1 \,D^{\tau_1}_{\eta_2}
+ \frac{1}{(2\pi i)^2 } \int^\tau_{i\infty} \dd \tau_1 \,D^{\tau_1}_{\eta_2} \int^{\tau_1}_{i\infty} \dd \tau_2 \,D^{\tau_2}_{\eta_2} + \ldots \Big)
Z^{i \infty}_{\eta_2}(\ast|1,2) \notag \\
&=  \sum_{n=0}^\infty  \frac{1}{(2\pi i)^{2n} }    \! \! \! \!  \! \! \! \!  \int \limits_{0<q_1<q_2<\ldots <q_n <q}    \! \! \! \!\! \! \! \! \frac{ \dd q_1 }{q_1} \,  \frac{ \dd q_2 }{q_2} \ldots  \frac{ \dd q_n }{q_n} \, D^{\tau_n}_{\eta_2}  \ldots  D^{\tau_2}_{\eta_2}  D^{\tau_1}_{\eta_2}  Z^{i \infty}_{\eta_2}(\ast|1,2) \, .
\label{preB3}
\end{align}
By the term $\sim \partial_{\eta_2}^2$ in (\ref{2ptA}), $D^\tau_{\eta_2}$ is a differential operator in $\eta_2$, so
$D^{\tau_i}_{\eta_2}$ and $D^{\tau_j}_{\eta_2}$ do not commute at different $\tau_i\neq \tau_j$ in
(\ref{preB3}). Instead, when all the $D^{\tau_i}_{\eta_2}$ on the right-hand side of (\ref{preB3}) are expanded
in terms of Eisenstein series via (\ref{2ptB}), each term can be identified as an iterated Eisenstein integral (\ref{preB5}),
\begin{align}
Z^\tau_{\eta_2}(\ast|1,2)&= \sum_{r=0}^{\infty}  \sum_{k_1,k_2,\ldots,k_r \atop{=0,4,6,8,\ldots} }  \prod_{j=1}^r (k_j{-}1)\, \gamma(k_1,k_2,\ldots,k_r|\tau)  r_{\eta_2}(\ep_{k_r} \ldots \ep_{k_2} \ep_{k_1} )
Z^{i\infty}_{\eta_2}(\ast|1,2) \, ,
  \label{intro12}
\end{align}
where $r_{\eta_2}(\ep_{k_i} \ep_{k_j} ) \equiv r_{\eta_2}(\ep_{k_i} )r_{\eta_2}(\ep_{k_j} )$.
This is the two-point instance of the general result (\ref{intro1}), where the derivations appear in a scalar
representation (as opposed to $(n{-}1)! \times (n{-}1)!$ matrices at $n$ points). Since each $r_{\eta_2}(\ep_k)$ in
(\ref{2ptC}) is linear in $s_{12}$, the summation variable $r$ in (\ref{intro12}) counts the powers of $\ap$ introduced
by the derivations. Note, however, that $Z^{i\infty}_{\eta_2}(\ast|1,2)$ may by itself depend on $\ap$, see the
discussion in the next sections.

Although the representations of $\ep_{k}$ are not yet matrix-valued in the two-point case, the
term $\sim \partial_{\eta_2}^2$ in $r_{\eta_2}(\ep_{0})$ prevents it from commuting. Accordingly, it is
nontrivial to check that relations (\ref{DER3}) are preserved by $r_{\eta_2}(\cdot)$,
e.g.\ that $[r_{\eta_2}( \ep_0), [r_{\eta_2}(\ep_0), [r_{\eta_2}( \ep_0),r_{\eta_2}(\ep_4)]]] {\cal F}(\eta_2)= 0$
for an arbitrary function ${\cal F}$ of $\eta_2$. However, since all the $r_{\eta_2}(\ep_{k})$ at positive $k>0$
are multiplicative, commutators $[r_{\eta_2}(\ep_{k_1}),r_{\eta_2}(\ep_{k_2})]$ with $k_1,k_2>0$ always vanish,
and relations such as $[r_{\eta_2}(\ep_{10}),r_{\eta_2}(\ep_4)]-3[ r_{\eta_2}(\ep_{8}),r_{\eta_2}(\ep_6)]=0$ (see
(\ref{DER4})) hold trivially. And the vanishing of $r_{\eta_2}( \ep_2)$ is of course compatible
with $\ep_2$ being a central element, cf.\ (\ref{poll1}).

As a consequence of $[r_{\eta_2}(\ep_{k_1}),r_{\eta_2}(\ep_{k_2})] = 0 \
\forall \ k_1,k_2>0$, iterated Eisenstein integrals of depth $\geq 2$ appear
in a constrained manner in (\ref{intro12}). For instance,
$\gamma(k_1,k_2|\tau)$ and $\gamma(k_2,k_1|\tau)$ with $k_1,k_2>0$ will always
have the same coefficient and conspire to the shuffle product
$\gamma(k_1,k_2|\tau)+\gamma(k_2,k_1|\tau) = \gamma(k_1|\tau) \cdot
\gamma(k_2|\tau)$. This is a genus-one analogue of the dropout of
depth-$(d\geq 2)$ MZVs from four-point disk integrals: The $\ap$-expansion of
the latter is still expressible via products of $\zeta_m$ whereas $(n\geq
5)$-point disk integrals involve irreducible higher-depth MZVs
\cite{Schlotterer:2012ny}.

On the other hand, since $[r_{\eta_2}(\ep_{k}),r_{\eta_2}(\ep_{0})] \neq 0 \
\forall \ k>2$, not all the iterated Eisenstein integrals at depth $\geq 2$ in
(\ref{intro12}) can be written as shuffles of depth-one instances: Cases like
$\gamma(4,4,0,0|\tau),\gamma(4,0,4,0|\tau)$ and $\gamma(4,0,0,4|\tau)$ with
entries $k_j=0$ appear in combinations that cannot be reduced to
$\gamma(4,0|\tau)^2$ and $\gamma(4|\tau)\cdot \gamma(4,0,0|\tau)$. Hence, in
spite of the vanishing commutators among the $r_{\eta_2}(\ep_{k>0})$, the
$\ap$-expansion of the two-point $A$-cycle integrals goes beyond the
coefficients of the meta-abelian quotient of the KZB associator \cite{metaab}.


\subsection{The initial value at the cusp: degenerating the integrand}
\label{sec:3.3}

\begin{figure}
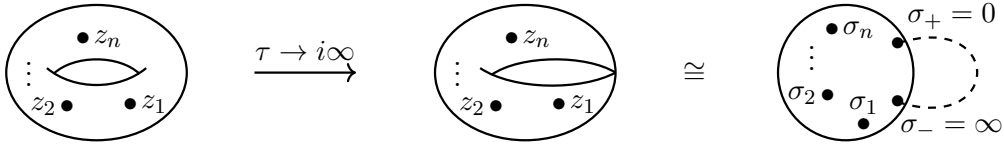

\begin{center}
\tikzpicture[scale=0.3,line width=0.30mm]
\draw(0,0) ellipse  (4cm and 3cm);
\draw(-2.2,0.2) .. controls (-1,-0.8) and (1,-0.8) .. (2.2,0.2);
\draw(-1.9,-0.05) .. controls (-1,0.8) and (1,0.8) .. (1.9,-0.05);
\draw(1.5,-1.4)node{$\bullet$} node[right]{$z_1$};
\draw(-1.3,-1.5)node{$\bullet$} node[left]{$z_2$};
\draw(-0.6,1.5)node{$\bullet$} node[right]{$z_n$};
\draw(-3.0,0.3)node{$\vdots$};
\draw[->](7,0) -- (11.5,0);
\draw(9.25,0.98)node{$\tau\rightarrow i \infty$};
\scope[xshift=19cm]
\draw(0,0) ellipse  (4cm and 3cm);
\draw(-2,0.2) .. controls (-0.5,-0.8) and (2.5,-0.8) .. (4,0);
\draw(-1.5,-0.10) .. controls (-0.2,0.8) and (2.7,0.8) .. (4,0);
\draw(1.5,-1.4)node{$\bullet$} node[right]{$z_1$};
\draw(-1.3,-1.5)node{$\bullet$} node[left]{$z_2$};
\draw(-0.6,1.5)node{$\bullet$} node[right]{$z_n$};
\draw(-3.0,0.3)node{$\vdots$};
\endscope
\draw(26.5,0)node{$\cong$};
\scope[xshift=33.2cm]
\draw(0,0)circle(3cm);
\draw(0.8,-2.3)node{$\bullet$} node[above]{$\sigma_1$};
\draw(-0.8,-1)node{$\bullet$} node[left]{$\sigma_2$};
\draw(-0.6,1.9)node{$\bullet$} node[right]{$\sigma_n$};
\draw(2.3,1.3)node{$\bullet$};
\draw(2.3,-1.3)node{$\bullet$};
\draw(4.7,2.5)node{$\sigma_+=0$};
\draw(4.7,-2.5)node{$\sigma_-=\infty$};
\draw(-1.5,0.3)node[rotate=180]{$\vdots$};
\draw[dashed](2.3,1.3) .. controls (7,2.8) and (7,-2.8) .. (2.3,-1.3);
\endscope
\endtikzpicture
\end{center}
\caption{The degeneration of the torus at $\tau \rightarrow i \infty$ pinches the $A$-cycle and
yields the topology of a Riemann sphere. In particular, the pinched $A$-cycle introduces a pair
of identified punctures $\sigma_{+}=0$ and $\sigma_{-}=\infty$ on the Riemann sphere.}
\label{nodsph}
\end{figure}

Given that the pattern of iterated Eisenstein integrals in the two-point integrals is fixed by (\ref{intro12}),
the next step is to find the explicit form of their initial value $Z^{i\infty}_{\eta_2}(\ast|1,2)$, i.e.\ the
degeneration of (\ref{B11}) at the cusp $\tau \rightarrow i \infty$. In this limit, both the Green function
and the Kronecker--Eisenstein series in the integrand are most conveniently described in the variables
\beq
\sigma_j= e^{2\pi i z_j} \ , \ \ \ \ \ \ \dd z_j = \frac{ \dd \sigma_j }{2\pi i \, \sigma_j} \, .
\label{G3}
\eeq
This parameterization of the punctures is tailored to the nodal Riemann sphere which arises from the degeneration
of a torus as visualized in figure \ref{nodsph}. The functions of $\sigma_j$ we will encounter can be interpreted as
${\rm SL}_{2}$-fixed expressions on a sphere, where the identified punctures from the pinching of the $A$-cycle in
figure \ref{nodsph} are $\sigma_+=0$ and $\sigma_-=\infty$, and the choice $z_1=0$ is mapped
to $\sigma_1=1$ under (\ref{G3}). The degeneration of
the Kronecker--Eisenstein series can be assembled from the $\tau \rightarrow i\infty$ behavior of the
functions $f^{(k)}(z,\tau)$ and $f^{(k)}(z{-}\tauh,\tau)$ studied in section 3.2.1 of \cite{Broedel:2017jdo}.
In the planar and non-planar cases with $v_{ij} \in \RR$, the results of the reference imply
\beq
\lim_{\tau \rightarrow i\infty}  \Omega(v_{ij},\eta,\tau) = \pi \cot(\pi \eta) + G_{ij}  \, , \ \ \ \ \ \
\lim_{\tau \rightarrow i\infty}  \Omega(v_{ij}+\tauh,\eta,\tau) = \frac{ \pi }{ \sin(\pi \eta) }
  \, .
\label{Gpl}
\eeq
Remarkably, the Kronecker--Eisenstein series $\Omega(v_{ij}{+}\tauh,\eta,\tau)$ on the non-planar
cycle no longer depends on the punctures at the cusp. In the planar case,
the only dependence of $ \Omega(v_{ij},\eta,\tau \rightarrow i \infty)$ on $\sigma_i$ and $\sigma_j$
occurs at the zero$^{\rm th}$ order in $\eta$ through the Green function on the nodal Riemann sphere\footnote{Note
the different fonts used for the letter 'G' in the open-string Green function ${\cal G}(z,\tau)$, the holomorphic
Eisenstein series ${\rm G}_k(\tau)$ and the Green function $G_{ij}$ on the nodal sphere in (\ref{shGij}).},
\beq
G_{ij}  \equiv 2\pi i \frac{\sigma_i+\sigma_j}{2(\sigma_i-\sigma_j)}  \, .
\label{shGij}
\eeq
The trigonometric functions of $\eta$ in (\ref{Gpl}) have the straightforward Laurent expansion
\begin{align}
 \pi \cot(\pi \eta) &=  \frac{1}{\eta} - 2 \zeta_2 \eta - 2 \zeta_4 \eta^3 - 2 \zeta_6 \eta^5 -\ldots \ \; \! =
 \frac{1}{\eta} - 2 \sum_{k=1}^{\infty} \zeta_{2k} \eta^{2k-1} \label{trigdeg} \\
 \frac{ \pi }{\sin(\pi \eta)} &=  \frac{1}{\eta} + \zeta_2 \eta + \frac{ 7}{4} \zeta_4 \eta^3 + \frac{31}{16} \zeta_6 \eta^5 +\ldots =
 \frac{1}{\eta} +  \sum_{k=1}^{\infty} \frac{2^{2k-1} {-}1 }{2^{2k-2}} \zeta_{2k}  \eta^{2k-1}\,. \notag
\end{align}
The degeneration of the planar Green function can for instance be determined by inserting the identity
$\lim_{\tau \rightarrow i \infty} \dd z' \,  f^{(1)}(z'-z_j,\tau) = \frac{ \dd \sigma' }{\sigma' - \sigma_j} -
\frac{ \dd \sigma' }{2\sigma'}$ at real values of $z'$ and $z_j$ into (\ref{B1}),
\beq
\lim_{\tau \rightarrow i\infty}  {\cal G}(v_{ij},\tau) = \frac{1}{2} \log (\sigma_i)
+  \frac{1}{2} \log (\sigma_j) -   \log (\sigma_j {-} \sigma_i) \, , \ \ \ \ \ \ v_i<v_j \, ,
\label{Gplother}
\eeq
where cases with $v_j<v_i$ give rise to $-\log (\sigma_i {-} \sigma_j)$ in the place of $-\log (\sigma_j {-} \sigma_i)$.
For the non-planar Green function (\ref{B2}), the reasoning in appendix \ref{app:C} yields the
regularized value
\beq
\lim_{\tau \rightarrow i\infty} {\cal G}(v_{ij}{+}\tauh,\tau)  = 0 \, .
\label{Gnpnew}
\eeq
In summary, the results of this subsection pinpoint the following degeneration of the
integrands of the $Z^\tau_{\eta_2}(\ast|1,2)$ in (\ref{B11}),
\begin{align}
\lim_{\tau \rightarrow i\infty}   \Omega(v_{12},\eta_2,\tau) \,e^{s_{12}{\cal G}(v_{12},\tau)}
&= ( \pi \cot(\pi \eta_2) + G_{12}  \big) \sigma_2^{s_{12}/2} (1-\sigma_2)^{-s_{12}}
\label{someintA}
\\
\lim_{\tau \rightarrow i\infty}   \Omega(v_{12}{+}\tauh,\eta_2,\tau) \,e^{s_{12}{\cal G}(v_{12}{+}\tauh,\tau)} &= \frac{ \pi }{\sin(\pi \eta_2)} \, .
\label{someintB}
\end{align}
In the non-planar case, the integrand at $\tau \rightarrow i \infty$ no longer depends on the puncture,
so one can immediately perform the integral $\int_{{\cal C}\big(\begin{smallmatrix} 2 \\ 1 \end{smallmatrix} \big)} \dd z_2=1$
and obtain the degeneration
\beq
Z^{i\infty}_{\eta_2}\big( \begin{smallmatrix} 2 \\ 1 \end{smallmatrix}|1,2\big) = \frac{ \pi }{\sin(\pi \eta_2)} \, .
\label{someintD}
\eeq
Upon insertion into the solution (\ref{intro12}) to the differential equation in $\tau$,
this completes the result for the $\ap$-expansion of the non-planar $A$-cycle integral,
\begin{align}
Z^\tau_{\eta_2}( \begin{smallmatrix} 2 \\ 1 \end{smallmatrix} |1,2)&= \sum_{r=0}^{\infty}  \sum_{k_1,k_2,\ldots,k_r \atop{=0,4,6,8,\ldots} }  \prod_{j=1}^r (k_j{-}1)\, \gamma(k_1,k_2,\ldots,k_r|\tau)  r_{\eta_2}(\ep_{k_r} \ldots \ep_{k_2} \ep_{k_1} )  \frac{ \pi }{\sin(\pi \eta_2)}\, .
\label{someintE}
 \end{align}
Since $ \frac{ \pi }{\sin(\pi \eta_2)}$ does not depend on $\ap$ and $r_{\eta_2}(\ep_{k})$ are linear in $s_{12}$, the
order of $(\ap)^r$ in (\ref{someintE}) exclusively features iterated Eisenstein integrals of length $r$. As will
be detailed in section \ref{sec:3.5}, the integral over a specific $f^{(k)}(v{-}\tauh,\tau)$ rather than
$\Omega(v{-}\tauh,\eta,\tau)$ can be obtained by expanding $\frac{ \pi }{\sin(\pi \eta_2)}$ as in (\ref{trigdeg}) and
isolating the desired power of $\eta_2$ after summing over the words in $r_{\eta_2}(\ep_{k})$. The analogous
statements in the planar case will be derived in the next section -- the dependence of (\ref{someintA}) on $\sigma_2$
will require extra care.

It is also worth highlighting that (\ref{someintE}) manifests the absence of
twisted eMZVs which are naively expected when the $\ap$-expansion of $\int^1_0
\dd v\, f^{(k)}(v{-}\tauh,\tau) e^{s_{12} {\cal G}(v-\tauh,\tau)}$ is performed by the
methods of \cite{Broedel:2017jdo}. Hence, half-odd integer powers $q^{m}, \ m
\in \NN{-}\frac{1}{2}$ are guaranteed to drop out at any $\ap$-order of
$q^{-s_{12}/8}Z^\tau_{\eta_2}( \begin{smallmatrix} 2 \\ 1 \end{smallmatrix}
|1,2)$, where the factor of $q^{-s_{12}/8}$ eliminates the contribution
$\frac{i\pi \tau}{4}$ to the non-planar Green function.


\subsection{The initial value at the cusp: deforming the integration contour}
\label{sec:3.4}

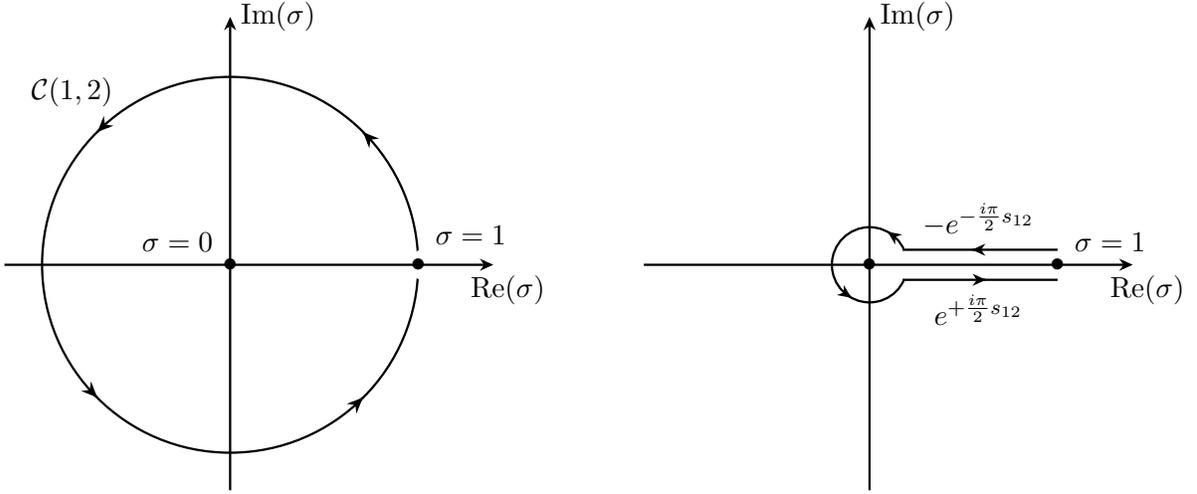
\begin{figure}
\begin{center}
\begin{tikzpicture}[line width=0.30mm]
\draw (2.5,0) arc (0:360:2.5cm);
\draw[white, fill=white] (2.6,0) circle (0.2cm);
\draw[arrows={-Stealth[width=1.8mm, length=2.1mm]}](1.767,1.765)--(1.766,1.766);
\draw[arrows={-Stealth[width=1.8mm, length=2.1mm]}](-1.767,-1.765)--(-1.766,-1.766);
\draw[arrows={-Stealth[width=1.8mm, length=2.1mm]}](-1.765,1.767)--(-1.766,1.766);
\draw[arrows={-Stealth[width=1.8mm, length=2.1mm]}](1.765,-1.767)--(1.766,-1.766);
\draw (0,0)node{$\bullet$} ;
\draw(-0.7,0.3)node{$\sigma=0$};
\draw (2.5,0)node{$\bullet$} ;
\draw(3.2,0.4)node{$\sigma=1$};
\draw [arrows={-Stealth[width=1.6mm, length=1.8mm]}] (-3,0) -- (3.5,0) node[below]{$\quad\te{Re}(\sigma)$};
\draw [arrows={-Stealth[width=1.6mm, length=1.8mm]}] (0,-3) -- (0,3.3) node[right]{$\te{Im}(\sigma)$};
\draw(-2.1,2.3)node{${\cal C}(1,2)$};
\begin{scope}[xshift=8.5cm]
  \draw[fill=white] (0,0) circle (0.5cm);
  \draw[white, fill=white] (0.5,0) circle (0.2cm);
  \draw(2.5,0.2) -- (0.44,0.2);
  \draw(2.5,-0.2) -- (0.44,-0.2);
  \draw[arrows={-Stealth[width=1.6mm, length=1.8mm]}](1.37,0.2)--(1.36,0.2);
  \draw[arrows={-Stealth[width=1.6mm, length=1.8mm]}](1.54,-0.2)--(1.55,-0.2);
  \draw[arrows={-Stealth[width=1.6mm, length=1.8mm]}](0.25,0.433)--(0.24,0.44);
  \draw[arrows={-Stealth[width=1.6mm, length=1.8mm]}](-0.25,-0.433)--(-0.24,-0.44);
  \draw (0,0)node{$\bullet$} ;
  \draw (2.5,0)node{$\bullet$} ;
  \draw(3.2,0.3)node{$\sigma=1$};
  \draw [arrows={-Stealth[width=1.6mm, length=1.8mm]}] (-3,0) -- (3.5,0) node[below]{$\quad\te{Re}(\sigma)$};
  \draw [arrows={-Stealth[width=1.6mm, length=1.8mm]}] (0,-3) -- (0,3.3) node[right]{$\te{Im}(\sigma)$};
  \draw(1.45,0.6)node{$-e^{-\frac{i \pi }{2} s_{12}}$};
  \draw(1.45,-0.6)node{$e^{+\frac{i \pi }{2} s_{12}}$};
  \end{scope}
\end{tikzpicture}
\end{center}
\caption{As depicted in the left panel, the integration contour ${\cal C}(1,2)$ in the $\sigma_2 =e^{2\pi i z_2}$ variable
is the unit circle $|\sigma_2|=1$. When replacing ${\cal C}(1,2)$ by the homotopy-equivalent combination
of paths visualized the right panel, the multivalued part of the integrand $\sigma_2^{s_{12}/2}$ introduces
phases $e^{\pm \frac{i \pi }{2} s_{12}}$ depending on the sign of the small imaginary part of $\sigma_2$.}
\label{figcirc}
\end{figure}

In this section, the degeneration $Z^{i\infty}_{\eta_2}(1,2|1,2)$ of the planar
two-point $A$-cycle integral will be reduced to a four-point disk integral. The integrand
(\ref{someintA}) yields expressions of the form
\beq
 I^{\te{tree}}(1,2| {\cal F}(\sigma_2)) = \int_{{\cal C}(1,2)} \frac{ \dd \sigma_2 }{2\pi i \sigma_2} \, \sigma_2^{s_{12}/2} (1-\sigma_2)^{-s_{12}} {\cal F}(\sigma_2)
 \label{someintF}
 \eeq
with ${\cal F}(\sigma_2) \rightarrow 1$ and ${\cal F}(\sigma_2) \rightarrow G_{12}$,
\begin{align}
Z^{i\infty}_{\eta_2}(1,2|1,2) &= \pi \cot(\pi \eta_2) I^{\te{tree}}(1,2| 1) +  I^{\te{tree}}(1,2|G_{12}) \, .
 \label{someintG}
\end{align}
The integrand of $ I^{\te{tree}}(1,2|G_{12}) $ is in fact a total derivative and integrates to zero,
\beq
\frac{  G_{12} }{2\pi i \sigma_2} \sigma_2^{s_{12}/2} (1-\sigma_2)^{-s_{12}}
= \frac{1}{s_{12}} \frac{\partial }{\partial \sigma_2}\big( \sigma_2^{s_{12}/2} (1-\sigma_2)^{-s_{12}} \big)
\ \ \ \Rightarrow \ \ \
I^{\te{tree}}(1,2|G_{12}) = 0 \, .
 \label{someintH}
\eeq
Hence, the leftover work is to simplify the expression (\ref{someintF}) for $ I^{\te{tree}}(1,2| 1)$.
In the $\sigma_2 = e^{2\pi i z_2}$ variable, the
integration contour ${\cal C}(1,2)$ is the unit circle visualized in the left panel of figure \ref{figcirc} instead of the
unit interval $z_2 \in (0,1)$. In order to connect with genus-zero techniques, it is convenient to deform the
unit circle ${\cal C}(1,2)$ to the homotopy-equivalent contour in the right panel of figure \ref{figcirc}. The contour
deformation must not cross the branch points $\sigma_2=0$ and $\sigma_2=1$ of the multivalued part of the
integrand $\sigma_2^{s_{12}/2} (1-\sigma_2)^{-s_{12}}$. Accordingly, one arrives at the two straight paths
parallel to the unit interval $\sigma_2\in (0,1)$ whose small displacement from the real axis indicates that
they are loaded with different branches of $\sigma_2^{s_{12}/2}$. The phases associated with the straight
paths above and below the real axis are $-e^{-\frac{i \pi }{2} s_{12}}$ and $e^{+\frac{i \pi }{2} s_{12}}$, respectively,
and the minus sign of the former stems from the negative orientation of the path.

The contour deformation in figure \ref{figcirc} has been used in \cite{Enriquez:Emzv} and \cite{Broedel:2017jdo}
to evaluate the $\tau\rightarrow i\infty$ degeneration of $A$-cycle eMZVs and twisted eMZVs, respectively.
In these references, the small circle around the origin introduces separate contributions to the (twisted) eMZVs
at the cusp. In our situation with the additional factor of $\sigma_2^{s_{12}/2} (1-\sigma_2)^{-s_{12}}$,
the net effect of the small circle is to attribute phases to the straight paths that differ by
$\frac{ (e^{2\pi i }\sigma_2)^{s_{12}/2} }{ (\sigma_2)^{s_{12}/2} } = e^{2\pi i \frac{s_{12}}{2}}$, and we arrive at
\begin{align}
\int_{{\cal C}(1,2)}  \dd \sigma_2 \, \sigma_2^{s_{12}/2} (1-\sigma_2)^{-s_{12}} &=
\bigg(e^{- \frac{i \pi }{2} s_{12}} \int_1^{0} \dd \sigma_2 + e^{\frac{i \pi }{2} s_{12}} \int_0^{1} \dd \sigma_2  \bigg)
\, \sigma_2^{s_{12}/2} (1-\sigma_2)^{-s_{12}}
\notag\\
&= 2 i   \sin \Big( \frac{ \pi s_{12}}{2} \Big) \int^1_0  \dd \sigma_2
\, \sigma_2^{s_{12}/2} (1-\sigma_2)^{-s_{12}}\, .
 \label{someintI}
\end{align}
Contour deformations and $s_{ij}$-dependent phases of this type are well-known from the Kawai--Lewellen--Tye
relations \cite{Kawai:1985xq} and monodromy relations \cite{BjerrumBohr:2009rd, Stieberger:2009hq} among string
tree amplitudes. More recently, these techniques have been connected \cite{Mizera:2017cqs, Mizera:2019gea} with
the framework of intersection theory (see e.g.\ \cite{intersection}), where the choices of branch for the Koba--Nielsen
factor $\sim \sigma_{ij}^{-s_{ij}}$
lead to the notion of twisted cycles. Accordingly, (\ref{someintI}) can be viewed as a relation between twisted
cycles which holds for any rational function of $\sigma_j$ in the integrand. Upon insertion into (\ref{someintF}),
the desired integral can be evaluated in terms of the well-known Euler Beta function
\begin{align}
 I^{\te{tree}}(1,2| 1) &= \frac{2i}{2\pi i}  \sin \Big( \frac{ \pi s_{12}}{2} \Big) \int^1_{0} \frac{ \dd \sigma_2}{\sigma_2}
  \, \sigma_2^{s_{12}/2} (1-\sigma_2)^{-s_{12}}  \label{KNdrft1}  \\
&= \frac{1}{\pi} \sin \Big( \frac{ \pi s_{12}}{2} \Big)  \frac{ \Gamma(\tfrac{s_{12}}{2})\Gamma(1-s_{12}) }{\Gamma(1-\tfrac{s_{12}}{2})}  = \frac{ \Gamma(1-s_{12}) }{ \big[ \Gamma(1-\tfrac{s_{12}}{2}) \big]^2}  \, ,\notag
\end{align}
or equivalently, a four-point disk integral (\ref{cocyc1}) at special arguments $(-\tfrac{s_{12}}{2}, s_{12})$ in
the place of $(s_{12},s_{23})$. In section \ref{sec:5}, we will spell out the general
dictionary between $(\tau\rightarrow i \infty)$-limits of planar $n$-point $A$-cycle integrals and $(n{+}2)$-point
disk integrals of Parke--Taylor type.

By (\ref{someintH}) and (\ref{KNdrft1}), the planar two-point $A$-cycle integral at the cusp takes the form
\begin{align}
Z^{i\infty}_{\eta_2}(1,2|1,2) &= \pi \cot(\pi \eta_2) \frac{ \Gamma(1-s_{12}) }{ \big[ \Gamma(1-\tfrac{s_{12}}{2}) \big]^2}  \, .
 \label{someintL}
\end{align}
When inserted into the general solution (\ref{intro12}) of the differential equation for
$Z^\tau_{\eta_2}(\ast |1,2)$, one can assemble the $\ap$-expansion of the planar
$A$-cycle integral from
\begin{align}
Z^\tau_{\eta_2}(1,2 |1,2)&=   \frac{ \Gamma(1-s_{12}) }{ \big[ \Gamma(1-\tfrac{s_{12}}{2}) \big]^2}  \sum_{r=0}^{\infty}  \sum_{k_1,k_2,\ldots,k_r \atop{=0,4,6,8,\ldots} }  \prod_{j=1}^r (k_j{-}1)\, \gamma(k_1,k_2,\ldots,k_r|\tau) \notag\\
&\ \ \ \ \ \ \ \ \ \ \ \ \ \ \ \ \ \ \ \ \ \times  r_{\eta_2}(\ep_{k_r} \ldots \ep_{k_2} \ep_{k_1} ) \pi \cot(\pi \eta_2) \, ,
 \label{someintM}
 \end{align}
along with the straightforward expansion of the $\Gamma$-functions in terms of Riemann zeta values
\begin{align}
\frac{ \Gamma(1-s_{12}) }{ \big[ \Gamma(1-\tfrac{s_{12}}{2}) \big]^2}&=
\exp \left( \sum_{k=2}^{\infty} \frac{ \zeta_k }{k}  ( 1- 2^{1-k}) s_{12}^k  \right)
\notag \\
&= 1 + \frac{1}{4} s_{12}^2 \zeta_2 + \frac{1}{4} s_{12}^3 \zeta_3 + \frac{19}{160} s_{12}^4 \zeta_2^2
+ \frac{ 1}{16} s_{12}^5 \zeta_2 \zeta_3  + \frac{ 3}{16} s_{12}^5 \zeta_5   \notag \\
& \ \ \ \ \ \
+ \frac{ 55}{896} s_{12}^6 \zeta_2^3 +\frac{ 1}{32} s_{12}^6 \zeta_3^2 + {\cal O}(\ap^7) \, .
 \label{KNdrft2}
\end{align}
The non-planar $\ap$-expansion in (\ref{someintE}) does not exhibit any analogue of the
$\ap$-dependent factor (\ref{KNdrft2}), and in fact, no MZVs at all. By the planar result
(\ref{someintM}), the coefficient of $(\ap)^r$ in $Z^\tau_{\eta_2}(1,2 |1,2)$ comprises
products $\gamma(k_1,k_2,\ldots,k_\ell|\tau) \zeta_{w_1}\zeta_{w_2}\ldots\zeta_{w_k}$
such that the length $\ell$ and the weights $w_j$ conspire to $r=\ell+w_1+w_2+\ldots+w_k$.


\subsection{Extracting component integrals}
\label{sec:3.5}

One-loop string amplitudes can be reduced to $A$-cycle integrals over products of $f_{ij}^{(m_k)}$ at fixed
overall weight $\sum_k m_k$ \cite{Tsuchiya:1988va, Dolan:2007eh, Broedel:2014vla, Bianchi:2015vsa, Berg:2016wux}
and do not involve the full $\eta$-dependent $\Omega(z_{ij},\eta,\tau)$ in the integrand. In order to apply
the results on the generating functions $Z^\tau_{\vec{\eta}}$ in (\ref{intro2}) to the integrals in
string amplitudes, one has to extract particular orders in $\eta_j$. In the two-point case, this
amounts to studying the following component integrals
(with the usual placeholder $\ast$ for planar or non-planar cycles)
\beq
Z^{\tau}_{(m)}(\ast) \equiv Z^{\tau}_{\eta_2}(\ast|1,2) \, \big|_{\eta_2^{m-1}} = \int_{{\cal C}(\ast)} \dd z_2 \, f^{(m)}(z_{12},\tau) \,e^{s_{12}{\cal G}(z_{12},\tau)} \, .
\label{compin1}
\eeq
As will be explained below, each order in $\ap$ and $\eta_2$ only receives a finite number of contributions
from the general formula (\ref{intro12}), and we will spell out detailed cutoffs on the relevant infinite sums.
Like this, one can efficiently obtain two-point $\ap$-expansions to high orders from the
differential-equation method of this work.

The $\ap$-expansion of $Z^{\tau}_{(0)}(1,2) $ and $Z^{\tau}_{(0)}(\begin{smallmatrix} 2 \\ 1 \end{smallmatrix})$
with a plain Koba--Nielsen factor in the integrand has already been studied in \cite{Broedel:2018izr}, and the
coefficients where dubbed planar and non-planar $A$-cycle graph functions\footnote{Following earlier
work on closed-string $\ap$-expansions \cite{Green:2008uj, DHoker:2015wxz}, the property
$\int^0_1 \dd z\, {\cal G}(z,\tau)=0$ of the open-string Green function has been exploited to organize
the $\ap$-expansion of $Z^{\tau}_{(0)}(1,2) $ and generalizations via one-particle irreducible graphs
$\Gamma$ \cite{Broedel:2018izr}.
The notation $A_w$ and $A_{\underline{w}}$ for $A$-cycle graph functions in (\ref{CC160}) refers to planar
banana graphs with $w$ edges, and the translation to the notation ${\bf A}[\Gamma]$ in \cite{Broedel:2018izr}
is exemplified as follows:

 \tikzpicture
\draw(0,0)node{$A_2= {\bf A}\Big[ \ \ \ \ \ \ \Big]$};
\draw(0.3,0)node{$\bullet$};
\draw(0.8,0)node{$\bullet$};
\draw(0.3,0) .. controls (0.45,0.25) and (0.65,0.25) .. (0.8,0);
\draw(0.3,0) .. controls (0.45,-0.25) and (0.65,-0.25) .. (0.8,0);
\scope[xshift=3.2cm]
\draw(0,0)node{$A_3= {\bf A}\Big[ \ \ \ \ \ \ \Big]$};
\draw(0.3,0)node{$\bullet$};
\draw(0.8,0)node{$\bullet$};
\draw(0.3,0) .. controls (0.45,0.25) and (0.65,0.25) .. (0.8,0);
\draw(0.3,0) .. controls (0.45,-0.25) and (0.65,-0.25) .. (0.8,0);
\draw(0.3,0) -- (0.8,0);
\endscope
\scope[xshift=6.4cm]
\draw(0,0)node{$A_4= {\bf A}\Big[ \ \ \ \ \ \ \Big]$};
\draw(0.3,0)node{$\bullet$};
\draw(0.8,0)node{$\bullet$};
\draw(0.3,0) .. controls (0.45,0.15) and (0.65,0.15) .. (0.8,0);
\draw(0.3,0) .. controls (0.45,-0.15) and (0.65,-0.15) .. (0.8,0);
\draw(0.3,0) .. controls (0.45,0.35) and (0.65,0.35) .. (0.8,0);
\draw(0.3,0) .. controls (0.45,-0.35) and (0.65,-0.35) .. (0.8,0);
\endscope
\scope[xshift=9.6cm]
\draw(0,0)node{$A_{\underline{2}}= {\bf A}\Big[ \ \ \ \ \ \ \Big]$};
\draw(0.3,0)node{$\bullet$};
\draw(0.8,0)node{$\bullet$};
\draw(0.3,0) .. controls (0.45,0.25) and (0.65,0.25) .. (0.8,0);
\draw(0.3,0) .. controls (0.45,-0.25) and (0.65,-0.25) .. (0.8,0);
\draw[dashed](0.55,0.28) -- (0.55,-0.28);
\endscope
\scope[xshift=12.8cm]
\draw(0,0)node{$A_{\underline{3}}= {\bf A}\Big[ \ \ \ \ \ \ \Big]$};
\draw(0.3,0)node{$\bullet$};
\draw(0.8,0)node{$\bullet$};
\draw(0.3,0) .. controls (0.45,0.25) and (0.65,0.25) .. (0.8,0);
\draw(0.3,0) .. controls (0.45,-0.25) and (0.65,-0.25) .. (0.8,0);
\draw(0.3,0) -- (0.8,0);
\draw[dashed](0.55,0.28) -- (0.55,-0.28);
\endscope
\endtikzpicture} $A_w$ and $A_{\underline{w}}$, respectively,
\beq
Z^{\tau}_{(0)}(1,2)= \sum_{w=0}^{\infty} \frac{ s_{12}^w }{w!} A_{w}(\tau) \, , \ \ \ \ \ \
Z^{\tau}_{(0)}\big(\begin{smallmatrix} 2 \\ 1 \end{smallmatrix} \big)= q^{s_{12}/8} \sum_{w=0}^{\infty} \frac{ s_{12}^w }{w!} A_{\underline{w}}(\tau)\, .
\label{CC160}
\eeq
In the expansion of $Z^{\tau}_{(0)}\big(\begin{smallmatrix} 2 \\ 1 \end{smallmatrix} \big)$,
the factor of $q^{s_{12}/8}$ has been separated to ensure that both $A_w$ and $A_{\underline{w}}$
have a Fourier expansion in $q$. Accordingly, $A$-cycle graph functions take a more compact
form\footnote{In case of $A_2$ and $A_3$, rewriting $\gamma(\ldots|\tau)$ in terms of $\gamma_0(\ldots|\tau)$
will absorb the contributions $\sim \zeta_4 \gamma(0, 0|\tau)$ and $\sim \zeta_6 \gamma(0, 0, 0|\tau)$ from
\begin{align*}
A_2(\tau) &= \frac{ \zeta_2}{2}  -12 \zeta_4 \gamma(0, 0|\tau)-6 \gamma(4, 0|\tau) \, , \ \ \ \ A_3(\tau) = \frac{3 \zeta_3}{2}  + 384 \zeta_6 \gamma(0, 0, 0|\tau) +
144 \zeta_2 \gamma(4, 0, 0|\tau)- 60 \gamma(6, 0, 0|\tau)  \, .
\end{align*}} when expressed in terms of the convergent iterated Eisenstein integrals $\gamma_0(k_1,k_2,\ldots|\tau)$
with $k_1{>}0$, see (\ref{preB5con}) for their definition and (\ref{qgamma1}) for their $q$-expansion. In the planar case,
we have $A_0(\tau)=1, \ A_1(\tau)=0$ and \cite{Broedel:2018izr}
\begin{align}
A_{2} (\tau)&= \frac{ \zeta_2}{2} - 6 \gamma_0(4, 0|\tau)
\label{CC18} \\
A_{3}(\tau) &= \frac{ 3 \zeta_3}{2} + 144 \zeta_2 \gamma_0(4, 0, 0|\tau) -
 60 \gamma_0(6, 0, 0|\tau)
\notag \\
A_{4} (\tau)&= \frac{57 \zeta_4}{8} - 18 \zeta_2 \gamma_0(4, 0|\tau)  - 3456 \zeta_4 \gamma_0(4, 0, 0, 0|\tau)
+ 216 \gamma_0(4, 0,4,0|\tau)  -  432 \gamma_0(4, 4, 0, 0|\tau)
 \notag \\
 &\ \ \ \ +  5760 \zeta_2 \gamma_0(6, 0, 0, 0|\tau) - 3024 \gamma_0(8, 0, 0, 0|\tau) \, .\notag
\end{align}
The simplest non-planar examples are $A_{\underline{0}}(\tau)=1, \ A_{\underline{1}}(\tau)=0$ and \cite{Broedel:2018izr}
\begin{align}
A_{\underline{2}}(\tau) &= - 6 \gamma_0(4, 0|\tau)
\label{CC17} \\
A_{\underline{3}}(\tau) &= -72 \zeta_2 \gamma_0(4, 0, 0|\tau) - 60 \gamma_0(6, 0, 0|\tau)
\notag \\
A_{\underline{4}}(\tau) &= -3456 \zeta_4 \gamma_0(4, 0, 0, 0|\tau) +
 216 \gamma_0(4, 0, 4, 0|\tau) - 432 \gamma_0(4, 4, 0, 0|\tau)\notag \\
 &\ \ \ \ -
 2880 \zeta_2 \gamma_0(6, 0, 0, 0|\tau) - 3024 \gamma_0(8, 0, 0, 0|\tau) \, .
\notag
\end{align}
In order to efficiently generate $A$-cycle graph functions from (\ref{intro12}), we reiterate that
\begin{itemize}
\item $Z^{\tau}_{(0)}(\ast)$ are the coefficients of $\eta_2^{-1}$ in $Z^\tau_{\eta_2}(\ast|1,2)$, and
$A_w,A_{\underline{w}}$ occur at their $\ap^w$-order
\item the initial values $Z^{i\infty}_{\eta_2}(\ast|1,2)$ depend on $\eta_2$ via
straightforward trigonometric series (\ref{trigdeg})
\item the derivations $r_{\eta_2}(\ep_k)$ in (\ref{2ptC}) are linear in $\ap$ and
change the order in $\eta_2$ by $\geq k{-}2$
\end{itemize}
The last point implies that the order-$(\ap)^w$ contribution $r_{\eta_2}(\ep_{k_w}\ldots \ep_{k_1})$ to (\ref{intro12})
cannot lower the order in $\eta_2$ by more than $2w$ units.
Hence, the $A$-cycle graph functions $A_w,A_{\underline{w}}$ along with $\ap^w$
are only sensitive to the orders $\eta_2^{\leq 2w-1}$ in the expansion of
$\pi \cot(\pi \eta_2)$ in $Z^{i\infty}_{\eta_2}(\ast|1,2)$.

Similarly, only a small number of length-$w$ words $\gamma(k_1,\ldots,k_w|\tau)
r_{\eta_2}(\ep_{k_w}\ldots \ep_{k_1})$ in (\ref{intro12}) contributes to a given $A_w$ or
$A_{\underline{w}}$: Since the initial values $Z^{i\infty}_{\eta_2}(\ast|1,2)$ have no
poles of higher order than $\frac{1}{\eta_2}$, the combined effect of the $r_{\eta_2}(\ep_k)$
must not raise the power of $\eta_2$ in order to contribute to $Z^\tau_{(0)}(\ast)$.
This leads to the bound $k_1+k_2+\ldots+k_w \leq 2w$
for the $\gamma(k_1,\ldots,k_w|\tau)$  that contribute to $A_w$ and $A_{\underline{w}}$,
in lines with the examples in (\ref{CC18}) and (\ref{CC17}).

So far, we did not take the MZVs in the planar $\ap$-expansion (\ref{someintM}) into account.
Since the series (\ref{KNdrft2}) in $\zeta_m$ can be factored out from $Z^{\tau}_{\eta_2}(1,2|1,2)$,
some of the terms in $A_w,A_{\underline{w}}$ are products of $\zeta_m$ multiplying $A$-cycle graph functions
of lower weight.

This kind of counting can be straightforwardly generalized to higher-order component integrals
 $Z^\tau_{(m>0)}(\ast)$ in (\ref{compin1}), resulting in
sharp bounds on the possible contributions to specific $\ap$-orders.
While integrals over odd functions $f^{(2k-1)}$ vanish identically,
\beq
Z^\tau_{(2k-1)}(1,2) = Z^{\tau}_{(2k-1)}\big(\begin{smallmatrix} 2 \\ 1 \end{smallmatrix} \big) = 0\, , \ \ \ \ \ \ k\geq 1\, ,
\label{compin2}
\eeq
the simplest examples beyond $A$-cycle graph functions read
\begin{align}
Z^\tau_{(2)}(1,2)  &=  -2 \zeta_2
+  3 s_{12} \gamma_0(4|\tau)
+  s_{12}^2 \Big({-} \frac{ 5 \zeta_4}{4 }
-  18   \zeta_2 \gamma_0(4, 0|\tau)
+ 10   \gamma_0(6, 0|\tau)  \Big) \notag \\
&+ s_{12}^3 \Big( {-}  \frac{ \zeta_2 \zeta_3}{2 }
+  \frac{ 3 \zeta_2}{4 }   \gamma_0(4|\tau)
  +  24   \zeta_4 \gamma_0(4, 0, 0|\tau) -
 9   \gamma_0(4, 0, 4|\tau)
 + 18   \gamma_0(4, 4, 0|\tau) \notag  \\
 & \ \ \ \ -
 220   \zeta_2 \gamma_0(6, 0, 0|\tau) +
 126   \gamma_0(8, 0, 0|\tau) \Big) + {\cal O}(\ap^4) \label{compin3}
\\
Z^\tau_{(4)}(1,2) &= -2 \zeta_4 + s_{12} \Big(  {-} 6  \zeta_2 \gamma_0(4|\tau) +  5   \gamma_0(6|\tau) \Big)  \notag \\
&
+ s_{12}^2 \Big( {-}  \frac{ 7 \zeta_6}{8 }  - 42   \zeta_4 \gamma_0(4, 0|\tau) +
 9   \gamma_0(4, 4|\tau) - 100   \zeta_2 \gamma_0(6, 0|\tau) +
 63   \gamma_0(8, 0|\tau) \Big) + {\cal O}(\ap^3) \notag
\\
Z^\tau_{(6)}(1,2)  &=  -2 \zeta_6 + s_{12} \Big( {-} 6  \zeta_4 \gamma_0(4|\tau) -
 10   \zeta_2 \gamma_0(6|\tau) + 7   \gamma_0(8|\tau) \Big)+ {\cal O}(\ap^2) \notag
\end{align}
as well as
\begin{align}
q^{-\frac{s_{12}}{8}}Z^{\tau}_{(2)}\big(\begin{smallmatrix} 2 \\ 1 \end{smallmatrix} \big)&=  \zeta_2 + 3 s_{12} \gamma_0(4|\tau)
+ s_{12}^2 \Big( 9 \zeta_2 \gamma_0(4, 0|\tau) + 10   \gamma_0(6, 0|\tau)  \Big) \notag \\
&+  s_{12}^3\Big( 114  \zeta_4 \gamma_0(4, 0, 0|\tau) -
 9   \gamma_0(4, 0, 4|\tau) + 18   \gamma_0(4, 4, 0|\tau)  \notag \\
 &\ \ \ \ +  110   \zeta_2 \gamma_0(6, 0, 0|\tau) +
 126  \gamma_0(8, 0, 0|\tau) \Big) + {\cal O}(\ap^4)  \label{compin3x}  \\
q^{-\frac{s_{12}}{8}}Z^{\tau}_{(4)}\big(\begin{smallmatrix} 2 \\ 1 \end{smallmatrix} \big)&=
   \frac{7 \zeta_4}{4 }
+ s_{12}\Big( 3  \zeta_2 \gamma_0(4|\tau) +   5   \gamma_0(6|\tau) \Big)  \notag \\
& + s_{12}^2 \Big(  \frac{ 147 \zeta_4}{4 } \gamma_0(4, 0|\tau) +
  9   \gamma_0(4, 4|\tau) + 50   \zeta_2 \gamma_0(6, 0|\tau) +
  63   \gamma_0(8, 0|\tau) \Big)+ {\cal O}(\ap^3) \notag
 \\
q^{-\frac{s_{12}}{8}}Z^{\tau}_{(6)}\big(\begin{smallmatrix} 2 \\ 1 \end{smallmatrix} \big)&= \frac{31 \zeta_6}{16 } + s_{12} \Big(  \frac{21 \zeta_4}{4}   \gamma_0(4|\tau)
+  5   \zeta_2 \gamma_0(6|\tau) + 7   \gamma_0(8|\tau)  \Big)+ {\cal O}(\ap^2) \, . \notag
\end{align}
There is no bottleneck in generating much high orders in $\ap$ from (\ref{intro12}), even without
optimizing the performance of a computer implementation.


\subsection{On $B$-cycle graph functions and modular graph functions}
\label{sec:3.6}

The motivation for defining $A$-cycle graph functions in \cite{Broedel:2018izr} was to find
an open-string analogue of the modular graph functions in closed-string $\ap$-expansions
\cite{Green:2008uj, DHoker:2015wxz}. Indeed, the $B$-cycle graph functions obtained from
modular $\tau \rightarrow - \frac{1}{\tau}$ transformation of their $A$-cycle analogues were
taken as a starting point to propose an elliptic single-valued map connecting genus-one integrals
in open- and closed-string one-loop amplitudes \cite{Broedel:2018izr}.

The asymptotics of various graph functions at the cusp $\tau \rightarrow i \infty$ has recently attracted
a lot of attention. In the case of $B$-cycle and modular graph functions, the behavior at the cusp is
governed by Laurent polynomials in $(\pi \tau)$ and $(\pi \Im \tau)$, respectively, whose coefficients are $\QQ$-linear
combinations of MZVs\footnote{For $B$-cycle graph functions, the appearance of $\QQ$-linear
rather than $\QQ[(2\pi i)^{-1}]$-linear combinations of MZVs has been proven in \cite{Broedel:2018izr}
and, with an improved bound on the degree of the Laurent polynomial, in \cite{Zerbini:2018hgs}.
For modular graph functions, the coefficients in the Laurent polynomial are proven to be $\QQ$-linear
combinations of cyclotomic MZVs and conjectured to be single-valued MZVs \cite{Zerbini:2015rss, Zerbini:2018sox}.}.
For the planar two-point $B$-cycle graph functions, i.e.\ the modular transformations of $A_w$,
the Laurent polynomials at all $w\in \NN$ have been expressed in terms of Riemann zeta values \cite{Zagier:2019eus}.
Similarly, the Laurent polynomials of all two-point modular graph functions were determined in terms of
odd (i.e.\ single-valued) Riemann zeta values \cite{DHoker:2019xef, Zagier:2019eus}.

While the recent results on $B$-cycle graph functions at the cusp settle the behavior of $A_w(\tau)$
at $\tau \rightarrow 0$, their behavior at $\tau\rightarrow i\infty$ has not yet been spelled out for all weights.
By isolating the $\eta^{-1}$ order of (\ref{someintM}) and exploiting that the iterated Eisenstein integrals do not
contribute at the cusp, we infer the generating function
\beq
\sum_{w=0}^{\infty} \frac{ s_{12}^w }{w!} A_{w}(i\infty) = \frac{ \Gamma(1-s_{12}) }{ \big[ \Gamma(1-\tfrac{s_{12}}{2}) \big]^2} \,,
\eeq
that of course reproduces the results $A_{w}(i\infty)=\frac{ \zeta_2}{2}, \ A_{3}(i\infty)= \frac{3 \zeta_3}{2}$
and $A_{4}(i\infty)=\frac{ 57 \zeta_4}{8}$ in (\ref{CC18}).
Similarly, the non-planar $\ap$-expansion (\ref{someintE}) implies the vanishing of all non-trivial non-planar
$A$-cycle graph functions $A_{\underline{w}}$ at the cusp
\beq
A_{\underline{w}}(i\infty) = 0 \ , \ \ \ \ \ \ w \geq 1\, .
\eeq
In later sections, similar generating series will be given for the asymptotics of higher-point $A$-cycle
graph functions. We hope that the techniques of this work can be helpful to determine the
much richer patterns of MZVs in the Laurent polynomials of $B$-cycle and modular graph functions
beyond two points. Also, it would
be interesting to study the generalization of $B$-cycle and modular graph functions to additional factors of $f^{(m)}_{ij}$
in the integrand, i.e.\ the $B$-cycle and closed-string analogues of the higher $Z^\tau_{(m>0)}(\ast)$
functions in (\ref{compin1}).


\section{Differential equations at $n$ points}
\label{sec:4}

In the previous section, we have illustrated the key ideas of this work by a two-point example.
The purpose of the remaining sections is to propose the generalization of all the steps to the $n$-point
$A$-cycle integrals in (\ref{intro2}). In this section, we explain the derivation of the homogeneous and linear
first-order differential equation (\ref{intro3}) at $n$ points and describe the explicit form of the
$(n{-}1)! \times (n{-}1)!$ matrix-valued differential operator $D^\tau_{\vec{\eta}}$. While the expressions
up to and including $n=5$ are based on rigorous calculations, the form of $D^\tau_{\vec{\eta}}$ at
$n\geq 6$ is conjectural.

The subsequent differential equations hold for any planar or non-planar integration cycle ${\cal C}(\ast)$
in (\ref{intro2}) that is realized in one-loop open-string amplitudes, cf.\ figure \ref{basiccyl}. We will
always set $z_1=0$ and use the notation
\beq
{\cal C}(1,2,3,\ldots,n) \equiv \{ z_2,z_3,\ldots,z_n \in \RR, \ 0=z_1<z_2<z_3<\ldots<z_n<1\}
 \label{allC1}
\eeq
in the planar case as well as
\begin{align}
{\cal C}(\begin{smallmatrix} r{+}1,\ldots,n \\ 1,2,3,\ldots,r \end{smallmatrix}) &\equiv{\cal C}(1,2,\ldots,r) \times
\Big\{  z_{j} = \tauh + v_{j} \ \forall \ j=r{+}1,\ldots,n ,  \label{allC2} \\
&\  \ \ \ \ \bigvee_{\rho \ \te{cyclic}(r{+}1,\atop{r{+}2,\ldots,n)}}  0< v_{\rho(r+1)}< v_{\rho(r+2)}<\ldots < v_{\rho(n)} <1 \Big\}
\notag
\end{align}
for non-planar integration cycles, in lines with the two-point cases in (\ref{2cycl}). The second line of
(\ref{allC2}) ensures that, for fixed $z_1=0$, all the cyclic arrangements of punctures $z_{r{+}1},\ldots,z_n$
on the upper boundary in figure \ref{basiccyl} are taken into account. The Koba--Nielsen
factors will henceforth be denoted by the shorthand ${\rm KN}^\tau_{12\ldots n}$ defined in (\ref{dfkn}).


\subsection{Three points}
\label{sec:4.1}

The three-point instances of the $A$-cycle integrals (\ref{intro2}) are given by
\beq
Z_{\eta_2,\eta_3}^{\tau}(\ast |1,2,3) = \int_{{\cal C}(\ast)} \dd z_2\, \dd z_3 \,\Omega(z_{12},\eta_2{+}\eta_3,\tau)
\Omega(z_{23},\eta_3,\tau) \, \te{KN}^\tau_{123}
\label{df3a}
\eeq
and allow for two inequivalent planar integration cycles ${\cal C}(1,2,3), {\cal C}(1,3,2)$ as well as three inequivalent
non-planar ones ${\cal C}\big( \begin{smallmatrix} 3 \\ 1,2 \end{smallmatrix} \big), {\cal C}\big( \begin{smallmatrix} 2 \\ 1,3 \end{smallmatrix} \big), {\cal C}\big( \begin{smallmatrix} 2,3 \\ 1 \end{smallmatrix} \big)$. The $\tau$-derivative
of (\ref{df3a}) can be evaluated by iteration of the steps in the two-point computation (\ref{B12}),
\begin{align}
2\pi i \partial_\tau &Z_{\eta_2,\eta_3}^{\tau}(\ast |1,2,3) =
 \int_{{\cal C}(\ast)} \dd z_2\, \dd z_3  \, \Big( -  2 \zeta_2 s_{123}  \Omega(z_{12}, \beta_1,\tau)
  \Omega(z_{23},\beta_2,\tau) \notag \\
& \ \ \
+ s_{12} \big[ f^{(1)}_{12} \partial_{\beta_1} \Omega(z_{12}, \beta_1,\tau)
- f^{(2)}_{12}   \Omega(z_{12}, \beta_1,\tau) \big] \Omega(z_{23},\beta_2,\tau)\notag \\
& \ \ \
+ s_{23} \big[ f^{(1)}_{23} \partial_{\beta_2} \Omega(z_{23},\beta_2,\tau)
- f^{(2)}_{23} \Omega(z_{23},\beta_2,\tau)  \big]   \Omega(z_{12},  \beta_1,\tau) \notag \\
&\ \ \ + s_{13} \big[ f^{(1)}_{13} \Omega(z_{23},\beta_2,\tau)  \partial_{\beta_1} \Omega(z_{12},  \beta_1,\tau)
+f^{(1)}_{13}   \Omega(z_{12}, \beta_1,\tau) \partial_{\beta_2} \Omega(z_{23},\beta_2,\tau)  \notag \\
& \ \ \ \ \ \ \ \
-f^{(2)}_{13}   \Omega(z_{12}, \beta_1,\tau) \Omega(z_{23}, \beta_2,\tau)
  \big]
 \Big) \, \te{KN}^\tau_{123}\, , \label{B32}
\end{align}
where we have again used the mixed heat equation (\ref{Aheat}), integration by parts to eliminate
$\partial_z\Omega(z_{ij},\ldots)$ as well as the Koba--Nielsen derivatives (\ref{dkn}).
The second arguments of the $\Omega(\ldots)$ are denoted by $\beta_1=\eta_2{+}\eta_3$ and $\beta_2=\eta_3$
in intermediate steps to make the scope of the $\partial_{\beta_j}$ more transparent. In the next step, we
express $f^{(1)}_{ij}$ and $f^{(2)}_{ij}$ in terms of the $\xi^0$-order of another Kronecker--Eisenstein
series $\Omega(z_{ji},\xi,\tau)$, see (\ref{B14}), and obtain
\begin{align}
2\pi i \partial_\tau  Z_{\eta_2,\eta_3}^{\tau}(\ast |1,2,3) &=   \int_{{\cal C}(\ast)} \dd z_2 \, \dd z_3 \,\Big(
 -2 \zeta_2 s_{123} \Omega(z_{12},\beta_1,\tau)\Omega(z_{23},\beta_2,\tau)
 \label{B33} \\
& \ \ \ - s_{12} \Omega(z_{23},\beta_2,\tau) ( \partial_{\beta_1}{+}\partial_\xi) \Omega(z_{12},\beta_1,\tau) \Omega(z_{21},\xi,\tau)
 \notag \\
 & \ \ \ - s_{23} \Omega(z_{12},\beta_1,\tau)  ( \partial_{\beta_2}{+}\partial_\xi)  \Omega(z_{23},\beta_2,\tau)\Omega(z_{32},\xi,\tau)
 \notag \\
 & \ \ \ - s_{13} ( \partial_{\beta_1}{+} \partial_{\beta_2}{+}\partial_\xi) \Omega(z_{12},\beta_1,\tau) \Omega(z_{23},\beta_2,\tau)\Omega(z_{31},\xi,\tau)
 \Big) \, \te{KN}^\tau_{123} \, \Big|_{\xi^0} \, .
 \notag
\end{align}
The main goal in the simplification of the $\tau$-derivative is to have at most two factors of $\Omega(\ldots)$ in
each term -- this is a necessary condition for writing the right-hand side in terms of integrals
$Z_{\eta_2,\eta_3}^{\tau}(\ldots)$. In the second and third line of (\ref{B33}), this can be readily achieved by
means of (\ref{A25DP}) whereas the last line requires the more general identity
\begin{align}
&(\partial_{\beta_1}+\partial_{\beta_2}+\partial_{\xi}) \Omega(z_{12},\beta_1,\tau)   \Omega(z_{23},\beta_2,\tau)  \Omega(z_{31},\xi,\tau)  \label{A27DP} \\
& \ \ =  - \wp(\beta_1,\tau)   \Omega(z_{23},\beta_2{-}\beta_1,\tau)    \Omega(z_{31},\xi{-}\beta_1,\tau)  \notag \\
& \ \ \ \ \; \, - \wp(\beta_2,\tau)   \Omega(z_{12},\beta_1{-}\beta_2,\tau)    \Omega(z_{31},\xi{-}\beta_2,\tau)  \notag \\
&  \ \ \ \ \; \, -  \wp(\xi,\tau)   \Omega(z_{12},\beta_1{-}\xi,\tau)    \Omega(z_{23},\beta_2{-}\xi,\tau) \notag
\end{align}
to be derived in appendix \ref{app:0.2}. After inserting (\ref{A25DP}) and (\ref{A27DP}) into (\ref{B33}) and
picking up the zero$^{\rm th}$ order in $\xi$, one arrives at
\begin{align}
2\pi i \partial_\tau Z_{\eta_2,\eta_3}^{\tau}(\ast |1,2,3) &=   \int_{{\cal C}(\ast)} \dd z_2 \, \dd z_3  \,\Big(
 -2 \zeta_2 s_{123} \Omega(z_{12},\beta_1,\tau)\Omega(z_{23},\beta_2,\tau)
 \label{B34D} \\
& \ \ \ + s_{12} \Omega(z_{23},\beta_2,\tau) \Big( \frac{1}{2} \partial_{\beta_1}^2 -  \wp(\beta_1,\tau) \Big) \Omega(z_{12},\beta_1,\tau)
 \notag \\
 & \ \ \ + s_{23} \Omega(z_{12},\beta_1,\tau)  \Big( \frac{1}{2} \partial_{\beta_2}^2 -  \wp(\beta_2,\tau) \Big)  \Omega(z_{23},\beta_2,\tau)
 \notag \\
 & \ \ \ + s_{13} \Big[  \big( \wp(\beta_1,\tau)  -\wp(\beta_2,\tau) \big)  \Omega(z_{13},\beta_1,\tau)  \Omega(z_{32},\beta_1{-}\beta_2,\tau)    \notag \\
& \ \ \ \ \ \ \ \ +\Big( \frac{1}{2}( \partial_{\beta_1}{+} \partial_{\beta_2})^2 -\wp(\beta_2,\tau) \Big)
  \Omega(z_{12},\beta_1,\tau)    \Omega(z_{23},\beta_2,\tau) \Big]
 \Big)
 \,  \te{KN}^\tau_{123} \, ,
 \notag
\end{align}
where the Fay identity (\ref{A21}) has been used to eliminate one of the three arrangements of the first
arguments $\Omega(z_{ij},\ldots)\Omega(z_{jk},\ldots)$:
\beq
\Omega(z_{12},\beta_1{-}\beta_2,\tau)    \Omega(z_{13},\beta_2,\tau) =
 \Omega(z_{12},\beta_1,\tau)    \Omega(z_{23},\beta_2,\tau)
+    \Omega(z_{13},\beta_1,\tau) \Omega(z_{32},\beta_1{-}\beta_2,\tau)
\, .
\label{B34C}
\eeq
At this point, one can see the benefit of rewriting $\beta_1=\eta_2{+}\eta_3$ and $\beta_2=\eta_3$ in
terms of the $\eta_j$ variables: The two contributions $\Omega(z_{12},\eta_2{+}\eta_3,\tau)
  \Omega(z_{23},\eta_3,\tau)$ and $\Omega(z_{13},\eta_2{+}\eta_3,\tau)  \Omega(z_{32},\eta_2,\tau)$ to
  the integrand of (\ref{B34D}) are permutations of each other w.r.t.\ $(z_2,\eta_2) \leftrightarrow (z_3,\eta_3)$.
Hence, the entire right-hand side of (\ref{B34D}) can be rewritten in terms of the original integral and one
permutation under $2\leftrightarrow 3$,
\begin{align}
2\pi i \partial_\tau Z_{\eta_2,\eta_3}^{\tau}(\ast |1,2,3) &=
\Big(
 s_{12} \big[ \tfrac{1}{2} \partial^2_{\eta_2}- \wp(\eta_2{+}\eta_3,\tau) \big]
+ s_{23}  \big[ \tfrac{1}{2} (\partial_{\eta_2}{-}\partial_{\eta_3})^2- \wp(\eta_3,\tau) \big]
\notag\\
& \ \ \ \ \ \ \ \ \ \ \ \ \
+ s_{13} \big[ \tfrac{1}{2} \partial^2_{\eta_3}- \wp(\eta_3,\tau) \big]  -2 \zeta_2 s_{123}  \Big)  Z_{\eta_2,\eta_3}^{\tau}(\ast |1,2,3)   \notag \\
&\ \ \ \ \ + s_{13} \big[ \wp(\eta_2{+}\eta_3,\tau) - \wp(\eta_3,\tau) \big]
Z_{\eta_2,\eta_3}^{\tau}(\ast|1,3,2)  \label{B38}
\\
&= \sum_{B \in S_2} D^{\tau}_{\eta_2,\eta_3}(2,3|b_2,b_3) Z_{\eta_2,\eta_3}^{\tau}(\ast |1,b_2,b_3)\, .  \notag
\end{align}
In the last step, $2\pi i \partial_\tau Z_{\eta_2,\eta_3}^{\tau}(\ast |1,2,3)$ has been lined up with the general
form (\ref{intro3}) of the differential equation. The words $B=b_2b_3\ldots b_n$ in a summation $B\in S_{n-1}$
are always understood to be permutations of $2,3,\ldots, n$. The first row of the $2\times 2$ differential
operator $D^{\tau}_{\eta_2,\eta_3}$ can be immediately read off from (\ref{B38}),
\begin{align}
D^{\tau}_{\eta_2,\eta_3}(2,3|2,3) &=
 s_{12} \big[ \tfrac{1}{2} \partial^2_{\eta_2}- \wp(\eta_2{+}\eta_3,\tau) \big] + s_{23}  \big[ \tfrac{1}{2} (\partial_{\eta_2}{-}\partial_{\eta_3})^2- \wp(\eta_3,\tau) \big]
\notag \\
& \ \ \ \ \ \ \ \ + s_{13} \big[ \tfrac{1}{2} \partial^2_{\eta_3}- \wp(\eta_3,\tau) \big] -2 \zeta_2 s_{123} \label{B40} \\
D^{\tau}_{\eta_2,\eta_3}(2,3|3,2) &=
s_{13} \big[ \wp(\eta_2{+}\eta_3,\tau) - \wp(\eta_3,\tau) \big]  \, , \notag
\end{align}
whereas the second row is obtained by relabeling $s_{12}\leftrightarrow s_{13}$ and $\eta_2\leftrightarrow \eta_3$,
\begin{align}
D^{\tau}_{\eta_2,\eta_3}(3,2|3,2)&=
s_{13} \big[ \tfrac{1}{2} \partial^2_{\eta_3}- \wp(\eta_2{+}\eta_3,\tau) \big]
+ s_{23}  \big[ \tfrac{1}{2} (\partial_{\eta_2}{-}\partial_{\eta_3})^2- \wp(\eta_2,\tau) \big]
\notag\\
& \ \ \ \ \ \ \ \
+ s_{12} \big[ \tfrac{1}{2} \partial^2_{\eta_2}- \wp(\eta_2,\tau) \big] -2 \zeta_2 s_{123}  \label{B40A} \\
D^{\tau}_{\eta_2,\eta_3}(3,2|2,3)&=
s_{12} \big[ \wp(\eta_2{+}\eta_3,\tau) - \wp(\eta_2,\tau) \big]  \, . \notag
\end{align}
The expansion (\ref{intro3a}) of the Weierstrass function casts the differential operator into the
form
\beq
D^\tau_{\eta_2,\eta_3} = \sum_{m=0}^{\infty} (1{-}2m) {\rm G}_{2m}(\tau) r_{\eta_2,\eta_3}(\ep_{2m}) \, ,
 \label{intro3y}
\eeq
and yields the following $(2\times 2)$-matrix representation of the derivations
\begin{align}
r_{\eta_2,\eta_3}(\ep_k) &= \delta_{k,0}\Big( 2 \zeta_2 s_{123} - \frac{1}{2} s_{12} \partial_{\eta_2}^2 - \frac{1}{2} s_{13} \partial_{\eta_3}^2
- \frac{1}{2} s_{23} (\partial_{\eta_2} - \partial_{\eta_3})^2 \Big) 1_{2\times 2}
\label{DER11} \\
&\ \ + \eta^{k-2}_{23} \ccb s_{12} &-s_{13} \\-s_{12}&s_{13} \cce
+ \eta^{k-2}_{2} \ccb 0 &0 \\ s_{12} &s_{12}{+}s_{23} \cce
+ \eta^{k-2}_{3} \ccb s_{13}{+}s_{23} & s_{13} \\0 &0 \cce\,,\quad k\neq2   \notag
\end{align}
with $r_{\eta_2,\eta_3}(\ep_{2})=0$ and shorthand $\eta_{23}\equiv \eta_2{+}\eta_3$. All the matrix entries
are linear in $s_{ij}$ and therefore in $\ap$. As a three-point instance of (\ref{intro1}), the solution of (\ref{B38})
via Picard iteration yields the following $\ap$-expansion
\begin{align}
Z^\tau_{\eta_2,\eta_3}(\ast |1,2,3)&=    \sum_{r=0}^{\infty}  \sum_{k_1,k_2,\ldots,k_r \atop{=0,4,6,8,\ldots} }  \prod_{j=1}^r (k_j{-}1)\, \gamma(k_1,k_2,\ldots,k_r|\tau)   \label{someint33}\\
&   \times \sum_{B \in S_2} r_{\eta_2,\eta_3}(\ep_{k_r} \ldots \ep_{k_2} \ep_{k_1} )_{23}{}^{B} Z^{i\infty}_{\eta_2,\eta_3}(\ast |1,B) \, .
\notag
 \end{align}
In contrast to the scalar representations (\ref{2ptC}) at two points, the $r_{\eta_2,\eta_3}(\ep_{k\geq 4})$
no longer commute. By $[r_{\eta_2,\eta_3}(\ep_{4}) , r_{\eta_2,\eta_3}(\ep_{6})] \neq 0$,
for instance, $\gamma(4,6|\tau)$ and $\gamma(6,4|\tau)$ enter (\ref{someint33}) with different coefficients
and introduce an irreducible iterated Eisenstein integral at depth two which is not expressible via
$\gamma(4|\tau)$ and $\gamma(6|\tau)$.

The initial values $Z^{i\infty}_{\eta_2,\eta_3}(\ast |1,b_2,b_3)$ will be later on assembled from $(n{\leq} 5)$-point
disk integrals, see section \ref{sec:5.4} for the planar cycles ${\cal C}(\ast)$ and section \ref{sec:6.2} for the non-planar ones.


\subsection{Four points}
\label{sec:4.2}

Starting from the four-point example of the $A$-cycle integrals (\ref{intro2}), we will
use the vector-notation $\vec{\eta}$ to refer to the $\eta_{j}$ in the subscript. I.e.\ we have
$\vec{\eta}=\eta_2,\eta_3,\eta_4$ in the definition
\beq
Z_{\vec{\eta}}^{\tau}(\ast |1,2,3,4) = \int_{{\cal C}(\ast)} \dd z_2\, \dd z_3 \, \dd z_4 \,\Omega(z_{12},\eta_2{+}\eta_3{+}\eta_4,\tau)
\Omega(z_{23},\eta_3{+}\eta_4,\tau) \Omega(z_{34},\eta_4,\tau) \, \te{KN}^\tau_{1234}
\label{th4pt}
\eeq
that applies to 3, 8 and 6 cyclically inequivalent integration cycles of the type
${\cal C}\big(\begin{smallmatrix} k,l \\ i,j \end{smallmatrix}\big),
{\cal C}\big(\begin{smallmatrix} l \\ i,j,k \end{smallmatrix}\big)$ and ${\cal C}(i,j,k,l)$, respectively.
The $\tau$-derivative of (\ref{th4pt}) can be computed by repeating the steps (\ref{B32}), (\ref{B33}) in the three-point
case\footnote{Note that two integrations by parts w.r.t.\ $z_3,z_4$ are required to remove the $z$-derivative from
$2\pi i \partial_\tau \Omega(z_{23},\beta_2,\tau)
= \partial_{z_2} \partial_{\beta_2}  \Omega(z_{23},\beta_2,\tau)$. More specifically, the Koba--Nielsen derivative
(\ref{dkn}) implies that
\begin{align*}
\partial_{z_2} \partial_{\beta_2}  \Omega(z_{23},\beta_2,\tau) \Omega(z_{34},\beta_3,\tau) {\rm KN}_{1234}^{\tau}
&=\partial_{\beta_2} \Omega(z_{23},\beta_2,\tau)  {\rm KN}_{1234}^{\tau} ( \partial_{z_3} {+} s_{13}f_{13}^{(1)}
{+} s_{23}f_{23}^{(1)}  {-} s_{34}f_{34}^{(1)})\Omega(z_{34},\beta_3,\tau)  \\
&= (  s_{13}f_{13}^{(1)}{+} s_{23}f_{23}^{(1)}  {+} s_{14}f_{14}^{(1)}{+} s_{24}f_{24}^{(1)}  )
\partial_{\beta_2}\Omega(z_{23},\beta_2,\tau) \Omega(z_{34},\beta_3,\tau) {\rm KN}_{1234}^{\tau}
\end{align*}
after discarding total derivatives $(\partial_{z_3}{+}\partial_{z_4})\big[ \partial_{\beta_2}  \Omega(z_{23},\beta_2,\tau) \Omega(z_{34},\beta_3,\tau) {\rm KN}_{1234}^{\tau}\big]$.}, and it is convenient to use the
variables $\beta_1=\eta_2{+}\eta_3{+}\eta_4, \ \beta_2=\eta_3{+}\eta_4$ and $\beta_3=\eta_4$
in intermediate steps:
\begin{align}
2\pi i \partial_\tau Z_{\vec{\eta}}^{\tau}(\ast |1,2,3,4) &=    \int_{{\cal C}(\ast)} \dd z_2 \, \dd z_3 \, \dd z_4 \, {\rm KN}_{1234}^{\tau}
\, \Big({-}2 \zeta_2 s_{1234} \Omega(z_{12},\beta_1,\tau)\Omega(z_{23},\beta_2,\tau)  \Omega(z_{34},\beta_3,\tau)
  \notag \\
&\! \! \! \!  \! \! \! \! - s_{12} \Omega(z_{23},\beta_2,\tau) \Omega(z_{34},\beta_3,\tau)  ( \partial_{\beta_1}{+}\partial_\xi) \Omega(z_{12},\beta_1,\tau) \Omega(z_{21},\xi,\tau)
\notag \\
 &\! \! \! \!  \! \! \! \! - s_{23} \Omega(z_{12},\beta_1,\tau) \Omega(z_{34},\beta_3,\tau)  ( \partial_{\beta_2}{+}\partial_\xi)  \Omega(z_{23},\beta_2,\tau)\Omega(z_{32},\xi,\tau)
 \notag \\
 &\! \! \! \!  \! \! \! \! - s_{34} \Omega(z_{12},\beta_1,\tau) \Omega(z_{23},\beta_2,\tau)  ( \partial_{\beta_3}{+}\partial_\xi)  \Omega(z_{34},\beta_3,\tau)\Omega(z_{43},\xi,\tau)
 \label{B43}\\
 &\! \! \! \!  \! \! \! \! - s_{13} \Omega(z_{34},\beta_3,\tau)  ( \partial_{\beta_1}{+} \partial_{\beta_2}{+}\partial_\xi) \Omega(z_{12},\beta_1,\tau) \Omega(z_{23},\beta_2,\tau)\Omega(z_{31},\xi,\tau)
\notag \\
 &\! \! \! \!  \! \! \! \! - s_{24} \Omega(z_{12},\beta_1,\tau)  ( \partial_{\beta_2}{+} \partial_{\beta_3}{+}\partial_\xi) \Omega(z_{23},\beta_2,\tau) \Omega(z_{34},\beta_3,\tau)\Omega(z_{42},\xi,\tau)
\notag \\
 &\! \! \! \!  \! \! \! \! - s_{14}  ( \partial_{\beta_1}{+} \partial_{\beta_2}{+} \partial_{\beta_3}{+}\partial_\xi) \Omega(z_{12},\beta_1,\tau)  \Omega(z_{23},\beta_2,\tau) \Omega(z_{34},\beta_3,\tau)\Omega(z_{41},\xi,\tau)
 \Big) \, \Big|_{\xi^0} \, .
 \notag
\end{align}
The cycles of Kronecker--Eisenstein series can be resolved using the earlier lemmata
(\ref{A25DP}) and (\ref{A27DP}) as well as the following identity explained in appendix \ref{app:0.3}
\begin{align}
&(\partial_{\beta_1}+\partial_{\beta_2}+\partial_{\beta_3}+\partial_{\xi})
 \Omega(z_{12},\beta_1,\tau)   \Omega(z_{23},\beta_2,\tau)  \Omega(z_{34},\beta_3,\tau) \Omega(z_{41},\xi,\tau)  \label{A28om} \\
 & \ \ = - \wp(\beta_1,\tau)   \Omega(z_{23},\beta_2{-}\beta_1,\tau)    \Omega(z_{34},\beta_3{-}\beta_1,\tau) \Omega(z_{41},\xi{-}\beta_1,\tau)  \notag \\
& \ \ \ \ \; \, - \wp(\beta_2,\tau)   \Omega(z_{12},\beta_1{-}\beta_2,\tau)    \Omega(z_{34},\beta_3{-}\beta_2,\tau) \Omega(z_{41},\xi{-}\beta_2,\tau)  \notag \\
& \ \ \ \ \; \, - \wp(\beta_3,\tau)   \Omega(z_{12},\beta_1{-}\beta_3,\tau)    \Omega(z_{23},\beta_2{-}\beta_3,\tau) \Omega(z_{41},\xi{-}\beta_3,\tau)  \notag \\
& \ \ \ \ \; \, -  \wp(\xi,\tau)   \Omega(z_{12},\beta_1{-}\xi,\tau)    \Omega(z_{23},\beta_2{-}\xi,\tau) \Omega(z_{34},\beta_3{-}\xi,\tau) \, .\notag
\end{align}
After extracting the $\xi^0$-order in (\ref{B43}), one can enforce by a
sequence of Fay identities (\ref{B34C}) that the first arguments of the
Kronecker--Eisenstein series match one of the six permutations of
$\Omega(z_{12},\ldots) \Omega(z_{23},\ldots) \Omega(z_{34},\ldots)$ in
$2,3,4$. This is analogous to repeated partial-fraction manipulations which
can be used to expand expressions of the form $\frac{1}{z_{ij}z_{jk} z_{kl}}$
and $\frac{1}{z_{ij}z_{ik} z_{il}}$ (with $i,j,k,l$ pairwise distinct) in a
six-dimensional basis. When the $\beta_j$ are rewritten in terms of the
$\eta_j$ variables, we arrive at a linear combination of various
$Z_{\vec{\eta}}^{\tau}(\ast |1,i,j,k)$ in (\ref{th4pt}):
\begin{align}
2\pi i \partial_\tau &Z_{\vec{\eta}}^{\tau}(\ast |1,2,3,4)  = \Big(   \frac{1}{2}
\sum_{j=2}^4 s_{1j} \partial_{\eta_j}^2
+ \frac{1}{2} \sum_{2\leq i<j}^4 s_{ij} (\partial_{\eta_i}{-}\partial_{\eta_j})^2
 - s_{12} \wp(\eta_2{+}\eta_3{+}\eta_4,\tau)
 \notag \\
& \ \ \ \ \ \ \ \ - (s_{13}{+}s_{23}) \wp(\eta_3{+}\eta_4,\tau)
-(s_{14}{+}s_{24}{+}s_{34}) \wp(\eta_4,\tau) - 2 \zeta_2 s_{1234}
\Big)  Z_{\vec{\eta}}^{\tau}(\ast  |1,2,3,4)   \notag \\
& \ \ \ \ + s_{13} \big[ \wp(\eta_2{+}\eta_3{+}\eta_4,\tau) - \wp(\eta_3{+}\eta_4,\tau)\big] \big(
Z_{\vec{\eta}}^{\tau}(\ast |1,3,2,4)
+ Z_{\vec{\eta}}^{\tau}(\ast  |1,3,4,2)  \big) \notag \\
& \ \ \ \ +s_{24} \big[ \wp(\eta_3{+}\eta_4,\tau) - \wp(\eta_4,\tau)\big]
Z_{\vec{\eta}}^{\tau}(\ast  |1,2,4,3) \label{B45} \\
& \ \ \ \ +s_{14} \big[ \wp(\eta_3{+}\eta_4,\tau) - \wp(\eta_2{+}\eta_3{+}\eta_4,\tau)\big]
Z_{\vec{\eta}}^{\tau}(\ast  |1,4,3,2)  \notag \\
& \ \ \ \ +s_{14}\big[ \wp(\eta_3{+}\eta_4,\tau) - \wp(\eta_4,\tau)\big]
\big(
Z_{\vec{\eta}}^{\tau}(\ast  |1,2,4,3)
+ Z_{\vec{\eta}}^{\tau}(\ast  |1,4,2,3)   \big)
\notag \\
&= \sum_{B\in S_3} D^{\tau}_{\vec{\eta}}(2,3,4|b_2,b_3,b_4) Z_{\vec{\eta}}^{\tau}(\ast |1,b_2,b_3,b_4)\, . \notag
\end{align}
The last line defines the first row of a $6\times 6$ differential operator $D^{\tau}_{\vec{\eta}}$,
\begin{align}
D^{\tau}_{\vec{\eta}}(2,3,4|2,3,4) &=
   \frac{1}{2}
\sum_{j=2}^4 s_{1j} \partial_{\eta_j}^2
+ \frac{1}{2} \sum_{2\leq i<j}^4 s_{ij} (\partial_{\eta_i}{-}\partial_{\eta_j})^2
 - s_{12} \wp(\eta_2{+}\eta_3{+}\eta_4,\tau)
 \notag \\
& \ \ \ - (s_{13}{+}s_{23}) \wp(\eta_3{+}\eta_4,\tau)
-(s_{14}{+}s_{24}{+}s_{34}) \wp(\eta_4,\tau)  - 2 \zeta_2 s_{1234}
 \notag \\
D^{\tau}_{\vec{\eta}}(2,3,4|2,4,3) &= (s_{14}{+}s_{24}) \big[ \wp(\eta_3{+}\eta_4,\tau) - \wp(\eta_4,\tau)\big] \notag \\
D^{\tau}_{\vec{\eta}}(2,3,4|3,2,4) &= s_{13} \big[ \wp(\eta_2{+}\eta_3{+}\eta_4,\tau) - \wp(\eta_3{+}\eta_4,\tau)\big] \label{4difop} \\
D^{\tau}_{\vec{\eta}}(2,3,4|3,4,2) &= s_{13} \big[ \wp(\eta_2{+}\eta_3{+}\eta_4,\tau) - \wp(\eta_3{+}\eta_4,\tau)\big] \notag \\
D^{\tau}_{\vec{\eta}}(2,3,4|4,2,3) &= s_{14}\big[ \wp(\eta_3{+}\eta_4,\tau) - \wp(\eta_4,\tau)\big]  \notag \\
D^{\tau}_{\vec{\eta}}(2,3,4|4,3,2) &= s_{14} \big[ \wp(\eta_3{+}\eta_4,\tau) - \wp(\eta_2{+}\eta_3{+}\eta_4,\tau)\big] \, ,\notag
\end{align}
and the remaining rows are obtained from relabelings of both $s_{ij}$ and $\eta_j$ w.r.t.\ 2,3,4.
The $6\times 6$ representation of the derivations resulting from the expansion (\ref{intro3b}) in terms
of Eisenstein series can be conveniently organized as
\begin{align}
&r_{\vec{\eta}}(\ep_k) = \delta_{k,0}  \Big( 2 \zeta_2 s_{1234} - \frac{1}{2} \sum_{j=2}^4 s_{1j} \partial_{\eta_j}^2 -  \frac{1}{2}\sum_{2\leq i <j}^4 s_{ij} (\partial_{\eta_i} {-} \partial_{\eta_j})^2 \Big) 1_{6\times 6}  + \eta_{234}^{k-2} r_{\vec{\eta}}(e_{234})
\label{DER14}\\
&\  + \eta_{23}^{k-2} r_{\vec{\eta}}(e_{23})
+ \eta_{24}^{k-2}  r_{\vec{\eta}}(e_{24})
+ \eta_{34}^{k-2}  r_{\vec{\eta}}(e_{34})
+ \eta_{2}^{k-2}  r_{\vec{\eta}}(e_{2})
+ \eta_{3}^{k-2}  r_{\vec{\eta}}(e_{3})
+ \eta_{4}^{k-2}  r_{\vec{\eta}}(e_{4})\,,\quad k\neq2 \notag
\end{align}
with $r_{\vec{\eta}}(\ep_2) =0$, and we used the shorthands
$\eta_{ij}\equiv \eta_i{+}\eta_j$ and $\eta_{234} \equiv \eta_2{+}\eta_3{+}\eta_4$. The
$r_{\vec{\eta}}(e_{\ldots})$ refer to $6 \times 6$ matrices whose entries are
linear in the $s_{ij}$. Representative examples for their explicit form are
\begin{align}
 r_{\vec{\eta}}(e_{234}) &= \left( \begin{array}{cccccc}
s_{12} &0 &-s_{13} &-s_{13} &0 &s_{14} \\
0 &s_{12} &0 &s_{13} &-s_{14} &-s_{14} \\
-s_{12} &-s_{12} &s_{13} &0 &s_{14} &0 \\
0 &s_{12} &0 &s_{13} &-s_{14} &-s_{14} \\
-s_{12} &-s_{12} &s_{13} &0 &s_{14} &0 \\
s_{12} &0 &-s_{13} &-s_{13} &0 &s_{14}
\end{array} \right) \notag \\
 r_{\vec{\eta}}(e_{34}) &= \left( \begin{array}{cccccc}
s_{13}{+}s_{23} &-s_{14}{-}s_{24} &s_{13} &s_{13} &-s_{14} &-s_{14} \\
{-}s_{13}{-}s_{23} &s_{14}{+}s_{24} &-s_{13} &-s_{13} &s_{14} &s_{14} \\
0 &0 &0 &0 &0 &0 \\
0 &0 &0 &0 &0 &0 \\
0 &0 &0 &0 &0 &0 \\
0 &0 &0 &0 &0 &0
\end{array} \right)\label{DER15} \\
r_{\vec{\eta}}(e_{4}) &= \left( \begin{array}{cccccc}
s_{14}{+}s_{24}{+}s_{34} &s_{14}{+}s_{24} &0 &0 &s_{14} &0 \\
0 &0 &0 &0 &0 &0 \\
0 &0 &s_{14}{+}s_{24}{+}s_{34} &s_{14}{+}s_{34} &0 &s_{14} \\
0 &0 &0 &0 &0 &0 \\
0 &0 &0 &0 &0 &0 \\
0 &0 &0 &0 &0 &0
\end{array} \right)
 \, , \notag
\end{align}
and the expressions for
$r_{\vec{\eta}}(e_{23})$, $r_{\vec{\eta}}(e_{24})$, $r_{\vec{\eta}}(e_{2})$ and
$r_{\vec{\eta}}(e_{3})$ obtained from relabelings are given in appendix \ref{app:B}.
The solution to (\ref{B45}) via Picard iteration
\begin{align}
Z^\tau_{\vec{\eta}}(\ast |1,2,3,4)&=    \sum_{r=0}^{\infty}  \sum_{k_1,k_2,\ldots,k_r \atop{=0,4,6,8,\ldots} }  \prod_{j=1}^r (k_j{-}1)\, \gamma(k_1,k_2,\ldots,k_r|\tau)   \label{someint44}\\
& \times \sum_{B \in S_3} r_{\vec{\eta}}(\ep_{k_r} \ldots \ep_{k_2} \ep_{k_1} )_{234}{}^{B} Z^{i\infty}_{\vec{\eta}}(\ast |1,B)
\notag
 \end{align}
will be completed by the discussion of the initial values
$Z^{i\infty}_{\vec{\eta}}(\ast |1,B)$  in sections
\ref{sec:5.5} and \ref{sec:6.3}.


\subsection{Five points}
\label{sec:4.3}

The five-point $A$-cycle integrals (\ref{intro2}) with $\vec{\eta}=\eta_2,\eta_3,\eta_4,\eta_5$ are given by
\begin{align}
Z_{\vec{\eta}}^{\tau}(\ast |1,2,3,4,5) = \int_{{\cal C}(\ast)} &\dd z_2\, \dd z_3 \, \dd z_4\, \dd z_5 \,\Omega(z_{12},\eta_2{+}\eta_3{+}\eta_4{+}\eta_5,\tau)
\Omega(z_{23},\eta_3{+}\eta_4{+}\eta_5,\tau)  \notag \\
& \ \ \ \times \Omega(z_{34},\eta_4{+}\eta_5,\tau)
\Omega(z_{45},\eta_5,\tau) \, \te{KN}^\tau_{12345}
\end{align}
and may be applied to 20, 30 and 24 cyclically inequivalent integration cycles
${\cal C}\big(\begin{smallmatrix} l,m \\ i,j,k \end{smallmatrix}\big)$,
${\cal C}\big(\begin{smallmatrix} m \\ i,j,k,l \end{smallmatrix}\big)$ and ${\cal C}( i,j,k,l,m)$, respectively.
Their $\tau$-derivatives can be computed by iterating the steps of the earlier sections and
inserting the five-point version of the lemma (\ref{A29om}) to resolve the five-cycle of Kronecker--Eisenstein
series in the last line of
\begin{align}
2\pi i \partial_{\tau } &Z_{\vec{\eta}}^{\tau}(\ast |1,2,3,4,5)  = \int_{{\cal C}(\ast)} \dd z_2 \, \dd z_3 \, \dd z_4 \,  \dd z_5 \, {\rm KN}_{12345}^{\tau} \label{B52}  \\
 &\times \Big( -2 \zeta_2 s_{12345} \Omega(z_{12},\beta_1)\Omega(z_{23},\beta_2)\Omega(z_{34},\beta_3)\Omega(z_{45},\beta_4)
 \notag \\
& \ \ \ - s_{12} \Omega(z_{23},\beta_2)\Omega(z_{34},\beta_3)\Omega(z_{45},\beta_4) ( \partial_{\beta_1}{+}\partial_\xi) \Omega(z_{12},\beta_1) \Omega(z_{21},\xi)
 \notag \\
 & \ \ \ - s_{23}\Omega(z_{12},\beta_1) \Omega(z_{34},\beta_3)\Omega(z_{45},\beta_4) ( \partial_{\beta_2}{+}\partial_\xi)  \Omega(z_{23},\beta_2)\Omega(z_{32},\xi)
 \notag \\
 & \ \ \ - s_{34} \Omega(z_{12},\beta_1) \Omega(z_{23},\beta_2)\Omega(z_{45},\beta_4) ( \partial_{\beta_3}{+}\partial_\xi) \Omega(z_{34},\beta_3)\Omega(z_{43},\xi)
 \notag \\
 & \ \ \ - s_{45} \Omega(z_{12},\beta_1)\Omega(z_{23},\beta_2)\Omega(z_{34},\beta_3) ( \partial_{\beta_4}{+}\partial_\xi)  \Omega(z_{45},\beta_4)\Omega(z_{54},\xi)
 \notag \\
 & \ \ \ - s_{13} \Omega(z_{34},\beta_3)\Omega(z_{45},\beta_4) ( \partial_{\beta_1}{+} \partial_{\beta_2}{+}\partial_\xi) \Omega(z_{12},\beta_1) \Omega(z_{23},\beta_2)\Omega(z_{31},\xi)
 \notag \\
  & \ \ \ - s_{24} \Omega(z_{12},\beta_1)\Omega(z_{45},\beta_4) ( \partial_{\beta_2}{+} \partial_{\beta_3}{+}\partial_\xi)  \Omega(z_{23},\beta_2) \Omega(z_{34},\beta_3)\Omega(z_{42},\xi)
 \notag \\
   & \ \ \ - s_{35} \Omega(z_{12},\beta_1) \Omega(z_{23},\beta_2) ( \partial_{\beta_3}{+} \partial_{\beta_4}{+}\partial_\xi)   \Omega(z_{34},\beta_3) \Omega(z_{45},\beta_4)\Omega(z_{53},\xi)
 \notag \\
 & \ \ \ - s_{14} \Omega(z_{45},\beta_4) ( \partial_{\beta_1}{+} \partial_{\beta_2}{+} \partial_{\beta_3}{+}\partial_\xi) \Omega(z_{12},\beta_1) \Omega(z_{23},\beta_2) \Omega(z_{34},\beta_3)\Omega(z_{41},\xi)
 \notag \\
 & \ \ \ - s_{25} \Omega(z_{12},\beta_1)( \partial_{\beta_2}{+} \partial_{\beta_3}{+} \partial_{\beta_4}{+}\partial_\xi)  \Omega(z_{23},\beta_2) \Omega(z_{34},\beta_3)  \Omega(z_{45},\beta_4)  \Omega(z_{52},\xi)
 \notag \\
 & \ \ \ - s_{15} ( \partial_{\beta_1}{+} \partial_{\beta_2}{+} \partial_{\beta_3}{+} \partial_{\beta_4}{+}\partial_\xi) \Omega(z_{12},\beta_1) \Omega(z_{23},\beta_2) \Omega(z_{34},\beta_3)  \Omega(z_{45},\beta_4)   \Omega(z_{51},\xi) \Big) \, \Big|_{\xi^0} \, ,
 \notag
\end{align}
where $\beta_i = \sum_{j=i+1}^5 \eta_{j}$ for $i=1,2,3,4$. A sequence of Fay relations (\ref{B34C})
allows to cast the products of $\Omega(z_{ij},\ldots)$ due to (\ref{A29om}) into a 24-dimensional basis.
In view of the $n$-point generalization, we present the $24\times 24$ differential operator $D^{\tau}_{\vec{\eta}}$
\begin{align}
2\pi i \partial_\tau Z_{\vec{\eta}}^{\tau}(\ast |1,2,3,4,5) &= \sum_{B \in S_4} D^{\tau}_{\vec{\eta}}(2,3,4,5|b_2,b_3,b_4,b_5) Z_{\vec{\eta}}^{\tau}(\ast |1,b_2,b_3,b_4,b_5) \label{B455}
\end{align}
from a slightly different angle and express the $\tau$-derivative
in terms of an overcomplete set of $n!$ integrals with leg $1$ in
an arbitrary position of the second entry:
\begin{align}
2\pi i &\partial_\tau Z_{\vec{\eta}}^{\tau}(\ast |1,2,3,4,5) = \Big(
\frac{1}{2} \sum_{j=2}^5  s_{1j} \partial_{\eta_j}^2
+ \frac{1}{2} \sum_{2\leq i <j}^5  s_{ij} (\partial_{\eta_i}{-}\partial_{\eta_j})^2 - 2 \zeta_2 s_{12345} \Big) Z_{\vec{\eta}}^{\tau}(\ast |1,2,3,4,5)
\notag \\
&\! \! - \wp(\eta_{2345},\tau) \Big( s_{12}  Z_{\vec{\eta}}^{\tau}(\ast |1,2,3,4,5)
- s_{13} \big( Z_{\vec{\eta}}^{\tau}(\ast |1,3,2,4,5)
{+} Z_{\vec{\eta}}^{\tau}(\ast |1,3,4,2,5)
{+} Z_{\vec{\eta}}^{\tau}(\ast |1,3,4,5,2) \big)
\notag \\
& \ \ \ \ \ \ \  + s_{14} \big(Z_{\vec{\eta}}^{\tau}(\ast |1,4,5,3,2)
{+}Z_{\vec{\eta}}^{\tau}(\ast |1,4,3,5,2)
{+}Z_{\vec{\eta}}^{\tau}(\ast |1,4,3,2,5) \big)
- s_{15}  Z_{\vec{\eta}}^{\tau}(\ast |1,5,4,3,2)
\Big) \notag \\
&\! \! - \wp(\eta_{345},\tau) \Big( s_{23}  Z_{\vec{\eta}}^{\tau}(\ast |1,2,3,4,5)
- s_{24} \big( Z_{\vec{\eta}}^{\tau}(\ast |1,2,4,3,5)
{+}Z_{\vec{\eta}}^{\tau}(\ast |1,2,4,5,3) \big)
+ s_{25}  Z_{\vec{\eta}}^{\tau}(\ast |1,2,5,4,3)
\notag \\
& \ \ \ \ \ \ \    -s_{13}  Z_{\vec{\eta}}^{\tau}(\ast |2,1,3,4,5)
+ s_{14} \big(Z_{\vec{\eta}}^{\tau}(\ast |2,1,4,3,5)
{+}Z_{\vec{\eta}}^{\tau}(\ast |2,1,4,5,3) \big)
- s_{15}  Z_{\vec{\eta}}^{\tau}(\ast |2,1,5,4,3)
\Big) \notag \\
&\! \! - \wp(\eta_{45},\tau) \Big( s_{34} Z_{\vec{\eta}}^{\tau}(\ast |1,2,3,4,5)
- s_{24} \big(Z_{\vec{\eta}}^{\tau}(\ast |1,3,2,4,5)
{+}Z_{\vec{\eta}}^{\tau}(\ast |3,1,2,4,5) \big)
 + s_{14}  Z_{\vec{\eta}}^{\tau}(\ast |3,2,1,4,5)
 \notag \\
& \ \ \ \ \ \ \   -s_{35} Z_{\vec{\eta}}^{\tau}(\ast |1,2,3,5,4)
+ s_{25} \big(Z_{\vec{\eta}}^{\tau}(\ast |1,3,2,5,4)
{+}Z_{\vec{\eta}}^{\tau}(\ast |3,1,2,5,4) \big)
 - s_{15} Z_{\vec{\eta}}^{\tau}(\ast |3,2,1,5,4)
  \Big) \notag \\
&\! \! - \wp(\eta_{5},\tau) \Big( s_{45} Z_{\vec{\eta}}^{\tau}(\ast |1,2,3,4,5)
 - s_{35} \big(Z_{\vec{\eta}}^{\tau}(\ast |1,2,4,3,5)
 {+}Z_{\vec{\eta}}^{\tau}(\ast |1,4,2,3,5)
 {+}Z_{\vec{\eta}}^{\tau}(\ast |4,1,2,3,5) \big)
  \notag \\
& \ \ \ \ \ \ \   + s_{25} \big(Z_{\vec{\eta}}^{\tau}(\ast |4,3,1,2,5)
{+}Z_{\vec{\eta}}^{\tau}(\ast |4,1,3,2,5)
{+}Z_{\vec{\eta}}^{\tau}(\ast |1,4,3,2,5) \big)
- s_{15}  Z_{\vec{\eta}}^{\tau}(\ast |4,3,2,1,5)
\Big)  \, . \label{B459}
\end{align}
In an $n$-point context it will be useful to define
\beq
\eta_1 \equiv - (\eta_2+\eta_3+\ldots+\eta_n)\,,
\label{n5ptA}
\eeq
since a repeated application of Fay identities can be used to show that
$Z_{\vec{\eta}}^{\tau}(\ast |P\shuffle Q)=0$ for $P,Q\neq\emptyset$.
Therefore,
the $n!$ integrands of $Z_{\vec{\eta}}^{\tau}(\ast |a_1,a_2,\ldots,a_n), \ A \in S_n,$ can be
reduced to an $(n{-}1)!$-element basis of $Z_{\vec{\eta}}^{\tau}(\ast |1,c_2,\ldots,c_n), \ C \in S_{n-1}$.
More precisely \cite{Schocker},
\beq
Z_{\vec{\eta}}^{\tau}(\ast |A,1,B) = (-1)^{|A|} Z_{\vec{\eta}}^{\tau}\big(\ast |1,(A^t \shuffle B) \big) \,,
\label{n5ptB}
\eeq
where $|A|$ is the number of letters in the word $A=a_1 a_2\ldots a_{|A|}$, and $A^t=a_{|A|}\ldots a_2
a_1$ is obtained by reversing the order of the letters. The shuffle symbol yields a formal sum over all words
that preserve the order of $A^t$ and $B$, and any object labelled by words is understood to obey a linearity
property such as $Z_{\vec{\eta}}^{\tau}(\ast |1,C{+}D )=Z_{\vec{\eta}}^{\tau}(\ast |1,C)+Z_{\vec{\eta}}^{\tau}(\ast |1, D )$.

Remarkably, (\ref{n5ptB}) is the same shuffle symmetry obeyed by the
tree-level functions
$\frac{1}{z_{12}z_{23}\ldots z_{k-1,k}}$.
This can be seen from
the fact that Fay identities among
$\Omega(z_{ij},\ldots)\Omega(z_{jk},\ldots)$ mirror the partial-fraction
manipulations among $\frac{1}{z_{ij} z_{jk}}$ when disregarding the second
argument. The latter in turn are fixed by requiring both sides of the Fay
identity to have the same behavior under antiholomorphic derivatives
$\partial_{\bar z} \Omega(z,\eta,\tau) = - \frac{ \pi \eta}{\Im \tau }
\Omega(z,\eta,\tau) $.

By virtue of the identity (\ref{n5ptB}), the right-hand side of
(\ref{B459}) can be uniquely expanded in the $24$-element basis of
$Z_{\vec{\eta}}^{\tau}\big(\ast |1,i,j,k,l \big)$. The result is noted in
appendix \ref{app:A}, and one can read off the first row of the $24\times 24$
differential operator $D^\tau_{\vec{\eta}}$ in (\ref{B455}). Similarly, the
expansion (\ref{intro3b}) in terms of Eisenstein series straightforwardly
yields the $24\times 24$ matrix representation of the derivations which enters
the solution of (\ref{B455}) via Picard iteration,
\begin{align}
Z^\tau_{\vec{\eta}}(\ast |1,2,3,4,5)&=    \sum_{r=0}^{\infty}  \sum_{k_1,k_2,\ldots,k_r \atop{=0,4,6,8,\ldots} }  \prod_{j=1}^r (k_j{-}1)\, \gamma(k_1,k_2,\ldots,k_r|\tau)   \label{someint55}\\
&    \times \sum_{B \in S_4} r_{\vec{\eta}}(\ep_{k_r} \ldots \ep_{k_2} \ep_{k_1} )_{2345}{}^{B} Z^{i\infty}_{\vec{\eta}}(\ast |1,B)  \, .
\notag
 \end{align}
The initial values $Z^{i\infty}_{\vec{\eta}}(\ast |1,B)$ are determined by the discussion in later sections.


\subsection{$n$ points}
\label{sec:4.4}

There is no obstacle to extending the rigorous computations of the previous sections to the $\tau$-derivatives
of higher-point $A$-cycle integrals (\ref{intro2}). The mixed heat equation and the Koba--Nielsen derivatives
lead to the intermediate step
\begin{align}
2\pi i &\partial_{\tau } Z^\tau_{\vec{\eta}}(\ast|1,2,\ldots,n) = \int_{{\cal C}(\ast)}\dd z_2 \, \dd z_3 \, \ldots\, \dd z_n \,
{\rm KN}_{12\ldots n}^\tau  \Big( -2 \zeta_2 s_{12\ldots n}
\prod_{k=1}^{n-1}\Omega(z_{k,k+1},\beta_k)
 \notag \\
& \ \ \ \ \ \ \ \ \ \ \ - \sum_{1\leq i<j}^n s_{ij}(\partial_{\beta_i}{+}
\partial_{\beta_{i+1}}{+}\ldots{+}\partial_{\beta_{j-1}} {+} \partial_{\xi}) \Omega(z_{ji},\xi)
\prod_{k=1}^{n-1}\Omega(z_{k,k+1},\beta_k)
 \Big) \, \Big|_{\xi^0}\, ,
 \label{tocheck}
\end{align}
where $\beta_i = \sum_{j=i+1}^n \eta_j$ for $i=1,2,\ldots,n{-}1$. And the leftover work in extracting the
form of the differential operator $D^\tau_{\vec{\eta}}$ in (\ref{intro3}) is to resolve the cycles of $\Omega(z_{ij},\ldots)$
through the lemma (\ref{A29om}) and to reduce the result to the $(n{-}1)!$ basis of
$Z^\tau_{\vec{\eta}}(\ast|1,B), \ B \in S_{n-1}$ via Fay identities.

Instead of performing these tedious but straightforward calculations on a case-by-case basis
at $n{\geq} 6$ points, we take inspiration from certain patterns in the $(n{\leq}5)$-point results
to propose a closed formula for
$ 2\pi i \partial_{\tau } Z^\tau_{\vec{\eta}}(\ast|1,2,\ldots,n)$. These patterns are
based on a combinatorial operation dubbed the ``S-map'' \cite{Mafra:2014oia, Mafra:2014gsa}.
For any object $X(A)$ labelled by words $A=a_1 a_2 \ldots a_{|A|}$, the S-map is defined by
\begin{align}
X(S[A,B]) &=  \sum_{i=1}^{|A|} \sum_{j=1}^{|B|} (-1)^{i-j+|A|-1}
s_{a_i b_j}  \label{B462} \\
&\ \ \times X\big((a_1 a_2\ldots a_{i-1} \shuffle a_{|A|} a_{|A|-1}\ldots a_{i+1}),a_i,b_j,
(b_{j-1}\ldots b_2 b_1 \shuffle b_{j+1} \ldots b_{|B|}) \big)\, , \notag
\end{align}
for instance
\beq
X(S[1,2]) = s_{12} X(1,2) \, , \ \ \ \
X(S[12,3]) = s_{23} X(1,2,3) - s_{13} X(2,1,3) \, .
  \label{B462ex}
 \eeq
The S-map is antisymmetric $X(S[A,B]) = - X(S[B,A])$ if $X$ obeys the shuffle
identity
$X(A,1,B)=(-1)^{|A|} X(1,(A^t \shuffle B))$ as in (\ref{n5ptB}).

In the previous section, we have chosen to present the $\tau$-derivative at five points
in the form (\ref{B459}) because it can be compactly written in terms of the above
S-map (\ref{B462}):
\begin{align}
2\pi i \partial_\tau &Z_{\vec{\eta}}^{\tau}(\ast |1,2,3,4,5) = \Big(
\frac{1}{2} \sum_{j=2}^5  s_{1j} \partial_{\eta_j}^2
+ \frac{1}{2}\sum_{2\leq i <j}^5   s_{ij}(\partial_{\eta_i}{-}\partial_{\eta_j})^2 - 2 \zeta_2 s_{12345} \Big)
 Z_{\vec{\eta}}^{\tau}(\ast |1,2,3,4,5)
\notag  \\
& \ \ \ \ \ - \wp(\eta_{2}{+}\eta_{3}{+}\eta_{4}{+}\eta_5,\tau) Z_{\vec{\eta}}^{\tau}(\ast |S[1,2345])
 - \wp(\eta_{3}{+}\eta_{4}{+}\eta_5,\tau) Z_{\vec{\eta}}^{\tau}(\ast | S[12,345])  \label{B459a} \\
 &\ \ \ \ \ - \wp(\eta_{4}{+}\eta_5,\tau) Z_{\vec{\eta}}^{\tau}(\ast | S[123,45] )
- \wp(\eta_{5},\tau) Z_{\vec{\eta}}^{\tau}(\ast | S[1234,5])
 \, .\notag
\end{align}
The structure of the right-hand side harmonizes with a rewriting of the $(n{\leq} 4)$-point results
(\ref{B16alt}), (\ref{B38}) and (\ref{B45}) as
\begin{align}
2\pi i \partial_\tau Z_{\eta_2}^{\tau}(\ast |1,2) &= \Big(
  \frac{1}{2}  s_{12}\partial_{\eta_2}^2 - 2 \zeta_2 s_{12}  \Big) Z_{\eta_2}^{\tau}(\ast |1,2) - \wp(\eta_2,\tau) Z_{\eta_2}^{\tau}(\ast |S[1,2])
\notag  \\
2\pi i \partial_\tau Z_{\vec{\eta}}^{\tau}(\ast |1,2,3) &= \Big(
 \frac{1}{2} \sum_{j=2}^3 s_{1j} \partial_{\eta_j}^2
+ \frac{1}{2}s_{23} (\partial_{\eta_2}{-}\partial_{\eta_3})^2 - 2 \zeta_2 s_{123} \Big) Z_{\vec{\eta}}^{\tau}(\ast |1,2,3)
\notag \\
&\ \ - \wp(\eta_{2}{+}\eta_3,\tau)  Z_{\vec{\eta}}^{\tau}(\ast | S[1,23] )
- \wp(\eta_{3},\tau) Z_{\vec{\eta}}^{\tau}(\ast | S[12,3])  \label{B456a} \\
2\pi i \partial_\tau Z_{\vec{\eta}}^{\tau}(\ast |1,2,3,4) &= \Big(
\frac{1}{2}  \sum_{j=2}^4 s_{1j} \partial_{\eta_j}^2
+ \frac{1}{2}\sum_{2\leq i <j}^4 s_{ij} (\partial_{\eta_i}{-}\partial_{\eta_j})^2 - 2 \zeta_2 s_{1234} \Big) Z_{\vec{\eta}}^{\tau}(\ast |1,2,3,4)
\notag \\
& \ \ - \wp(\eta_{2}{+}\eta_{3}{+}\eta_4,\tau) Z_{\vec{\eta}}^{\tau}(\ast |S[1,234])
- \wp(\eta_{3}{+}\eta_{4},\tau) Z_{\vec{\eta}}^{\tau}(\ast |S[12,34])  \notag  \\
& \ \
 - \wp(\eta_{4},\tau) Z_{\vec{\eta}}^{\tau}(\ast |S[123,4]) \, ,\notag
\end{align}
where the equivalence to earlier results can be checked by means of
(\ref{n5ptB}). By inspecting the pattern in (\ref{B456a}) and (\ref{B459a}), it is natural to propose
the following generalization to $n$ points
\begin{align}
2\pi i \partial_\tau Z_{\vec{\eta}}^{\tau}(\ast |1,2,\ldots,n) &= \Big(
\frac{1}{2}  \sum_{j=2}^n s_{1j}  \partial_{\eta_j}^2
+ \frac{1}{2}  \sum_{2\leq i <j}^n s_{ij}  (\partial_{\eta_i}{-}\partial_{\eta_j})^2 - 2 \zeta_2 s_{123\ldots n}  \Big)
Z_{\vec{\eta}}^{\tau}(\ast |1,2,\ldots,n)
\notag \\
& \! \! - \sum_{j=2}^n \wp(\eta_{j}{+}\eta_{j+1}{+}\ldots{+}\eta_{n},\tau)
Z_{\vec{\eta}}^{\tau}(\ast | S[12\ldots j{-}1\, ,\, j(j{+}1)\ldots n])
  \, ,
 \label{B460a}
\end{align}
which is conjectural at $n{\geq}6$. There is no bottleneck in checking (\ref{B460a}) at fixed multiplicities $n=6,7,\ldots$
by starting from the intermediate step (\ref{tocheck}) and proceeding as outlined above. By carrying
out the S-map (\ref{B462}) and applying Fay identities (\ref{n5ptB}) to the result, (\ref{B460a}) yields a well-defined
proposal for the differential operator $D^\tau_{\vec{\eta}}$ at $n$ points. Accordingly, its expansion (\ref{intro3b}) in
Eisenstein series leads to concrete $(n{-}1)! \times (n{-}1)!$ representations of the derivations $r_{\vec{\eta}}(\ep_k)$.
Like this, the appearance of iterated Eisenstein integrals in the $\ap$-expansion of
$Z_{\vec{\eta}}^{\tau}(\ast |1,A)$ can be made fully explicit from (\ref{intro1}), and it remains to fix the
initial values $Z_{\vec{\eta}}^{i \infty}(\ast |1,B)$ that introduce MZVs.

The $r_{\vec{\eta}}(\ep_k)$ resulting from (\ref{B460a}) are linear in $\ap$ since each term in the first line
and the expansion of the S-map $Z_{\vec{\eta}}^{\tau}(\ast | S[12\ldots j{-}1\, ,\, j(j{+}1)\ldots n])$ features one
factor of $s_{ij}$, see (\ref{B462}). This will be important in section \ref{sec:others1}, where the structure of the
$\ap$-expansion (\ref{intro1}) is argued to imply uniform transcendentality of the $Z_{\vec{\eta}}^{\tau}(\ast |1,A)$.

Finally, the absence of the twisted Eisenstein series $f^{(k)}(\tauh,\tau)$ on the right-hand side of (\ref{B460a})
implies that the $\ap$-expansion (\ref{intro1}) of non-planar $A$-cycle integrals cannot involve any twisted eMZVs
(which would naively arise from the expansion method in \cite{Broedel:2017jdo}).
As a hallmark of twisted eMZVs, their $\tau$-derivatives involve Eisenstein series of congruence
subgroups of ${\rm SL}_2(\ZZ)$ \cite{Broedel:2017jdo}, i.e.\ $f^{(k)}(\tauh,\tau)$ in case of the twists
of the non-planar Green function. Still, the absence of twisted eMZVs by itself does not guarantee that
the iterated Eisenstein integrals in non-planar $\ap$-expansions (\ref{intro1}) conspire to eMZVs, see
the discussion in the next subsection.


\subsection{Representations of the derivation algebra}
\label{sec:4.5}

We shall now clarify to which extent the above matrices $r_{\vec{\eta}}(\ep_k)$ are known to
preserve the commutation relations of the derivations, i.e.\ to which extent they qualify to be
called a ``representation''.

The reasoning will rely on the fact that
the planar instances of the $A$-cycle integrals (\ref{intro2}) can be $\ap$-expanded in terms of eMZVs whose
coefficients are $\QQ$-linear combinations of MZVs. This can be seen by applying the methods of
\cite{Broedel:2014vla, Broedel:2019vjc} to each component integral over
$f^{(m_1)}_{12}f^{(m_2)}_{23}\ldots f^{(m_{n-1})}_{n-1,n}, \ m_j \in \NN_0 $ that arises
in the $\eta_j$-expansion: When integrating one puncture after the other, Fay identities can be used to have
only one instance of the integration variable among the subscripts of the $f^{(m_k)}_{ij}$ in each step. Regardless
on the accompanying monomials in planar Green functions (\ref{B1}), the integral over each puncture can therefore be
performed within the space of elliptic iterated integrals (\ref{elliter})\footnote{See \cite{Broedel:2014vla} for details on
the ``z-removal'' in intermediate steps and \cite{Broedel:2019vjc} for examples on the handling of kinematic poles.}.
The integration over the last puncture sets the argument
of the elliptic iterated integrals $\Gamma(\ldots;z)$ to $z=1$ and yields eMZVs by (\ref{A2emzv}).
At higher multiplicity or order in $\ap$, this method may be slowed down by a nesting of
kinematic poles and the multitude of relations among eMZVs. Still, it clarifies that the iterated Eisenstein integrals
in the $\ap$-expansion (\ref{intro1}) of planar $n$-point $A$-cycle integrals must conspire to eMZVs,
with admixtures of MZVs through ``z-removal'' identities\footnote{The appearance of MZVs in ``z-removal'' identities
is exemplified by \cite{Broedel:2014vla}
\[
\Gamma\left( \begin{smallmatrix} 1 &1  \\ z&0  \end{smallmatrix} ;  z |\tau\right)
=2\Gamma\left( \begin{smallmatrix} 0 &2  \\ 0&0  \end{smallmatrix} ;  z |\tau\right)
+\Gamma\left( \begin{smallmatrix} 2 &0  \\ 0&0  \end{smallmatrix} ;  z |\tau\right)
-2 \Gamma\left( \begin{smallmatrix} 1 &1  \\ 0&0  \end{smallmatrix} ;  z |\tau\right)+\zeta_2\, .
\]}.

We can now demonstrate that, when acting on the initial value $Z^{i \infty}_{\vec{\eta}}$ of planar $A$-cycle integrals,
any failure of the $r_{\vec{\eta}}(\ep_k)$ to satisfy the commutation relations
of the derivations would lead to a contradiction: As detailed in \cite{Broedel:2015hia}, each relation in the
derivation algebra selects a linear combination of iterated Eisenstein integrals that cannot be realized as
an eMZV, see section \ref{sec:2.5}. If one of these relations was violated by the $r_{\vec{\eta}}(\ep_k)$
acting on $Z^{i \infty}_{\vec{\eta}}$, then (\ref{intro1}) would introduce such a non-eMZV
combination of iterated Eisenstein integrals (say an isolated $\gamma(6,8|\tau)$) into
the $\ap$-expansion of some component integral over $f^{(m_1)}_{12}f^{(m_2)}_{23}\ldots f^{(m_{n-1})}_{n-1,n} $,
cf.\ section 4.3 of \cite{Broedel:2015hia}.
Since all the $\tau$-dependence in the $\ap$-expansion must stem from eMZVs by the previous paragraph,
we conclude that setting $\ep_k \rightarrow r_{\vec{\eta}}(\ep_k)$ in vanishing commutators
must annihilate the planar initial values $Z^{i \infty}_{\vec{\eta}}$. I.e.\ when restricted to act on the
subspace of functions of $\eta_j$ set by the initial values, the
$r_{\vec{\eta}}(\ep_k)$ furnish a matrix representation of the derivation algebra.

One might wonder whether this annihilation is a peculiarity of the $\eta_j$-dependence
of $Z^{i\infty}_{\vec{\eta}}$, say the factor of $\cot(\pi \eta_2)$
in the planar two-point case (\ref{someintL}). By the commutativity
$r_{\eta_2}([\ep_{k_1},\ep_{k_2}]) = 0 \ \forall \ k_1,k_2>0$ at two points, most of the nontrivial checks of
$r_{\vec{\eta}}(\ep_k)$ preserving the derivation algebra have been performed
at $n{\geq}3$ points, based on the matrix representations in (\ref{DER11}) and (\ref{DER14}).
For instance, we have verified that arbitrary functions of $\eta_j$
are annihilated when setting $\ep_k \rightarrow r_{\vec{\eta}}(\ep_k)$ in vanishing
combinations of commutators of the form
\begin{itemize}
\item $[\ep_{k_1},\ep_{k_2}]$ at $k_1{+}k_2\leq 30$ and $n=3,4$,
\item $[\ep_{\ell_1},[\ep_{\ell_2},\ep_{\ell_3}]]$ at $\ell_1{+}\ell_2{+}\ell_3\leq 30$ and $n=3,4$,
\item $[\ep_{p_1},[\ep_{p_2},[\ep_{p_3},\ep_{p_4}]]]$ at $p_1{+}p_2{+}p_3{+}p_4\leq 26,\ n=3$
and $p_1{+}p_2{+}p_3{+}p_4\leq 18,\ n=4$,
\end{itemize}
where some of the four-point checks rely on numerical methods.

For non-planar $A$-cycle integrals, the integration methods of \cite{Broedel:2017jdo, Broedel:2019vjc}
guarantee that their $\ap$-expansions are expressible in terms of twisted eMZVs. The result (\ref{intro1}) of
our differential-equation method implies these twisted eMZVs to
conspire to iterated Eisenstein integrals of ${\rm SL}_2(\ZZ)$, and one could wonder if they are necessarily
eMZVs on these grounds. However, at the time of writing, we cannot rule out the following loophole beyond
the range of our explicit checks: Setting $\ep_k \rightarrow r_{\vec{\eta}}(\ep_k)$ in some vanishing combination
of $\ep_k$-commutators might fail to annihilate the initial value
$Z^{i\infty}_{\vec{\eta}}\big( \begin{smallmatrix} Q \\ P \end{smallmatrix} |\cdot \big)$ of
a {\it non-planar} $A$-cycle integral with $P,Q\neq \emptyset$. This is related to the open question
whether all iterated Eisenstein integrals expressible in terms of twisted eMZVs are necessarily eMZVs.


\section{Planar genus-one integrals at the cusp}
\label{sec:5}

Starting from this section, the proposal (\ref{B460a}) for the $\tau$-derivatives of $n$-point $A$-cycle integrals will
be supplemented by an initial value at the cusp. We will now determine the degenerations
$Z^{i\infty}_{\vec{\eta}}(P|\cdot)$ in case of planar integration cycles ${\cal C}(P)$ and express the results in
terms of $(n{+}2)$-point Parke--Taylor integrals (\ref{cocyc1}) over disk boundaries:
\beq
Z^{i\infty}_{\vec{\eta}}(1,A|1,P) = \frac{1}{(2\pi i)^{n-1}} \sum_{B,Q \in S_{n-1}} {\cal H}_{\alpha'}(A|B) {\cal K}_{\vec{\eta}}(P|Q)
Z^{\rm tree}(+,B,1,- | +,Q,-,1)\, .
\label{inival1}
\eeq
The explicit form of the $(n{-}1)! \times (n{-}1)!$ matrices ${\cal H}_{\alpha'}(A|B)$ and ${\cal K}_{\vec{\eta}}(P|Q)$
will be given below. The normalization by powers of $2\pi i$ is due to the change of variables
$\prod_{j=2}^n \dd z_j = \frac{1}{(2\pi i)^{n-1}} \prod_{j=2}^n \frac{\dd \sigma_j}{\sigma_j}$, cf.\ (\ref{G3}).
The extra legs $+,-$ in the $(n{+}2)$-point
disk integrals $Z^{\rm tree}$ are associated with the identified punctures $\sigma_+=0$ and $\sigma_-=\infty$
due to the pinching of the $A$-cycle at $\tau \rightarrow i\infty$, see figure \ref{nodsph}. The
Mandelstam invariants associated with the extra legs are determined by the degeneration of the
planar Koba--Nielsen factor that follows from (\ref{Gplother})
\beq
{\rm KN}_{12\ldots n}^{i \infty}  \, \big|_{{\cal C}(1,a_2,a_3,\ldots,a_n)}
= \prod_{j=2}^n \sigma_j^{\frac{1}{2} \sum^n_{i\neq j}s_{ij}}
  \prod_{j=2}^n (1{-}\sigma_j)^{-s_{1j}}
  \prod_{2\leq i<j}^n (\sigma_{a_j} {-} \sigma_{a_i} )^{-s_{a_i a_j}}\, ,
\label{inival2}
\eeq
i.e.\ one can read off\footnote{We leave it as an open problem whether the identification of Mandelstam
variables $\{s_{ij}, \ 1\leq i <j\leq n\}$ with dot products of momenta can be extended to the formal
variables (\ref{inival3}) via suitable choices of $k_{\pm}$.
We will use $s_{j\pm}$ and $s_{+,-}$ as auxiliary variables that are uniquely determined
by ${\rm SL}_2$-invariance and whose virtue for the subsequent calculations does not rely
on a physical interpretation.}
\beq
s_{j+} = s_{j-} = - \frac{1}{2} \sum^n_{1\leq i\neq j} s_{ij} \, , \ \ \ \ \ \ s_{+,-} = \sum_{1\leq i <j }^n s_{ij}
\label{inival3}
\eeq
by identifying (\ref{inival2}) as an ${\rm SL}_2$-fixed version of $\sigma_{+,-}^{-s_{+,-}}\prod_{j=1}^n
\sigma_{j+}^{-s_{j+}} \sigma_{j-}^{-s_{j-}}\prod_{1\leq i<j}^{n}\sigma_{a_ja_i}^{-s_{a_ia_j}}$.

The first matrix $ {\cal H}_{\alpha'}(A|B)$ in (\ref{inival1}) translates the twisted cycles defined
by ${\cal C}(1,A)$ and (\ref{inival2}) into disk orderings
in an ${\rm SL}_2$-frame\footnote{Note that
(\ref{inival4}) takes the following form in a generic ${\rm SL}_2$-frame:
\[{\cal D}(a_1,a_2,\ldots,a_k) = \{ \sigma_1,\sigma_2,\ldots, \sigma_k \in \RR, \
-\infty  < \sigma_{a_1}< \sigma_{a_2}<  \ldots < \sigma_{a_k}< \infty\} \, .\]}
with $(\sigma_+,\sigma_1,\sigma_-)=(0,1,\infty)$,
\beq
{\cal D}_B \equiv {\cal D}(+,b_2,b_3,\ldots,b_n,1,-) = \{\sigma_{b_2},\sigma_{b_3},\ldots, \sigma_{b_n} \in \RR, \
0 {<} \sigma_{b_2} {<} \sigma_{b_3} {<}  \ldots {<} \sigma_{b_k}{<} 1\}\, .
\label{inival4}
\eeq
The relation (\ref{someintI}) among the twisted
cycles at two points amounts to ${\cal H}_{\alpha'}(2|2)= 2i \sin ( \frac{ \pi }{2}s_{12} )$, and the entries of
${\cal H}_{\alpha'}(\cdot|\cdot)$ at higher multiplicity will be determined in sections \ref{sec:5.1}
and \ref{sec:5.2} in terms of $s_{ij}$-dependent phases.

The second matrix ${\cal K}_{\vec{\eta}}(P|Q)$ in (\ref{inival1})
tracks the transformation between the basis integrands at genus zero \& genus one
and carries the dependence on $\eta_j$ through the $\pi \cot(\pi \eta)$ function in the
degeneration (\ref{Gpl}) of the Kronecker--Eisenstein series. At two points, the
fact that $\frac{ 1}{\sigma_2}$ descends from the Parke--Taylor factor
$(\sigma_{+2} \sigma_{2-} \sigma_{-1} \sigma_{1+})^{-1}$ in an ${\rm SL}_2$-frame
with $(\sigma_+,\sigma_1,\sigma_-)=(0,1,\infty)$ yields ${\cal K}_{\eta_2}(2|2)=\pi \cot(\pi \eta_2)$.
The extension to higher multiplicity will be the topic of section \ref{sec:5.3}.


\subsection{Recovering twisted cycles on the disk boundary up to five points}
\label{sec:5.1}

The strategy of section \ref{sec:3.4} to deform the integration contour from the unit circle
$|\sigma_j| = 1$ to the interval $0<\sigma_j<1$ carries over to higher multiplicity. The only
additional feature at $n\geq 3$ points concerns the relative ordering of $\sigma_2,\sigma_3,\ldots,\sigma_n$
on the unit circle with relative phases according to $0{<}z_2{<}z_3{<}\ldots{<}z_n{<}1$ in case of ${\cal C}(1,2,\ldots,n)$.

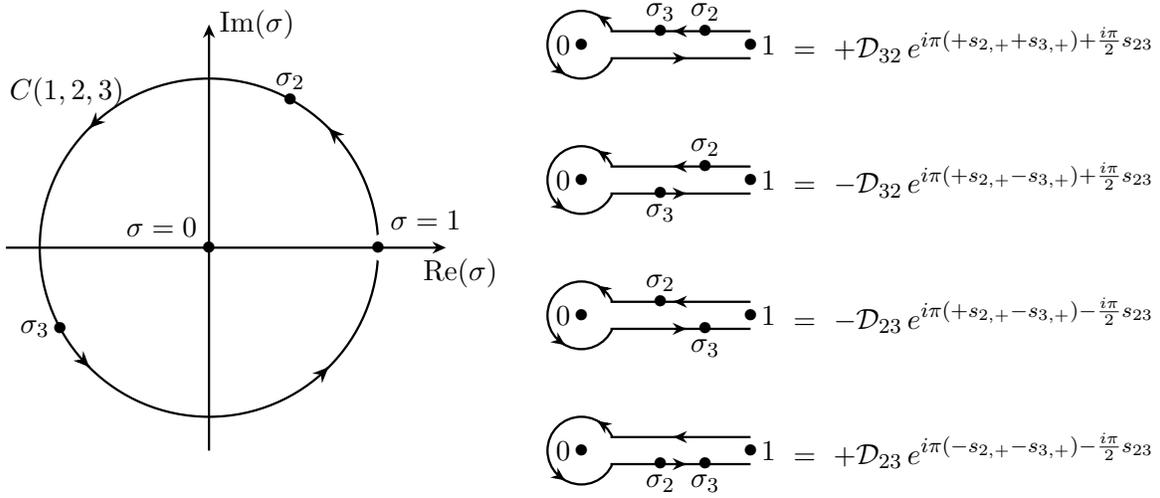
\begin{figure}
\begin{center}
\begin{tikzpicture}[line width=0.30mm, scale=0.9]
\draw (2.5,0) arc (0:360:2.5cm);
\draw[white, fill=white] (2.6,0) circle (0.2cm);
\draw[arrows={-Stealth[width=1.8mm, length=2.1mm]}](1.767,1.765)--(1.766,1.766);
\draw[arrows={-Stealth[width=1.8mm, length=2.1mm]}](-1.767,-1.765)--(-1.766,-1.766);
\draw[arrows={-Stealth[width=1.8mm, length=2.1mm]}](-1.765,1.767)--(-1.766,1.766);
\draw[arrows={-Stealth[width=1.8mm, length=2.1mm]}](1.765,-1.767)--(1.766,-1.766);
\draw (0,0)node{$\bullet$} ;
\draw(-0.7,0.3)node{$\sigma=0$};
\draw(1.2,2.18)node{$\bullet$}node[above]{$\sigma_2$};
\draw(-2.2,-1.2)node{$\bullet$}node[left]{$\sigma_3$};
\draw (2.5,0)node{$\bullet$} ;
\draw(3.2,0.4)node{$\sigma=1$};
\draw [arrows={-Stealth[width=1.6mm, length=1.8mm]}] (-3,0) -- (3.5,0) node[below]{$\quad\te{Re}(\sigma)$};
\draw [arrows={-Stealth[width=1.6mm, length=1.8mm]}] (0,-3) -- (0,3.3) node[right]{$\te{Im}(\sigma)$};
\draw(-2.1,2.3)node{$C(1,2,3)$};
\scope[xshift=5.5cm,yshift=3cm]
  \draw[fill=white] (0,0) circle (0.5cm);
  \draw[white, fill=white] (0.5,0) circle (0.2cm);
  \draw[arrows={-Stealth[width=1.6mm, length=1.8mm]}](0.25,0.433)--(0.24,0.44);
  \draw[arrows={-Stealth[width=1.6mm, length=1.8mm]}](-0.25,-0.433)--(-0.24,-0.44);
  \draw(2.5,0.2) -- (0.43,0.2);
  \draw(2.5,-0.2) -- (0.43,-0.2);
  \draw[arrows={-Stealth[width=1.6mm, length=1.8mm]}](1.37,0.2)--(1.36,0.2);
  \draw[arrows={-Stealth[width=1.6mm, length=1.8mm]}](1.54,-0.2)--(1.55,-0.2);
  \draw (0,0)node{$\bullet$}node[left]{$0$};
  \draw (2.5,0)node{$\bullet$}node[right]{$1$};
  \draw(1.17,0.2)node{$\bullet$}node[above]{$\sigma_3$};
  \draw(1.83,0.2)node{$\bullet$}node[above]{$\sigma_2$};
  \draw(5.8,0)node{$= \ +{\cal D}_{32}  \, e^{i\pi(+s_{2,+}+s_{3,+})+ \frac{i\pi}{2}s_{23}} $};
  \endscope
\scope[xshift=5.5cm, yshift=-1cm]
  \draw[fill=white] (0,0) circle (0.5cm);
  \draw[white, fill=white] (0.5,0) circle (0.2cm);
  \draw[arrows={-Stealth[width=1.6mm, length=1.8mm]}](0.25,0.433)--(0.24,0.44);
  \draw[arrows={-Stealth[width=1.6mm, length=1.8mm]}](-0.25,-0.433)--(-0.24,-0.44);
  \draw(2.5,0.2) -- (0.43,0.2);
  \draw(2.5,-0.2) -- (0.43,-0.2);
  \draw[arrows={-Stealth[width=1.6mm, length=1.8mm]}](1.37,0.2)--(1.36,0.2);
  \draw[arrows={-Stealth[width=1.6mm, length=1.8mm]}](1.54,-0.2)--(1.55,-0.2);
  \draw (0,0)node{$\bullet$}node[left]{$0$};
  \draw (2.5,0)node{$\bullet$}node[right]{$1$};
  \draw(1.17,0.2)node{$\bullet$}node[above]{$\sigma_2$};
  \draw(1.83,-0.2)node{$\bullet$}node[below]{$\sigma_3$};
  \draw(5.8,0)node{$= \  -{\cal D}_{23}  \, e^{i\pi(+s_{2,+}-s_{3,+}) - \frac{i\pi}{2}s_{23}}$};
\endscope
\scope[xshift=5.5cm,yshift=1cm]
  \draw[fill=white] (0,0) circle (0.5cm);
  \draw[white, fill=white] (0.5,0) circle (0.2cm);
  \draw[arrows={-Stealth[width=1.6mm, length=1.8mm]}](0.25,0.433)--(0.24,0.44);
  \draw[arrows={-Stealth[width=1.6mm, length=1.8mm]}](-0.25,-0.433)--(-0.24,-0.44);
  \draw(2.5,0.2) -- (0.43,0.2);
  \draw(2.5,-0.2) -- (0.43,-0.2);
  \draw[arrows={-Stealth[width=1.6mm, length=1.8mm]}](1.37,0.2)--(1.36,0.2);
  \draw[arrows={-Stealth[width=1.6mm, length=1.8mm]}](1.54,-0.2)--(1.55,-0.2);
  \draw (0,0)node{$\bullet$}node[left]{$0$};
  \draw (2.5,0)node{$\bullet$}node[right]{$1$};
  \draw(1.17,-0.2)node{$\bullet$}node[below]{$\sigma_3$};
  \draw(1.83,0.2)node{$\bullet$}node[above]{$\sigma_2$};
  \draw(5.8,0)node{$= \ -{\cal D}_{32}  \, e^{i\pi(+s_{2,+}-s_{3,+}) + \frac{i\pi}{2}s_{23}} $};
\endscope
\scope[xshift=5.5cm,yshift=-3cm]
  \draw[fill=white] (0,0) circle (0.5cm);
  \draw[white, fill=white] (0.5,0) circle (0.2cm);
  \draw[arrows={-Stealth[width=1.6mm, length=1.8mm]}](0.25,0.433)--(0.24,0.44);
  \draw[arrows={-Stealth[width=1.6mm, length=1.8mm]}](-0.25,-0.433)--(-0.24,-0.44);
  \draw(2.5,0.2) -- (0.43,0.2);
  \draw(2.5,-0.2) -- (0.43,-0.2);
  \draw[arrows={-Stealth[width=1.6mm, length=1.8mm]}](1.37,0.2)--(1.36,0.2);
  \draw[arrows={-Stealth[width=1.6mm, length=1.8mm]}](1.54,-0.2)--(1.55,-0.2);
  \draw (0,0)node{$\bullet$}node[left]{$0$};
  \draw (2.5,0)node{$\bullet$}node[right]{$1$};
  \draw(1.17,-0.2)node{$\bullet$}node[below]{$\sigma_2$};
  \draw(1.83,-0.2)node{$\bullet$}node[below]{$\sigma_3$};
  \draw(5.8,0)node{$= \ +{\cal D}_{23}  \, e^{i\pi(-s_{2,+}-s_{3,+}) - \frac{i\pi}{2}s_{23}} $};
\endscope
\end{tikzpicture}
\end{center}
\caption{In the $\sigma_j =e^{2\pi i z_j}$ variables, the integration contour ${\cal C}(1,2,3)$ is the unit circle in
the left panel, where the phases of $\sigma_2$ and $\sigma_3$ are ordered according to $z_2<z_3$.
Similar to figure \ref{figcirc}, we
replace ${\cal C}(1,2,3)$ by the homotopy-equivalent combination of paths visualized in the right panel. For
each of the four inequivalent relative positions of $\sigma_2$ and $\sigma_3$, the Koba--Nielsen
factor (\ref{inival2}) introduces a different phase that depends on $s_{2+},s_{3+}$ and $s_{23}$.}
\label{figcirc3pt}
\end{figure}

This key ideas become clear from the twisted three-point cycle defined by ${\cal C}(1,2,3)$ and
${\rm KN}_{123}^{i\infty}=\sigma_2^{-s_{2+}}\sigma_3^{-s_{3+}}(\sigma_3{-}\sigma_2)^{-s_{23}}$
with $s_{2+}=-\frac{1}{2}(s_{12}+s_{23})$ and $s_{3+}=-\frac{1}{2}(s_{13}+s_{23})$, see (\ref{inival2})
and (\ref{inival3}). When deforming ${\cal C}(1,2,3)$ to the homotopy-equivalent contour in the right panel
of figure \ref{figcirc3pt}, there are four different scenarios for the phases introduced by ${\rm KN}_{123}^{i\infty}$
depending on the relative positions of $\sigma_2$ and $\sigma_3$. The rules for determining these phases
can be easily written down in an $n$-point context:
\begin{itemize}
\item As in the two-point case, $\sigma_j^{-s_{j+}}$ with $j=2,3,\ldots,n$ introduce
phases $e^{+i\pi s_{j+}}$ and $e^{-i\pi s_{j+}}$ when $\sigma_j$ is slightly above and below
the real axis, respectively. This realizes the desired phase difference $e^{-2\pi i s_{j+}}$ when
transporting $\sigma_+$ on a circle around the origin.
\item Starting from three points, factors of $(\sigma_j{-}\sigma_i)^{-s_{ij}}$ with $2\leq i<j\leq n$ introduce
phases $e^{+\frac{i\pi}{2}s_{ij}}$ and $e^{- \frac{i\pi}{2}s_{ij}}$ when $\Re(\sigma_j)< \Re(\sigma_i)$
and $\Re(\sigma_i)< \Re(\sigma_j)$, respectively. This realizes the desired phase difference
$e^{i\pi  s_{ij}}$ when transporting $\sigma_j$ on a semicircle around~$\sigma_i$.
\end{itemize}
By adding up the four contributions in figure \ref{figcirc3pt} and adjoining a factor of $(-1)$ for the negative
path orientation of each puncture above the real line, we find the following relation among twisted cycles,
\begin{align}
{\cal C}(1,2,3)&=2i \sin(\pi s_{3+}) e^{i \pi (s_{2+}+\frac{1}{2}s_{23})} {\cal D}_{32}
- 2i \sin(\pi s_{2+}) e^{i \pi (-s_{3+}-\frac{1}{2}s_{23})} {\cal D}_{23}
  \label{s0jA} \\
&= 2i \sin\Big(\frac{\pi}{2} (s_{12}{+}s_{23}) \Big) e^{ \frac{i \pi}{2} s_{13}} {\cal D}_{23}
- 2i \sin\Big(\frac{\pi}{2} (s_{13}{+}s_{23}) \Big) e^{- \frac{i \pi}{2} s_{12}} {\cal D}_{32} \, ,  \notag
\end{align}
where ${\cal D}_{23}$ refers to $0<\sigma_2<\sigma_3<1$ according to the general notation (\ref{inival4}).
For the sake of brevity, we employ the abusive notation in (\ref{s0jA}) and later equations
to keep the Koba--Nielsen factor (\ref{inival2}) defining the twisted cycles implicit.

The rules for assigning phases can be straightforwardly applied to ${\cal C}(1,2,3,4)$, where
the iteration of the usual path deformation yields eight different scenarios for the phases,
depending on the relative positions of $\sigma_2, \sigma_3,\sigma_4$ in figure \ref{figcirc4pt},
\begin{align}
{\cal C}(1,2,3,4) &= - 2i \sin(\pi s_{2+}) \, e^{i\pi(-s_{3+}-s_{4+} )} \, e^{\frac{ i \pi }{2} (-s_{23}-s_{24}-s_{34})} {\cal D}_{234} \notag \\
&\ \ \ \ + 2i \sin(\pi s_{3+}) \, e^{i\pi(+s_{2+}-s_{4+} )} \, e^{\frac{ i \pi }{2} (+s_{23}-s_{24}-s_{34})} {\cal D}_{324}\notag \\
&\ \ \ \ + 2i \sin(\pi s_{3+}) \, e^{i\pi(+s_{2+}-s_{4+} )} \, e^{\frac{ i \pi }{2} (+s_{23}+s_{24}-s_{34})} {\cal D}_{342}\notag \\
&\ \ \ \ -  2i \sin(\pi s_{4+}) \, e^{i\pi(+s_{2+}+s_{3+} )} \, e^{\frac{ i \pi }{2} (+s_{23}+s_{24}+s_{34})} {\cal D}_{432} \notag \\
&= 2i \sin\Big(\frac{\pi}{2}(s_{12}{+}s_{23}{+}s_{24}) \Big) \, e^{\frac{ i \pi }{2}s_{134}} {\cal D}_{234} \label{s0jD} \\
&\ \ \ \ - 2i \sin\Big(\frac{\pi}{2}(s_{13}{+}s_{23}{+}s_{34}) \Big) \,  e^{\frac{ i \pi }{2}(-s_{12}+s_{14}-s_{24} )} {\cal D}_{324}\notag \\
&\ \ \ \ - 2i \sin\Big(\frac{\pi}{2}(s_{13}{+}s_{23}{+}s_{34}) \Big) \, e^{\frac{ i \pi }{2}(-s_{12}+s_{14}+s_{24} )}{\cal D}_{342}\notag \\
&\ \ \ \ +  2i  \sin\Big(\frac{\pi}{2}(s_{14}{+}s_{24}{+}s_{34}) \Big) \,  e^{-\frac{ i \pi }{2}s_{123}} {\cal D}_{432} \, .
\notag
\end{align}

\begin{figure}
\begin{center}
\begin{tikzpicture}[line width=0.30mm]
  \draw[fill=white] (0,0) circle (0.5cm);
  \draw[white, fill=white] (0.5,0) circle (0.2cm);
  \draw[arrows={-Stealth[width=1.6mm, length=1.8mm]}](0.1,0.5)--(-0.1,0.5);
  \draw[arrows={-Stealth[width=1.6mm, length=1.8mm]}](-0.1,-0.5)--(0.1,-0.5);
  \draw(3,0.2) -- (0.43,0.2);
  \draw(3,-0.2) -- (0.43,-0.2);
  \draw (0,0)node{$\bullet$}node[left]{$0$};
  \draw (3,0)node{$\bullet$}node[right]{$1$};
  \draw(2.36,0.2)node{$\bullet$}node[above]{$\sigma_2$};
  \draw(1.72,0.2)node{$\bullet$}node[above]{$\sigma_3$};
  \draw(1.08,0.2)node{$\bullet$}node[above]{$\sigma_4$};
  \draw(7.5,0)node{$= \ \ -{\cal D}_{432}  \, e^{i\pi(+s_{2+}+s_{3+}+s_{4+})} \, e^{\frac{i \pi}{2}( +s_{23}+s_{24}+s_{34})}$};
\scope[yshift=-2cm]
  \draw[fill=white] (0,0) circle (0.5cm);
  \draw[white, fill=white] (0.5,0) circle (0.2cm);
  \draw[arrows={-Stealth[width=1.6mm, length=1.8mm]}](0.1,0.5)--(-0.1,0.5);
  \draw[arrows={-Stealth[width=1.6mm, length=1.8mm]}](-0.1,-0.5)--(0.1,-0.5);
  \draw(3,0.2) -- (0.43,0.2);
  \draw(3,-0.2) -- (0.43,-0.2);
  \draw (0,0)node{$\bullet$}node[left]{$0$};
  \draw (3,0)node{$\bullet$}node[right]{$1$};
  \draw(1.08,-0.2)node{$\bullet$}node[below]{$\sigma_4$};
  \draw(1.72,0.2)node{$\bullet$}node[above]{$\sigma_3$};
  \draw(2.36,0.2)node{$\bullet$}node[above]{$\sigma_2$};
  \draw(7.5,0)node{$= \ \ +{\cal D}_{432}  \, e^{i\pi(+s_{2+}+s_{3+}-s_{4+})} \, e^{\frac{i \pi}{2}( +s_{23}+s_{24}+s_{34})}$};
\endscope
\scope[yshift=-4cm]
  \draw[fill=white] (0,0) circle (0.5cm);
  \draw[white, fill=white] (0.5,0) circle (0.2cm);
  \draw[arrows={-Stealth[width=1.6mm, length=1.8mm]}](0.1,0.5)--(-0.1,0.5);
  \draw[arrows={-Stealth[width=1.6mm, length=1.8mm]}](-0.1,-0.5)--(0.1,-0.5);
  \draw(3,0.2) -- (0.43,0.2);
  \draw(3,-0.2) -- (0.43,-0.2);
  \draw (0,0)node{$\bullet$}node[left]{$0$};
  \draw (3,0)node{$\bullet$}node[right]{$1$};
  \draw(1.08,0.2)node{$\bullet$}node[above]{$\sigma_3$};
  \draw(1.72,-0.2)node{$\bullet$}node[below]{$\sigma_4$};
  \draw(2.36,0.2)node{$\bullet$}node[above]{$\sigma_2$};
  \draw(7.5,0)node{$= \ \ +{\cal D}_{342}  \, e^{i\pi(+s_{2+}+s_{3+}-s_{4+})} \, e^{\frac{i \pi}{2}( +s_{23}+s_{24}-s_{34})}$};
\endscope
\scope[yshift=-6cm]
  \draw[fill=white] (0,0) circle (0.5cm);
  \draw[white, fill=white] (0.5,0) circle (0.2cm);
  \draw[arrows={-Stealth[width=1.6mm, length=1.8mm]}](0.1,0.5)--(-0.1,0.5);
  \draw[arrows={-Stealth[width=1.6mm, length=1.8mm]}](-0.1,-0.5)--(0.1,-0.5);
  \draw(3,0.2) -- (0.43,0.2);
  \draw(3,-0.2) -- (0.43,-0.2);
  \draw (0,0)node{$\bullet$}node[left]{$0$};
  \draw (3,0)node{$\bullet$}node[right]{$1$};
  \draw(1.08,0.2)node{$\bullet$}node[above]{$\sigma_3$};
  \draw(1.72,0.2)node{$\bullet$}node[above]{$\sigma_2$};
  \draw(2.36,-0.2)node{$\bullet$}node[below]{$\sigma_4$};
  \draw(7.5,0)node{$= \ \ +{\cal D}_{324}  \, e^{i\pi(+s_{2+}+s_{3+}-s_{4+})} \, e^{\frac{i \pi}{2}( +s_{23}-s_{24}-s_{34})}$};
\endscope
\scope[yshift=-8cm]
  \draw[fill=white] (0,0) circle (0.5cm);
  \draw[white, fill=white] (0.5,0) circle (0.2cm);
  \draw[arrows={-Stealth[width=1.6mm, length=1.8mm]}](0.1,0.5)--(-0.1,0.5);
  \draw[arrows={-Stealth[width=1.6mm, length=1.8mm]}](-0.1,-0.5)--(0.1,-0.5);
  \draw(3,0.2) -- (0.43,0.2);
  \draw(3,-0.2) -- (0.43,-0.2);
  \draw (0,0)node{$\bullet$}node[left]{$0$};
  \draw (3,0)node{$\bullet$}node[right]{$1$};
  \draw(1.08,-0.2)node{$\bullet$}node[below]{$\sigma_3$};
  \draw(1.72,-0.2)node{$\bullet$}node[below]{$\sigma_4$};
  \draw(2.36,0.2)node{$\bullet$}node[above]{$\sigma_2$};
  \draw(7.5,0)node{$= \ \ -{\cal D}_{342}  \, e^{i\pi(+s_{2+}-s_{3+}-s_{4+})} \, e^{\frac{i \pi}{2}( +s_{23}+s_{24}-s_{34})}$};
\endscope
\scope[yshift=-10cm]
  \draw[fill=white] (0,0) circle (0.5cm);
  \draw[white, fill=white] (0.5,0) circle (0.2cm);
  \draw[arrows={-Stealth[width=1.6mm, length=1.8mm]}](0.1,0.5)--(-0.1,0.5);
  \draw[arrows={-Stealth[width=1.6mm, length=1.8mm]}](-0.1,-0.5)--(0.1,-0.5);
  \draw(3,0.2) -- (0.43,0.2);
  \draw(3,-0.2) -- (0.43,-0.2);
  \draw (0,0)node{$\bullet$}node[left]{$0$};
  \draw (3,0)node{$\bullet$}node[right]{$1$};
  \draw(1.08,-0.2)node{$\bullet$}node[below]{$\sigma_3$};
  \draw(1.72,0.2)node{$\bullet$}node[above]{$\sigma_2$};
  \draw(2.36,-0.2)node{$\bullet$}node[below]{$\sigma_4$};
  \draw(7.5,0)node{$= \ \ -{\cal D}_{324}  \, e^{i\pi(+s_{2+}-s_{3+}-s_{4+})} \, e^{\frac{i \pi}{2}( +s_{23}-s_{24}-s_{34})}$};
\endscope
\scope[yshift=-12cm]
  \draw[fill=white] (0,0) circle (0.5cm);
  \draw[white, fill=white] (0.5,0) circle (0.2cm);
  \draw[arrows={-Stealth[width=1.6mm, length=1.8mm]}](0.1,0.5)--(-0.1,0.5);
  \draw[arrows={-Stealth[width=1.6mm, length=1.8mm]}](-0.1,-0.5)--(0.1,-0.5);
  \draw(3,0.2) -- (0.43,0.2);
  \draw(3,-0.2) -- (0.43,-0.2);
  \draw (0,0)node{$\bullet$}node[left]{$0$};
  \draw (3,0)node{$\bullet$}node[right]{$1$};
  \draw(1.08,0.2)node{$\bullet$}node[above]{$\sigma_2$};
  \draw(1.72,-0.2)node{$\bullet$}node[below]{$\sigma_3$};
  \draw(2.36,-0.2)node{$\bullet$}node[below]{$\sigma_4$};
  \draw(7.5,0)node{$= \ \ -{\cal D}_{234}  \, e^{i\pi(+s_{2+}-s_{3+}-s_{4+})} \, e^{\frac{i \pi}{2}( -s_{23}-s_{24}-s_{34})}$};
\endscope
\scope[yshift=-14cm]
  \draw[fill=white] (0,0) circle (0.5cm);
  \draw[white, fill=white] (0.5,0) circle (0.2cm);
  \draw[arrows={-Stealth[width=1.6mm, length=1.8mm]}](0.1,0.5)--(-0.1,0.5);
  \draw[arrows={-Stealth[width=1.6mm, length=1.8mm]}](-0.1,-0.5)--(0.1,-0.5);
  \draw(3,0.2) -- (0.43,0.2);
  \draw(3,-0.2) -- (0.43,-0.2);
  \draw (0,0)node{$\bullet$}node[left]{$0$};
  \draw (3,0)node{$\bullet$}node[right]{$1$};
  \draw(1.08,-0.2)node{$\bullet$}node[below]{$\sigma_2$};
  \draw(1.72,-0.2)node{$\bullet$}node[below]{$\sigma_3$};
  \draw(2.36,-0.2)node{$\bullet$}node[below]{$\sigma_4$};
  \draw(7.5,0)node{$= \ \ +{\cal D}_{234}  \, e^{i\pi(-s_{2+}-s_{3+}-s_{4+})} \, e^{\frac{i \pi}{2}( -s_{23}-s_{24}-s_{34})}$};
\endscope
\end{tikzpicture}
\end{center}
\caption{Genus-zero contributions ${\cal D}_{ijk}$ to the degenerate genus-one integral over the
four-point cycle ${\cal C}_{1234}= \{ 0 <z_2<z_3<z_4<1\}$.}
\label{figcirc4pt}
\end{figure}

In the five-point generalization of the above bookkeeping, one has to take sixteen different scenarios into account
for the phases (depending on the relative positions of $\sigma_2,\sigma_3,\sigma_4,\sigma_5$), and we simply
quote the end result of adding these contributions:
\begin{align}
{\cal C}(1,2,3,4,5) &= 2i \sin \Big(\frac{\pi}{2}( s_{12}{+}s_{23}{+}s_{24}{+}s_{25}) \Big) \, e^{+\frac{i\pi}{2} s_{1345} } {\cal D}_{2345} \notag \\
&\ \ \ \ -  2i \sin \Big(\frac{\pi}{2}( s_{13}{+}s_{23}{+}s_{34}{+}s_{35}) \Big)\, e^{\frac{i\pi}{2} (-s_{12}+s_{14}+s_{15} - s_{24}-s_{25}+s_{45}) } {\cal D}_{3245} \notag \\
&\ \ \ \ -  2i \sin\Big(\frac{\pi}{2}( s_{13}{+}s_{23}{+}s_{34}{+}s_{35}) \Big)\,  e^{\frac{i\pi}{2} (-s_{12}+s_{14}+s_{15} + s_{24}-s_{25}+s_{45}) }  {\cal D}_{3425} \notag \\
&\ \ \ \ -  2i \sin\Big(\frac{\pi}{2}( s_{13}{+}s_{23}{+}s_{34}{+}s_{35}) \Big)\,  e^{\frac{i\pi}{2} (-s_{12}+s_{14}+s_{15} + s_{24}+s_{25}+s_{45}) }  {\cal D}_{3452} \notag \\
&\ \ \ \ +  2i \sin\Big(\frac{\pi}{2}( s_{14}{+}s_{24}{+}s_{34}{+}s_{45}) \Big)\,  e^{\frac{i\pi}{2} (-s_{12}-s_{13}+s_{15} - s_{23}-s_{25}-s_{35}) }  {\cal D}_{4325} \notag \\
&\ \ \ \ +  2i \sin\Big(\frac{\pi}{2}( s_{14}{+}s_{24}{+}s_{34}{+}s_{45}) \Big)\,e^{\frac{i\pi}{2} (-s_{12}-s_{13}+s_{15} - s_{23}+s_{25}-s_{35}) }{\cal D}_{4352} \notag \\
&\ \ \ \ +  2i \sin\Big(\frac{\pi}{2}( s_{14}{+}s_{24}{+}s_{34}{+}s_{45}) \Big)\, e^{\frac{i\pi}{2} (-s_{12}-s_{13}+s_{15} - s_{23}+s_{25}+s_{35}) } {\cal D}_{4532} \notag \\
&\ \ \ \ -  2i \sin\Big(\frac{\pi}{2}( s_{15}{+}s_{25}{+}s_{35}{+}s_{45}) \Big)\, e^{-\frac{i\pi}{2} s_{1234} } {\cal D}_{5432} \, .
\label{s0jG}
\end{align}
%


\subsection{Recovering twisted cycles on the disk boundary at $n$ points}
\label{sec:5.2}

In order to extend the above relations between twisted genus-one and genus-zero integration cycles
to higher multiplicity, we start by introducing some notation: It is convenient to absorb the exponentials
of (\ref{s0jA}), (\ref{s0jD}) and (\ref{s0jG}) into
\begin{align}
\widehat {\cal D}\Big(2,\begin{smallmatrix} \varepsilon_3 \\ 3\end{smallmatrix} \Big) &\equiv e^{ \frac{ i \pi }{2} \varepsilon_3 s_{13} } {\cal D}_{23} \notag \\
\widehat {\cal D}\Big(2,\begin{smallmatrix} \varepsilon_3 &\varepsilon_4 \\ 3 &4\end{smallmatrix} \Big) &\equiv e^{ \frac{ i \pi }{2} (\varepsilon_3 (s_{13}+s_{34}) + \varepsilon_4 s_{14}) } {\cal D}_{234} \label{s0jH} \\
\widehat {\cal D}\Big(2,\begin{smallmatrix} \varepsilon_3 &\varepsilon_4 &\varepsilon_5 \\ 3 &4 &5\end{smallmatrix} \Big) &\equiv e^{ \frac{ i \pi }{2} (\varepsilon_3 (s_{13}+s_{34}+s_{35}) + \varepsilon_4 (s_{14} + s_{45}) + \varepsilon_5 s_{15}) } {\cal D}_{2345}\, ,
\notag
\end{align}
where $\varepsilon_j\in \{1,-1\}$ or in short $\varepsilon_j = \pm$. These definitions allow to compactly rewrite
\begin{align}
{\cal C}(1,2,3) &=  2i \sin\Big(\frac{\pi}{2} (s_{12}{+}s_{23}) \Big) \widehat {\cal D}\Big(2,\begin{smallmatrix} + \\ 3\end{smallmatrix} \Big)
- 2i \sin\Big(\frac{\pi}{2} (s_{13}{+}s_{23}) \Big) \widehat {\cal D}\Big(3,\begin{smallmatrix} - \\ 2\end{smallmatrix} \Big)  \notag \\
{\cal C}(1,2,3,4) &=  2i \sin\Big(\frac{\pi}{2}(s_{12}{+}s_{23}{+}s_{24}) \Big) \,
\widehat {\cal D}\Big(2,\begin{smallmatrix} + &+ \\ 3 &4\end{smallmatrix} \Big)
\notag \\
&\ \ \ \ - 2i \sin\Big(\frac{\pi}{2}(s_{13}{+}s_{23}{+}s_{34}) \Big) \,  \widehat {\cal D}\Big(3,\Big(
\begin{smallmatrix} - \\ 2\end{smallmatrix} \shuffle
\begin{smallmatrix} + \\ 4\end{smallmatrix}
\Big)\Big)  \notag \\
&\ \ \ \ +  2i  \sin\Big(\frac{\pi}{2}(s_{14}{+}s_{24}{+}s_{34}) \Big) \,  \widehat {\cal D}\Big(4,\begin{smallmatrix} - &- \\ 3 &2\end{smallmatrix} \Big)
\label{s0jI} \\
{\cal C}(1,2,3,4,5) &= 2i \sin \Big(\frac{\pi}{2}( s_{12}{+}s_{23}{+}s_{24}{+}s_{25}) \Big) \,
\widehat {\cal D}\Big(2,\begin{smallmatrix} + &+ &+ \\ 3 &4 &5 \end{smallmatrix} \Big)\notag \\
&\ \ \ \ -  2i \sin \Big(\frac{\pi}{2}( s_{13}{+}s_{23}{+}s_{34}{+}s_{35}) \Big)\,
\widehat {\cal D}\Big(3,\Big(
\begin{smallmatrix} - \\ 2\end{smallmatrix} \shuffle
\begin{smallmatrix} + &+ \\ 4 &5\end{smallmatrix}
\Big)\Big)  \notag \\
&\ \ \ \ +  2i \sin\Big(\frac{\pi}{2}( s_{14}{+}s_{24}{+}s_{34}{+}s_{45}) \Big)\,
\widehat {\cal D}\Big(4,\Big(
\begin{smallmatrix} - &- \\ 3 &2\end{smallmatrix} \shuffle
\begin{smallmatrix} + \\ 5\end{smallmatrix}
\Big)\Big)  \notag \\
&\ \ \ \ -  2i \sin\Big(\frac{\pi}{2}( s_{15}{+}s_{25}{+}s_{35}{+}s_{45}) \Big)\,
\widehat {\cal D}\Big(5,\begin{smallmatrix} - &- &- \\ 4 &3 &2 \end{smallmatrix} \Big)\, .\notag
\end{align}
The shuffle symbol is understood to act on the combined letters $\begin{smallmatrix} \varepsilon_j \\ j\end{smallmatrix}$, e.g.\
\beq
\widehat {\cal D}\Big(3,\Big(
\begin{smallmatrix} - \\ 2\end{smallmatrix} \shuffle
\begin{smallmatrix} + \\ 4\end{smallmatrix} \Big)\Big) =
\widehat {\cal D}\Big(3,\begin{smallmatrix} - &+ \\ 2 &4\end{smallmatrix}\Big)
+\widehat {\cal D}\Big(3,\begin{smallmatrix} + &- \\ 4 &2\end{smallmatrix}\Big) \, ,
\label{s0jJ}
\eeq
and one can formally align the two-point case into the same pattern with
$\widehat {\cal D}(2) = {\cal D}_2$ and ${\cal C}(1,2) = 2i \sin\big( \frac{\pi }{2}s_{12} \big) \widehat {\cal D}(2)$.
Based on the obvious $n$-point generalization of (\ref{s0jH}),
\begin{align}
\widehat {\cal D}\Big(2,\begin{smallmatrix} \varepsilon_3 &\varepsilon_4 &\ldots &\varepsilon_n \\ 3 &4 &\ldots &n\end{smallmatrix} \Big) &\equiv e^{ \frac{ i \pi }{2} (\varepsilon_3 (s_{13}+s_{34}+\ldots+s_{3n}) + \varepsilon_4 (s_{14} + s_{45}+\ldots+s_{4n})+ \ldots
+ \varepsilon_{n-1}(s_{1,n-1}+s_{n-1,n}) + \varepsilon_n s_{1,n}) } {\cal D}_{23\ldots n} \notag \\
&= \prod_{j=3}^n  e^{ \frac{ i \pi }{2} \varepsilon_j \big( s_{1j}+ \sum_{m=j+1}^n s_{jm}  \big)}  {\cal D}(+,2,3,\ldots,n,1,-)  \, ,
\label{s0jK}
\end{align}
the patterns in (\ref{s0jI}) can be extended to the following all-multiplicity proposal
(which is conjectural at $n\geq 6$ points):
\begin{align}
{\cal C}(1,2,3,\ldots, n) &= 2i \sum_{j=2}^n (-1)^j \sin \Big( \frac{\pi}{2}\sum_{i=1 \atop{i\neq j}}^n s_{ij} \Big)
\widehat {\cal D}\Big(j,\Big(
\begin{smallmatrix} - &- &\ldots &- &- \\ j{-}1 &j{-}2 &\ldots &3 &2\end{smallmatrix} \shuffle
\begin{smallmatrix} + &+ &\ldots &+ \\ j{+}1 &j{+}2 &\ldots &n\end{smallmatrix} \Big)\Big)\, .
\label{s0jL}
\end{align}
These relations among twisted cycles determine the first row of the $(n{-}1)! \times (n{-}1)!$
matrices ${\cal H}_{\alpha'}(A|B)$ in (\ref{inival1}) at all multiplicities by matching (\ref{s0jK}) and (\ref{s0jL}) with
\beq
{\cal C}(1,2,3,\ldots, n) = \sum_{B \in S_{n-1}}  {\cal H}_{\alpha'}(2,3,\ldots,n|b_2,b_3,\ldots,b_n)
{\cal D}(+,b_2,b_3,\ldots,b_n,1,-) \, .
\label{s0jZ}
\eeq
The remaining rows of ${\cal H}_{\alpha'}(A|B)$ can be obtained by relabelings as for instance
seen in the three-point case, cf.\ (\ref{s0jA})
\begin{align}
{\cal H}_{\ap}(2,3|2,3) &= 2i \sin\Big(\frac{\pi}{2} (s_{12}{+}s_{23}) \Big) e^{\frac{ i \pi}{2}  s_{13}}
\, , \ \ \ \  {\cal H}_{\ap}(2,3|3,2) =  - 2i \sin\Big(\frac{\pi}{2} (s_{13}{+}s_{23}) \Big) e^{-\frac{ i \pi}{2} s_{12}}
\\
{\cal H}_{\ap}(3,2|3,2) &= 2i \sin\Big(\frac{\pi}{2} (s_{13}{+}s_{23}) \Big) e^{ \frac{ i \pi}{2}  s_{12}}
\, , \ \ \ \ {\cal H}_{\ap}(3,2|2,3) =  - 2i \sin\Big(\frac{\pi}{2} (s_{12}{+}s_{23}) \Big) e^{- \frac{ i \pi}{2} s_{13}} \, . \notag
\end{align}
Four-point examples of the entries ${\cal H}_{\alpha'}(A|B)$ can be found in appendix \ref{app:G1}.

\paragraph{Symmetrized cycles:}
The definition of planar $A$-cycle graph functions at $n\geq 3$ points \cite{Broedel:2018izr} is based on symmetrized
combinations of the integration cycles ${\cal C}(1,a_2,a_3,\ldots,a_n)$,
\beq
{\cal C}^{(n)}_{\rm symm} \equiv \sum_{A\in S_{n-1}} {\cal C}(1,a_2,a_3,\ldots,a_n) = \{ 0<z_j<1 , \ j=2,3,\ldots,n\} \, .
\label{Csym}
\eeq
Instead of adding up permutations of $Z^\tau_{\vec{\eta}}(1,a_2,\ldots,a_n|\ast)$ and separately
applying (\ref{s0jL}) to each twisted cycle, it is rewarding to simplify (\ref{Csym}) via standard
trigonometric identities\footnote{More specifically, one can straightforwardly derive
(\ref{CsymA}) by (possibly repeated) use of
\[e^{\pm ix} = \cos(x)\pm i\sin(x)\, , \ \ \ \,
\cos(a{\pm} b) = \cos(a)\cos(b)\mp \sin(a)\sin(b) \, , \ \ \ \,
\sin(a{\pm} b) = \sin(a)\cos(b)\pm \sin(b)\cos(a)\,.
\]}. For instance, by adding up 2, 6 and 24 permutations of
(\ref{s0jA}), (\ref{s0jD}) and (\ref{s0jG}), respectively, one arrives at
\begin{align}
{\cal C}^{(3)}_{\rm symm}   &= -4  \sin\Big(\frac{\pi}{2} (s_{12}{+}s_{23}) \Big)
 \sin \Big( \frac{ \pi }{2} s_{13}\Big) {\cal D}(+,2,3,1,-)
+ (2\leftrightarrow 3) \notag
\\
{\cal C}^{(4)}_{\rm symm}   &= - 8 i \sin\Big( \frac{ \pi}{2} (s_{12}{+}s_{23}{+}s_{24})\Big)
 \sin\Big( \frac{ \pi}{2} (s_{13}{+}s_{34})\Big)  \notag \\
&\ \ \ \ \times \sin\Big( \frac{ \pi}{2}  s_{14}\Big)  \, {\cal D}(+,2,3,4,1,-)
 + {\rm perm}(2,3,4)
 \label{CsymA}
 \\
 {\cal C}^{(5)}_{\rm symm}   &= 16 \sin\Big( \frac{ \pi}{2} (s_{12}{+}s_{23}{+}s_{24}{+}s_{25})\Big)
\sin\Big( \frac{ \pi}{2} (s_{13}{+}s_{34}{+}s_{35})\Big) \notag \\
&\ \ \ \ \times
\sin\Big( \frac{ \pi}{2} (s_{14}{+}s_{45})\Big)\sin\Big( \frac{ \pi}{2} s_{15}\Big)
 \, {\cal D}(+,2,3,4,5,1,-)
+ {\rm perm}(2,3,4,5) \, ,\notag
\end{align}
where the permutations of $2,3,\ldots,n$ affect the labels of both $s_{ij}$ and
${\cal D}(+,a_2,a_3,\ldots,a_n,1,-)$.
The representations of ${\cal C}^{(n)}_{\rm symm}$ in (\ref{CsymA}) manifest that $n{-}1$
powers of $\pi s_{ij}$ can be factored out from the $\ap$-expansion of each $\sin(\pi x)=
\pi x(1-\zeta_2 x^2 + \frac{3}{4} \zeta_4 x^4+\ldots)$. Extrapolating
the patterns in (\ref{CsymA}) leads us to propose the general formula
\beq
 {\cal C}^{(n)}_{\rm symm}   = (2i)^{n-1} \prod_{j=2}^n \sin \Big( \frac{ \pi }{2} \Big( s_{1j}
+ \! \! \! \sum_{m=j+1}^n  \! \! \! s_{j,m}\Big) \Big) \, {\cal D}(+,2,3,\ldots ,n,1,-) + {\rm perm}(2,3,\ldots,n)
 \label{CsymB}
\eeq
which reproduces ${\cal C}^{(2)}_{\rm symm} = {\cal C}(1,2) = 2i \sin( \frac{ \pi }{2} s_{12}) {\cal D}(+,2,1,-)$
and is conjectural at $n\geq 6$. Note that the product of sine functions in (\ref{CsymB})
matches the diagonal entry $S_{\alpha'/2}(n,\ldots,3,2|$
$n,\ldots,3,2)_1$ of the
string-theory KLT kernel at $\ap \rightarrow \ap/2$ which also plays a role in the simplification
of symmetrized disk integrals $\sum_{A\in S_{n-1}} Z^{\rm tree}(1,A|\ast)$ \cite{Carrasco:2016ygv}.


\subsection{Recovering Parke--Taylor disk integrands}
\label{sec:5.3}

In order to complete the dictionary between the initial values $Z^{i\infty}_{\vec{\eta}}$ of the
$A$-cycle integrals and disk integrals $Z^{\rm tree}$, it remains to recover ${\rm SL}_2$-fixed
Parke--Taylor factors from the degeneration of the Kronecker--Eisenstein integrands.
Based on the expression (\ref{Gpl}) for the planar $\Omega(v,\eta,i\infty)$ at $v \in \RR$, we will determine
the $\eta_j$-dependent entries of the $(n{-}1)! \times (n{-}1)!$ matrix ${\cal K}_{\vec{\eta}}(P|Q)$
in (\ref{inival1}).

As a first step, we generalize the definition (\ref{someintF}) to planar $n$-point cycles $A \in S_n$
\beq
I^{\rm tree}(A| {\cal F}(\sigma_j)) \equiv \int_{{\cal C}(A)} \frac{ \dd \sigma_2 \, \dd \sigma_3\, \ldots \, \dd \sigma_n }{(2\pi i)^{n-1} \,\sigma_2\,\sigma_3\,\ldots \, \sigma_n} \, {\rm KN}_{12\ldots n}^{i\infty}\,{\cal F}(\sigma_j)\, ,
\label{rePT1}
\eeq
where ${\cal F}(\sigma_j)$ may denote an arbitrary rational function of $\sigma_2,\ldots,\sigma_n$,
and the degenerate Koba--Nielsen factor ${\rm KN}_{12\ldots n}^{i\infty}$ can be found in (\ref{inival2}).
In any instance of (\ref{rePT1}) that arises from the degeneration $Z^{i\infty}_{\vec{\eta}}$ of an $n$-point
$A$-cycle integral (\ref{intro2}), the rational function ${\cal F}(\sigma_j)$ is a product of up to $n{-}1$
factors of $G_{ij}=i\pi  \frac{\sigma_i{+}\sigma_j}{\sigma_i-\sigma_j}$, the Green function on the nodal sphere.
The three- and four-point generalizations of the two-point result (\ref{someintG}) are given by
\begin{align}
Z^{i\infty}_{\vec{\eta}}(1,a_2,a_3|1,2,3)&=\pi^2 \cot(\pi \eta_{23})\cot(\pi \eta_3) I^{\rm tree}(1,a_2,a_3|1) +I^{\rm tree}(1,a_2,a_3|G_{12} G_{23}) \notag \\
& \! \! \! \! \! \! \! + \pi \cot(\pi \eta_{23}) I^{\rm tree}(1,a_2,a_3|G_{23})
+ \pi \cot(\pi \eta_{3}) I^{\rm tree}(1,a_2,a_3|G_{12})
 \label{rePT2}
\end{align}
as well as
\begin{align}
&Z^{i\infty}_{\vec{\eta}}(1,a_2,a_3,a_4|1,2,3,4)=\pi^3 \cot(\pi \eta_{234})\cot(\pi \eta_{34}) \cot(\pi \eta_{4})
 I^{\rm tree}(1,a_2,a_3,a_4|1)  \notag \\
&\ \ + \pi \cot(\pi \eta_{234})  I^{\rm tree}(1,a_2,a_3,a_4|G_{23}G_{34})
+ \pi\cot(\pi \eta_{34})  I^{\rm tree}(1,a_2,a_3,a_4|G_{12}G_{34}) \notag \\
&\ \ + \pi\cot(\pi \eta_{4}) I^{\rm tree}(1,a_2,a_3,a_4|G_{12} G_{23})
+ \pi^2  \cot(\pi \eta_{234}) \cot(\pi \eta_{34}) I^{\rm tree}(1,a_2,a_3,a_4|G_{34}) \notag \\
&\ \  +\pi^2 \cot(\pi \eta_{234}) \cot(\pi \eta_{4}) I^{\rm tree}(1,a_2,a_3,a_4|G_{23})
+\pi^2 \cot(\pi \eta_{34}) \cot(\pi \eta_{4}) I^{\rm tree}(1,a_2,a_3,a_4|G_{12}) \notag  \\
&\ \ +I^{\rm tree}(1,a_2,a_3,a_4|G_{12} G_{23}G_{34}) \, ,
\label{rePT3}
\end{align}
where the expansion of the cotangent can be found in (\ref{trigdeg}), and we use the shorthand
 $\eta_{ij\ldots l} =\eta_i{+}\eta_j{+}\ldots{+}\eta_l$ throughout this section.

The degeneration of the $n$-point integrand $\prod_{j=2}^n \Omega(z_{j-1,j},\eta_{j,j+1\ldots n},\tau)$
of $Z^{i\infty}_{\vec{\eta}}(\ast|1,2,\ldots,n)$ can be described by summing over all the $2^{n-1}$
possibilities to distribute the $n{-}1$ factors into two distinguishable and disjoint sets $P$ and $Q$.
For each label $j$ in set $P$ and $Q$, the factor of $\Omega(z_{j-1,j},\eta_{j,j+1\ldots n},i\infty)$ is mapped to its contribution
$\pi \cot(\pi \eta_{j,j+1\ldots n})$ and $G_{j-1,j}$, respectively:
\begin{align}
Z^{i\infty}_{\vec{\eta}}(1,A|1,2,3,\ldots,n) &= \! \! \! \! \! \sum_{\{2,3,\ldots,n\}  = P \cup Q\atop{ P \cap Q = \emptyset}} \! \! \!
\Big(\prod_{i \in P} \pi \cot(\pi \eta_{i,i+1\ldots n-1,n})  \Big)
I^{\rm tree}\big(1,A| \, \prod_{j \in Q}  G_{j-1,j} \big)\, .
\label{rePT4}
\end{align}
Since the integrands in (\ref{rePT1}) only depend on the punctures via $(\sigma_2 \sigma_3\ldots \sigma_n)^{-1}$
and products of $G_{ij}$, one can apply the algorithmic method of \cite{He:2017spx} to recover combinations of
${\rm SL}_2$-fixed Parke--Taylor factors in $n{+}2$ punctures,
\beq
{\rm PT}(a_1,a_2,a_3,\ldots,a_{n+1},a_{n+2}) \equiv \frac{1}{\sigma_{a_1a_2} \sigma_{a_2a_3} \sigma_{a_3a_4}
\ldots \sigma_{a_{n+1}a_{n+2}} \sigma_{a_{n+2}a_1}}\, .
\label{inival5}
\eeq
Genus-one integrands $ I^{\rm tree}(\ast|1)$ without any insertion of $G_{ij}$ descend from\footnote{This
can be seen from the following corollary of repeated partial-fraction identities:
\[
\frac{1}{\sigma_2 \sigma_3\ldots \sigma_n} = \frac{(-1)^n}{ \sigma_{2} \sigma_{23} \sigma_{34}\ldots\sigma_{n-2,n-1}\sigma_{n-1,n}} + {\rm perm}(2,3,\ldots,n)\, .
\]}
\begin{align}
\prod_{j=2}^n \frac{1}{\sigma_j} &=  (-1)^{n} \lim_{\sigma_{-} \rightarrow \infty \atop {\sigma_+ \rightarrow 0}}  |\sigma_-|^2
\sum_{A \in S_{n-1}} {\rm PT}(+,a_2,a_3,\ldots,a_n,-,1)
\notag\\
&=  (-1)^{n-1} \lim_{\sigma_{-} \rightarrow \infty \atop {\sigma_+ \rightarrow 0}}  |\sigma_-|^2
\sum_{B \in S_n} {\rm PT}(+,b_1,b_2,\ldots,b_n,-)  \, , \label{cocyc8}
\end{align}
where Kleiss--Kuijf relations \cite{Kleiss:1988ne} have been used to trade $(n{-}1)!$ permutations $A$ of $2,3,\ldots,n$
for $n!$ permutations $B$ of $1,2,\ldots,n$. The insertion of $|\sigma_-|^2$ stems from the Jacobian $|\sigma_{1,+}
\sigma_{1,-}\sigma_{+,-}|$ of the ${\rm SL}_2(\RR)$ frame $(\sigma_+,\sigma_1,\sigma_-)=(0,1,\infty)$.

The structure of the $n!$-term Parke--Taylor expansion in (\ref{cocyc8}) is preserved when
adjoining factors of $G_{ij}$: As one can anticipate from the example
\begin{align}
\frac{ G_{12} }{\sigma_2 \sigma_3} &= i \pi \lim_{\sigma_{-} \rightarrow \infty \atop {\sigma_+ \rightarrow 0}}  |\sigma_-|^2
\Big( {-} {\rm PT}(+,1,2,3,-) - {\rm PT}(+,1,3,2,-) - {\rm PT}(+,3,1,2,-) \notag \\
& \ \ \ \ \ \ \ \ \ \ \ \ \ \ \ \ \
+{\rm PT}(+,2,1,3,-) + {\rm PT}(+,2,3,1,-)+ {\rm PT}(+,3,2,1,-) \Big) \, ,
 \label{cocyc21}
\end{align}
the net effect of the factor $G_{ij}$ is to attribute relative minus signs to individual Parke--Taylor
factors, depending on the relative position of legs $i$ and $j$ in its cyclic ordering. These signs
can be compactly encoded in the shorthand
\beq
\te{sgn}_{ij}^B = \left\{ \begin{array}{rl} +1 &: \ i \ \te{is right of} \ j \ \te{in} \ B=b_1,b_2,\ldots,b_n
\\
 -1 &: \ i \ \te{is left of} \ j \ \te{in} \ B=b_1,b_2,\ldots,b_n   \end{array}\right. \, ,
 \label{cocyc22}
\eeq
which enters the following all-multiplicity formula for $k\leq n{-}1$ factors of $G_{ij}\rightarrow i\pi \, {\rm sgn}_{ij}^B$ \cite{He:2017spx}:
\begin{align}
\Big(\prod_{j=2}^n \frac{1}{\sigma_j} \Big)&G_{p_1 q_1} G_{p_2 q_2} \ldots G_{p_k q_k} = (-1)^{n-1} (i \pi)^k \lim_{\sigma_{-} \rightarrow \infty \atop {\sigma_+ \rightarrow 0}}  |\sigma_-|^2 \label{cocyc24} \\
&\times
\sum_{B \in S_n} {\rm PT}(+,b_1,b_2,\ldots,b_n,-) {\rm sgn}^{B}_{p_1 q_1}  {\rm sgn}^{B}_{p_2 q_2} \ldots
{\rm sgn}^{B}_{p_k q_k} \, .
\notag
\end{align}
This result cannot be applied to products $G_{p_1 q_1} G_{p_2 q_2} \ldots G_{p_k q_k}$ whose labels form
a cycle, e.g.\ $G_{12}^2$ and $G_{12}G_{23}G_{31}$ do not admit a Parke--Taylor decomposition
via (\ref{cocyc24}). Cycles of this type do not appear in the integrands of $I^{\rm tree}(\cdot|\cdot)$ in
(\ref{rePT4}) -- the arrangement of $G_{ij}$ inherits the chain structure from the integrands
$\Omega(z_{12},\ldots)\Omega(z_{23},\ldots)\ldots \Omega(z_{n-1,n},\ldots)$ of $Z^\tau_{\vec{\eta}}(\cdot|\cdot)$.

The main results of this subsection follow from applying the integrand manipulations (\ref{cocyc8}) and
(\ref{cocyc24}) to the general definition of $I^{\rm tree}(\cdot|\cdot)$ in (\ref{rePT1}).
In absence of $G_{ij}$, each term in the Parke--Taylor decomposition
\begin{align}
I^{\rm tree}(1,A|1) &=(-1)^{n-1} \int_{{\cal C}(A)} \frac{ \dd \sigma_+ \, \dd \sigma_- \, \dd \sigma_1 \, \ldots \,
\dd \sigma_n}{(2\pi i)^{n-1}\, \te{vol} \, {\rm SL}_2(\RR)} \, {\rm KN}_{12\ldots n}^{i\infty} \sum_{B \in S_n}
{\rm PT}(+,B,-) \notag \\
&= \frac{1}{(-2\pi i)^{n-1}} \sum_{P\in S_{n-1}} {\cal H}_{\ap}(A|P) \sum_{B\in S_n}
Z^{\rm tree}(+,P,1,- | +,B,-)
\label{rePT11}
\end{align}
reproduces the definition (\ref{cocyc1}) of disk integrals, with $s_{j+},s_{j-}$ and $s_{+,-}$ given by (\ref{inival3}).
The matrix ${\cal H}_{\ap}(A|P)$ arises from the translation of planar genus-one cycles ${\cal C}(1,A)$
into disk orderings via (\ref{s0jZ}). The same reasoning
applies to products of $G_{ij}$ without subcycles,
\begin{align}
I^{\rm tree}(1,A| &G_{p_1 q_1} G_{p_2 q_2} \ldots G_{p_k q_k} ) = \frac{(i\pi)^k}{(-2\pi i)^{n-1}} \sum_{P\in S_{n-1}} {\cal H}_{\ap}(A|P) \label{reprePT12} \\
&\times \sum_{B\in S_n}  {\rm sgn}^{B}_{p_1 q_1}  {\rm sgn}^{B}_{p_2 q_2} \ldots
{\rm sgn}^{B}_{p_k q_k}  Z^{\rm tree}(+,P,1,- | +,B,-)\, .
\notag
\end{align}
Upon insertion into (\ref{rePT4}), this implies that any initial value $Z^{i \infty}_{\vec{\eta}}(\cdot|\cdot)$
is expressible in terms of $(n{+}2)$-point Parke--Taylor integrals. Note that each factor of $G_{ij}$ introduces
a minus sign into the expressions (\ref{reprePT12}) when trading the original integration cycle $A=a_2a_3\ldots a_n$
for its reversal $A^t= a_n\ldots a_3 a_2$. Hence, integrands of (\ref{rePT1}) with a fixed number $k$ of $G_{ij}$ factors
integrate to zero on ${\cal C}(1,A) - (-1)^k {\cal C}(1,A^t)$, and one may replace
$ {\cal H}_{\ap}(A|P) $ by the parity-weighted combination
$\frac{1}{2} \big[  {\cal H}_{\ap}(A|P)
+ (-1)^k  {\cal H}_{\ap}(A^t|P) \big] $ in (\ref{reprePT12}),
%
\begin{align}
I^{\rm tree}(1,A| G_{p_1 q_1} G_{p_2 q_2} \ldots &G_{p_k q_k} ) = (-1)^{n-1} \frac{ (i\pi)^{k-n+1} }{2^n} \sum_{P\in S_{n-1}}
 \big[  {\cal H}_{\ap}(A|P) + (-1)^k  {\cal H}_{\ap}(A^t|P) \big]
\notag \\
&\times \sum_{B\in S_n}  {\rm sgn}^{B}_{p_1 q_1}  {\rm sgn}^{B}_{p_2 q_2} \ldots
{\rm sgn}^{B}_{p_k q_k}  Z^{\rm tree}(+,P,1,- | +,B,-)\, .
 \label{rePT12}
\end{align}
As one can see from the examples in appendix \ref{app:G2}, the coefficients in the $\ap$-expansion of
${\cal H}_{\ap}(A|P)+ (-1)^k  {\cal H}_{\ap}(A^t|P)$ are real and imaginary if $n{-}k$ is odd and even,
respectively. Hence, by the prefactor $(i\pi)^{k-n+1}$ in its first line, (\ref{rePT12}) always yields
real linear combinations of disk integrals $Z^{\rm tree}$.

The combination of (\ref{rePT4}) and
(\ref{rePT12}) determines the transformation matrix ${\cal K}_{\vec{\eta}}(P|Q)$ between the basis integrands
in (\ref{inival1}) once the $n!$ Parke--Taylor factors of $Z^{\rm tree}(\ast| +,B,-)$ are reduced to
an $(n{-}1)!$-element basis of $Z^{\rm tree}(\ast | +,Q,-,1)$ via BCJ relations \cite{BCJ, Zfunctions}. One
can also attain this BCJ basis by removing any appearance of $i,j=1$ from $G_{ij}$
via Fay relations
\beq
G_{12}G_{23}+G_{23}G_{31}+G_{31}G_{12}= \pi^2
\label{faygij}
\eeq
combined with integration by parts \cite{He:2017spx}
\beq
\partial_{v_i} \te{KN}^{i\infty}_{12\ldots n} \, \big|_{{\cal C}(A)}  = -  \sum_{j=1 \atop{j\neq i}}^n s_{ij} G_{ij}  \te{KN}^{i\infty}_{12\ldots n} \, .
\label{rePT13}
\eeq
The latter follows from the limit $\tau \rightarrow i \infty$ of (\ref{B5}) on a planar integration cycle.

The techniques of this subsection will now be illustrated by various examples.
The two-point instance of (\ref{rePT11}) with ${\cal H}_{\ap}(2|2)=2i   \sin  ( \frac{ \pi }{2} s_{12})$
reproduces the earlier result (\ref{KNdrft1}) via
\begin{align}
I^{\rm tree}(1,2|1)  &= - \frac{1}{\pi} \sin \Big( \frac{ \pi }{2}  s_{12}\Big) \big[Z^{\rm tree}(+,2,1,-|+,1,2,-)
+ Z^{\rm tree}(+,2,1,-|+,2,1,-) \big] \notag \\
&=  \frac{1}{\pi} \sin \Big( \frac{ \pi }{2}  s_{12}\Big) Z^{\rm tree}(+,2,1,-|+,2,-,1)   = \frac{ \Gamma(1-s_{12}) }{ \big[ \Gamma(1-\tfrac{s_{12}}{2}) \big]^2} \, .  \label{KNdeg162}
\end{align}
Several three- and four-point examples can be found in the next subsections.


\subsection{Three points}
\label{sec:5.4}

As a first application of the general results above, we will now relate the constituents
of $Z^{i\infty}_{\eta_2,\eta_3}(1,2,3|1,2,3)$
in (\ref{rePT2}) to five-point disk integrals. In absence of $G_{ij}$ in the integrand
at $\tau \rightarrow i\infty$, one can symmetrize the cycle by $ I^{\rm tree}(1,2,3|1)= I^{\rm tree}(1,3,2|1)$,
and (\ref{rePT11}) implies
\begin{align}
 I^{\rm tree}(1,2,3|1)&=   \frac{1}{2\pi^2}
 \Big[ \sin \Big( \frac{ \pi }{2} (s_{12}{+}s_{23}) \Big) \sin \Big( \frac{ \pi }{2} s_{13} \Big) \sum_{B \in S_3}
 Z^{\rm tree}(+,2,3,1,-| +,B,-)  \notag \\
 & \ \ \ \ \ \ \ \  \ \ +  \sin \Big( \frac{ \pi }{2} (s_{13}{+}s_{23}) \Big) \sin \Big( \frac{ \pi }{2} s_{12} \Big)
 \sum_{B \in S_3} Z^{\rm tree}(+,3,2,1,-| +,B,-)    \Big]  \label{rePT17} \\
 &= - \frac{1}{2\pi^2}
 \Big[ \sin \Big( \frac{ \pi }{2} (s_{12}{+}s_{23}) \Big) \sin \Big( \frac{ \pi }{2} s_{13} \Big)
 Z^{\rm tree}(+,2,3,1,-| +,(2\shuffle 3),-,1)  \notag \\
 & \ \ \ \ \ \ \ \  \ \ +  \sin \Big( \frac{ \pi }{2} (s_{13}{+}s_{23}) \Big) \sin \Big( \frac{ \pi }{2} s_{12} \Big)
 Z^{\rm tree}(+,3,2,1,-| +,(2\shuffle 3),-,1)    \Big] \, ,
\notag
\end{align}
see (\ref{s0jC}) for the trigonometric functions in $\frac{1}{2}\big[{\cal H}_{\ap}(2,3|P)
{+}   {\cal H}_{\ap}(3,2|P)\big]$. The disk integrals have been rewritten in a BCJ basis
in passing to the last two lines. Integrals over a single $G_{ij}$ in turn are asymmetric in the cycle,
$ I^{\rm tree}(1,2,3|G_{23}) =- I^{\rm tree}(1,3,2|G_{23})$, i.e.\
\small
\begin{align}
 I^{\rm tree}&(1,2,3|G_{23})=  \frac{1}{2\pi}   \label{rePT18}  \\
 &\! \! \! \! \times \Big[ \sin \Big( \frac{ \pi }{2} (s_{12}{+}s_{23}) \Big) \cos \Big( \frac{ \pi }{2} s_{13} \Big)
 \Big( Z^{\rm tree}(+,2,3,1,-| +,2,3,-,1)  - Z^{\rm tree}(+,2,3,1,-| +,3,2,-,1) \Big)  \notag \\
 &  -  \sin \Big( \frac{ \pi }{2} (s_{13}{+}s_{23}) \Big) \cos \Big( \frac{ \pi }{2} s_{12} \Big)
 \Big( Z^{\rm tree}(+,3,2,1,-| +,2,3,-,1) - Z^{\rm tree}(+,3,2,1,-| +,3,2,-,1) \Big)    \Big]  \notag
\end{align}
\normalsize
after reduction to a BCJ basis. The remaining configurations of $G_{ij}$
in (\ref{rePT2}) follow from the Fay identity (\ref{faygij}) and integration by parts (\ref{rePT13}) at the level of the $G_{ij}$,
\begin{align}
s_{12} I^{\rm tree}(1,a_2,a_3|G_{12}) &= s_{23}  I^{\rm tree}(1,a_2,a_3|G_{23})  \label{rePT180}
\\
 I^{\rm tree}(1,a_2,a_3|G_{12}G_{23}) &= \frac{ \pi^2 s_{13}}{s_{123}}  I^{\rm tree}(1,a_2,a_3|1)\, , \notag
\end{align}
so (\ref{rePT2}) can be decomposed as follows in terms of (\ref{rePT17}) and (\ref{rePT18}):
\begin{align}
Z^{i\infty}_{\eta_2,\eta_3}(1,a_2,a_3|1,2,3)&= \Big( \pi^2 \cot(\pi \eta_{23})\cot(\pi \eta_3) +\frac{ \pi^2 s_{13}}{s_{123}} \Big) I^{\rm tree}(1,a_2,a_3|1)  \notag \\
& \! \! \! \! \! \! \! + \Big( \pi \cot(\pi \eta_{23})  +  \frac{ s_{23}}{s_{12}} \pi \cot(\pi \eta_{3}) \Big)
 I^{\rm tree}(1,a_2,a_3|G_{23})\, .
 \label{rePT2new}
\end{align}
Since the $I^{\rm tree}(\cdot | \cdot)$ have been expanded in the $2\times 2$ basis of
 $Z^{\rm tree}(+,P,1,-|+,Q,-,1)$, one can read off the entries of the transformation matrix
 ${\cal K}_{\eta_2,\eta_3}(\cdot|\cdot)$ for the integrands in~(\ref{inival1}):
\begin{align}
{\cal K}_{\eta_2,\eta_3}(2,3|2,3)&= - \Big( \pi^2 \cot(\pi \eta_{23})\cot(\pi \eta_3) +\frac{ \pi^2 s_{13}}{s_{123}} \Big)
+ i \pi \Big( \pi \cot(\pi \eta_{23})  +  \frac{ s_{23}}{s_{12}} \pi \cot(\pi \eta_{3}) \Big)
 \label{rePT3new} \\
 {\cal K}_{\eta_2,\eta_3}(2,3|3,2)&= - \Big( \pi^2 \cot(\pi \eta_{23})\cot(\pi \eta_3) +\frac{ \pi^2 s_{13}}{s_{123}} \Big)
- i \pi \Big( \pi \cot(\pi \eta_{23})  +  \frac{ s_{23}}{s_{12}} \pi \cot(\pi \eta_{3}) \Big) \, . \notag
\end{align}

\paragraph{$\ap$-expansions:}

One can now insert the $\ap$-expansion of the disk integrals (see for instance \cite{wwwMZV, Mafra:2016mcc})
into (\ref{rePT17}) and (\ref{rePT18}). By rewriting $s_{j+}=s_{j-}=-\frac{1}{2}(s_{1j}{+}s_{23})$ and
$\pi^2 = 6 \zeta_2$, we arrive at the following permutation symmetric series in $s_{ij}$,
\begin{align}
 I^{\rm tree}(1,2,3|1)&= \frac{1}{2} + \frac{1}{8} \zeta_2 (s_{12}^2 + s_{13}^2 + s_{23}^2)
+  \frac{1}{8}  \zeta_3 (s_{12}^3 + s_{13}^3  + s_{23}^3 + s_{12} s_{13} s_{23}) \notag \\
&
+ \frac{1}{320} \zeta_2^2 \Big[19 s_{12}^4
+ 10 s_{12}^2 s_{13}^2
+ 12 s_{12}^2 s_{13} s_{23}  +  {\rm cyc}(1,2,3) \Big] \notag\\
&+  \frac{1}{32} \zeta_5 \Big[ 3 s_{12}^5
+ 2 s_{12}^3 s_{13} s_{23}
+ 3 s_{12}^2 s_{13}^2 s_{23} +  {\rm cyc}(1,2,3) \Big]   \label{KNdeg178} \\
&+  \frac{1}{32}  \zeta_2  \zeta_3 (s_{12}^2 + s_{13}^2 + s_{23}^2) (s_{12}^3 + s_{13}^3 +  s_{23}^3+ s_{12} s_{13} s_{23} )
+ {\cal O}(\ap^6) \notag
\end{align}
as well as
\begin{align}
s_{23} &I^{\rm tree}(1,2,3|G_{23})= 1
+  \frac{1}{4} \zeta_2 (s_{12} {+} s_{13} {+} s_{23})^2  \notag \\
&+  \frac{1}{4} \zeta_3 \Big\{ \Big[s_{12}^3
+ 3 s_{12}^2 s_{13} + 3 s_{12} s_{13}^2    +  {\rm cyc}(1,2,3) \Big]
+  7 s_{12} s_{13} s_{23} \Big\}
   \notag \\
&+ \frac{1}{160} \zeta_2^2 (s_{12} {+} s_{13} {+} s_{23}) \Big\{ \Big[
19 s_{12}^3 + 57 s_{12}^2 s_{13} + 57 s_{12} s_{13}^2 +  {\rm cyc}(1,2,3) \Big]
+ 126 s_{12} s_{13} s_{23}   \Big\} \notag \\
&+ \frac{  1}{16} \zeta_5  \Big[
3 s_{12}^5
  + 15 s_{12}^4 s_{13}  + 15 s_{12} s_{13}^4
  + 30 s_{12}^3 s_{13}^2 + 30 s_{12}^2 s_{13}^3
    + 62 s_{12}^3 s_{13} s_{23}
\label{KNdeg179}\\
 & \ \ \ \ \ \ \  \ \ \ \ \ \ \  \ \ \ \ \ \ \  \ \ \ \ \ \ \  +  93 s_{12}^2 s_{13}^2 s_{23}   +  {\rm cyc}(1,2,3) \Big]  \notag \\
&+ \frac{ 1}{16}  \zeta_2  \zeta_3  (s_{12} {+} s_{13} {+} s_{23})^2
 \Big\{ \Big[
s_{12}^3 + 3 s_{12}^2 s_{13} + 3 s_{12} s_{13}^2    +  {\rm cyc}(1,2,3) \Big]
+  7 s_{12} s_{13} s_{23} \Big\} + {\cal O}(\ap^6) \, .
 \notag
\end{align}
At higher orders in $\ap$, the irreducible depth-$(d{\geq} 2)$ MZVs $\zeta_{3,5},\zeta_{3,7}$ and $\zeta_{3,5,3}$ of the
five-point $Z^{\rm tree}(\cdot|\cdot)$ drop out from both (\ref{KNdeg178}) and (\ref{KNdeg179}). This
is a peculiarity of the arguments $s_{j+ }=s_{j-}=-\frac{1}{2}(s_{1j}{+}s_{23})$ which cause the
$I^{\rm tree}(\cdot|\cdot)$ to be functions of three variables instead of the five Mandelstam
invariants of a five-point disk integral.

\paragraph{Component integrals:}
By inserting (\ref{KNdeg178}) and (\ref{KNdeg179}) into the initial value (\ref{rePT2new}),
one can access any order in the $\ap$-expansion of the three-point $A$-cycle integrals via (\ref{someint33}).
In particular, inspection of specific orders in $\eta_2{+}\eta_3$ and $\eta_3$ yields explicit
results for the component integrals
\begin{align}
Z^{\tau}_{(m_1,m_2)}(A|1,2,3) &\equiv Z^{\tau}_{\eta_2,\eta_3}(A|1,2,3) \, \big|_{(\eta_{2}{+}\eta_3)^{m_1-1}\eta_3^{m_2-1}}
\label{compin71}\\
&= \int_{{\cal C}(A)} \dd z_2 \, \dd z_3 \, f^{(m_1)}(z_{12},\tau)f^{(m_2)}(z_{23},\tau) \, {\rm KN}_{123}^\tau \, .
\notag
\end{align}
Since all the $r_{\eta_2,\eta_3}(\ep_k)$ in (\ref{DER11}) are even under $(\eta_2,\eta_3) \rightarrow -(\eta_2,\eta_3)$,
one can take a convenient shortcut in extracting component integrals from the generating series:
Any $Z^{\tau}_{(m_1,m_2)}(A|1,2,3)$ with even parity $m_1{+}m_2 \in 2\NN_0$
(odd parity $m_1{+}m_2 \in 2\NN_0{+}1$) is determined by the even (odd) part of the initial value
$Z^{i\infty}_{\eta_2,\eta_3}$ in (\ref{rePT2new}) w.r.t.\ $\eta_j \rightarrow - \eta_j$.
Hence, $I^{\te{tree}}(A|G_{ij})$ with odd coefficients $\sim \cot(\pi \eta_{23})$ or $\cot(\pi \eta_{j})$ do
not contribute to $Z^{\tau}_{(m_1,m_2)}(A|1,2,3)\big|_{m_1+m_2 \ \te{even}}$, and one can similarly
disregard $I^{\te{tree}}(A|1)$ in (\ref{rePT2new}) when computing
$Z^{\tau}_{(m_1,m_2)}(A|1,2,3)\big|_{m_1+m_2 \ \te{odd}}$.

Similar to the two-point case (\ref{CC160}), the simplest integrand $(m_1,m_2)=(0,0)$ yields a generating
series of planar $A$-cycle graph functions involving up to three vertices\footnote{In the graphical notation
of \cite{Broedel:2018izr}, the three-vertex graph functions in (\ref{CC33pl}) are represented as follows:

\begin{center}
 \tikzpicture
\draw(-0.135,0)node{$A_{111}= {\bf A}\Big[ \ \ \ \ \ \ \Big]$};
\draw(0.3,-0.2)node{$\bullet$};
\draw(0.8,-0.2)node{$\bullet$};
\draw(0.55,0.22)node{$\bullet$};
\draw(0.3,-0.2)--(0.8,-0.2);
\draw(0.55,0.22)--(0.8,-0.2);
\draw(0.3,-0.2)--(0.55,0.22);
\scope[xshift=5cm]
\draw(-0.135,0)node{$A_{211}= {\bf A}\Big[ \ \ \ \ \ \ \Big]$};
\draw(0.3,-0.2)node{$\bullet$};
\draw(0.8,-0.2)node{$\bullet$};
\draw(0.55,0.22)node{$\bullet$};
\draw(0.3,-0.2) .. controls (0.45,-0.1) and (0.65,-0.1) .. (0.8,-0.2);
\draw(0.3,-0.2) .. controls (0.45,-0.3) and (0.65,-0.3) .. (0.8,-0.2);
\draw(0.55,0.22)--(0.8,-0.2);
\draw(0.3,-0.2)--(0.55,0.22);
\endscope
\endtikzpicture \end{center}},
\begin{align}
Z^{\tau}_{(0,0)}&(1,2,3|1,2,3) + Z^{\tau}_{(0,0)}(1,3,2|1,2,3) = 1 + \frac{1}{2} (s_{12}^2 {+} s_{13}^2{+}s_{23}^2)   A_2(\tau)
 \notag \\
& \ \ \ \ + s_{12}s_{13}s_{23} A_{111}(\tau)  + \frac{1}{3!}   ( s_{12}^3 {+}s_{13}^3{+}s_{23}^3) A_{3}(\tau)
+ \frac{1}{4} ( s_{12}^2 s_{13}^2{+}s_{12}^2s_{23}^2{+}s_{13}^2s_{23}^2) A_2(\tau)^2
 \notag \\
&\ \ \ \ +\frac{1}{2} s_{12}s_{13}s_{23} (s_{12}{+}s_{13}{+}s_{23}) A_{211}(\tau) + \frac{1}{4!}   (s_{12}^4 {+}s_{13}^4{+}s_{23}^4) A_{4}(\tau) + {\cal O}(\ap^5)
\label{CC33pl} \, ,
\end{align}
see (\ref{CC18}) for the planar two-vertex graph functions $A_w(\tau)$.
The $\ap$-expansion (\ref{someint33}) has been checked to reproduce
all the planar three-vertex $A$-cycle graph functions
\beq
A_{ijk}(\tau) = i! j! k!
\big[ Z^{\tau}_{(0,0)}(1,2,3|1,2,3)
+ Z^{\tau}_{(0,0)}(1,3,2|1,2,3)  \big] \, \big|_{s_{12}^i s_{23}^j s_{13}^k } = 0 \, ,
\ \ \ \ \ \ i,j,k\geq 1
\label{inshere}
\eeq
of weight $i{+}j{+}k\leq 6$ known from \cite{Broedel:2018izr}, e.g.
\begin{align}
A_{111}(\tau)&= \frac{ \zeta_3}{4} + 36 \zeta_2 \gamma_0(4, 0, 0|\tau)
-  60 \gamma_0(6, 0, 0|\tau)
\notag\\
A_{211}(\tau)&=
\frac{3 \zeta_4}{8} -144 \zeta_4 \gamma_0(4, 0, 0, 0|\tau) -
 36 \gamma_0(4, 4, 0, 0|\tau)  \label{CC36}  \\
 &\ \ \ \ + 1680 \zeta_2 \gamma_0(6, 0, 0, 0|\tau) -
 756 \gamma_0(8, 0, 0, 0|\tau)\, . \notag
\end{align}
Moreover, the differential-equation method of this work directly yields the fully
simplified iterated-Eisenstein-integral representation of $A_{ijk}(\tau)$ which is particularly helpful
at higher weights $i{+}j{+}k$ beyond the current reach of the eMZV datamine \cite{WWWe}. Moreover,
the degeneration of planar three-vertex $A$-cycle
graph functions at the cusp follows from
\begin{align}
Z^{i\infty}_{(0,0)}&(1,2,3|1,2,3) + Z^{i\infty}_{(0,0)}(1,3,2|1,2,3) = 2  I^{\rm tree}(1,2,3|1)  \label{gen3v} \\
&= -\frac{1}{\pi^2}
\sin \Big( \frac{ \pi }{2} (s_{12}{+}s_{23}) \Big) \sin \Big( \frac{ \pi }{2} s_{13} \Big)
 Z^{\rm tree}(+,2,3,1,-| +,(2\shuffle 3),-,1)
 + (2\leftrightarrow 3) \notag
\end{align}
upon insertion into (\ref{inshere}).

The expansion of integrals (\ref{compin71}) over non-constant $f^{(m_1)}_{12}f^{(m_2)}_{23}$
with $m_j\neq 0$ has not yet been discussed in the literature, and the leading orders of some
representative examples read
\begin{align}
Z^{\tau}_{(2,0)}&(1,2,3|1,2,3) = -\zeta_2
+ \frac{3}{2} s_{12} \gamma_0(4|\tau)
- \frac{1}{4} (s_{12}^2 {+} s_{13}^2 {+} s_{23}^2) \zeta_2^2 \notag \\
&\! \! \! \! +3(   s_{13}^2 {+}   s_{23}^2  {-}  2 s_{13} s_{23}  {-} 3 s_{12}^2  ) \zeta_2 \gamma_0(4, 0|\tau)
  + 5 (s_{12}^2 {+} 2 s_{13} s_{23}) \gamma_0(6, 0|\tau)
   + {\cal O}(\ap^3)  \notag \\
Z^{\tau}_{(3,3)}&(1,2,3|1,2,3) =
21 s_{13} \zeta_4 \gamma_0(4|\tau)
+ 5 s_{13} \zeta_2 \gamma_0(6|\tau) -
\frac{ 7}{2} s_{13} \gamma_0(8|\tau)  \label{smalleq} \\
 &\! \! \! \! +  27 s_{13} (5 s_{12} {-} 8 s_{13} {+} 5 s_{23}) \zeta_6   \gamma_0(4, 0|\tau)
+  15 s_{13} (17 s_{12} {+} 10 s_{13} {+} 17 s_{23}) \zeta_4  \gamma_0(6, 0|\tau) \notag\\
&\! \! \! \! + 9 s_{13} (3 s_{12} {+} s_{13} {+} 3 s_{23}) \zeta_2 \gamma_0(4, 4|\tau)
+21 s_{13} (5 s_{12} {+} 7 s_{13} {+} 5 s_{23}) \zeta_2 \gamma_0(8, 0|\tau)   - \frac{15}{2} s_{13}^2 \gamma_0(4, 6|\tau)  \notag \\
&\! \! \! \!   -\frac{15}{2} s_{13} (3 s_{12} {+} s_{13} {+} 3 s_{23})   \gamma_0(6, 4|\tau)
 -\frac{9}{2} s_{13} (19 s_{12} {+} 20 s_{13} {+} 19 s_{23})  \gamma_0(10, 0|\tau) + {\cal O}(\ap^3)\, . \notag
\end{align}
Integrals over $f^{(1)}_{ij}f^{(m_2)}_{jk}$ with even $m_2$ introduce kinematic poles that stem from the initial value
(\ref{rePT2new}) when computing the $\ap$-expansion via (\ref{someint33}), e.g.\
\begin{align}
Z^{\tau}_{(1,0)}&(1,2,3|1,2,3) = \frac{1}{s_{12}} \Big\{ 1
+(s_{12} {+} s_{13} {+} s_{23})^2 \Big( \frac{ \zeta_2}{4}  - 3 \gamma_0(4,0|\tau)   \Big)  \notag \\
& +(s_{12} {+} s_{13} {+} s_{23})^3 \Big( \frac{ \zeta_3}{4} -  10  \gamma_0(6,0,0|\tau)
+  24  \zeta_2 \gamma_0(4,0,0|\tau)  \Big)\label{medeq} \\
&+  s_{12} s_{13} s_{23} \Big( \frac{ \zeta_3}{4}   - 90  \gamma_0(6,0,0|\tau)   \Big) +{\cal O}(\ap^4) \Big\}\, . \notag
\end{align}
Integrals over $f^{(1)}_{ij}f^{(m_2)}_{jk}$ with odd $m_2\geq 3$ in turn are regular as $s_{ij}\rightarrow 0$,
and the $\ap^{\leq 3}$-orders of the example $Z^{\tau}_{(1,3)}(1,2,3|1,2,3)$ can be found in (\ref{bigeq}).
Finally, integration over $f^{(1)}_{ij}f^{(1)}_{jk}$ introduces the pole structure $\sim \frac{1}{s_{123}}$ as
one can anticipate from the behavior (\ref{rePT180}) of the $\tau \rightarrow i\infty$ degeneration.

Note that (\ref{smalleq}) and (\ref{bigeq}) exemplify some of the simplest situations
to encounter irreducible iterated Einstein integrals of depth two with different entries: Neither $ \gamma_0(6, 4|\tau)$
at the $\ap^2$-order of $Z^{\tau}_{(3,3)}(1,2,3|1,2,3)$ nor $\gamma_0(4, 4, 0|\tau)$
or $\gamma_0(4, 6, 0|\tau)$ at the $\ap^3$-order of $Z^{\tau}_{(1,3)}(1,2,3|1,2,3)$ are expressible as
shuffle products of depth-one representatives.


\subsection{Four points}
\label{sec:5.5}

All the constituents of $Z^{i\infty}_{\vec{\eta}}(1,A|1,2,3,4)$ in (\ref{rePT3}) can be expressed in
terms of six-point disk integrals via (\ref{rePT12}), for instance
\begin{align}
I^{\rm tree}&(1,2,3,4|1)= - \frac{1}{4\pi^3}  \! \! \sum_{B\in S_3} \! \!
 \Big[
 \sin \Big( \frac{ \pi }{2} (s_{12}{+}s_{23}{+}s_{24}) \Big) \cos \Big( \frac{ \pi }{2} s_{134} \Big)  Z^{\rm tree}(+,2,3,4,1,-|+,B,-,1) \notag \\
 & \ \ \ \ \ \ \ \
 - \sin \Big( \frac{ \pi }{2} (s_{13}{+}s_{23}{+}s_{34}) \Big) \cos \Big( \frac{ \pi }{2}({-}s_{12}{+}s_{14}{-}s_{24}) \Big)
 Z^{\rm tree}(+,3,2,4,1,-|+,B,-,1)  \notag \\
 & \ \ \ \ \ \ \ \
  - \sin \Big( \frac{ \pi }{2} (s_{13}{+}s_{23}{+}s_{34}) \Big) \cos \Big( \frac{ \pi }{2}({-}s_{12}{+}s_{14}{+}s_{24}) \Big)
  Z^{\rm tree}(+,3,4,2,1,-|+,B,-,1)\notag \\
 & \ \ \ \ \ \ \ \
 + \sin \Big( \frac{ \pi }{2} (s_{14}{+}s_{24}{+}s_{34}) \Big) \cos \Big( \frac{ \pi }{2}s_{123} \Big)
 Z^{\rm tree}(+,4,3,2,1,-|+,B,-,1)
    \Big]  \label{4ptlim0}
\end{align}
as well as
\begin{align}
 I^{\rm tree}&(1,2,3,4|G_{23})=
  \frac{1}{4\pi^2} \sum_{B \in S_3} {\rm sgn}_{23}^{B} \notag \\
 &\times \Big[
 \sin \Big( \frac{ \pi }{2} (s_{12}{+}s_{23}{+}s_{24}) \Big) \sin \Big( \frac{ \pi }{2} s_{134} \Big)
 Z^{\rm tree}(+,2,3,4,1,-| +,B,- ,1)\notag \\
 & \ \ \ \ \ \
- \sin \Big( \frac{ \pi }{2} (s_{13}{+}s_{23}{+}s_{34}) \Big) \sin \Big( \frac{ \pi }{2} ({-}s_{12}{+}s_{14}{-}s_{24}) \Big)
 Z^{\rm tree}(+,3,2,4,1,-| +,B,- ,1)\notag \\
 & \ \ \ \ \ \
- \sin \Big( \frac{ \pi }{2} (s_{13}{+}s_{23}{+}s_{34}) \Big) \sin \Big( \frac{ \pi }{2} ({-}s_{12}{+}s_{14}{+}s_{24}) \Big)
 Z^{\rm tree}(+,3,4,2,1,-| +,B,-,1 )\notag \\
 & \ \ \ \ \ \
- \sin \Big( \frac{ \pi }{2} (s_{14}{+}s_{24}{+}s_{34}) \Big) \sin \Big( \frac{ \pi }{2} s_{123} \Big)
 Z^{\rm tree}(+,4,3,2,1,-| +,B,-,1 )
   \Big]  \label{4ptlimX}
\end{align}
and
\begin{align}
 I^{\rm tree}&(1,2,3,4|G_{12}G_{34})=
- \frac{1}{4\pi}  \sum_{B \in S_4}  {\rm sgn}_{12}^B {\rm sgn}_{34}^B  \notag \\
&\times \Big[
 \sin \Big( \frac{ \pi }{2} (s_{12}{+}s_{23}{+}s_{24}) \Big) \cos \Big( \frac{ \pi }{2} s_{134} \Big)  Z^{\rm tree}(+,2,3,4,1,-|+,B,-) \notag \\
 & \ \ \ \
 - \sin \Big( \frac{ \pi }{2} (s_{13}{+}s_{23}{+}s_{34}) \Big) \cos \Big( \frac{ \pi }{2}({-}s_{12}{+}s_{14}{-}s_{24}) \Big)
 Z^{\rm tree}(+,3,2,4,1,-|+,B,-)  \notag \\
 & \ \ \ \
  - \sin \Big( \frac{ \pi }{2} (s_{13}{+}s_{23}{+}s_{34}) \Big) \cos \Big( \frac{ \pi }{2}({-}s_{12}{+}s_{14}{+}s_{24}) \Big)
  Z^{\rm tree}(+,3,4,2,1,-|+,B,-)\notag \\
 & \ \ \ \
 + \sin \Big( \frac{ \pi }{2} (s_{14}{+}s_{24}{+}s_{34}) \Big) \cos \Big( \frac{ \pi }{2}s_{123} \Big)
 Z^{\rm tree}(+,4,3,2,1,-|+,B,-)
    \Big]\, .
\label{cocyc30}
\end{align}
When the integrand of $I^{\rm tree}(\cdot|\cdot)$ in (\ref{rePT3}) is $G_{12}G_{23}G_{34}$
or one of $G_{12}G_{23}, \, G_{23}G_{34}$ with overlapping labels, repeated use of
the Fay identity (\ref{faygij}) and integration by parts (\ref{rePT13}) can be used to reduce these cases to (permutations of)
(\ref{4ptlim0}), (\ref{4ptlimX}), and (\ref{cocyc30}), for instance
\begin{align}
 I^{\rm tree}&(A|G_{12}G_{23}) =
\frac{ \pi^2 s_{13}}{s_{123} }   I^{\rm tree}(A|1)
+ \frac{s_{34} }{s_{123}}  I^{\rm tree}(A|G_{12}G_{34})
- \frac{ s_{14} }{s_{123}} I^{\rm tree}(A|G_{14}G_{23})
 \label{4ptlim10}
 \end{align}
as well as
\begin{align}
s_{1234} I^{\rm tree}(A|G_{12}G_{13}G_{14}) &= - \pi^2
\big[ s_{34} I^{\rm tree}(A|G_{12})
+s_{24} I^{\rm tree}(A|G_{13})
+s_{23} I^{\rm tree}(A|G_{14}) \big] \!  \label{extraIBP} \\
s_{1234} I^{\rm tree}(A|G_{12}G_{23}G_{34}) &=  \pi^2
\big[ s_{124} I^{\rm tree}(A|G_{12})
+s_{134} I^{\rm tree}(A|G_{34})
+(s_{14}{-}s_{23}) I^{\rm tree}(A|G_{23}) \big] \, . \notag
\end{align}
\paragraph{$\ap$-expansions:}

Given the $\ap$-expansion of six-point disk integrals (see for instance \cite{wwwMZV, Mafra:2016mcc}), we
arrive at expressions such as
\begin{align}
I^{\rm tree}&(1,2,3,4|1)= \frac{1}{6} + \frac{ \zeta_3}{4\pi^2}
(s_{12} {-} 2 s_{13} {+} s_{14} {+} s_{23} {-} 2 s_{24} {+} s_{34}) \notag\\
&\! \! \! \! \! +\zeta_2 \Big( \Big[ \frac{13 s_{12}^2}{240}  - \frac{ s_{12}s_{14}}{120}+ \frac{ s_{12}(s_{13}{+}s_{24}) }{240} + {\rm cyc}(1,2,3,4) \Big] \notag  \\
&\! \! \! \! \!  \ \ \ \ + \frac{ s_{13}s_{24} }{60} - \frac{ (s_{12}s_{34}{+}s_{14}s_{23}) }{120}
+ \frac{ (s_{13}^2{+}s_{24}^2) }{60} \Big) \notag \\
&\! \! \! \! \! +\frac{ \zeta_3}{96} \Big( \Big[
 5 s_{12}^3 {+} 3 s_{12}^2 s_{14} {+} 3 s_{14}^2 s_{23} {+} 3 s_{14}^2 s_{34} {-} 3 s_{12} s_{13}^2 {-}
 3 s_{13}^2 s_{14} {+} 4 s_{12} s_{13} s_{23}  \label{4ptlim1}  \\
 &\! \! \! \! \!  \ \ \ \ {+} 6 s_{12} s_{14} s_{34} {-} 6 s_{12} s_{13} s_{24}
 + {\rm cyc}(1,2,3,4) \Big] +2 s_{13}^3 {+} 2 s_{24}^3 {-} 6 s_{13}^2 s_{24} {-} 6 s_{13} s_{24}^2
\Big) \notag \\
&\! \! \! \! \! +\frac{ \zeta_5 }{16\pi^2} \Big( \Big[
2 s_{12}^3 {-} 2 s_{12}^2 s_{13} {-} 2 s_{13} s_{23}^2 {+} 7 s_{12} s_{13}^2 {+} 7 s_{13}^2 s_{23} {-}
 5 s_{12}^2 s_{23} {-} 5 s_{12} s_{23}^2 {-} 5 s_{12}^2 s_{34} \notag \\
 &\! \! \! \! \!  \ \ \ \  {-} 11 s_{12} s_{23} s_{34} {+}  11 s_{12} s_{13} s_{24} + {\rm cyc}(1,2,3,4) \Big]
  - 4 s_{13}^3 {-} 4 s_{24}^3 {+} 10 s_{13}^2 s_{24} {+} 10 s_{13} s_{24}^2 \Big)
+{\cal O}(\ap^4) \, . \notag
\end{align}
Similar expansions for $I^{\rm tree}(1,2,3,4|\ast)$ with integrands $G_{12},\, G_{13},\, G_{12}G_{34},\,
G_{13}G_{24}$ and $G_{12}G_{13}$ are displayed in appendix \ref{app:H4}.

The inverse powers of $\pi$ in the combinations $\frac{ \zeta_3}{\pi^2}$ and $\frac{ \zeta_5}{\pi^2}$
at the $\ap$- and $\ap^3$-orders of (\ref{4ptlim1}) are a generic feature of $I^{\rm tree}(\cdot|\cdot)$
at $n{\geq} 4$ points: The MZVs in the $\ap$-expansion of $n$-point $I^{\rm tree}(\cdot|\cdot)$
may be accompanied by up to $n{-}2$ inverse powers of $\pi$ (up to $n{-}3$ inverse powers
if $n$ is odd). This can be seen from the following properties of their decomposition (\ref{rePT12})
into disk integrals:
\begin{itemize}
\item the overall prefactors $(i\pi)^{k-n+1}$, where $k$ is the number of $G_{ij}$ in the integrand
\item the sine functions in (\ref{s0jL}) render the $\ap$-expansion of any ${\cal H}_\ap(A|P)$ proportional to $\pi$
\item in case of odd multiplicity $n\in 2\NN +1$ and $k=0$ powers of $i\pi$ due to $G_{ij}$,
the combination $\frac{1}{2} ({\cal H}_\ap(A|P) {+}{\cal H}_\ap(A^t|P) )$ allows to factorize $\pi^2$,
leading to $3{-}n$ powers of $\pi$ in (\ref{rePT12})
\item the $\ap$-expansion of generic $Z^{\rm tree}(\cdot| \cdot)$ involves $\QQ$-linear combinations of MZVs
\end{itemize}
In the limit of on-shell kinematics $(s_{13},s_{14},s_{24},s_{34}) \rightarrow
(-s_{12}{-}s_{23},s_{23},-s_{12}{-}s_{23},s_{12})$ of four massless particles, the
$\ap$-expansion of $I^{\rm tree}(1,2,3,4|1)$ governs the $\tau \rightarrow i \infty$ contribution
to the one-loop four-point amplitude of the open superstring \cite{Green:1982sw}.
The on-shell limit of the six Mandelstam invariants $s_{ij}$ in (\ref{4ptlim1})
is compatible with the all-order result \cite{Green:1981ya}
\begin{align}
I^{\rm tree}&(1,2,3,4|1) \rightarrow
 - \frac{1}{2\pi^2 s_{12}s_{23} } \alpha' \frac{ \partial }{\partial \alpha'} \frac{ \Gamma(1-s_{12}) \Gamma(1-s_{23}) }{\Gamma(1-s_{12}-s_{23}) } \, . \label{4ptlim2}
\end{align}
Still, it is beneficial to retain the dependence of various $I^{\rm tree}(1,2,3,4|B)$ on
six $s_{ij}$ since this will turn out to yield important building blocks for non-planar six-point $A$-cycle integrals.

\paragraph{Component integrals:}

The $\ap$-expansion (\ref{someint44}) of the generating function $Z^{\tau}_{\eta_2,\eta_3,\eta_4}$
allows to extract component integrals
\begin{align}
Z^{\tau}_{(m_1,m_2,m_3)}(A|1,2,3,4) &\equiv Z^{\tau}_{\eta_2,\eta_3,\eta_4}(A|1,2,3,4) \, \big|_{(\eta_{2}{+}\eta_3{+}\eta_4)^{m_1-1}(\eta_{3}{+}\eta_4)^{m_2-1}\eta_4^{m_3-1}}
\label{compin171}\\
&= \int_{{\cal C}(A)} \dd z_2 \, \dd z_3 \, \dd z_4 \, f^{(m_1)}(z_{12},\tau)
f^{(m_2)}(z_{23},\tau) f^{(m_3)}(z_{34},\tau) \, {\rm KN}_{1234}^\tau \, ,
\notag
\end{align}
see (\ref{compin71}) for the analogous definition at three points. Again, since the derivations
in the four-point representation (\ref{DER14}) are even under $\eta_j \rightarrow - \eta_j$,
component integrals with even and odd values of $m_1{+}m_2{+}m_3$ decouple in the following
sense: $Z^{\tau}_{(m_1,m_2,m_3)}$ with $m_1{+}m_2{+}m_3\in 2\NN_0$
($m_1{+}m_2{+}m_3\in 2\NN_0{+}1$) only receive contributions from the $I^{\rm tree}$
in the initial value (\ref{rePT3}) with an even (odd) number of $G_{ij}$ in the integrand, respectively.
Note that even and odd powers of $G_{ij}$ do not mix under Fay relations (\ref{faygij})
and integration by parts (\ref{rePT13}).

The $\ap$-expansion of the simplest component integral (\ref{compin171}) starts with
\begin{align}
Z^{\tau}_{(0,0,0)}&(1,2,3,4|1,2,3,4) = I^{\rm tree}(1,2,3,4|1) +6 (s_{12} {-} 2 s_{13} {+} s_{14} {+} s_{23} {-} 2 s_{24} {+} s_{34}) \gamma_0(4,0,0|\tau) \notag \\
& + 60 \Big[ s_{12}^2 {-} s_{13}^2 {+} s_{13}(s_{12}{+}s_{23}) {-} 2s_{12}s_{23} {+} s_{13} s_{24} {-} s_{12}s_{34}
+ {\rm cyc}(1,2,3,4) \Big] \gamma_0(6,0,0,0|\tau)  \notag \\
&  - \frac{1}{2} \Big(\sum_{1\leq i <j}^4 s_{ij}^2 \Big) \gamma_0(4,0|\tau) +
{\cal O}(\ap^3) \,,
\label{compin271}
\end{align}
in agreement with \cite{Broedel:2014vla}.

The symmetrization of (\ref{4ptlim1}) w.r.t.\ planar integration cycles no longer involves inverse powers of
$\pi$ since the permutation sum $\sum_{A \in S_3}{\cal H}_\ap(A|P)$ yields products of three sine
functions by (\ref{CsymA}) and therefore an overall factor of $\pi^3$,
\begin{align}
\sum_{A \in S_3} I^{\rm tree}(1,A|1)=  \frac{1}{\pi^3}  &
 \sin \Big( \frac{ \pi }{2} (s_{12}{+}s_{23}{+}s_{24}) \Big)
 \sin \Big( \frac{ \pi }{2} (s_{13}{+}s_{34}) \Big)
 \sin \Big( \frac{ \pi }{2} s_{14} \Big)  \label{4ptsy}\\
 &\times \sum_{B\in S_3}  Z^{\rm tree}(+,2,3,4,1,-|+,B,-,1)
  + {\rm perm}(2,3,4) \, .  \notag
\end{align}
As will be detailed in appendix \ref{app:qexp}, symmetrized combinations $\sum_{A \in S_{n-1}} I^{\rm tree}(1,A|\ast)$
at any multiplicity (possibly involving $G_{ij}$ in the integrand) yield $\ap$-expansions with $\QQ$-linear as
opposed to $\QQ[(2\pi i)^{-1}]$-linear combinations of MZVs in their coefficients.

At the leading orders in $\ap$, (\ref{4ptsy}) amounts to
\begin{align}
\sum_{A \in S_3} &I^{\rm tree}(1,A|1) = 1 + \frac{ \zeta_2}{4} \sum_{1\leq i<j}^4 s_{ij}^2
+ \frac{ \zeta_3}{4}
 \sum_{1\leq i<j}^4 s_{ij}^3
 \label{KNdeg191}\\
&+ \frac{ \zeta_3}{4} (  s_{12}s_{13}s_{23} {+}
 s_{12}s_{14}s_{24} {+}
  s_{13}s_{14}s_{34} {+}
   s_{23}s_{24}s_{34} )
 + {\cal O}(\ap^4) \, .
\notag
\end{align}
This is in fact the $\tau \rightarrow i \infty$ degeneration of the simplest $A$-cycle graph functions which
are generated by the component integral $\sum_{A\in S_3}  Z^{\tau}_{(0,0,0)}(1,A|1,2,3,4) $.
We have checked the differential-equation method to reproduce the simplest example of
four-point $A$-cycle graph functions \cite{Broedel:2018izr}
\beq
A_{ijkl}(\tau) = i! j! k! l!  \sum_{A\in S_3}  Z^{\tau}_{(0,0,0)}(1,A|1,2,3,4)  \, \big|_{s_{12}^i s_{23}^j s_{34}^k s_{41}^l} \, ,
\label{defA4}
\eeq
namely
\beq
A_{1111}(\tau) = \frac{1}{8} \zeta_4 + 960 \zeta_2 \gamma_0(6,0,0,0|\tau) - 840 \gamma_0(8,0,0,0|\tau )
 \, . \label{CC66}
\eeq
It will be shown in appendix \ref{app:qexp} that the coefficients in the $q$-expansion of $A$-cycle graph functions
at any multiplicity are $\QQ$-linear as opposed to $\QQ[(2\pi i)^{-1}]$-linear combinations of~MZVs.

Four-point one-loop amplitudes with half- or quarter-maximal spacetime supersymmetry
involve moduli-space integrals over $f_{ij}^{(2)}$ and $f_{ij}^{(1)} f_{pq}^{(1)}$ \cite{Bianchi:2015vsa, Berg:2016wux}.
Hence, their $\ap$-expansions will require the on-shell limit of
$Z^{\tau}_{(2,0,0)}(1,2,3,4|1,b_2,b_3,4)$ and
$Z^{\tau}_{(1,0,1)}(1,2,3,4|1,b_2,b_3,4)$ with $(b_2,b_3)\in S_2$,
which address the cyclically inequivalent integrands $f_{12}^{(2)},f_{13}^{(2)},f_{12}^{(1)} f_{34}^{(1)}$
and $f_{13}^{(1)} f_{24}^{(1)}$. Integrals over $f_{12}^{(1)} f_{23}^{(1)}$ with overlapping labels again
follow from Fay relations and integration by parts.


\section{Non-planar genus-one integrals at the cusp}
\label{sec:6}

We shall now extend the above discussion of initial values $Z^{i \infty}_{\vec{\eta}}(A|\ast)$
associated with planar integration cycles ${\cal C}(A)$ to non-planar ones ${\cal C}\big(\begin{smallmatrix} Q \\ P \end{smallmatrix} \big)$.
The $\ap$-dependence of
the desired initial values $Z^{i\infty}_{\vec{\eta}} \big(\begin{smallmatrix} Q \\ P \end{smallmatrix} |\ast\big)$ will
be shown to reside in products $I^{\rm tree}(P|\ast) I^{\rm tree}(Q|\ast) $ of lower-multiplicity integrals (\ref{rePT1})
that boil down to products of disk integrals by virtue of (\ref{rePT12}).

One of the main reasons for this simplified $\ap$-dependence is the factorization of the
degenerate Koba--Nielsen on a non-planar cycle into
\beq
{\rm KN}_{12\ldots n}^{i\infty} \, \big|_{{\cal C}\big(\begin{smallmatrix} Q \\ P \end{smallmatrix} \big)}
= \big( {\rm KN}_{P}^{i\infty}  \, \big|_{{\cal C}(P)} \big) \times \big( {\rm KN}_{Q}^{i\infty}  \, \big|_{{\cal C}(Q)}  \big)\, ,
\label{CC10A}
\eeq
see (\ref{inival2}) for the degeneration of the planar Koba--Nielsen factors on the right-hand side.
This factorization follows from (\ref{Gnpnew}) and can be intuitively understood from figure \ref{basiccyl}:
The cylinder worldsheet becomes infinitely long as $\tau \rightarrow i \infty$, so its boundaries decouple
in this limit and give rise to separate Koba--Nielsen factors.


\subsection{General result}
\label{sec:6.1}

The combinations and coefficients of $I^{\rm tree}(P|\ast) I^{\rm tree}(Q|\ast) $
in the initial values $Z^{i\infty}_{\vec{\eta}} \big(\begin{smallmatrix} Q \\ P \end{smallmatrix} |\ast\big)$
can be conveniently described by means of the following notation: For a given integrand
$\Omega(z_{12},\eta_{23\ldots n},\tau) \Omega(z_{23},\eta_{3\ldots n},\tau)\ldots
\Omega(z_{n-1,n},\eta_n,\tau)$ associated with $\ast \rightarrow 1,2,\ldots,n$, we define
the following collection of integrated punctures $z_j$ such that both $z_{j-1}$ and $z_j$
belong to the same boundary:
\begin{align}
X^{12\ldots n}_P = \{j \in \{2,3,\ldots,n\}, \ j \in P \ \wedge \  j{-}1 \in P \} \, .
\label{CC10B}
\end{align}
By the parameterization in (\ref{allC2}), each $z_{j-1,j}$ with $j \in X^{12\ldots n}_P$ or $j \in X^{12\ldots n}_Q$
is real on ${\cal C}\big(\begin{smallmatrix} Q \\ P \end{smallmatrix} |\ast\big)$. Similarly,
the complement in the sense of $X^{12\ldots n}_P \cup X^{12\ldots n}_Q \cup Y^{12\ldots n}_{P,Q}=\{2,3,\ldots,n\}$
is given by
\beq
Y^{12\ldots n}_{P,Q} = \{j \in \{ 2,3,\ldots,n \}, \ (j \in P  \wedge  j{-}1 \in Q) \ \vee \ (j \in Q  \wedge  j{-}1 \in P)   \} \, ,
\label{CC10C}
\eeq
such that $\Im z_{j-1,j}= \pm \Im(\tauh) \ \forall \ j \in Y^{12\ldots n}_{P,Q}$ on ${\cal C}\big(\begin{smallmatrix}
Q \\ P \end{smallmatrix} |\ast\big)$. Given that the degeneration $\lim_{\tau \rightarrow i \infty} \Omega(v_{ij}{\pm}\tauh,\eta,\tau)
= \frac{ \pi }{\sin(\pi \eta)}$ does not depend on $v_{ij} \in \RR$, the factorization (\ref{CC10A})
of the non-planar Koba--Nielsen factor propagates to
\begin{align}
Z^{i\infty}_{\vec{\eta}} \big(\begin{smallmatrix} Q \\ P \end{smallmatrix} |1,2,\ldots,n\big) &= \prod_{i \in Y^{12\ldots n}_{P,Q}} \frac{ \pi }{\sin(\pi \eta_{i,i+1 \ldots n-1,n})}  \bigg( \int_{{\cal C}(P)} {\rm KN}_P^{i\infty} \! \prod_{j \in X^{12\ldots n}_P} \! \Omega(z_{j-1,j},\eta_{j,j+1\ldots n},i\infty)  \bigg) \notag \\
& \ \  \times
 \bigg( \int_{{\cal C}(Q)} {\rm KN}_Q^{i\infty}  \! \prod_{k \in X^{12\ldots n}_{Q}} \! \Omega(z_{k-1,k},\eta_{k,k+1\ldots n},i\infty)  \bigg) \, \dd z_2\, \dd z_3\, \ldots \, \dd z_n\, .
 \label{CC10D}
\end{align}
Once we insert the degeneration (\ref{Gpl}) of the remaining factors $\Omega(\ldots)$,
the two decoupled integrals in the first and second line both follow the degeneration
(\ref{rePT4}) in the planar case,
\begin{align}
Z^{i\infty}_{\vec{\eta}} &\big(\begin{smallmatrix} Q \\ P \end{smallmatrix} |1,2,\ldots,n\big) = \prod_{i \in Y^{12\ldots n}_{P,Q}} \frac{ \pi }{\sin(\pi \eta_{i,i+1\ldots n-1,n})} \notag \\
&\times \sum_{X^{12\ldots n}_P= A \cup B \atop{A \cap B = \emptyset} }\Big( \prod_{j \in A} \pi \cot(\pi \eta_{j,j+1\ldots n}) \Big) I^{\rm tree}(P| \prod_{\bar j \in B} G_{\bar j-1,\bar j} ) \label{CC10E} \\
&\times \sum_{X^{12\ldots n}_Q= C \cup D \atop{C \cap D = \emptyset}} \Big( \prod_{k \in C} \pi \cot(\pi \eta_{k,k+1\ldots n}) \Big) I^{\rm tree}(Q| \prod_{\bar k \in D} G_{\bar k-1,\bar k} )\, .
\notag
\end{align}
For $P,Q$ with one or two elements (say $P=i$ or $P=i,j$), one can readily simplify (\ref{CC10E}) via,
\beq
 I^{\rm tree}(i | 1 )=1 \, , \ \ \ \ \ \  I^{\rm tree}(i,j | G_{ij} ) = 0 \, , \ \ \ \ \ \
I^{\rm tree}(i,j | 1 ) = \frac{ \Gamma(1-s_{ij}) }{ \big[ \Gamma(1-\tfrac{s_{ij}}{2}) \big]^2} \, ,
 \label{CC10F}
\eeq
cf.\ section \ref{sec:3.4}. On these grounds, the non-planar two-point result (\ref{someintD}) is recovered
from (\ref{CC10E}) with $P=1, \ Q=2$ as well as $X^{12}_1= X^{12}_2= \emptyset$ and $Y^{12}_{1,2}=\{2\}$.
For permutations of the integrand in (\ref{CC10E}), say
$\Omega(z_{e_{j-1}e_j},\eta_{e_je_{j+1}\ldots e_n},\tau)$ with $E=e_1 e_2 \ldots e_n$,
the sets in (\ref{CC10B}) and (\ref{CC10C}) have to be adjusted to $Y^{12\ldots n}_{P,Q} \rightarrow Y^{E}_{P,Q}$
and similarly for $X_P^{\cdot},X_Q^{\cdot}$,
\begin{align}
Z^{i\infty}_{\vec{\eta}} &\big(\begin{smallmatrix} Q \\ P \end{smallmatrix} |E\big) = \prod_{i \in Y^{E}_{P,Q}} \frac{ \pi }{\sin(\pi \eta_{e_i,e_{i+1}\ldots e_n})} \notag \\
&\times \sum_{X^{E}_P= A \cup B \atop{A \cap B = \emptyset} }\Big( \prod_{j \in A} \pi \cot(\pi \eta_{e_j,e_{j+1}\ldots e_n}) \Big) I^{\rm tree}(P| \prod_{\bar j \in B} G_{e_{\bar j-1},e_{\bar j}} ) \label{CC10G} \\
&\times \sum_{X^{E}_Q= C \cup D \atop{C \cap D = \emptyset}} \Big( \prod_{k \in C} \pi \cot(\pi \eta_{e_k,e_{k+1}\ldots e_n}) \Big) I^{\rm tree}(Q| \prod_{\bar k \in D} G_{e_{\bar k-1},e_{\bar k}} )\, .
\notag
\end{align}
In this way, the labels of $G_{e_{\bar j-1},e_{\bar j}} $ and $G_{e_{\bar k-1},e_{\bar k}} $
are contained in the first entry $P$ of $I^{\rm tree}(P|\ast )$ and $Q$ of $I^{\rm tree}(Q|\ast )$, respectively.
Given the general dictionary (\ref{rePT12}) between $I^{\rm tree}(P | \cdot)$ and $(|P|{+}2)$-point disk
integrals, (\ref{CC10G}) reduces the initial values for non-planar cycles ${\cal C}\big(\begin{smallmatrix}
Q \\ P \end{smallmatrix} |\ast\big)$ to products of disk integrals. As an additional simplification in
comparison to the planar case, these disk integrals have multiplicities $\leq {\rm max}(|P|,|Q|){+}2$
instead of $n{+}2=|P|{+}|Q|{+}2$.


\subsection{Three points}
\label{sec:6.2}

At three points, the general prescription (\ref{CC10G}) for non-planar initial values
yields the following two inequivalent cases
\begin{align}
Z^{i\infty}_{\eta_2,\eta_3} \big(\begin{smallmatrix} 3 \\ 1,2 \end{smallmatrix} |1,2,3\big) &
=  \frac{ \pi^2 \cot(\pi \eta_{23}) }{\sin(\pi \eta_3)}   I^{\rm tree}(1,2|1)
=  \frac{ \pi^2 \cot(\pi \eta_{23}) }{\sin(\pi \eta_3)}  \frac{ \Gamma(1-s_{12}) }{ \big[ \Gamma(1-\tfrac{s_{12}}{2}) \big]^2}
 \label{in3np} \\
%
Z^{i\infty}_{\eta_2,\eta_3} \big(\begin{smallmatrix} 3 \\ 1,2 \end{smallmatrix} |1,3,2\big) &
=  \frac{ \pi^2}{ \sin(\pi \eta_{23}) \sin(\pi \eta_2)}  I^{\rm tree}(1,2|1)
=  \frac{ \pi^2}{ \sin(\pi \eta_{23}) \sin(\pi \eta_2)}  \frac{ \Gamma(1-s_{12}) }{ \big[ \Gamma(1-\tfrac{s_{12}}{2}) \big]^2}
 \,.\notag
\end{align}
We have exploited the vanishing of $I^{\rm tree}(1,2|G_{12})$, and the $\ap$-expansion of the gamma functions
in terms of Riemann zeta values $\zeta_{m\geq 2}$ can be found in (\ref{KNdrft2}). The $\ap$-expansion of non-planar
$A$-cycle integrals at three points is completely determined by (\ref{someint33}) and (\ref{in3np}). As before,
any order in the $\ap$-expansion of the component integrals
\begin{align}
\widehat Z^{\tau}_{(m_1,m_2)}\big(\begin{smallmatrix} k \\ i,j \end{smallmatrix} |1,2,3\big) &\equiv q^{-\frac{1}{8}(s_{ik}{+}s_{jk}) } Z^\tau_{\eta_2 ,\eta_3}\big(\begin{smallmatrix} k \\ i,j \end{smallmatrix} |1,2,3\big) \, \big|_{(\eta_{2}{+}\eta_3)^{m_1-1}\eta_3^{m_2-1}}
\label{npcompin71}
\\
&\phantom{:}= q^{-\frac{1}{8}(s_{ik}{+}s_{jk}) } \int_{{\cal C}\big(\begin{smallmatrix} k \\ i,j \end{smallmatrix} \big) } \dd z_2 \, \dd z_3 \, f^{(m_1)}(z_{12},\tau)f^{(m_2)}(z_{23},\tau) \, {\rm KN}_{123}^\tau
\notag
\end{align}
can be assembled from a finite number of elementary operations. By the extra factor
$q^{-\frac{1}{8}(s_{ik}{+}s_{jk}) }$ in comparison to the planar component integrals (\ref{compin71}),
any $\ap$-order of $\widehat Z^{\tau}_{(m_1,m_2)}\big(\begin{smallmatrix} k \\ i,j \end{smallmatrix} |1,2,3\big)$
admits a Fourier expansion w.r.t.\ $q$.

The component integral $\widehat Z^{\tau}_{(0,0)} (  \begin{smallmatrix} 3 \\ 1,2 \end{smallmatrix}   |1,2,3 ) $ generates
non-planar $A$-cycle graph functions\footnote{In the graphical notation
of \cite{Broedel:2018izr}, the three-vertex graph functions in (\ref{CC33}) are represented as follows:

\begin{center}
 \tikzpicture
\draw(-0.135,0)node{$A_{1\underline{11}}= {\bf A}\Big[ \ \ \ \ \ \ \Big]$};
\draw(0.3,-0.2)node{$\bullet$};
\draw(0.8,-0.2)node{$\bullet$};
\draw(0.55,0.22)node{$\bullet$};
\draw(0.3,-0.2)--(0.8,-0.2);
\draw(0.55,0.22)--(0.8,-0.2);
\draw(0.3,-0.2)--(0.55,0.22);
\draw[dashed](0.3,0)--(0.8,0);
\scope[xshift=4cm]
\draw(-0.135,0)node{$A_{2\underline{11}}= {\bf A}\Big[ \ \ \ \ \ \ \Big]$};
\draw(0.3,-0.2)node{$\bullet$};
\draw(0.8,-0.2)node{$\bullet$};
\draw(0.55,0.22)node{$\bullet$};
\draw(0.3,-0.2) .. controls (0.45,-0.1) and (0.65,-0.1) .. (0.8,-0.2);
\draw(0.3,-0.2) .. controls (0.45,-0.3) and (0.65,-0.3) .. (0.8,-0.2);
\draw(0.55,0.22)--(0.8,-0.2);
\draw(0.3,-0.2)--(0.55,0.22);
\draw[dashed](0.3,0)--(0.8,0);
\endscope
\scope[xshift=8cm]
\draw(-0.135,0)node{$A_{1\underline{21}}= {\bf A}\Big[ \ \ \ \ \ \ \Big]$};
\draw(0.3,-0.2)node{$\bullet$};
\draw(0.8,-0.2)node{$\bullet$};
\draw(0.55,0.22)node{$\bullet$};
\draw(0.3,-0.2) .. controls (0.45,-0.1) and (0.65,-0.1) .. (0.8,-0.2);
\draw(0.3,-0.2) .. controls (0.45,-0.3) and (0.65,-0.3) .. (0.8,-0.2);
\draw(0.55,0.22)--(0.8,-0.2);
\draw(0.3,-0.2)--(0.55,0.22);
\draw[dashed](0.42,-0.32)--(0.8,0.15);
\endscope
\endtikzpicture \end{center}} with two and three vertices \cite{Broedel:2018izr}
\begin{align}
\widehat Z^{\tau}_{(0,0)} &\big(  \begin{smallmatrix} 3 \\ 1,2 \end{smallmatrix}   |1,2,3 \big) = 1 + \frac{1}{2} \big[ s_{12}^2 A_2(\tau) + (s_{13}^2{+}s_{23}^2) A_{\underline{2}}(\tau) \big]
+ \frac{1}{3!} \big[ s_{12}^3 A_3(\tau) + (s_{13}^3{+}s_{23}^3) A_{\underline{3}}(\tau) \big]  \notag \\
& \! \! \! \!  \! \! \! \!
+ s_{12}s_{13}s_{23} A_{1\underline{11}}(\tau)
+\frac{1}{2} s_{12}s_{13}s_{23} (s_{13}{+}s_{23}) A_{1\underline{21}}(\tau)
+ \frac{1}{2} s_{12}^2s_{13}s_{23} A_{2\underline{11}}(\tau)
\label{CC33}\\
& \! \! \! \!  \! \! \! \!
+ \frac{1}{4} \big[ s_{12}^2 (s_{13}^2{+}s_{23}^2) A_2(\tau)   A_{\underline{2}}(\tau)
+ s_{13}^2 s_{23}^2 A_{\underline{2}}(\tau)^2 \big]
+ \frac{1}{4!} \big[ s_{12}^4 A_4(\tau) + (s_{13}^4{+}s_{23}^4) A_{\underline{4}}(\tau) \big]
+ {\cal O}(\ap^5)\, ,
\notag
\end{align}
and the methods of this work have been used to reproduced all examples of
\beq
A_{i\underline{jk}}(\tau) = i! j! k!
\widehat Z^{\tau}_{(0,0)} \big(  \begin{smallmatrix} 3 \\ 1,2 \end{smallmatrix}   |1,2,3 \big) \, \big|_{s_{12}^i s_{23}^j s_{13}^k } = 0 \, ,
\ \ \ \ \ \ i,j,k\geq 1
 \label{in3npA}
\eeq
computed in \cite{Broedel:2018izr}, namely
\begin{align}
A_{1\underline{11}}(\tau)&= -60 \gamma_0(6, 0, 0 |\tau)
\label{CC34} \\
A_{2\underline{11}}(\tau)&= - 144 \zeta_4 \gamma_0(4, 0, 0, 0|\tau) -
 36 \gamma_0(4, 4, 0, 0|\tau) + 960 \zeta_2 \gamma_0(6, 0, 0, 0|\tau) -
 756 \gamma_0(8, 0, 0, 0|\tau)
\notag \\
A_{1\underline{21}}(\tau)&= - 144 \zeta_4 \gamma_0(4, 0, 0, 0|\tau) -
 36 \gamma_0(4, 4, 0, 0|\tau) - 480 \zeta_2 \gamma_0(6, 0, 0, 0|\tau) -
 756 \gamma_0(8, 0, 0, 0|\tau)\, .
\notag
\end{align}
Given that the initial value (\ref{in3np}) is independent on $s_{13}$ and $s_{23}$,
the generating function (\ref{in3npA}) implies that any non-planar three-vertex graph function vanishes at the cusp,
\beq
A_{i\underline{jk}}(i \infty) = 0 \, ,
\ \ \ \ \ \ i,j,k\geq 1 \, .  \label{in3npB}
\eeq
Both the initial values in (\ref{in3np}) and the three-point derivations (\ref{DER11}) are even functions
under $(\eta_2,\eta_3) \rightarrow - (\eta_2,\eta_3)$. Hence, any component integral (\ref{npcompin71})
with an integrand of odd weight vanishes at any value of $\tau$,
\beq
\widehat Z^{\tau}_{(m_1,m_2)} \big(  \begin{smallmatrix} 3 \\ 1,2 \end{smallmatrix}   |1,a_2,a_3 \big) \, \Big|_{m_1+m_2 \ \te{odd}} = 0 \, .
 \label{in3npC}
\eeq
In particular, there are no kinematic poles in the non-planar component integrals
$\widehat Z^{\tau}_{(m_1,m_2)}(\cdot|\cdot)$, as one can also see
from the absence of $I^{\rm tree}(\ast|G_{ij})$ in (\ref{in3np}). The simplest example of a component integral
beyond (\ref{CC33}) is given by,
\begin{align}
\widehat Z^{\tau}_{(2,0)} &\big(  \begin{smallmatrix} 3 \\ 1,2 \end{smallmatrix}   |1,2,3 \big)
= -2 \zeta_2  + 3 s_{12} \gamma_0( 4|\tau) - \frac{ s_{12} }{2}    \zeta_2^2  \label{in3npD} \\
&+6 ( s_{13}^2  {+} s_{23}^2 {+} 4 s_{13} s_{23} {-} 3 s_{12}^2)  \zeta_2 \gamma_0(4, 0|\tau)
 +10 ( s_{12}^2  {+} 2s_{13} s_{23} ) \gamma_0(6, 0|\tau)+{\cal O}(\ap^3) \, , \notag
\end{align}
see (\ref{smalleq}) for its planar analogue $Z^{\tau}_{(2,0)}(1,2,3|1,2,3)$.


\subsection{Four points}
\label{sec:6.3}

At four points, the general prescription (\ref{CC10G}) for initial values yields
\begin{align}
Z^{i\infty}_{\vec{\eta}} \big(\begin{smallmatrix} 3,4 \\ 1,2 \end{smallmatrix} |1,2,3,4\big) &=  \frac{ \pi^3 \cot(\pi \eta_{234}) \cot(\pi \eta_4)}{\sin(\pi \eta_{34})}  I^{\rm tree}(1,2|1) I^{\rm tree}(3,4|1) \notag \\
Z^{i\infty}_{\vec{\eta}} \big(\begin{smallmatrix} 3,4 \\ 1,2 \end{smallmatrix} |1,3,2,4\big) &=  \frac{ \pi^3}{ \sin(\pi \eta_{234}) \sin(\pi \eta_{24}) \sin(\pi \eta_4)}  I^{\rm tree}(1,2|1) I^{\rm tree}(3,4|1)   \label{in3npE} \\
Z^{i\infty}_{\vec{\eta}} \big(\begin{smallmatrix} 3,4 \\ 1,2 \end{smallmatrix} |1,3,4,2\big) &=  \frac{ \pi^3 \cot(\pi \eta_{24}) }{\sin(\pi \eta_{234}) \sin(\pi \eta_2)}  I^{\rm tree}(1,2|1) I^{\rm tree}(3,4|1)  \notag
\end{align}
in case of a ``2+2'' cycle ${\cal C}\big(  \begin{smallmatrix} k,l \\ i,j \end{smallmatrix} \big)$
as well as
\begin{align}
Z^{i\infty}_{\vec{\eta}} &\big(\begin{smallmatrix} 2,3,4 \\ 1 \end{smallmatrix} |1,2,3,4\big) =  \frac{ \pi }{\sin(\pi \eta_{234})}
\Big\{ \pi^2 \cot(\pi \eta_{34}) \cot(\pi \eta_4) I^{\rm tree}(2,3,4|1)   \label{in3npF} \\
&\! \! \! \! \! \! \! + \pi \cot( \pi \eta_{34})  I^{\rm tree}(2,3,4|G_{34}) + \pi \cot(\pi \eta_4) I^{\rm tree}(2,3,4|G_{23}) +  I^{\rm tree}(2,3,4|G_{23} G_{34}) \Big\} \notag
\end{align}
in case of a ``3+1'' cycle ${\cal C}\big(  \begin{smallmatrix} l \\ i,j,k \end{smallmatrix} \big)$.
The latter can be further simplified by importing the integration-by-parts relations (\ref{rePT180})
from the planar three-point case,
\begin{align}
Z^{i\infty}_{\vec{\eta}} \big(\begin{smallmatrix} 2,3,4 \\ 1 \end{smallmatrix} |1,2,3,4\big) =  \frac{ \pi }{\sin(\pi \eta_{234})}
&\Big\{ \Big( \pi^2 \cot(\pi \eta_{34}) \cot(\pi \eta_4) + \frac{ \pi^2 s_{24} }{s_{234}} \Big) I^{\rm tree}(2,3,4|1)   \label{in3npFsimp} \\
&\ + \Big( \pi \cot( \pi \eta_{34}) + \frac{ s_{34} }{s_{23}}  \pi \cot(\pi \eta_4)  \Big)
 I^{\rm tree}(2,3,4|G_{34}) \Big\} \, .\notag
\end{align}
The remaining permutations of the integrands follow from (\ref{in3npE}) upon relabeling in
$3\leftrightarrow 4$ and from (\ref{in3npFsimp}) upon relabeling $2,3,4$ on the respective
right-hand sides. Non-planar component integrals are defined by
\begin{align}
\widehat Z^{\tau}_{(m_1,m_2,m_3)}\big(\begin{smallmatrix} k,l \\ i,j \end{smallmatrix} |1,2,3,4\big) &\equiv q^{-\frac{1}{8}(s_{ik}{+}s_{jk}{+}s_{il}{+}s_{jl})}
Z^{\tau}_{\eta_2,\eta_3,\eta_4}\big(\begin{smallmatrix} k,l \\ i,j \end{smallmatrix} |1,2,3,4\big) \, \big|_{\eta_{234}^{m_1-1}\eta_{34}^{m_2-1}\eta_4^{m_3-1}} \notag
\\
\widehat Z^{\tau}_{(m_1,m_2,m_3)}\big(\begin{smallmatrix} l \\ i,j,k \end{smallmatrix} |1,2,3,4\big) &\equiv q^{-\frac{1}{8}(s_{il}{+}s_{jl}{+}s_{kl})}
Z^{\tau}_{\eta_2,\eta_3,\eta_4}\big(\begin{smallmatrix} l \\ i,j,k \end{smallmatrix} |1,2,3,4\big) \, \big|_{\eta_{234}^{m_1-1}\eta_{34}^{m_2-1}\eta_4^{m_3-1}}\, ,  \label{in3npG}
\end{align}
where the extra factors $q^{-\frac{1}{8} s_{ij}}$ in comparison to the planar component integrals (\ref{compin171})
ensure that any $\ap$-order admits a Fourier expansion w.r.t.\ $q$.

The on-shell limits $(s_{13},s_{14},s_{24},s_{34}) \rightarrow
(-s_{12}{-}s_{23},s_{23},-s_{12}{-}s_{23},s_{12})$ of the component integrals $\widehat Z^{\tau}_{(0,0,0)}$
enter the four-point one-loop amplitude of the open superstring \cite{Green:1982sw}, see the
discussion around (\ref{4ptlim2}) for its planar sector. As we will see, the contributions to the non-planar amplitude
from the cusp can be related to the above initial values:
\begin{itemize}
\item Any component integral on the ``2+2 cycle'' ${\cal C}\big(\begin{smallmatrix} 3,4\\ 1,2 \end{smallmatrix}\big)$
is proportional to the universal factor of
\beq
I^{\rm tree}(1,2|1) I^{\rm tree}(3,4|1)= \frac{   \Gamma( 1 - s_{12} ) \Gamma( 1 - s_{34} )   }{ \big( \Gamma( 1 - \tfrac{ s_{12} }{2 })  \big)^2 \big( \Gamma( 1 - \tfrac{ s_{34} }{2 })  \big)^2 }
\label{in3npH}
\eeq
seen in each line of (\ref{in3npE}). In the on-shell limit $s_{34} \rightarrow s_{12}$, this
reproduces the $\tau \rightarrow i \infty$ behavior of the non-planar
``2+2 cycle''-amplitude determined in \cite{Hohenegger:2017kqy}:
\beq
I^{\rm tree}(1,2|1) I^{\rm tree}(3,4|1)\rightarrow
\frac{ \big( \Gamma( 1 - s_{12} )  \big)^2 }{ \big( \Gamma( 1 - \tfrac{ s_{12} }{2 })  \big)^4 }
= \frac{2^{-s_{12}} }{\pi} \left( \frac{  \Gamma( \tfrac{1}{2}-\tfrac{ s_{12} }{2 }) }{\Gamma( 1 - \tfrac{ s_{12} }{2 }) } \right)^2
\label{in3npI}
\eeq
\item On a ``3+1 cycle'', component integrals $\widehat Z^{\tau}_{(m_1,m_2,m_3)}\big(\begin{smallmatrix} 2,3,4 \\ 1 \end{smallmatrix} | \ast \big)$ at even $m_1{+}m_2{+}m_3 \in 2\ZZ$ are determined by the odd part
of (\ref{in3npFsimp}) w.r.t.\ $\eta_j \rightarrow - \eta_j$ which can be isolated by discarding
$I^{\rm tree}(2,3,4|G_{34})$,
\begin{align}
Z^{i\infty}_{\vec{\eta}} &\big(\begin{smallmatrix} 2,3,4 \\ 1 \end{smallmatrix} |1,2,3,4\big) \, \big|^{\te{odd in}}_{\eta_j \rightarrow - \eta_j} =  \frac{ \pi^3  I^{\rm tree}(2,3,4|1)}{\sin(\pi \eta_{234})}
\Big\{  \cot(\pi \eta_{34}) \cot(\pi \eta_4)  + \frac{s_{24}}{s_{234}} \Big\} \, .   \label{in3npFF}
\end{align}
Hence, the non-planar ``3+1 cycle'' amplitude determined by
$\widehat Z^\tau_{(0,0,0)}\big(\begin{smallmatrix} 2,3,4 \\ 1 \end{smallmatrix} |1,2,3,4\big)$ is proportional to the
on-shell limit of $I^{\rm tree}(2,3,4|1)$. In order to match the known $\tau \rightarrow i\infty$ degeneration of the
``3+1 cycle'' amplitude \cite{Green:1981ya}, the on-shell limit $s_{234} \rightarrow 0$ should yield
\beq
I^{\rm tree}(2,3,4|1) \rightarrow
 -   \frac{1}{\pi^2} \Big(
\frac{ \Gamma(-s_{23}) \Gamma(-s_{34} ) }{\Gamma(1{-}s_{23}{-}s_{34})}
+ \frac{\Gamma(-s_{24}) \Gamma(-s_{34} ) }{\Gamma(1{-}s_{24}{-}s_{34})}
+ \frac{\Gamma(-s_{23}) \Gamma(-s_{24} ) }{\Gamma(1{-}s_{23}{-}s_{24})} \Big) \, \Big|_{s_{234}=0}\, .
\label{in3npJ}
\eeq
This is consistent with the $\ap$-orders in (\ref{KNdeg178}), and it would be interesting
to find a general proof based on the relation (\ref{rePT17}) with five-point disk integrals.
\end{itemize}
At generic values of $\tau$, the $\ap$-expansions of $\widehat Z^\tau_{(0,0,0)}\big(
\begin{smallmatrix} 2,3,4 \\ 1 \end{smallmatrix} |\ast\big)$ and $\widehat Z^\tau_{(0,0,0)}
\big(\begin{smallmatrix} 3,4 \\ 1,2 \end{smallmatrix} |\ast\big)$ generate planar and non-planar
$A$-cycle graph functions \cite{Broedel:2017jdo, Broedel:2018izr}.
As a consistency check for
the differential-equation method of this work, (\ref{someint44}) with initial values (\ref{in3npE})
and (\ref{in3npFsimp}) has been verified to reproduce all $A$-cycle graph functions at the orders of $\ap^{\leq 4}$
including the simplest instances
\begin{align}
A_{11\underline{11}}(\tau) &=
240 \zeta_2 \gamma_0(6,0,0,0|\tau)- 840 \gamma_0(8,0,0,0|\tau)
= A_{1\underline{1}1\underline{1}}(\tau)\notag
\\
A_{\underline{1111}}(\tau) &=
{-}480 \zeta_2 \gamma_0(6,0,0,0|\tau)- 840 \gamma_0(8,0,0,0|\tau)
 \label{CC67}
\end{align}
of non-planar four-vertex graph functions
\begin{align}
A_{ij\underline{kl}}(\tau) &= 2 i! j! k! l!   \widehat Z^{\tau}_{(0,0,0)}\big(
\begin{smallmatrix} 2,3,4 \\ 1 \end{smallmatrix} |1,2,3,4\big)  \, \big|_{ s_{23}^i s_{34}^j s_{41}^k s_{12}^l}  \notag \\
A_{i\underline{j} k \underline{l}}(\tau) &=  i! j! k! l!   \widehat Z^{\tau}_{(0,0,0)}\big(
\begin{smallmatrix} 3,4 \\ 1,2 \end{smallmatrix} |1,2,3,4\big)  \, \big|_{s_{12}^i s_{23}^j s_{34}^k s_{41}^l}  \label{in3npK}  \\
A_{\underline{ijkl}}(\tau) &=  i! j! k! l!   \widehat Z^{\tau}_{(0,0,0)}\big(
\begin{smallmatrix} 3,4 \\ 1,2 \end{smallmatrix} |1,2,3,4\big)  \, \big|_{s_{13}^i s_{23}^j s_{24}^k s_{41}^l}  \, ,\notag
\end{align}
see (\ref{defA4}) and (\ref{CC66}) for their planar counterparts. One
can easily check via (\ref{qgamma1}) that the non-planar $(n{\leq} 4)$-point $A$-cycle graph functions
in (\ref{CC17}), (\ref{CC34}) and (\ref{in3npK}) do not exhibit any negative powers of $\pi$ in
their $q$-expansion, in lines with the $n$-point discussion in appendix~\ref{app:qexp}.


\subsection{Higher points}
\label{sec:6.4}

Non-planar initial values $Z^{i\infty}_{\vec{\eta}} (\begin{smallmatrix} Q\\ P \end{smallmatrix} |1,2,\ldots,n)$
at $n{\geq} 5$ points lead to a rapidly growing number of inequivalent cases.
Instead of spelling out an exhaustive list of their reduction to $I^{\rm tree}(\cdot|\cdot)$, we content
ourselves to representative examples of the general prescription (\ref{CC10G}). At five
points, cycles of ``3+2'' and ``4+1'' type give rise to initial values such as
\begin{align}
Z^{i\infty}_{\vec{\eta}} &\big(\begin{smallmatrix} 2,4 \\ 1,3,5 \end{smallmatrix} |1,2,3,4,5\big)  =
\frac{ \pi^4 I^{\rm tree}(1,3,5|1) I^{\rm tree}(2,4|1) }{\sin(\pi \eta_{2345}) \sin(\pi \eta_{345}) \sin(\pi \eta_{45}) \sin(\pi \eta_5)}
\label{in3npN}\\
Z^{i\infty}_{\vec{\eta}} &\big(\begin{smallmatrix} 3 \\ 1,2,4,5 \end{smallmatrix} |1,2,3,4,5\big)  =
\frac{ \pi^2  }{  \sin(\pi \eta_{345}) \sin(\pi \eta_{45}) } \Big\{ \pi^2 \cot(\pi \eta_{2345}) \cot(\pi \eta_5)
I^{\rm tree}(1,2,4,5|1)  \notag \\
&\! \! \! \! \! + \pi \cot(\pi \eta_{2345}) I^{\rm tree}(1,2,4,5|G_{45})
{+} \pi  \cot(\pi \eta_5)I^{\rm tree}(1,2,4,5|G_{12})
{+} I^{\rm tree}(1,2,4,5|G_{12}G_{45})  \Big\}\, , \notag
\end{align}
and representative six-point analogues read
\begin{align}
Z^{i\infty}_{\vec{\eta}} \big(\begin{smallmatrix} 2,5,6 \\ 1,3,4 \end{smallmatrix} |1,2,3,4,5,6\big)  &= \frac{ \pi^3  }{  \sin(\pi \eta_{23456}) \sin(\pi \eta_{3456}) \sin(\pi \eta_{56}) }  \notag \\
&\! \! \! \! \! \! \! \! \! \! \! \! \! \! \! \!  \times \Big\{ \pi \cot(\pi \eta_{456}) I^{\rm tree}(1,3,4|1) + I^{\rm tree}(1,3,4|G_{34}) \Big\}  \notag\\
&\! \! \! \! \! \! \! \! \! \! \! \! \! \! \! \! \times \Big\{ \pi \cot(\pi \eta_{6}) I^{\rm tree}(2,5,6|1) + I^{\rm tree}(2,5,6|G_{56}) \Big\}
\label{in3npP}
\\
Z^{i\infty}_{\vec{\eta}} \big(\begin{smallmatrix} 3,5 \\ 1,2,4,6 \end{smallmatrix} |1,2,3,4,5,6\big)  &= \frac{ \pi^4  \,  I^{\rm tree}(3,5|1)   }{  \sin(\pi \eta_{3456}) \sin(\pi \eta_{456}) \sin(\pi \eta_{56}) \sin(\pi \eta_{6}) }\notag\\
&\! \! \! \! \! \! \! \! \! \! \! \! \! \! \! \!  \times \Big\{
\pi \cot(\pi \eta_{23456}) I^{\rm tree}(1,2,4,6|1)  + I^{\rm tree}(1,2,4,6|G_{12})
\Big\} \, . \notag
\end{align}
Note that even the on-shell limit of $Z^{i\infty}_{\vec{\eta}} (\begin{smallmatrix} 3,5 \\ 1,2,4,6 \end{smallmatrix} |\ast)$
(with nine independent Mandelstam invariants instead of fifteen $s_{ij}$) requires the full expressions
for $I^{\rm tree}(1,2,4,6|\ast)$ in terms of the six Mandelstam invariants $s_{ij}$ with $i,j \in \{1,2,4,6\}$.
This is a major motivation for performing the computations in this work without assuming any relations among the $s_{ij}$.
We reiterate that non-planar $Z^{i\infty}_{\vec{\eta}}(\begin{smallmatrix} Q\\ P \end{smallmatrix}|\cdot)$ characterized
by $|P|$- and $|Q|$-point cycles can be reduced to disk integrals $Z^{\rm tree}(\cdot|\cdot)$ at
multiplicities $\leq \te{max}(|P|,|Q|){+}2$ via (\ref{rePT12}).

Higher-point analogues of the component integrals (\ref{npcompin71}) and (\ref{in3npG})
can be generated from
\beq
\widehat Z^{\tau}_{\vec{\eta}} \big(\begin{smallmatrix} Q \\ P \end{smallmatrix} |1,2,\ldots,n\big)
\equiv \Big( \prod_{i \in P} \prod_{j \in Q} q^{-\frac{1}{8}s_{ij}} \Big)Z^{\tau}_{\vec{\eta}} \big(\begin{smallmatrix} Q \\ P \end{smallmatrix} |1,2,\ldots,n\big) \, ,
 \label{in3npL}
\eeq
where the extra factors of $q^{-\frac{1}{8}s_{ij}}$ ensure that
each order in $\ap$ admits a Fourier expansion w.r.t.\ $q$. However, these
extra factors modify the differential equation, i.e.\ the $\tau$-derivative of
$\widehat Z^{\tau}_{\vec{\eta}} (\cdot|\cdot)$ can no longer be described by
a universal differential operator $D_{\vec{\eta}}^\tau$ as in (\ref{intro3}):
The cycle-dependent redefinition in (\ref{in3npL}) yields
\beq
2\pi i \partial_\tau   \widehat Z^{\tau}_{\vec{\eta}} \big(\begin{smallmatrix} Q \\ P \end{smallmatrix} |1,A\big)
=\sum_{B\in S_{n-1}}
\widehat D_{\vec{\eta},P,Q}^\tau(A|B)
   \widehat Z^{\tau}_{\vec{\eta}} \big(\begin{smallmatrix} Q \\ P \end{smallmatrix} |1,B\big)
   \label{in3npO}
\eeq
instead of (\ref{intro3}), with $P$- and $Q$-dependent modifications in the diagonal entries,
\beq
\widehat D_{\vec{\eta},P,Q}^\tau(A|B) = D_{\vec{\eta}}^\tau(A|B) +
3\delta_{A,B} \zeta_2 \sum_{i \in P} \sum_{j \in Q}  s_{ij}\, .
\label{in3npM}
\eeq
The two-point instance of (\ref{in3npL}) obeys $2\pi i \partial_\tau \widehat Z^\tau_{\eta_2}
(\begin{smallmatrix} 2 \\1 \end{smallmatrix} |1,2 )= s_{12}  ( \frac{1}{2} \partial_{\eta_2}^2 - \wp(\eta_2,\tau)
+ \zeta_2) \widehat Z^\tau_{\eta_2}(\begin{smallmatrix} 2 \\1 \end{smallmatrix} |1,2 )$ with $+\zeta_2$ in place
of $-2\zeta_2$ in (\ref{B16alt}).


\section{Formal properties}
\label{sec:others}

In this section, we take advantage of the new representations and structures of
$n$-point $A$-cycle integrals to derive some of their formal properties.


\subsection{Uniform transcendentality}
\label{sec:others1}

As one of the central results of this work, we can prove on the basis of the differential-equation method
that the $\ap$-expansion of the integrals
\beq
\int_{{\cal C}(\ast)} \dd z_2\, \dd z_3\,\ldots\, \dd z_n \, f^{(m_1)}_{12} f^{(m_2)}_{23} f^{(m_3)}_{34}\ldots f^{(m_{n-1})}_{n-1,n}
 \, {\rm KN}_{12\ldots n}^\tau \, , \ \ \ \ \ m_1,m_2,\ldots,m_{n-1} \geq 0
 \label{in3npR}
\eeq
is uniformly transcendental with weight $k{+}m_1{+}m_2{+}\ldots{+}m_{n-1}$ at the order of $\ap^k$ -- for
any planar or non-planar integration cycle ${\cal C}(P)$ or ${\cal C}(\begin{smallmatrix} Q \\P \end{smallmatrix})$.
We will derive the equivalent claim
\beq
\begin{array}{c}
\te{The $\ap^k$-order of the $A$-cycle integrals $Z^\tau_{\vec{\eta}}(\cdot|\cdot)$ in (\ref{intro2})} \\
\te{is uniformly transcendental with weight $k{+}n{-}1$.}
\end{array}
\label{in3npS}
\eeq
where the $\eta_j$ are assigned transcendental weight $-1$. With this choice,
the expansion (\ref{trigdeg}) of each trigonometric factor $\pi \cot(\pi \eta)$ or $\frac{ \pi }{\sin(\pi \eta)}$
carries transcendental weight $+1$. In the subsequent derivation of (\ref{in3npS}), it is convenient to
introduce the shorthand
\beq
\Psi^\tau_{\vec{\eta}} \equiv \sum_{r=0}^{\infty}  \sum_{k_1,k_2,\ldots,k_r \atop{=0,4,6,8,\ldots} }  \prod_{j=1}^r (k_j{-}1)\, \gamma(k_1,k_2,\ldots,k_r|\tau) r_{\vec{\eta}}(\ep_{k_r} \ep_{k_{r-1}} \ldots \ep_{k_2} \ep_{k_1} )
\label{in3npT}
\eeq
for the path-ordered exponential of $D^\tau_{\vec{\eta}}$ in the $\ap$-expansion (\ref{intro1}) of
$Z^\tau_{\vec{\eta}}(\cdot|\cdot)$, i.e.\ we will determine the transcendentality properties of
both ingredients on the right-hand side of
\beq
Z^{\tau}_{\vec{\eta}} \big(\begin{smallmatrix} Q \\ P \end{smallmatrix} |1,A\big) =
  \sum_{B \in S_{n-1}} \Psi^\tau_{\vec{\eta}}(A|B) Z^{i\infty}_{\vec{\eta}} \big(\begin{smallmatrix} Q \\ P \end{smallmatrix} |1,B\big) \, .
\label{in3npU}
\eeq
\begin{itemize}
\item The expressions (\ref{rePT4}) and (\ref{CC10G}) for the initial values
$Z^{i\infty}_{\vec{\eta}} (A|B)$ and
$Z^{i\infty}_{\vec{\eta}} \big(\begin{smallmatrix} Q \\ P \end{smallmatrix} |B\big)$
are linear or quadratic in $I^{\rm tree}(\cdot|\cdot)$. The simplest contributions
due to $I^{\rm tree}(\cdot|1)$ with constant integrand involve $\QQ[(2\pi i)^{-1}]$-linear combinations
of MZVs of overall weight $k$ at the order of $\ap^k$. This follows from
(\ref{rePT11}) and the genus-zero result that the combinations $(2\pi i)^{1-n}Z^{\rm tree}(\cdot|\cdot)$
with $(n{+}2)$-point disk integrals (\ref{cocyc1}) are uniformly transcendental with weight $k$ at the order of $\ap^k$
(and the same is true for the trigonometric entries of ${\cal H}_{\ap}(A|P)$).
Any appearance of $I^{\rm tree}(\cdot|1)$ is accompanied by
$n{-}1$ trigonometric factors $\pi \cot(\pi \eta)$ or $\frac{ \pi }{\sin(\pi \eta)}$ of weight one each,
resulting in overall weight $k{+}n{-}1$ at the order of $\ap^k$ in the respective contribution
to $n$-point $Z^{i\infty}_{\vec{\eta}}$.

The same transcendentality counting holds in presence of $0\leq \ell \leq n{-}1$ factors of $G_{ij}$
 in the integrands of $I^{\rm tree}(\cdot|\cdot)$: Each insertion of
$G_{ij}\rightarrow i\pi \, {\rm sgn}_{ij}^B$ in the expansion (\ref{rePT12}) of
$I^{\rm tree}(\cdot|\cdot)$ increases the weight by one, but at the same time
suppresses one of the trigonometric weight-one prefactors, cf.\ (\ref{Gpl}).
This leads to the conclusion that
\beq
\begin{array}{c}
\te{Initial values $Z^{i\infty}_{\vec{\eta}} (\cdot|\cdot)$ for planar or non-planar $n$-point $A$-cycle integrals}  \\
\te{are uniformly transcendental with weight $k{+}n{-}1$ at the order of $\ap^k$.}
\end{array}
\label{in3npV}
\eeq
\item For the $\tau$-dependent factor of $\Psi^\tau_{\vec{\eta}}(A|B)$ in (\ref{in3npU}), the transcendentality
properties are determined by the combinations of $\gamma(k_1,k_2,\ldots,k_r|\tau)
 r_{\vec{\eta}}(\ep_{k_r} \ep_{k_{r-1}} \ldots \ep_{k_2} \ep_{k_1} )$
in (\ref{in3npT}). The transcendental weight of the above iterated Eisenstein integral is $k_1{+}k_2{+}\ldots{+}k_r{-}r$, see
section \ref{sec:2.4}, and the accompanying derivations are of homogeneity degree
$r_{\vec{\eta}}(\ep_{k_j})\sim s_{pq} \eta^{k_j-2}$ if $k_j>0$. The exceptional term $2\zeta_2 s_{12\ldots n}$
in $r_{\vec{\eta}}(\ep_{0})$ follows the same transcendentality counting as its remaining terms
$\sim s_{ij} \eta^{-2}$ or $s_{ij} \partial_{\eta}^{2}$.

Hence, each product $\gamma(k_1,k_2,\ldots,k_r|\tau)  r_{\vec{\eta}}(\ep_{k_r} \ep_{k_{r-1}} \ldots \ep_{k_2} \ep_{k_1} )$
contributes transcendental weight $k_1{+}k_2{+}\ldots{+}k_r{-}r$ and $2r{-}k_1{-}k_2{-}\ldots{-}k_r$ from the
iterated Eisenstein integrals and the derivations, respectively. The combined transcendental weight $r$ matches
the order of $\ap$ due to the linearity of each $r_{\vec{\eta}}(\ep_{k_j\geq 0})$ in $s_{ij}$. As a result,
\beq
\begin{array}{c}
\te{The path-ordered exponentials $\Psi^\tau_{\vec{\eta}}$ in (\ref{in3npT}) are uni-}  \\
\te{formly transcendental with weight $k$ at the order of $\ap^k$.}
\end{array}
\label{in3npW}
\eeq
\end{itemize}
Uniform transcendentality (\ref{in3npS}) of the $A$-cycle integrals follows from combining the
observations (\ref{in3npV}) and (\ref{in3npW}) on the two ingredients in their $\ap$-expansion
(\ref{in3npU}). Therefore, by isolating the coefficient of $\prod_{j=1}^{n-1} \eta^{m_{j}-1}_{j+1 \ldots n}$
in $Z^\tau_{\vec{\eta}}(\ast|1,2,\ldots,n)$, the $\ap$-expansion of the integral (\ref{in3npR})
over $f^{(m_1)}_{12}f^{(m_2)}_{23}\ldots f^{(m_{n-1})}_{n-1,n}$ is shown to be uniformly transcendental
with weight $k{+}m_1{+}m_2{+}\ldots{+}m_{n-1}$ at the order of $\ap^k$.


\subsection{Coaction}
\label{sec:others2}

As will be shown in this section, $A$-cycle integrals and in particular their $\tau$-dependent
building blocks $\Psi^\tau_{\vec{\eta}} $ in (\ref{in3npU}) are preserved under the coaction.
This generalizes the behavior of $N$-point disk integrals (\ref{cocyc1}): The result of applying
the motivic coaction $\Delta$ \cite{Goncharov:2005sla, BrownTate, Brown:2011ik, Brown:ICM14}
order by order to the MZVs in the $\ap$-expansion can be resumed to
\cite{Schlotterer:2012ny, Drummond:2013vz}
\beq
\Delta Z^{\rm tree}(A|B) = \sum_{C,D \in S_{N-3}} S_0(C|D)Z^{\rm tree}(A|1,C,N,N{-}1) \otimes
Z^{\rm tree}_{\mathfrak{dr}}(1,D,N{-}1,N|B)\, ,
\label{coac1}
\eeq
where $S_0(C|D)$ is a local representation\footnote{At four and five points, for instance,
the KLT matrix is a scalar $S_0(2|2)=-s_{12}$ and a $2\times 2$ matrix with entries
such as $S_0(23|23)=s_{12}(s_{13}{+}s_{23})$ and
$S_0(23|32)=s_{12}s_{13}$, respectively.} of the $(N{-}3)! \times (N{-}3)!$ KLT
matrix in field theory \cite{Kawai:1985xq, Bern:1998sv, momentumKernel} and $A,B \in S_N$.
Generalizations of the coaction formula to genus-zero integrals with additional fixed punctures
have been recently studied in a physics context of Feynman integrals
\cite{Abreu:2017enx, Abreu:2017mtm, otherAbreu:2018, Abreu:2018nzy} and
a mathematics context of Lauricella hypergeometric functions \cite{Brown:2019jng}.

In view of the unsettled transcendentality properties of MZVs, only their motivic versions allow
for a well-defined coaction. Motivic MZVs $\zeta^{\mathfrak m}_{n_1,n_2,\ldots,n_r}$ are objects in
algebraic geometry whose elaborate definition can for instance be found in \cite{Goncharov:2005sla, BrownTate,
Brown:ICM14, Brown:2013gia}. All the $\QQ$-relations known from MZVs $\zeta_{n_1,n_2,\ldots,n_r}$ (see
e.g.\ \cite{Blumlein:2009cf, freetext}) by definition hold for motivic MZVs, and (\ref{coac1}) is understood
to apply to the motivic version of disk integrals,
where all the MZVs in their $\ap$-expansion are replaced by the respective $\zeta^{\mathfrak m}_{n_1,n_2,\ldots,n_r}$.

The subscript $_{\mathfrak{dr}}$ in the second factor on the right-hand side of (\ref{coac1}) instructs to
replace each MZV in the $\ap$-expansion by the associated de Rham period $\zeta^{\mathfrak dr}_{n_1,n_2,\ldots,n_r}$.
Loosely speaking, the net effect of passing to de Rham periods is to mod out by
$\zeta_{2}$ or $\zeta^{\mathfrak m}_{2}$, see e.g.\ \cite{Francislecture} for the mathematical background.

The genus-one analogue of (\ref{coac1}) is based on the following coaction of iterated Eisenstein
integrals \cite{Francislecture, Broedel:2018iwv}:
\beq
\Delta \gamma(k_1,k_2,\ldots,k_r|\tau) = \sum_{j=0}^r \gamma(k_1,k_2,\ldots,k_j|\tau)
\otimes \mathfrak{S}[\gamma(k_{j+1},\ldots,k_r|\tau)]\, .
\label{coac2}
\eeq
In the same way as the second entry of the motivic coaction of MZVs involves de Rham periods,
the iterated Eisenstein integrals in the second entry of (\ref{coac2}) are replaced by
abstract symbols
\beq
\mathfrak{S}[\gamma(k_{1},k_2,\ldots,k_r|\tau)] = \Big[ \, {-} \frac{ \dd \tau}{2\pi i } {\rm G}_{k_1}(\tau) \, \Big| \,
 {-} \frac{ \dd \tau}{2\pi i } {\rm G}_{k_2}(\tau) \, \Big| \, \ldots
 \, \Big| \,  {-} \frac{ \dd \tau}{2\pi i } {\rm G}_{k_r}(\tau) \, \Big] \, ,
 \label{coac3}
\eeq
which leads to examples like $\Delta (\tau)=\tau \otimes 1 + 1 \otimes [\dd \tau]$ or
$\Delta \gamma(k|\tau)= \gamma(k|\tau)\otimes 1 + 1 \otimes \big[{-} \frac{ \dd \tau \, {\rm G}_k}{2\pi i} \big]$.
When applying (\ref{coac2}) to each term in the $\ap$-expansion of (\ref{in3npT}), we obtain
\begin{align}
\Delta \Psi^\tau_{\vec{\eta}}(A|B) &=
 \sum_{r=0}^{\infty}  \sum_{k_1,k_2,\ldots,k_r \atop{=0,4,6,8,\ldots} }  \prod_{j=1}^r (k_j{-}1)\,  r_{\vec{\eta}}(\ep_{k_r} \ep_{k_{r-1}} \ldots \ep_{k_2} \ep_{k_1} )_A{}^B \notag\\
 &\ \ \ \ \times \sum_{j=0}^r  \gamma(k_1,k_2,\ldots,k_j|\tau)
\otimes \mathfrak{S}[\gamma(k_{j+1},\ldots,k_r|\tau)]
 \label{coac4} \\
 &= \sum_{r=0}^{\infty}  \sum_{k_1,k_2,\ldots,k_r \atop{=0,4,6,8,\ldots} }  \sum_{C \in S_{n-1}}
 r_{\vec{\eta}}(   \ep_{k_r}  \ldots \ep_{k_2} \ep_{k_1} )_A{}^C
  \sum_{s=0}^{\infty}  \sum_{\ell_1,\ell_2,\ldots,\ell_s \atop{=0,4,6,8,\ldots} }
 r_{\vec{\eta}}( \ep_{\ell_s}  \ldots \ep_{\ell_2} \ep_{\ell_1} )_C{}^B  \notag \\
 & \ \ \ \ \times  \prod_{j=1}^s (\ell_j{-}1) \gamma(\ell_1,\ell_2,\ldots,\ell_s|\tau)
 \otimes  \prod_{j=1}^r (k_j{-}1)
 \mathfrak{S}[ \gamma(k_1,k_2,\ldots,k_r|\tau)] \, .\notag
\end{align}
Since both entries of the tensor product can be recombined to the series-representation
of $\Psi^\tau_{\vec{\eta}}$, one may compactly rewrite (\ref{coac4}) as
\beq
\Delta \Psi^\tau_{\vec{\eta}}(A|B) = \sum_{C \in S_{n-1}} \Psi^\tau_{\vec{\eta}}(C|B) \otimes \mathfrak{S} \big[\overleftarrow{\Psi}^\tau_{\vec{\eta}}(A|C)\big]  \, .
 \label{coac5}
\eeq
The notation $\overleftarrow{\Psi}^\tau_{\vec{\eta}}$ indicates that the derivations in the second
entry (in particular the $\partial_{\eta_j}$ in the entries of $r_{\vec{\eta}}(\ep_0)$) act from the
left on the derivations in the first entry. This reflects the
relative positions of the $r_{\vec{\eta}}(\ep_{k_r} \ldots \ep_{k_1} )_A{}^C $ and
$r_{\vec{\eta}}(\ep_{\ell_s}  \ldots \ep_{\ell_1} )_C{}^B$ in (\ref{coac4}).
The order of the derivations within $\overleftarrow{\Psi}^\tau_{\vec{\eta}}$ is unchanged in
comparison to $\Psi^\tau_{\vec{\eta}}$.

The coaction of the full $A$-cycle integrals (\ref{in3npU}) requires an extension of (\ref{coac5})
to also incorporate the MZVs of the initial values $Z^{i\infty}_{\vec{\eta}}(\cdot|\cdot)$. In the planar case,
their decomposition (\ref{inival1}) into disk integrals and the genus-zero coaction (\ref{coac1}) imply
a result of the schematic form
\beq
\Delta Z^{i\infty}_{\vec{\eta}}(1,A|1,P) = \sum_{B,C \in S_{n-1}} T_{\vec{\eta}}(B|C) Z^{i\infty}_{\vec{\eta}}(1,A|1,B)
\otimes Z^{i\infty}_{\vec{\eta},\mathfrak{dr}}(1,C|1,P)\, .
\label{coac6}
\eeq
The $(n{-}1)! \times (n{-}1)!$ matrix $T_{\vec{\eta}}(B|C)$ may be reconstructed from the KLT matrix $S_0$ as well as the
objects ${\cal H}_{\ap},{\cal K}_{\vec{\eta}}$ in (\ref{inival1}). It depends on $\eta_j,s_{ij}$ and does not affect
the second entry through its coaction $\Delta T_{\vec{\eta}}(B|C) = T_{\vec{\eta}}(B|C) \otimes 1$ since it comprises
no MZVs other than powers of $\pi$. On these grounds,
one can combine (\ref{coac5}) and (\ref{coac6}) to find
\begin{align}
\Delta Z^{\tau}_{\vec{\eta}}(1,A|1,B) &= \sum_{C \in S_{n-1}} \Delta \Psi^\tau_{\vec{\eta}}(B|C) \cdot \Delta Z^{i\infty}_{\vec{\eta}}(1,A|1,C)\notag\\
&= \! \! \! \sum_{C,D,E,F\in S_{n-1}} \! \! \!  \Psi^\tau_{\vec{\eta}}(D|C) T_{\vec{\eta}}(E|F) Z^{i\infty}_{\vec{\eta}}(1,A|1,E)
\otimes \mathfrak{S} \big[\overleftarrow{\Psi}^\tau_{\vec{\eta}}(B|D)\big]  Z^{i\infty}_{\vec{\eta},\mathfrak{dr}}(1,F|1,C)
\notag
\\
&= \! \! \! \sum_{C,D,E,F,G\in S_{n-1}} \! \! \!  \Psi^\tau_{\vec{\eta}}(D|C)  T_{\vec{\eta}}(E|F) \big[
\Psi^\tau_{\vec{\eta}}(E|G)^{-1} Z^{\tau}_{\vec{\eta}}(1,A|1,G) \big]  \label{coac7} \\
&\ \ \ \ \ \ \ \ \ \ \ \ \ \ \ \ \ \ \otimes \mathfrak{S} \big[\overleftarrow{\Psi}^\tau_{\vec{\eta}}(B|D)\big]  Z^{i\infty}_{\vec{\eta},\mathfrak{dr}}(1,F|1,C)\, ,
\notag
\end{align}
where the inverse $(\Psi^\tau_{\vec{\eta}})^{-1}$ of the path-ordered exponential features an expansion with
alternating signs $(-1)^r$ and a reversed order of composing the derivations in comparison to $\Psi^\tau_{\vec{\eta}}$:
\beq
\Psi^\tau_{\vec{\eta}}(A|B)^{-1} = \sum_{r=0}^{\infty} (-1)^r  \sum_{k_1,k_2,\ldots,k_r \atop{=0,4,6,8,\ldots} }  \prod_{j=1}^r (k_j{-}1)\, \gamma(k_1,k_2,\ldots,k_r|\tau) r_{\vec{\eta}}(\ep_{k_1} \ep_{k_{2}} \ldots  \ep_{k_r} )_A{}^B\, .
\label{coac8}
\eeq
The derivations in the object
$\mathfrak{S}\big[\overleftarrow{\Psi}^\tau_{\vec{\eta}}(B|D)\big]$ in the
second entry of (\ref{coac7}) are understood to act from the left on all the
functions of $\eta_j$ in the first entry as well as
$Z^{i\infty}_{\vec{\eta},\mathfrak{dr}}(1,F|1,C)$ in the second entry.
Similarly, the derivations in $\Psi^\tau_{\vec{\eta}}(D|C)$ act on all the
$\eta_j$ in $T_{\vec{\eta}}(E|F),
Z^{i\infty}_{\vec{\eta},\mathfrak{dr}}(1,F|1,C)$ and
$Z^{i\infty}_{\vec{\eta}}(1,A|1,E)$ (or equivalently $\sum_{G \in
S_{n-1}}\Psi^\tau_{\vec{\eta}}(E|G)^{-1} Z^{\tau}_{\vec{\eta}}(1,A|1,G)$).
However, as indicated by $[\ldots]$ in the third line of (\ref{coac7}), the
scope of the derivations in $\Psi^\tau_{\vec{\eta}}(E|G)^{-1}$ is limited to
the factor $Z^{\tau}_{\vec{\eta}}(1,A|1,G)$ enclosed by the bracket.

As a bottom line of (\ref{coac7}), the coaction of the planar $A$-cycle
integral $Z^{\tau}_{\vec{\eta}}(1,A|1,B)$ is expressible via linear operations
acting on an $(n{-}1)!$ basis of $A$-cycle integrals
$Z^{\tau}_{\vec{\eta}}(1,A|1,G)$, $G\in S_{n-1}$ with the same integration
cycle ${\cal C}(1,A)$. Said linear operations include infinite series in
$\gamma(k_1,\ldots,k_r|\tau)$ and $\eta_j$-derivatives that enter through the
factor of $\Psi^\tau_{\vec{\eta}}$ and its inverse.

The same kind of conclusion holds for the coaction of non-planar $A$-cycle integrals
$Z^\tau_{\vec{\eta}}(\begin{smallmatrix} Q \\ P \end{smallmatrix} |\cdot)$:
Given that the initial values $Z^{i\infty}_{\vec{\eta}}(\begin{smallmatrix} Q
\\ P \end{smallmatrix} |\cdot)$ are expressible via products of disk
integrals, see (\ref{CC10G}) and (\ref{rePT12}), their coaction is still
determined by (\ref{coac1}). With a bit of additional bookkeeping effort to
track the information of $P$ and $Q$, one can adjust the details of
(\ref{coac6}) and (\ref{coac7}) to the non-planar setting.


\section{Conclusions}
\label{sec:concl}

In this work, we have studied iterated integrals over $A$-cycles on a torus that generate the contributions
of a cylinder or M\"obius-strip surface to $n$-point one-loop open-string amplitudes.
These $A$-cycle integrals were shown to satisfy linear and homogeneous
first-order differential equations w.r.t.\ the modular parameter $\tau$. Moreover,
their degeneration as $\tau \rightarrow i \infty$ is reduced to explicitly known combinations of
genus-zero integrals from open-string tree amplitudes.

The solution to this initial-value problem via standard Picard iteration exposes the
structure of the $\ap$-expansion of the $A$-cycle integrals, see (\ref{intro1}):
Any order in $\ap$ is expressible via iterated Eisenstein integrals and MZVs
whose composition follows from elementary operations -- matrix multiplications and
differentiation in auxiliary variables $\eta_j$. Then, by isolating specific orders in
the Laurent expansion w.r.t.\ $\eta_2,\eta_3,\ldots,\eta_n$, one recovers the
$\ap$-expansions relevant to one-loop open-string amplitudes.

The form of the differential equations is universal to any integration cycle
in planar- and non-planar one-loop amplitudes and shared by the $A$-cycle
component of the elliptic KZB associator \cite{KZB, EnriquezEllAss, Hain}. The
differential equations of $n$-point $A$-cycle integrals induce conjectural
$(n{-}1)!\times (n{-}1)!$ matrix representations of Tsunogai's derivations
\cite{Tsunogai}, reflecting the fact that the iterated Eisenstein integrals in
the $\ap$-expansion correspond to eMZVs \cite{Enriquez:Emzv, Broedel:2014vla,
Broedel:2015hia}. In particular, the appearance of twisted eMZVs in the
$\ap$-expansion of non-planar cylinder integrals \cite{Broedel:2017jdo} is
completely bypassed, corroborating and generalizing the four-point observations of the
reference.

The homogeneity of the differential equations is tied to considering a
collection of $(n{-}1)!$ Kronecker--Eisenstein-type integrands. The latter are believed to span the
twisted cohomology at genus one, i.e.\ to generalize the basis of $(n{-}3)!$
Parke--Taylor factors for the integration-by-parts inequivalent genus-zero
integrands \cite{Aomoto87}. Accordingly, the $A$-cycle integrals
$Z^\tau_{\vec{\eta}}(\cdot|\cdot)$ are proposed to be the genus-one generalization
of the disk- or $Z$-integrals that are universal to open-string tree
amplitudes \cite{Mafra:2011nv, Zfunctions, Huang:2016tag, Azevedo:2018dgo}.
The $\ap$-expansion of $Z$-integrals at genus zero can be interpreted in terms
of effective-field-theory amplitudes of (bi-)colored scalars
\cite{Carrasco:2016ldy, Carrasco:2016ygv} and allows for Berends--Giele
recursions \cite{Mafra:2016mcc}. Hence, it would be interesting to investigate
a genus-one echo of these features at the level of $A$-cycle integrals
$Z^\tau_{\vec{\eta}}(\cdot|\cdot)$.

The differential equations of $A$-cycle integrals presented in this work suggest a variety of follow-up investigations:
\begin{itemize}
\item The techniques of this work call for an extension to closed strings. As will be demonstrated
elsewhere \cite{inprogr}, similar methods can be used to derive Cauchy-Riemann
and Laplace equations for modular graph forms (see e.g.\ \cite{DHoker:2015foa,
DHoker:2015wxz, DHoker:2016mwo, DHoker:2016quv, Kleinschmidt:2017ege, Basu:2019idd})
at the level of generating series. Closed-string differential equations of this type should harbor
valuable input on relations between open- and closed-string one loop amplitudes through an elliptic
single-valued map \cite{Broedel:2018izr, Gerken:2018jrq, Zagier:2019eus}. These research directions
are hoped to clarify the role the non-holomorphic modular forms and single-valued iterated Eisenstein
integrals of Brown \cite{Brown:2017qwo, Brown:2017qwo2} in closed-string $\ap$-expansions.
\item The solution (\ref{intro1}) to the differential equations of $A$-cycle integrals
only depends on the integration cycle through the initial value at $\tau \rightarrow i \infty$.
Hence, relations between $A$-cycle integrals for different planar and non-planar orderings should be
encoded in the genus-zero information at the cusp. It would be interesting connect the monodromy
relations among open-string tree-level amplitudes \cite{BjerrumBohr:2009rd, Stieberger:2009hq}
with their loop-level extensions \cite{Tourkine:2016bak, Hohenegger:2017kqy} along these lines.
\item In the final expressions for one-loop open-string amplitudes, $A$-cycle integrals are integrated
over the modular parameter $\tau$, at least at the level of individual orders in their expansion w.r.t.\ $\eta_j$.
The differential equations in this work should be instrumental for performing the desired $\tau$-integrals,
for instance by identifying various contributions as total derivatives in $\tau$. Also, it would be interesting to
relate techniques for $\tau$-integrations in open-string amplitudes with recent progress on the closed-string
side \cite{Green:2008uj, DHoker:2015foa, DHoker:2019mib, DHoker:2019blr}.
\item The strategy of this work to infer $\ap$-expansions from the solution
of an initial-value problem should be universally applicable at various genera.
Similar linear and homogeneous differential equations
in the complex-structure moduli are expected to yield recursions for moduli-space integrals in string amplitudes
w.r.t.\ the loop order. This is analogous to the recursions for disk integrals w.r.t.\ the number of legs that descend
from KZ equations in a puncture on a genus-zero surface \cite{Broedel:2013aza}.

At genus two, for instance, a promising intermediate goal is to find differential equations for moduli-space
integrals w.r.t.\ off-diagonal entry of the $2\times 2$ period matrix. The separating degeneration of the surface
could then provide an initial value, where the dependence on the diagonal entries of the period
matrix enter via products of genus-one integrals. We
hope that these directions contribute to understanding the structure \& explicit examples
of $\ap$-expansions beyond one loop and to advancing the study of modular graph functions
at higher genus \cite{DHoker:2017pvk, DHoker:2018mys}.
\end{itemize}


\section*{Acknowledgments}

We are grateful to Johannes Broedel, Jan Gerken, Axel Kleinschmidt, Nils
Matthes and Federico Zerbini for inspiring discussions and collaboration on
related topics. Moreover, Claude Duhr, Hermann Nicolai, Albert Schwarz and in
particular Sebastian Mizera are thanked for valuable discussions. We are
grateful to the organizers of the programme ``Modular forms, periods and
scattering amplitudes'' at the ETH Institute for Theoretical Studies for
providing a stimulating atmosphere and financial support. CRM is supported
by a University Research Fellowship from the Royal Society. OS thanks the
organizers of the workshop ``Automorphic Structures in String Theory'' at the
Simons Center in Stony Brook and those of the workshop ``String Theory from a
Worldsheet Perspective'' at the GGI Florence for setting up inspiring
meetings. OS is supported by the European Research Council under
ERC-STG-804286 UNISCAMP.


\appendix


\section{Resolving cycles of Kronecker--Eisenstein series}
\label{app:0}

This appendix is dedicated to a series of lemmata to simplify cyclic products of
Kronecker--Eisenstein series. Many steps in the subsequent calculations are based
on the meromorphic Kronecker--Eisenstein series
\begin{align}
F(z,\eta,\tau) \equiv \frac{ \theta_1'(0,\tau) \theta_1(z+\eta,\tau) }{\theta_1(z,\tau)  \theta_1(\eta,\tau)} \, ,
\label{1.2mer}
\end{align}
which does not share the double-periodicity of its non-holomorphic counterpart (\ref{1.2}).
Its expansion coefficient at the order of $\eta^0$ is denoted by $g^{(1)}(z,\tau)$ and satisfies
\beq
g^{(1)}(z,\tau) = \partial_z \log \theta_1(z,\tau) \, , \ \ \ \ \ \ \partial_z g^{(1)}(z,\tau) = - \wp(z,\tau) - {\rm G}_2(\tau)\, .
\label{defg1}
\eeq


\subsection{Two points}
\label{app:0.1}

We start by proving the two-point identity\footnote{This identity is
equivalent to the identities
(3.3) from \cite{BrownLev} and (20) from \cite{Enriquez:Emzv}.} \eqref{A25DP} which plays a key
role in the derivation of the differential equation (\ref{B16alt}) of two-point $A$-cycle integrals.
The starting point is the following version of the Fay identity (\ref{A21}):
\beq
F(z_{12},\eta_2,\tau) F(z_{21}{+}\varepsilon,\xi,\tau) = F(\varepsilon,\eta_2,\tau) F(z_{21}{+}\varepsilon,\xi{-}\eta_2,\tau)
+F(\varepsilon,\xi,\tau) F(z_{12},\eta_2{-}\xi,\tau) \, .
\label{A21pr}
\eeq
Then, we evaluate the limit $\varepsilon\rightarrow 0$ based on
$F(\varepsilon,\eta,\tau)= \frac{1}{\varepsilon}+g^{(1)}(\eta,\tau) + {\cal O}(\varepsilon)$, where
$\frac{1}{\varepsilon}F(z_{21}{+}\varepsilon,\xi{-}\eta_2,\tau)$
introduces the $z$-derivative $\partial_z F(z_{21},\xi{-}\eta_2)$ at the zero$^{\rm th}$ order in $\varepsilon$:
\beq
F(z_{12},\eta_2,\tau) F(z_{21},\xi,\tau) = F(z_{12},\eta_2{-}\xi,\tau)
 \big( g^{(1)}(\xi,\tau) -   g^{(1)}(\eta_2,\tau) \big) + \partial_z F(z_{21},\xi{-}\eta_2) \, .
\label{A22}
\eeq
The next step is to apply the differential operator $(\partial_{\eta_2}+\partial_{\xi})$ to both sides,
which does not affect functions of $\eta_2{-}\xi$ and only needs to be applied to the $g^{(1)}$:
\beq
(\partial_{\eta_2}+\partial_{\xi}) F(z_{12},\eta_2,\tau) F(z_{21},\xi,\tau) =  \big( \wp(\eta_2,\tau) -  \wp(\xi,\tau) \big) F(z_{12},\eta_2{-}\xi,\tau)  \, .
\label{A22a}
\eeq
Note the cancellation of ${\rm G}_2$ from $(\partial_{\eta_2}{+}\partial_{\xi}) (g^{(1)}(\xi,\tau) {-}   g^{(1)}(\eta_2,\tau))
= \wp(\eta_2,\tau) {-} \wp(\xi,\tau)$, cf.\ (\ref{defg1}). Finally, we arrive at (\ref{A25DP}) by multiplying
both sides of (\ref{A22a}) by $\exp(2\pi i (\eta_2-\xi) \frac{ \Im z_{12} }{\Im \tau})$. This exponential commutes
with $(\partial_{\eta_2}+\partial_{\xi}) $ and promotes all instances of the meromorphic Kronecker--Eisenstein
series (\ref{1.2mer}) to the doubly-periodic one (\ref{1.2}).

When applied to (\ref{B15}), corollaries of (\ref{A22a}) include
$f^{(2)}_{12} f^{(2)}_{12}  - 2 f^{(3)}_{12} f^{(1)}_{12} = 3 {\rm G}_4 - 2 f^{(4)}_{12} $
as well as
\begin{align}
f^{(3)}_{12} f^{(2)}_{12}  - 3 f^{(4)}_{12} f^{(1)}_{12} &= 3 {\rm G}_4 f^{(1)}_{12} - 5 f^{(5)}_{12}  \notag \\
f^{(4)}_{12} f^{(2)}_{12}  - 4 f^{(5)}_{12} f^{(1)}_{12} &= 3 {\rm G}_4 f^{(2)}_{12} + 5 {\rm G}_6 - 9 f^{(6)}_{12}  \label{B13} \\
f^{(5)}_{12} f^{(2)}_{12}  - 5 f^{(6)}_{12} f^{(1)}_{12} &= 3 {\rm G}_4 f^{(3)}_{12} + 5 {\rm G}_6  f^{(1)}_{12} - 14 f^{(7)}_{12}  \notag \\
f^{(6)}_{12} f^{(2)}_{12}  - 6 f^{(7)}_{12} f^{(1)}_{12} &= 3 {\rm G}_4 f^{(4)}_{12} + 5 {\rm G}_6  f^{(2)}_{12}+ 7 {\rm G}_8 - 20 f^{(8)}_{12} \notag
\end{align}
and more generally ($n \in \NN_0$)
\beq
f^{(n)}_{12} f^{(2)}_{12}  - n f^{(n+1)}_{12} f^{(1)}_{12} = \sum_{k=4}^{n+2}(k{-}1) {\rm G}_k f_{12}^{(n+2-k)}
 - \frac{1}{2}(n{-}1)(n{+}2)  f^{(n+2)}_{12}  \,.
\label{B13gen}
\eeq


\subsection{Three points}
\label{app:0.2}

The identity (\ref{A22}) to resolve a two-cycle of Kronecker--Eisenstein series can be generalized
to longer cycles of the form $F(z_{12},\beta_1,\tau) F(z_{23},\beta_2,\tau) \ldots F(z_{n-1,n},\beta_{n-1},\tau)
F(z_{n,1},\xi,\tau)$. Three-cycles can be reduced to (\ref{A22}) by applying the Fay identity (\ref{A21}) in a first step,
\begin{align}
&F(z_{12},\beta_1,\tau)   F(z_{23},\beta_2,\tau)  F(z_{31},\xi,\tau)
\label{A26} \\
& \ \ = \big[
F(z_{13},\beta_1,\tau) F(z_{23},\beta_2{-}\beta_1,\tau)
+F(z_{13},\beta_2,\tau) F(z_{12},\beta_1{-}\beta_2,\tau)
\big]  F(z_{31},\xi,\tau)  \notag \\
& \ \ = F(z_{23},\beta_2{-}\beta_1,\tau)  \big[
F(z_{13},\beta_1{-}\xi,\tau)
 \big( g^{(1)}(\xi,\tau) -   g^{(1)}(\beta_1,\tau) \big) + \partial_z F(z_{31},\xi{-}\beta_1)
 \big] \notag \\
 &\ \ \ \ + F(z_{12},\beta_1{-}\beta_2,\tau)  \big[
F(z_{13},\beta_2{-}\xi,\tau)
 \big( g^{(1)}(\xi,\tau) -   g^{(1)}(\beta_2,\tau) \big) + \partial_z F(z_{31},\xi{-}\beta_2)
 \big] \, . \notag
\end{align}
When acting with $(\partial_{\beta_1}{+}\partial_{\beta_2}{+}\partial_{\xi})$, the contributions
from $\partial_z F(\ldots)$ again vanish:
\begin{align}
&(\partial_{\beta_1}+\partial_{\beta_2}+\partial_{\xi}) F(z_{12},\beta_1,\tau)   F(z_{23},\beta_2,\tau)  F(z_{31},\xi,\tau)  \label{A27} \\
& \ \ = F(z_{13},\beta_1{-}\xi,\tau) F(z_{23},\beta_2{-}\beta_1,\tau)
 \big(    \wp(\beta_1,\tau) - \wp(\xi,\tau) \big)  \notag \\
 &\ \ \ \ + F(z_{12},\beta_1{-}\beta_2,\tau)  F(z_{13},\beta_2{-}\xi,\tau)
 \big(   \wp(\beta_2,\tau)  - \wp(\xi,\tau)  \big)   \notag \\
 & \ \ = - \wp(\beta_1,\tau)   F(z_{23},\beta_2{-}\beta_1,\tau)    F(z_{31},\xi{-}\beta_1,\tau)  \notag \\
& \ \ \ \ \; \, - \wp(\beta_2,\tau)   F(z_{12},\beta_1{-}\beta_2,\tau)    F(z_{31},\xi{-}\beta_2,\tau)  \notag \\
& \ \ \ \ \; \, -  \wp(\xi,\tau)   F(z_{12},\beta_1{-}\xi,\tau)    F(z_{23},\beta_2{-}\xi,\tau) \, .\notag
\end{align}
The Fay identity has been used in the last step to manifest the cyclic symmetry of
the result under simultaneous exchange of $z_j$ and $\beta_1,\beta_2,\xi$.
Multiplication of both sides by $\exp( \frac{2\pi i}{\Im \tau}(\beta_1 z_{12}+\beta_2 z_{23} + \xi z_{31}))$
which is inert under the combination $\partial_{\beta_1}{+}\partial_{\beta_2}{+}\partial_{\xi}$
replaces the $F(\ldots)$ in (\ref{A27}) by $\Omega(\ldots)$ and implies the lemma (\ref{A27DP}).


\subsection{Higher points}
\label{app:0.3}

The methods of the previous subsections can be straightforwardly uplifted to simplify the four-cycle
analogue of (\ref{A27}),
\begin{align}
&(\partial_{\beta_1}+\partial_{\beta_2}+\partial_{\beta_3}+\partial_{\xi})
 F(z_{12},\beta_1,\tau)   F(z_{23},\beta_2,\tau)  F(z_{34},\beta_3,\tau) F(z_{41},\xi,\tau)  \label{A28} \\
 & \ \ = - \wp(\beta_1,\tau)   F(z_{23},\beta_2{-}\beta_1,\tau)    F(z_{34},\beta_3{-}\beta_1,\tau) F(z_{41},\xi{-}\beta_1,\tau)  \notag \\
& \ \ \ \ \; \, - \wp(\beta_2,\tau)   F(z_{12},\beta_1{-}\beta_2,\tau)    F(z_{34},\beta_3{-}\beta_2,\tau) F(z_{41},\xi{-}\beta_2,\tau)  \notag \\
& \ \ \ \ \; \, - \wp(\beta_3,\tau)   F(z_{12},\beta_1{-}\beta_3,\tau)    F(z_{23},\beta_2{-}\beta_3,\tau) F(z_{41},\xi{-}\beta_3,\tau)  \notag \\
& \ \ \ \ \; \, -  \wp(\xi,\tau)   F(z_{12},\beta_1{-}\xi,\tau)    F(z_{23},\beta_2{-}\xi,\tau) F(z_{34},\beta_3{-}\xi,\tau) \, .\notag
\end{align}
Upon multiplication by $\exp( \frac{2\pi i}{\Im \tau}(\beta_1 z_{12}+\beta_2 z_{23} + \beta_3 z_{34} + \xi z_{41}))$,
one arrives at the analogous statement (\ref{A28om}) for a cycle of $\Omega(\ldots)$. The same logic can
be repeated at higher multiplicity and is expected to yield the cyclic result
\begin{align}
&(\partial_{\beta_1}{+}\partial_{\beta_2}{+}\ldots{+}\partial_{\beta_{n-1}}{+}\partial_{\xi})
 F(z_{12},\beta_1,\tau)   F(z_{23},\beta_2,\tau) \ldots F(z_{n-1,n},\beta_{n-1},\tau) F(z_{n,1},\xi,\tau)  \label{A29} \\
& \ \ = - \sum_{j=1}^{n-1} \wp(\beta_j,\tau) F(z_{n,1},\xi{-}\beta_j,\tau) \prod_{i=1 \atop{i\neq j}}^{n-1} F(z_{i,i+1},\beta_{i}{-}\beta_j,\tau)    -  \wp(\xi,\tau)  \prod_{i=1}^{n-1} F(z_{i,i+1},\beta_{i}{-}\xi,\tau)
\, ,\notag
\end{align}
which is tested up to and including $n=5$ and conjectural at $n\geq 6$. Note that both sides have the same
poles in $\xi,\beta_1,\beta_2,\ldots,\beta_{n-1}$ and monodromies as $z_j \rightarrow z_j{+}\tau$. It
would be interesting to prove (\ref{A29}) by induction.

As before, multiplication by $\exp( \frac{2\pi i}{\Im \tau}( \sum_{j=1}^{n-1}\beta_j z_{j,j+1} + \xi z_{n,1}))$
promotes all the meromorphic Kronecker--Eisenstein series in (\ref{A29}) to the doubly-periodic ones,
\begin{align}
&(\partial_{\beta_1}{+}\partial_{\beta_2}{+}\ldots{+}\partial_{\beta_{n-1}}{+}\partial_{\xi})
 \Omega(z_{12},\beta_1,\tau)   \Omega(z_{23},\beta_2,\tau) \ldots \Omega(z_{n-1,n},\beta_{n-1},\tau) \Omega(z_{n,1},\xi,\tau)  \label{A29om} \\
& \ \ = - \sum_{j=1}^{n-1} \wp(\beta_j,\tau) \Omega(z_{n,1},\xi{-}\beta_j,\tau) \prod_{i=1 \atop{i\neq j}}^{n-1} \Omega(z_{i,i+1},\beta_{i}{-}\beta_j,\tau)    -  \wp(\xi,\tau)  \prod_{i=1}^{n-1} \Omega(z_{i,i+1},\beta_{i}{-}\xi,\tau)
\, .\notag
\end{align}
%


\section{The non-planar Green function at the cusp}
\label{app:C}

The purpose of this appendix is to justify the regularized value (\ref{Gnpnew}), i.e.\ the vanishing
of the non-planar Green function $ {\cal G}(v_{ij}{+}\tauh,\tau)$ at the cusp. In order to do so,
we first establish a result that does not require any regularization:
\beq
\lim_{\tau \rightarrow i\infty} \Big[ {\cal G}(v_{ij}{+}\tauh,\tau) - \frac{ i \pi \tau}{4} \Big] = 0 \, .
\label{Gnp}
\eeq
The vanishing of this limit follows from two observations on the
difference ${\cal G}(v_{ij}{+}\tauh,\tau) - \frac{ i \pi \tau}{4}$:
\begin{itemize}
\item Its derivatives (\ref{B5}) w.r.t.\ $v_i$
and $\tau$ vanish at the cusp since $\lim_{\tau \rightarrow i \infty} f^{(1)}(z{-}\tauh,\tau) = 0$
and $\lim_{\tau \rightarrow i \infty} f^{(2)}(z{-}\tauh,\tau) = \zeta_2$. Hence, the right-hand side
of (\ref{Gnp}) must be a constant.
\item The representation (\ref{B2}) of ${\cal G}(v_{ij}{+}\tauh,\tau)$ in terms of elliptic iterated integrals
implies that $\int^1_0 \dd v_j \, \big[{\cal G}(v_{ij}{+}\tauh,\tau) - \frac{ i \pi \tau}{4} \big]=0$,
so the constant in the previous step must be zero.
\end{itemize}
The degeneration of the non-planar Green function itself (rather than ${\cal G}(v_{ij}{+}\tauh,\tau)
 - \frac{ i \pi \tau}{4}$) can be obtained from the regularized value $\lim_{\tau \rightarrow i \infty}  \tau = 0$.
 This choice lines up with the net effect $\log q = \int^q_0 \frac{ \dd q_1 }{q_1} =
 2\pi i \lim_{\tau' \rightarrow i \infty} (\tau - \tau')$ of the tangential-base-point regularization
 \cite{Brown:mmv} quoted below (\ref{preB5}). This assigns a vanishing degeneration to both terms
in (\ref{Gnp}), and we arrive at the desired result (\ref{Gnpnew}).


\section{More on the $\tau$-derivatives of $A$-cycle integrals}
\label{app:AB}


\subsection{The $6\times6$ representation of the derivations at four points}
\label{app:B}

In this appendix, we complete the expression (\ref{DER14}) for the $6\times6$ representation of the derivations
due to the four-point $A$-cycle integrals in (\ref{th4pt}). While the matrices $r_{\vec{\eta}}(e_{234}),r_{\vec{\eta}}(e_{34})$
and $r_{\vec{\eta}}(e_{4})$ can be found
in (\ref{DER15}), we still have to supply the expressions for $r_{\vec{\eta}}(e_{23}),
 r_{\vec{\eta}}(e_{24}),  r_{\vec{\eta}}(e_{2})$ and
$r_{\vec{\eta}}(e_{3})$: As one can reconstruct from permutations of (\ref{4difop}) in $2,3,4$, they are given by
\begin{align}
 r_{\vec{\eta}}(e_{23}) &= \left( \begin{array}{cccccc}
0 &0 &0 &0 &0 &0 \\
0 &0 &0 &0 &0 &0 \\
0 &0 &0 &0 &0 &0 \\
0 &0 &0 &0 &0 &0 \\
s_{12} &s_{12} &-s_{13} &-s_{13} &s_{12}{+}s_{24} &-s_{13}{-}s_{34} \\
-s_{12} &-s_{12} &s_{13} &s_{13} &-s_{12}{-}s_{24} &s_{13}{+}s_{34}
\end{array} \right)  \notag \\
r_{\vec{\eta}}(e_{24}) &= \left( \begin{array}{cccccc}
0 &0 &0 &0 &0 &0 \\
0 &0 &0 &0 &0 &0 \\
s_{12} &s_{12} &s_{12}{+}s_{23} &-s_{14}{-}s_{34}&-s_{14} &-s_{14}\\
-s_{12} &-s_{12} &-s_{12}{-}s_{23} &s_{14}{+}s_{34}&s_{14} &s_{14}\\
0 &0 &0 &0 &0 &0 \\
0 &0 &0 &0 &0 &0
\end{array} \right) \label{DER16}
\end{align}
as well as
\begin{align}
 r_{\vec{\eta}}(e_{2}) &= \left( \begin{array}{cccccc}
0 &0 &0 &0 &0 &0 \\
0 &0 &0 &0 &0 &0 \\
0 &0 &0 &0 &0 &0 \\
s_{12} &0 &s_{12}{+}s_{23} &s_{12}{+}s_{23}{+}s_{24} &0 &0 \\
0 &0 &0 &0 &0 &0 \\
0 &s_{12} &0 &0 &s_{12}{+}s_{24} &s_{12}{+}s_{23}{+}s_{24}
\end{array} \right) \notag \\
r_{\vec{\eta}}(e_{3}) &= \left( \begin{array}{cccccc}
0 &0 &0 &0 &0 &0 \\
s_{13}{+}s_{23} &s_{13}{+}s_{23}{+}s_{34} &s_{13} &0 &0 &0 \\
0 &0 &0 &0 &0 &0 \\
0 &0 &0 &0 &0 &0 \\
0 &0 &0 &s_{13} &s_{13}{+}s_{23}{+}s_{34} &s_{13}{+}s_{34}\\
0 &0 &0 &0 &0 &0
\end{array} \right) \label{DER17}  \, .
\end{align}
%


\subsection{The $\tau$-derivative at five points in a 24-element basis}
\label{app:A}

The purpose of this appendix is to give the minimal form of (\ref{B459}),
where all the integrals $Z_{\vec{\eta}}^{\tau}(\ast |i,j,k,l,m)$ on the right-hand side
are in the 24-element basis with $i=1$:
\begin{align}
2\pi i \partial_\tau &Z_{\vec{\eta}}^{\tau}(\ast |1,2,3,4,5)  = \Big(   \frac{1}{2}
\sum_{j=2}^5 s_{1j} \partial_{\eta_j}^2
+ \frac{1}{2} \sum_{2\leq i<j}^5 s_{ij} (\partial_{\eta_i}{-}\partial_{\eta_j})^2
 \notag \\
& \ \ \ \ \ \ \ \  \ \ \ \ \ \ \ \
 - s_{12} \wp(\eta_{2345},\tau) - (s_{13}{+}s_{23}) \wp(\eta_{345},\tau)
-(s_{14}{+}s_{24}{+}s_{34}) \wp(\eta_{45},\tau)  \notag \\
& \ \ \ \ \ \ \ \ \ \ \ \ \ \ \ \
-(s_{15}{+}s_{25}{+}s_{35}{+}s_{45}) \wp(\eta_5,\tau) - 2 \zeta_2 s_{12345}
\Big)  Z_{\vec{\eta}}^{\tau}(\ast  |1,2,3,4,5) \notag \\
&+ s_{13} \big[ \wp(\eta_{2345},\tau) - \wp(\eta_{345},\tau)\big]
\big( Z_{\vec{\eta}}^{\tau}(\ast  |1,3,2,4,5) + Z_{\vec{\eta}}^{\tau}(\ast  |1,3,4,2,5) + Z_{\vec{\eta}}^{\tau}(\ast  |1,3,4,5,2)   \big) \notag \\
&+ s_{24} \big[ \wp(\eta_{345},\tau) - \wp(\eta_{45},\tau)\big] \big(
Z_{\vec{\eta}}^{\tau}(\ast  |1,2,4,3,5) + Z_{\vec{\eta}}^{\tau}(\ast  |1,2,4,5,3)   \big) \notag \\
&+ s_{35} \big[ \wp(\eta_{45},\tau) - \wp(\eta_{5},\tau)\big]
Z_{\vec{\eta}}^{\tau}(\ast  |1,2,3,5,4)  \notag \\
&+ s_{14} \big[ \wp(\eta_{345},\tau) - \wp(\eta_{2345},\tau)\big]
\big( Z_{\vec{\eta}}^{\tau}(\ast  |1,4,3,2,5) +Z_{\vec{\eta}}^{\tau}(\ast  |1,4,3,5,2)+Z_{\vec{\eta}}^{\tau}(\ast  |1,4,5,3,2)\big)  \notag \\
&+ s_{14} \big[ \wp(\eta_{345},\tau) - \wp(\eta_{45},\tau)\big]
\big( Z_{\vec{\eta}}^{\tau}(\ast  |1,2,4,3,5) +Z_{\vec{\eta}}^{\tau}(\ast  |1,2,4,5,3)  \label{B4559}\\
& \ \ \ \ \ \ \ \ \ \ \ \ \ \ \ \ +Z_{\vec{\eta}}^{\tau}(\ast  |1,4,2,3,5)
+ Z_{\vec{\eta}}^{\tau}(\ast  |1,4,2,5,3) +Z_{\vec{\eta}}^{\tau}(\ast  |1,4,5,2,3)\big)  \notag \\
&+ s_{25} \big[ \wp(\eta_{45},\tau) - \wp(\eta_{345},\tau)\big]   Z_{\vec{\eta}}^{\tau}(\ast  |1,2,5,4,3)   \notag \\
&+ s_{25} \big[ \wp(\eta_{45},\tau) - \wp(\eta_{5},\tau)\big]  \big( Z_{\vec{\eta}}^{\tau}(\ast  |1,2,3,5,4)  + Z_{\vec{\eta}}^{\tau}(\ast  |1,2,5,3,4) \big) \notag \\
&+ s_{15} \big[ \wp(\eta_{2345},\tau) - \wp(\eta_{345},\tau)\big] Z_{\vec{\eta}}^{\tau}(\ast  |1,5,4,3,2) \notag \\
&+ s_{15} \big[ \wp(\eta_{45},\tau) - \wp(\eta_{345},\tau)\big]
\big( Z_{\vec{\eta}}^{\tau}(\ast  |1,2,5,4,3) + Z_{\vec{\eta}}^{\tau}(\ast  |1,5,2,4,3) + Z_{\vec{\eta}}^{\tau}(\ast  |1,5,4,2,3)   \big) \notag \\
&+ s_{15} \big[ \wp(\eta_{45},\tau) - \wp(\eta_{5},\tau)\big]
\big( Z_{\vec{\eta}}^{\tau}(\ast  |1,2,3,5,4) + Z_{\vec{\eta}}^{\tau}(\ast  |1,2,5,3,4) + Z_{\vec{\eta}}^{\tau}(\ast  |1,5,2,3,4)   \big) \, .
\notag
\end{align}
%


\section{Transformation matrices between twisted cycles}
\label{app:G}

In this appendix, we give more details on the transformation matrices ${\cal H}_\ap(A|B)$
between planar genus-one cycles and disk orderings ${\cal D}_B$ in (\ref{inival4}). We follow
the slightly abusive notation in the main text and relate twisted cycles without explicitly referring to
the degenerate Koba--Nielsen factor (\ref{inival2}).


\subsection{Four-point example}
\label{app:G1}

Given the definition of ${\cal H}_\ap(\cdot|\cdot)$ via ${\cal C}(1,A) = \sum_{B \in S_{n-1}}{\cal H}_\ap(A|B) {\cal D}_B$,
the four-point relation (\ref{s0jD}) allows to read off
\begin{align}
{\cal H}_{\alpha'}(2,3,4|2,3,4) &= 2i \sin\Big(\frac{\pi}{2}(s_{12}{+}s_{23}{+}s_{24}) \Big) \, e^{\frac{ i \pi }{2}s_{134}}
\notag \\
{\cal H}_{\alpha'}(2,3,4|3,2,4) &= - 2i \sin\Big(\frac{\pi}{2}(s_{13}{+}s_{23}{+}s_{34}) \Big) \,  e^{\frac{ i \pi }{2}(-s_{12}+s_{14}-s_{24} )}
\notag \\
{\cal H}_{\alpha'}(2,3,4|3,4,2) &= - 2i \sin\Big(\frac{\pi}{2}(s_{13}{+}s_{23}{+}s_{34}) \Big) \, e^{\frac{ i \pi }{2}(-s_{12}+s_{14}+s_{24} )}
\label{apG1} \\
{\cal H}_{\alpha'}(2,3,4|4,3,2) &=2i  \sin\Big(\frac{\pi}{2}(s_{14}{+}s_{24}{+}s_{34}) \Big) \,  e^{-\frac{ i \pi }{2}s_{123}}
\notag \\
{\cal H}_{\alpha'}(2,3,4|2,4,3) &={\cal H}_{\alpha'}(2,3,4|4,2,3) = 0 \, . \notag
\end{align}
The remaining rows of ${\cal H}_\ap(A|B)$ are obtained by relabeling
$2,3,4$ in $s_{ij}$, $A$ and $B$.


\subsection{Weighted combinations}
\label{app:G2}

Depending on the number of $G_{ij}$ insertions in the integrands of (\ref{rePT12}), one
encounters combinations ${\cal H}_\ap(A|B) \pm {\cal H}_\ap(A^t|B)$
with the reversed genus-one cycle ${\cal C}(1,A^t)$. The weighted three-point cycles are related by
\begin{align}
\frac{1}{2}\big[{\cal C}(1,2,3)+{\cal C}(1,3,2) \big] &= -2 \Big[ \sin\Big(\frac{\pi}{2} (s_{12}{+}s_{23}) \Big)
 \sin \Big( \frac{ \pi }{2} s_{13}\Big) {\cal D}_{23}
+ (2\leftrightarrow 3)
\Big] \notag \\
\frac{1}{2} \big[{\cal C}(1,2,3)-{\cal C}(1,3,2)\big]&= 2i \Big[ \sin\Big(\frac{\pi}{2} (s_{12}{+}s_{23}) \Big)
\cos \Big( \frac{ \pi }{2} s_{13}\Big) {\cal D}_{23}
- (2\leftrightarrow 3)
 \Big] \, , \label{s0jC}
\end{align}
and the analogous four-point identity is
\begin{align}
\frac{1}{2}\big[ {\cal C}(1,2,3,4) + {\cal C}(1,4,3,2 )\big] &=2i \Big[  \sin \Big( \frac{ \pi }{2} (s_{12}{+}s_{23}{+}s_{24}) \Big)
\cos   \Big( \frac{ \pi }{2} s_{134} \Big) {\cal D}_{234} \notag \\
& \ \ \ \ -\sin \Big( \frac{ \pi }{2} (s_{13}{+}s_{23}{+}s_{34}) \Big)
\cos   \Big( \frac{ \pi }{2} ({-}s_{12}{+}s_{14}{-}s_{24}) \Big) {\cal D}_{324} \notag \\
& \ \ \ \ -\sin \Big( \frac{ \pi }{2} (s_{13}{+}s_{23}{+}s_{34}) \Big)
\cos   \Big( \frac{ \pi }{2} ({-}s_{12}{+}s_{14}{+}s_{24}) \Big) {\cal D}_{342} \notag \\
& \ \ \ \ + \sin \Big( \frac{ \pi }{2} (s_{14}{+}s_{24}{+}s_{34}) \Big)
\cos   \Big( \frac{ \pi }{2} s_{123} \Big) {\cal D}_{432} \Big] \notag
\\
\frac{1}{2}\big[ {\cal C}(1,2,3,4) - {\cal C}(1,4,3,2 )\big] &= -2 \Big[  \sin \Big( \frac{ \pi }{2} (s_{12}{+}s_{23}{+}s_{24}) \Big)
\sin  \Big( \frac{ \pi }{2} s_{134} \Big) {\cal D}_{234} \label{s0jE} \\
& \ \ \ \ -\sin \Big( \frac{ \pi }{2} (s_{13}{+}s_{23}{+}s_{34}) \Big)
\sin   \Big( \frac{ \pi }{2} ({-}s_{12}{+}s_{14}{-}s_{24}) \Big) {\cal D}_{324} \notag \\
& \ \ \ \ -\sin \Big( \frac{ \pi }{2} (s_{13}{+}s_{23}{+}s_{34}) \Big)
\sin  \Big( \frac{ \pi }{2} ({-}s_{12}{+}s_{14}{+}s_{24}) \Big) {\cal D}_{342} \notag \\
& \ \ \ \ - \sin \Big( \frac{ \pi }{2} (s_{14}{+}s_{24}{+}s_{34}) \Big)
\sin   \Big( \frac{ \pi }{2} s_{123} \Big) {\cal D}_{432} \Big] \, . \notag
\end{align}
The corresponding entries of ${\cal H}_\ap(A|B)$ can be read off by matching these
expressions with ${\cal C}(1,A) = \sum_{B \in S_{n-1}}{\cal H}_\ap(A|B) {\cal D}_B$, e.g.\
\begin{align}
\frac{1}{2}\big[{\cal H}_{\alpha'}(2,3|2,3)+{\cal H}_{\alpha'}(3,2|2,3) \big] &=
-2 \sin\Big(\frac{\pi}{2} (s_{12}{+}s_{23}) \Big) \sin \Big( \frac{ \pi }{2} s_{13}\Big)
\notag \\
\frac{1}{2}\big[{\cal H}_{\alpha'}(2,3|2,3)-{\cal H}_{\alpha'}(3,2|2,3) \big] &=
2i \sin\Big(\frac{\pi}{2} (s_{12}{+}s_{23}) \Big)
\cos \Big( \frac{ \pi }{2} s_{13}\Big) \, . \label{apG2}
\end{align}


\section{Examples of $\ap$-expansions}
\label{app:H}


\subsection{Three points: integrating $f^{(1)}_{12}f^{(3)}_{23}$}
\label{app:H3}

Among the three-point component integrals (\ref{compin71}) over $f^{(m_1)}_{12} f^{(m_2)}_{23}$,
examples of their $\ap$-expansions at $(m_1,m_2)=(0,0), (2,0),(3,3)$ and $(1,0)$ have been given
in (\ref{CC33pl}), (\ref{smalleq}) and (\ref{medeq}). The following $\ap$-expansion at $(m_1,m_2)=(1,3)$
contains further examples of irreducible iterated Eisenstein integrals at depth two such as
$\gamma_0(4, 4, 0|\tau)$ and $\gamma_0(4, 6, 0|\tau)$:
\begin{align}
Z^{\tau}_{(1,3)}&(1,2,3|1,2,3) = s_{13} \Big( 6  \zeta_2 \gamma_0(4|\tau) - \frac{5}{2}  \gamma_0(6|\tau) \Big) \notag \\
&+ s_{13}(s_{12}{+}s_{13}) \Big( -36 \zeta_4 \gamma_0(4, 0|\tau) - \frac{9}{2} \gamma_0(4, 4|\tau) +
 60 \zeta_2 \gamma_0(6, 0|\tau) - \frac{63}{2} \gamma_0(8, 0|\tau) \Big) \notag \\
 &+ s_{13} \Big\{
  \frac{15}{4} (s_{12}^2 {+} s_{13}^2 {+} s_{23}^2) \zeta_4 \gamma_0(4|\tau)
 -\frac{5}{8} (s_{12}^2 {+} s_{13}^2 {+} s_{23}^2) \zeta_2 \gamma_0(6|\tau) \notag \\
 &\ \ \ +9 (36 s_{12}^2 {-} 42 s_{12} s_{13} {+} 36 s_{13}^2 {-} 31 s_{12} s_{23} {-} 31 s_{13} s_{23} {-}
   34 s_{23}^2) \zeta_6 \gamma_0(4, 0, 0|\tau)\notag \\
 &\ \ \
   -30 (57 s_{12} s_{13} {+} 28 s_{12} s_{23} {+} 28 s_{13} s_{23} {-} 5 s_{23}^2) \zeta_4  \gamma_0(6, 0, 0|\tau) \notag \\
   &\ \ \ +42 (36 s_{12}^2 {+} 18 s_{12} s_{13} {+} 36 s_{13}^2 {+} 29 s_{12} s_{23} {+} 29 s_{13} s_{23} {-}
   34 s_{23}^2) \zeta_2 \gamma_0(8, 0, 0|\tau)\notag \\
 &\ \ \
 -9 (90 s_{12}^2 {+} 21 s_{12} s_{13} {+} 90 s_{13}^2 {+} 59 s_{12} s_{23} {+} 59 s_{13} s_{23} {-}
   85 s_{23}^2) \gamma_0(10, 0, 0|\tau)\label{bigeq} \\
 &\ \ \
+18 (s_{12}^2 {+} s_{13}^2 {+} s_{12} s_{23} {+} s_{13} s_{23} {-} s_{23}^2)  \zeta_2 \gamma_0(4, 0, 4|\tau)\notag \\
 &\ \ \
+18 (6 s_{12}^2 {+} 2 s_{12} s_{13} {+} 6 s_{13}^2 {+} 3 s_{12} s_{23} {+} 3 s_{13} s_{23} {-}
   8 s_{23}^2) \zeta_2 \gamma_0(4, 4, 0|\tau)\notag \\
 &\ \ \
-\frac{15}{2} (9 s_{12}^2 {-} 2 s_{12} s_{13} {+} 9 s_{13}^2 {+} 4 s_{12} s_{23} {+} 4 s_{13} s_{23} {-}
   15 s_{23}^2) \gamma_0(6, 4, 0|\tau)
 \notag \\
 &\ \ \
-15 (s_{12}^2 {+} s_{13}^2 {+} 2 s_{12} s_{23} {+} 2 s_{13} s_{23}) \gamma_0(6, 0, 4|\tau)
+\frac{15}{2} (s_{12}^2 {+} s_{13}^2 {+} s_{23}^2) \gamma_0(4, 0, 6|\tau)\notag \\
 &\ \ \  -\frac{15}{2} (9 s_{12}^2 {+} 4 s_{12} s_{13} {+} 9 s_{13}^2 {+} 2 s_{12} s_{23} {+} 2 s_{13} s_{23} {-}
   5 s_{23}^2) \gamma_0(4, 6, 0|\tau) \Big\} + {\cal O}(\ap^4) \notag
\end{align}


\subsection{Four points: MZVs for the initial values}
\label{app:H4}

This appendix contains some further samples of the $\ap$-expansions of
$I^{\rm tree}(1,2,3,4|\ast)$ with one or two powers of $G_{ij}$ in the integrand
(see (\ref{4ptlim1}) for $I^{\rm tree}(1,2,3,4|1)$ at the orders of $\ap^{\leq 3}$).
For a given planar integration cycle, there are two cyclically inequivalent representatives
to consider for one and two powers of $G_{ij}$:
\begin{align}
I^{\rm tree}(1,2,3,4|G_{12})
 &= \frac{1}{2s_{12}} + \frac{ \zeta_2 }{8}
 \Big( \frac{ s_{123}^2 + (s_{14}{+}s_{24})^2 + s_{34}^2 }{s_{12}} + 2(s_{24}{-} s_{23})\Big)
 \label{4ptlim5}
 \\
&\! \! \! \!\! \! \! \!\! \! \! \!\! \! \! \!\! \! \! \!\! \! \! \!  +\frac{ \zeta_3}{8} \Big( \frac{ s_{123}^3 + (s_{14}{+}s_{24})^3 + s_{34}^3 +s_{123}(s_{14}{+}s_{24}) s_{34}}{s_{12}}
-3 s_{12} s_{23} - s_{13} s_{23} + s_{14} s_{23} \notag
    \\
 &\! \! \! \!\! \! \! \!\! \! \! \!\! \! \! \!\! \! \! \!\! \! \! \! \ \ \ \  - 3 s_{23}^2 + 3 s_{12} s_{24} -
    s_{13} s_{24} + 5 s_{14} s_{24} + 3 s_{24}^2 + s_{23} s_{34} - s_{24} s_{34}\Big) + {\cal O}(\ap^3) \notag \\
I^{\rm tree}(1,2,3,4|G_{13})
&= \frac{ \zeta_2}{4} (s_{14}-s_{12}-s_{23}+s_{34}) \label{4ptlim6}  \\
&\! \! \! \!\! \! \! \!\! \! \! \!\! \! \! \!\! \! \! \!\! \! \! \! + \frac{ \zeta_3}{8} \Big(-3 s_{12}^2 - 3 s_{12} s_{13} + 3 s_{13} s_{14} + 3 s_{14}^2 - 4 s_{12} s_{23}
 -    3 s_{13} s_{23} - 3 s_{23}^2 + s_{12} s_{24}  \notag \\
 &\! \! \! \!\! \! \! \!\! \! \! \!\! \! \! \!\! \! \! \!\! \! \! \! \ \ \ \ - s_{14} s_{24} + s_{23} s_{24} + 3 s_{13} s_{34} +
   4 s_{14} s_{34} - s_{24} s_{34} + 3 s_{34}^2 \Big)  + {\cal O}(\ap^3) \notag \\
I^{\rm tree}(1,2,3,4|G_{12}G_{34})
 &= \frac{1}{s_{12}s_{34}}
+ \frac{ \zeta_2}{4} \frac{ s_{1234}^2 }{s_{12} s_{34}} \label{4ptlim7}  \\
&\! \! \! \!\! \! \! \!\! \! \! \! \! \! \! \!\! \! \! \!\! \! \! \! +\frac{ \zeta_{3}}{4}\Big( \frac{ s_{1234}^3 }{s_{12}s_{34}}
+ \frac{ (s_{13}{+}s_{14})(s_{23}{+}s_{24}) }{s_{34}}
+ \frac{ (s_{13}{+}s_{23})(s_{14}{+}s_{24}) }{s_{12}}
+ s_{14}+s_{23} \Big) + {\cal O}(\ap^2) \notag  \\
I^{\rm tree}(1,2,3,4|G_{13}G_{24})
&= \frac{ \zeta_{3}}{4} (s_{12} - s_{14} - s_{23} + s_{34}) + {\cal O}(\ap^2) \, .
\label{4ptlim8}
\end{align}
Integrands $G_{ij} G_{jk}$ with overlapping labels $i,j,k$ can be reduced to non-overlapping cases
via integration by parts (\ref{4ptlim10}). They exhibit three-particle poles $s_{ijk}^{-1}$ as for instance
seen in
\begin{align}
 I^{\rm tree}&(1,2,3,4|G_{12}G_{13})= \frac{1}{s_{12}s_{123} }
- \frac{ \zeta_2 s_{23}}{s_{123}}
+ \frac{ \zeta_2 }{4} \frac{ ( s_{123}{+}s_{14}{+}s_{24}{+}s_{34})^2 }{s_{12}s_{123}}
+ \frac{\zeta_3 s_{13} s_{23}}{s_{123}}
\notag \\
&\ \ \ \ + \frac{ \zeta_3}{4}  \frac{ ( s_{123}{+}s_{14}{+}s_{24}{+}s_{34})^2 }{s_{12}s_{123}}
+ \frac{ \zeta_3(s_{14}{+}s_{24}) s_{34} }{4 s_{12}}  + \frac{ \zeta_3 }{4} (s_{24}{-}s_{23}) + {\cal O}(\ap^2)\, .
\label{s123ex}
\end{align}


\section{Fourier expansion of $A$-cycle graph functions}
\label{app:qexp}

In this appendix, we demonstrate on the basis of the conjecture (\ref{CsymB}) that the coefficients in the
$q$-expansion of planar and non-planar $A$-cycle graph functions
are $\QQ$-linear as opposed to $\QQ[(2\pi i)^{-1}]$-linear combinations of MZVs.
By their definition in \cite{Broedel:2018izr}, $A$-cycle graph functions are symmetrized
integrals $\prod_{j=2}^n \int^1_0 \dd v_j$
over products of the planar Green functions ${\cal G}(v_{ij},\tau)$ in (\ref{B1}) and non-planar Green functions
${\cal G}(v_{ij}{+}\tauh,\tau) - \frac{i \pi \tau}{4}$ without the additive $\frac{i \pi \tau}{4}$ in (\ref{B2}).
This definition implies that they are expressible in terms of $\QQ$-linear combinations of $A$-cycle eMZVs and their twisted
counterparts. However, the $q$-expansion of generic (twisted) eMZVs may involve inverse powers of
$\pi$, so it remains to show that $A$-cycle graph functions only realize those combinations of (twisted) eMZVs,
where negative powers of $\pi$ drop out. The subsequent arguments are tailored to $\ell$ punctures on the
$A$-cycle and $n{-}\ell$ punctures on its displacement by $\tauh$.

In this work, $A$-cycle graph functions are generated from the $\eta_j^{1-n}$-order of
symmetrized $A$-cycle integrals
$\sum_{P \in S_{\ell-1}} \sum_{Q \in S_{n-\ell-1}} Z^{\tau}_{\vec{\eta}}(\begin{smallmatrix}n,Q \\ 1,P\end{smallmatrix}|\ast)$,
which does not depend on the choice of the integrand $\ast$.
We will proceed in three steps and establish the absence of negative powers of $\pi$
first for the symmetrizations $\sum_{A \in S_{n-1}} I^{\rm tree}(1,A|\cdot)$, then for the initial values
$Z^{i\infty}_{\vec{\eta}}$ of the above generating series and finally for the entire $q$-dependence
of $A$-cycle graph functions.
\begin{itemize}
\item The absence of $\pi^{-1}$ in the symmetrization of $I^{\rm tree}(1,A|\cdot )$
can be understood from (\ref{rePT12}): Its prefactor $(i\pi )^{1-n}$ is compensated
by the symmetrization in $\sum_{A \in S_{n-1}} {\cal H}_\ap(A|P)$ that yields $n{-}1$ sine factors of the schematic form
$\sin( \frac{\pi}{2}\sum_{j}s_{ij})$ and therefore $n{-}1$ overall powers of $\pi$ in its $\ap$-expansion.
The appearance of the sine functions follows from the relation (\ref{CsymB}) between twisted
cycles which is conjectural at $n\geq 6$.
\item The initial values $\sum_{P \in S_{\ell-1}} \sum_{Q \in S_{n-\ell-1}}
Z^{i\infty}_{\vec{\eta}}(\begin{smallmatrix}n,Q \\ 1,P\end{smallmatrix}|\ast)$
relevant for $A$-cycle graph functions boil down to products of
symmetrized $I^{\rm tree}(1,A|\cdot )$ by (\ref{CC10G}). These products
do not involve any inverse powers of $\pi$ by the previous step, and their
coefficients are built from $\pi \cot(\pi \eta)$ and $\frac{\pi}{ \sin(\pi \eta)}$ (with $\eta$
referring to generic sums of $\eta_2,\eta_3,\ldots,\eta_n$).

Whenever the integrands of the $I^{\rm tree}(1,A|\cdot )$ in (\ref{CC10G})
involve $k$ powers of $G_{ij}$ in total, then the number of trigonometric factors is
$n{-}1{-}k$, and they have an expansion of the form $\eta^{k+1-n} \sum_{p=0}^{\infty} c_p (\pi \eta)^p$
with $c_p\in \QQ$, see (\ref{trigdeg}). Since each $G_{ij}$ in the integrand of $I^{\rm tree}(1,A|\cdot )$
introduces an extra power of $i\pi$ by (\ref{rePT12}), the counting of relative powers of $\pi$ and $\eta$
in the entire initial value is governed by the $k=0$ case: The $\ap$-expansion of
$\sum_{P \in S_{\ell-1}} \sum_{Q \in S_{n-\ell-1}} Z^{i\infty}_{\vec{\eta}}(\begin{smallmatrix}n,Q \\ 1,P\end{smallmatrix}|\ast)$
has the schematic form of $\eta^{1-n} \sum_{p=0}^{\infty}\hat c_p (\pi \eta)^p, \ \hat c_p\in \QQ$,
multiplying series in $\ap$ with $\QQ$-linear combinations of MZVs from the symmetrized $I^{\rm tree}$.
\item $A$-cycle graph functions arise from the $\eta^{1-n}$-order of the full $\tau$-dependent
$Z^{\tau}_{\vec{\eta}}(\cdot|\cdot)$ upon symmetrization. It remains to check that the operator
$\Psi^\tau_{\vec{\eta}}$ defined in (\ref{in3npT}) does not introduce any inverse powers of $\pi$
through the $q$ series of $ \gamma(k_1,k_2,\ldots,k_r|\tau) r_{\vec{\eta}}(\ep_{k_r} \ldots   \ep_{k_{2}}\ep_{k_1} )$.
The two types of Eisenstein series $\gamma(\ldots)$ and $\gamma_0(\ldots)$ are equivalent w.r.t.\ powers
of $\pi$ since all coefficients of $q^0$ and $q^{n\geq 1}$ in ${\rm G}_k$ are rational multiples of $\pi^k$, cf.\ (\ref{A14}).
Hence, the counting of $\pi$ in $\Psi^\tau_{\vec{\eta}}$ is captured by
replacing $\gamma(\ldots)\rightarrow \gamma_0(\ldots)$ and inserting the
$q$-expansion (\ref{qgamma1}) which determines each Fourier coefficient in
$ \gamma_0(k_1,k_2,\ldots,k_r|\tau)$ to be a rational multiple of $\pi^{-2r+k_1+k_2+\ldots+k_r}$.

The derivations $r_{\vec{\eta}}(\ep_{k} )$ are of homogeneity degree $k{-}2$ in $\eta_j$,
with the exception of the term $2\zeta_2 s_{123\ldots n}$ in $r_{\vec{\eta}}(\ep_{0} )$ which can
be viewed as introducing an extra power $(\pi \eta)^2$.
Hence, each term $ \gamma_0(k_1,k_2,\ldots,k_r|\tau)
r_{\vec{\eta}}(\ep_{k_r} \ldots   \ep_{k_{2}}\ep_{k_1} )$ in $\Psi^\tau_{\vec{\eta}}$ has a homogeneity
degree $\sim (\pi \eta)^{-2r+k_1+k_2+\ldots+k_r}$
up to tentative extra factors of $(\pi \eta)^2$ due to the exceptional term in~$r_{\vec{\eta}}(\ep_{0} )$.

Contributions with negative powers of $\pi$ (say $( \pi \eta)^{-2}$ due to $\gamma_0(4,0,0|\tau)
r_{\vec{\eta}}(\ep_{0} \ep_{0}\ep_{4} )$) can only affect the $\eta^{1-n}$-order relevant to $A$-cycle graph functions
if the initial value $Z^{i\infty}_{\vec{\eta}}$ contributes with higher orders in $\eta$ as compared to the
leading term $\sim \eta^{1-n}$. As argued above, the $\eta$-dependence of $Z^{i\infty}_{\vec{\eta}}$
has the schematic form $\eta^{1-n} \sum_{p=0}^{\infty}\hat c_p (\pi \eta)^p$ with $\hat c_p \in \QQ$.
Hence, any negative power $( \pi \eta)^{-p}$ due to $\Psi^\tau_{\vec{\eta}}$
must be compensated by a positive power $( \pi \eta)^{p}$ from the initial value in order to contribute to the
order $\eta^{1-n} $ relevant to $A$-cycle graph functions.
\end{itemize}


\bibliographystyle{JHEP}

\providecommand{\href}[2]{#2}\begingroup\raggedright\endgroup


\end{document}